\documentclass[conference]{IEEEtran}
\IEEEoverridecommandlockouts
\usepackage{cite}
\usepackage{amsmath,amssymb,amsfonts}
\usepackage{algorithmic}
\usepackage{graphicx}
\usepackage{textcomp}
\usepackage{xcolor}
\usepackage{caption}
\usepackage{subcaption}
\usepackage{lscape}
\usepackage{longtable}
\usepackage{csvsimple}
\usepackage{booktabs}
\usepackage{etoolbox}
\usepackage{adjustbox}
\usepackage{array}
\usepackage{xcolor,colortbl}
\usepackage{hyperref}
\usepackage{siunitx}
\usepackage{import}
\usepackage{natbib}
\usepackage{numprint}
\npthousandsep{\,}
\usepackage{xstring}

\def\BibTeX{{\rm B\kern-.05em{\sc i\kern-.025em b}\kern-.08em
    T\kern-.1667em\lower.7ex\hbox{E}\kern-.125emX}}

\makeatletter
\let\pragma@iinput=\@iinput
\def\@iinput#1{\xdef\@pragmafile{#1}\pragma@iinput{#1}}
\def\@pragmafile{default}
\def\pragmaonce{%
   \csname pragma@\@pragmafile\endcsname
   \global\expandafter\let \csname pragma@\@pragmafile\endcsname = \endinput
}
\makeatother

\ifdefined\mydraft
\mydraft
\fi

\begin{document}

\title{Best-Effort Communication Improves Performance and Scales Robustly on Conventional Hardware}

\author{\IEEEauthorblockN{Matthew Andres Moreno}
\IEEEauthorblockA{\textit{Computer Science and Engineering} \\
\textit{Michigan State University}\\
\texttt{mmore500@msu.edu} \\
\url{https://orcid.org/0000-0003-4726-4479}
}
\and
\IEEEauthorblockN{Charles Ofria}
\IEEEauthorblockA{\textit{Computer Science and Engineering} \\
\textit{Michigan State University}\\
\url{https://orcid.org/0000-0003-2924-1732}
}
}

\maketitle

\begin{abstract}
Exponential advances in HPC hardware enables profound scientific and industrial innovation, but performance overhead from synchronization and error recovery has become increasingly challenging.
``Best-effort'' approaches that improve efficiency by relaxing guarantees of correctness and determinism have emerged as a promising remedy.
Here, we test the performance and scalability of fully-asynchronous, best-effort communication on existing, commercially-available HPC hardware.

A first set of experiments tested whether best-effort communication strategies can benefit performance compared to the traditional perfect communication model.
At high CPU counts, best-effort communication improved both the number of computational steps executed per unit time and the solution quality achieved within a fixed-duration run window.
Computation-heavy benchmark workloads yielded the strongest scaling efficiency, achieving at 64 processes 92\% the update-rate of single-process execution.
We observed a relative speedup of up to $7.8\times$ under communication-heavy workloads.

Under the best-effort model, characterizing the distribution of quality of service across processing components and over time is critical to understanding the actual computation being performed.
Additionally, a complete picture of scalability under the best-effort model requires analysis of how such quality of service fares at scale.
To answer these questions, we designed and measured a suite of quality of service metrics: simulation update period, message latency, message delivery failure rate, and message delivery coagulation.
Under a lower communication-intensivity benchmark parameterization, we found that median values for all quality of service metrics were stable when scaling from 64 to 256 process.
Under maximal communication intensivity, we found only minor --- and, in most cases, nil --- degradation in median quality of service.

In an additional set of experiments, we tested the effect of an apparently faulty compute node on performance and quality of service.
Despite extreme quality of service degradation among that node and its clique, median performance and quality of service remained stable.

We used the Conduit C++ library for best-effort communication to perform reported experiments.
Development of this library stemmed from a practical need for an general-purpose, prepackaged framework for best-effort communication.
We hope that free availability of this library, which includes built-in tools to measure quality of service metrics, can facilitate broader incorporation of best-effort communication into HPC applications.
\end{abstract}

\providecommand{\dissertationonly}[1]{%
\ifdefined\DISSERTATION%
#1%
\else%
\fi
}

\section{Introduction}
\label{sec:introoduction;ch:conduit}

The parallel and distributed processing capacity of high-performance computing (HPC) clusters continues to grow rapidly and enable profound scientific and industrial innovations \citep{gagliardi2019international}.
These advances in hardware capacity and economy afford great opportunity, but also pose a serious challenge: developing approaches to effectively harness it.
As HPC systems scale, it becomes increasingly difficult to write software that makes efficient use of available hardware and also provides reproducible results (or even near-perfectly reproducible results --- i.e., up to effects from floating point non-transitivity) consistent with models of computation as being performed a reliable digital machine \citep{heroux2014toward}.

The bulk synchronous parallel (BSP) model, which is prevalent among HPC applications \citep{dongarra2014applied}, illustrates the challenge.
This model segments fragments of computation into sequential global supersteps, with fragments at superstep $i$ depending only on data from strictly preceding fragments $<i$, often just $i-1$.
Computational fragments are assigned across a pool of available processing components.
The BSP model assumes perfectly reliable messaging: all dispatched messages between computational fragments are faithfully delivered.
In practice, realizing this assumption introduces overhead costs: secondary acknowledgment messages to confirm delivery and mechanisms to dispatch potential resends as the need arises.
Global synchronization occurs between supersteps, with computational fragments held until their preceding superstep has completed \citep{valiant1990bridging}.
This ensures that computational fragments will have at hand every single expected input, including those required from fragments located on other processing elements, before proceeding.
So, supersteps only turn over once the entire pool of processing components have completed their work for that superstep.
Put another way, all processing components stall until the most laggardly component catches up.
In a game of double dutch with several jumpers, this would be like slowing the tempo to whoever is most slow-footed each particular turn of the rope.

Heterogeneous computational fragments, with some easy to process and others much slower, would result in poor efficiency under a naive approach where each processing element handled just one fragment.
Some processing elements with easy tasks would finish early then idle while more difficult tasks carry on.
To counteract such load imbalances, programmers can allow for ``parallel slack'' by ensuring computational fragments greatly outnumber processing elements or even performing dynamic load balancing at runtime \citep{valiant1990bridging}.

Unfortunately, hardware factors on the underlying processing elements ensure that inherent global superstep jitter will persist: memory access time varies due to cache effects, message delivery time varies due to network conditions, extra processing due to error detection and recovery, delays due to unfavorable process scheduling by the operating system, etc. \citep{dongarra2014applied}.
Power management concerns on future machines will likely introduce even more variability \citep{gropp2013programming}.
Worse yet, as we work with more and more processes, the expected magnitude of the worst-sampled jitter grows and grows --- and in lockstep with it, our expected superstep duration.
In the double dutch analogy, with enough jumpers, at almost every turn of the rope someone will need to stop and tie their shoe.
The global synchronization operations underpinning the BSP model further hinder its scalability.
Irrespective of time to complete computational fragments within a superstep, the cost of performing a global synchronization operation increases with processor count \citep{dongarra2014applied}.

Efforts to recover scalability by relaxing superstep synchronization fall under two banners.
The first approach, termed ``Relaxed Bulk-Synchronous Programming'' (rBSP), hides latency by performing collective operations asynchronously, essentially allowing useful computation to be performed at the same time as synchronization primitives for a single superstep percolate through the collective \citep{heroux2014toward}.
So, the time cost required to perform that synchronization can be discounted, up to the time taken up by computational work at one superstep.
Likewise, individual processes experiencing heavier workloads or performance degradation due to hardware factors can fall behind by up to a single superstep without slowing the entire collective.
However, this approach cannot mask synchronization costs or cumulative performance degradation exceeding a single superstep's duration.
The second approach, termed relaxed barrier synchronization, forgoes global synchronization entirely \citep{kim1998relaxed}.
Instead, computational fragments at superstep $i$ only wait on expected inputs from the subset of superstep $i-1$ fragments that they directly interact with.
Imagine a double-dutch routine where each jumper exchanges patty cakes with both neighboring jumpers at every turn of the rope.
Relaxed barrier synchronization would dispense entirely with the rope.
Instead, players would be free to proceed to their next round of patty cakes as soon as they had successfully patty-caked both neighbors.
With $n$ players, player 0 could conceivably advance $n$ rounds ahead of player $n-1$ (each player would be one round ahead of their right neighbor).
Assuming fragment interactions form a graph structure that persists across supersteps, in the general case before causing the entire collective to slow an individual fragment can fall behind at most a number of supersteps equal to the graph diameter \citep{gamell2015local}.
Even though this approach can shield the collective from most one-off performance degradations of a single fragment (especially in large-diameter cases), persistently laggard hardware or extreme one-off degradations will ultimately still hobble efficiency.
Dynamic task scheduling and migration aim to address this shortcoming, redistributing work in order to ``catch up'' delinquent fragments \citep{acun2014parallel}.
With our double-dutch analogy, we could think of this something like a team coach temporarily benching a jumper who skinned their knee and instructing the other jumpers to pick up their roles in the routine.

In addition to concerns over efficiency, resiliency poses another inexorable problem to massive HPC systems.
In small scales, it can suffice to assume that failures occur negligibly, with any that do transpire likely to cause an (acceptably rare) global interruption or failure.
At large scales, however, software crashes and hardware failures become the rule rather than the exception \citep{dongarra2014applied} --- running a simulation to completion could even require so many retries as to be practically infeasible.
A typical contemporary approach to improve resiliency is checkpointing: the system periodically records global state then, when a failure arises, progress is rolled back to the most recent global known-good state and runtime restarts \citep{hursey2007design}.
Global checkpoint-based recovery is expensive, especially at scale due to overhead associated with regularly recording global state, losing progress since the most recent checkpoint, and actually performing a global teardown and restart procedure.
In fact, at large enough scales global recovery durations could conceivably exceed mean time between failures, making any forward simulation progress all but impossible \citep{dongarra2014applied}.
The local failure, local recovery (LFLR) paradigm eschews global recovery by maintaining persistent state on a process-wise basis and providing a recovery function to initialize a step-in replacement process \citep{heroux2014toward,teranishi2014toward}.
In practice, such an approach can require keeping running logs of all messaging traffic in order to replay them for the benefit of any potential step-in replacement \citep{chakravorty2004fault}.
Returning once more to the double dutch analogy, LFLR would transpire as something like a handful teammates pulling a stricken teammate aside to catch them up after an amnesia attack (rather than starting the entire team's routine back at the top of the current track).
The intervening jumpers would have to remind the stricken teammate of a previously recorded position then discreetly re-feign some of their moves that the stricken teammate had cued off of between that recorded position and the amnesia episode.

The possibility of multiple simultaneous failure (perhaps, for example, of dozens of processes resident on a single node) poses an even more difficult, although not insurmountable, challenge for LFLR that would likely necessitate even greater overhead.
On approach involves pairing up with a remote ``buddy'' process.
The ``buddy'' hangs to the focal process' snapshots and is carbon-copied on all of that process' messages in order to ensure an independently survivable log.
Unfortunately, this could potentially require forwarding all messaging traffic between simulation elements coresident on the focal process to its buddy, dragging inter-node communication into some otherwise trivial simulation operations \citep{chakravorty2007fault}.
Efforts to ensure resiliency beyond single-node failures currently appear unnecessary \citep[p. 12]{ni2016mitigation}.
Even though LFLR saves the cost of global spin-down and spin-up, all processes will potentially have to wait for work lost since the last checkpoint to be recompleted, although in some cases this could be helped along by tapping idle hardware to take over delinquent work from the failed process and help catch it up \citep{dongarra2014applied}.

Still more insidious to the reliable digital machine model, though, are soft errors --- events where corruption of data in memory occurs, usually do to environmental interference (i.e., ``cosmic rays'') \citep{karnik2004characterization}.
Further miniaturization and voltage reduction, which are assumed as a likely vehicle for continuing advances in hardware efficiency and performance, could conceivably worsen susceptibility to such errors \citep{dongarra2014applied,kajmakovic2020challenges}.
What makes soft errors so dangerous is their potential indetectability.
Unlike typical hardware or software failures, which result in an explicit, observable outcome (i.e., an error code, an exception, or even just a crash), soft errors can transpire silently and lead to incorrect computational results without leaving anyone the wiser.
Luckily, soft errors occur rarely enough to be largely neglected in most single-processor applications (except the most safety-critical settings); however, at scale soft errors occur at a non-trivial rate \citep{sridharan2015memory,scoles2018cosmic}.
Redundancy (be it duplicated hardware components or error correction codes) can reduce the rate of uncorrected (or at least undetected) soft errors, although at a non-trivial cost \citep{vankeirsbilck2015soft,sridharan2015memory}.
In some application domains with symmetries or conservation principles, the rate of soft errors (or, at least, silent soft errors) could be also reduced through so-called ``skeptical'' assertions at runtime \citep{dongarra2014applied}, although this too comes at a cost.

Even if soft errors can be effectively eradicated --- or at least suppressed to a point of inconsequentiality --- the nondeterministic mechanics of fault recovery and dynamic task scheduling could conceivably make guaranteeing bitwise reproducibility at exascale effectively impossible, or at least an unreasonable engineering choice \citep{dongarra2014applied}.
However, the assumption of the reliable digital machine model remains near-universal within parallel and distributed algorithm design \citep{chakradhar2010best}.
Be it just costly or simply a practical impossibility, the worsening burden of synchronization, fault recovery, and error correction begs the question of whether it is viable to maintain, or even to strive to maintain, the reliable digital machine model at scale.
Indeed, software and hardware that relaxes guarantees of correctness and determinism --- a so-called ``best-effort model'' --- have been shown to improve speed \citep{chakrapani2008probabilistic}, energy efficiency \citep{chakrapani2008probabilistic,bocquet2018memory}, and scalability \citep{meng2009best}.
Discussion around ``approximate computing'' overlaps significantly with ``best-effort computing,'' although focusing more heavily on using algorithm design to shirk non-essential computation (i.e., reducing floating point precision, inexact memoization, etc.) \citep{mittal2016survey}.
As technology advances, computing is becoming more distributed and we are colliding with physical limits for speed and reliability.
Massively distributed systems are becoming inevitable, and indeed if we are to truly achieve ``indefinite scalability'' \citep{ackley2011pursue} we must shift from guaranteed accuracy to best-effort methods that operate asynchronously and degrade gracefully under hardware failure.

The suitability of the best-effort model varies from application to application.
Some domains are clear cut in favor of the reliable digital machine model --- for example, due to regulatory issues \citep{dongarra2014applied}.
However, a subset of HPC applications can tolerate --- or even harness --- occasionally flawed or even fundamentally nondeterministic computation \citep{chakradhar2010best}.
Various approximation algorithms or heuristics fall into this category, with notable work being done on best-effort stochastic gradient descent for artificial neural network applications \citep{dean2012large,zhao2019elastic,niu2011hogwild,noel2014dogwild,rhodes2020real}.
Best-effort, real-time computing approaches have also been used in some artificial life models \citep{ray1995proposal}.
Likewise, algorithms relying on pseudo-stochastic methods that tend to exploit noise (rather than destabilize due to it) also make good candidates \citep{chakrapani2008probabilistic,chakradhar2010best}.
Real-time control systems that cannot afford to pause or retry, by necessity, fall into the best-effort category \citep{rahmati2011computing, rhodes2020real}.
\dissertationonly{%
For this dissertation we will, of course, focus on this latter case of systems well-suited to best-effort methods, as evolving systems already require noise to fuel variation.
}

This work distills best-effort communication from the larger issue of best-effort computing, paying it special attention and generally pretermiting the larger issue.
Specifically, we investigate the implications of relaxing synchronization and message delivery requirements.
Under this model, the runtime strives to minimize message latency and loss, but guarantees elimination of neither.
Instead, processes continue their compute work unimpeded and incorporate communication from collaborating processes as it happens to become available.
We still assume that messages, if and when they are delivered, retain contentual integrity.

We see best-effort communication as a particularly fruitful target for investigation.
Firstly, synchronization constitutes the root cause of many contemporary scaling bottlenecks, well below the mark of thousands or millions of cores where runtime failures and soft errors become critical considerations.
Secondly, future HPC hardware is expected to provide more heterogeneous, more variable (i.e., due to power management), and generally lower (relative to compute) communication bandwidth \citep{gropp2013programming, acun2014parallel}; a best-effort approach suits these challenges.
A best-effort communication model presents the possibility of runtime adaptation to effectively utilize available resources given the particular ratio of compute and communication capability at any one moment in any one rack.

Complex biological organisms exhibit characteristic best-effort properties: trillions of cells interact asynchronously while overcoming all but the most extreme failures in a noisy world.
As such, bio-inspired algorithms present strong potential to benefit from best-effort communication strategies.
For example, evolutionary algorithms commonly use guided stochastic methods (i.e., selection and mutation operators) resulting in a search process that does not guarantee optimality, but typically produces a diverse range of high-quality results.
Indeed, island model genetic algorithms are easy to parallelize and have been shown to perform well with asynchronous migration \citep{izzo2009parallel}.
Likewise, artificial life simulations commonly rely on a bottom-up approach and seek to model life-as-it-could-be evolving in a noisy environment akin to the natural world, yet distinct from it \citep{bonabeau1994we}. %
Although perfect reproducibility and observability have uniquely enabled digital evolution experiments to ask and answer otherwise intractable questions \citep{pontes2020evolutionary,lenski2003evolutionary,grabowski2013case,dolson2020interpreting,fortuna2019coevolutionary,goldsby2014evolutionary,covert2013experiments,zaman2011rapid,bundy2021footprint,dolson2017spatial}, the reliable digital machine model is not strictly necessary for all such work.
Issues of distributed and parallel computing are of special interest within the the artificial life subdomain of open-ended evolution (OEE) \citep{ackley2014indefinitely}, which studies long-term dynamics of evolutionary systems in order to understand factors that affect potential to generate ongoing novelty \citep{taylor2016open}.
Recent evidence suggests that the generative potential of at least some model systems are --- at least in part --- meaningfully constrained by available compute resources \citep{channon2019maximum}.

Much exciting work on best-effort computing has incorporated bespoke experimental hardware \citep{chippa2014scalable, ackley2011homeostatic, cho2012ersa, chakrapani2008probabilistic, rhodes2020real}.
However, here, we focus on exploring best-effort communication among parallel and distributed elements within existing, commercially-available hardware.
Existing software libraries, though, do not explicitly expose a convenient best-effort communication interface for such work.
As such, best-effort approaches remain rare in production software and efforts to study best-effort communication must make use of a combination of limited existing support and the development of new software tools.

The Message Passing Interface (MPI) standard \citep{gropp1996high} represents the mainstay for high-performance computing applications.
This standard exposes communication primitives directly to the end user.
MPI's nonblocking communication primitives, in particular, are sufficient to program distributed computations with relaxed synchronization requirements.
Although its explicit, the imperative nature of the MPI protocols enables precise control over execution; unfortunately it also poses significant expense in terms of programmability.
This cost manifests in terms of reduced programmer productivity and software quality, while increasing domain knowledge requirements and the effort required to tune for performance due to program brittleness \citep{gu2019comparative, tang2014mpi}.

In response to programmability concerns, many frameworks have arisen to offer useful parallel and distributed programming abstractions.
Task-based frameworks such as Charm++ \citep{kale1993charm++}, Legion \citep{bauer2012legion}, Cilk \citep{blumofe1996cilk}, and Threading Building Blocks (TBB) \citep{reinders2007intel} describe the dependency relationships among computational tasks and associated data and relies on an associated runtime to automatically schedule and manage execution.
These frameworks assume a deterministic relationship between tasks.
In a similar vein, programming languages and extensions like Unified Parallel C (UPC) \citep{el2006upc} and Chapel \citep{chamberlain2007parallel} rely on programmers to direct execution, but equips them with powerful abstractions, such as global shared memory.
However, Chapel's memory model explicitly forbids data races and UPC ultimately relies on a barrier model for data transfer.

To bridge these shortcomings, we employ a new software framework, the Conduit C++ Library for Best-Effort High Performance Computing \citep{moreno2021conduit}.
The Conduit library provides tools to perform best-effort communication in a flexible, intuitive interface and uniform inter-operation of serial, parallel, and distributed modalities.
Although Conduit currently implements distributed functionality via MPI intrinsics, in future work we will explore lower-level protocols like InfiniBand Unreliable Datagrams \citep{kashyap2006ip, koop2007high}.

Here, we present a set of on-hardware experiments to empirically characterize Conduit's best-effort communication model.
In order to survey across workload profiles, we tested performance under both a communication-intensive graph coloring solver and a compute-intensive artificial life simulation.

First, we determine whether best-effort communication strategies can benefit performance compared to the traditional perfect communication model.
We considered two measures of performance: computational steps executed per unit time and solution quality achieved within a fixed-duration run window.

We compare the best-effort and perfect-computation strategies across processor counts, expecting to see the marginal benefit from best-effort communication increase at higher processor counts.
We focus on weak scaling, growing overall problem size proportional to processor count.
Put another way, we hold problem size per processor constant.%
\footnote{
As opposed to strong scaling, where the problem size is held fixed while processor count increases.
}
This approach prevents interference from shifts in processes' workload profiles in observation of the effects of scaling up processor count.

To survey across hardware configurations, we tested scaling CPU count via threading on a single node and scaling CPU count via multiprocessing with each process assigned to a distinct node.
In addition to a fully best-effort mode and a perfect communication mode, we also tested two intermediate, partially synchronized modes: one where the processor pool completed a global barrier (i.e., they aligned at a synchronization point) at predetermined, rigidly scheduled timepoints and another where global barriers occurred on a rolling basis spaced out by fixed-length delays from the end of the last synchronization.%
\footnote{
Our motivation for these intermediate synchronization modes was interest in the effect of clearing any potentially-unbounded accumulation of message backlogs on laggard processes.
}

Second, we sought to more closely characterize variability in message dispatch, transmission, and delivery under the best-effort model.
Unlike perfect communication, real-time volatility affects the outcome of computation under the best-effort model.
Because real-time processing speed degradations and message latency or loss alters inputs to simulation elements, characterizing the distribution of these phenomena across processing components and over time is critical to understanding the actual computation being performed.
For example, consistently faster execution or lower messaging latency for some subset of processing elements could violate uniformity or symmetry assumptions within a simulation.
It is even possible to imagine reciprocal interactions between real-time best-effort dynamics and simulation state.
In the case of a positive feedback loop, the magnitude of effects might become extreme.
For example, in artificial life scenarios, agents may evolve strategies that selectively increase messaging traffic so as to encumber neighboring processing elements or even cause important messages to be dropped.

We monitor five aspects of real-time behavior, which we refer to as quality of service metrics \citep{karakus2017quality},
\begin{itemize}
  \item wall-time simulation update rate (``simstep period''),
  \item simulation-time message latency,
  \item wall-time message latency,
  \item steadiness of message inflow (``delivery clumpiness''), and
  \item delivery failure rate.
\end{itemize}

In an initial set of experiments, we use the graph coloring problem to test this suite of quality of service metrics across runtime conditions expected to strongly influence them.
We compare
\begin{itemize}
  \item increasing compute workload per simulation update step,
  \item within-node versus between-node process placement, and
  \item multithreading versus multiprocessing.
\end{itemize}
We perform these experiments using a graph coloring solver configured to maximize communication relative to computation (i.e., just one simulation element per CPU) in order to maximize sensitivity of quality of service to the runtime manipulations.

Finally, we extend our understanding of performance scaling from the preceding experiments by analyzing how each quality of service metric fares as problem size and processor count grow together, a ``weak scaling'' experiment.
This analysis would detect a scenario where raw performance remains stable under weak scaling, but quality of service (and, therefore, potentially quality of computation) degrades.

\section{Methods}

We performed two benchmarks to compare the performance of Conduit's best-effort approach to a traditional synchronous model.
We tested our benchmarks across both a multithread, shared-memory context and a distributed, multinode context.
In each hardware context, we assessed performance on two algorithmic contexts: a communication-intensive distributed graph coloring problem (Section \ref{sec:graph-coloring-benchmark}) and a compute-intensive digital evolution simulation (Section \ref{sec:digital_evolution_benchmark}).
The latter benchmark --- presented in Section \ref{sec:digital_evolution_benchmark} --- grew out of the original work developing the Conduit library to support large-scale experimental systems to study open-ended evolution.
The former benchmark --- presented in Section \ref{sec:graph-coloring-benchmark} --- complements the first by providing a clear definition of solution quality.
Metrics to define solution quality in the open-ended digital evolution context remain a topic of active research.

\subsection{ Digital Evolution Benchmark } \label{sec:digital_evolution_benchmark}

The digital evolution benchmark runs the DISHTINY (DIStributed Hierarchical Transitions in Individuality) artificial life framework.
This system is designed to study major transitions in evolution, events where lower-level organisms unite to form a self-replicating entity.
The evolution of multicellularity and eusociality exemplify such transitions.
Previous work with DISHTINY has explored methods for selecting traits characteristic of multicellularity such as reproductive division of labor, resource sharing within kin groups, resource investment in offspring, and adaptive apoptosis \citep{moreno2019toward}.

DISHTINY simulates a fixed-size toroidal grid populated by digital cells.
Cells can sense attributes of their immediate neighbors, can communicate with those neighbors through arbitrary message passing, and can interact with neighboring cells cooperatively through resource sharing or competitively through antagonistic competition to spawn daughter cells into limited space.
This cell behavior is controlled by SignalGP event-driven linear genetic programs \citep{lalejini2018evolving}.
Full details of the DISHTINY simulation are available in \citep{moreno2021exploring}.

We use Conduit-based messaging channels to manage all interactions between neighboring cells.
Conduit models messaging channels as independent objects.
However, support is provided for behind-the-scenes consolidation of communication along these channels between pairs of processes.
Pooling joins together exactly one message per messaging channel to create a fixed-size consolidated message.
Aggregation joins together arbitrarily many messages per channel  to  create a variable-size consolidated message.

During a computational update, each cell advances its internal state and pushes information about its current state to neighbor cells.
Several independent messaging layers handle disparate aspects of cell-cell interaction, including
\begin{itemize}
  \item Cell spawn messages, which contain variable-length genomes (seeded at 100 12-byte instructions with a hard cap of 1000 instructions). These are handled every 16 updates and use Conduit's built-in aggregation support for inter-process transfer.
  \item Resource transfer messages, consisting of a 4-byte float value. These are handled every update and use Conduit's built-in pooling support for inter-process transfer.
  \item Cell-cell communication messages, consisting of arbitrarily many 20-byte packets dispatched by genetic program execution. These are handled every 16 updates and use Conduit's built-in aggregation support for inter-process transfer.
  \item Environmental state messages, consisting of a 216-byte struct of data. These are handled every 8 updates and use conduit's built-in pooling support for inter-process transfer.
  \item Multicellular kin-group size detection messages, consisting of a 16-byte bitstring. These are handled every update and use Conduit's built-in pooling support for inter-process transfer.
\end{itemize}

Implementing all cell-cell interaction via Conduit-based messaging channels allows the simulation to be parallelized down to the granularity, potentially, of individual cells.
These messaging channels allow cells to communicate using the same interface whether they are placed within the same thread, across different threads, or across diffferent processes.
However, in practice, for this benchmarking we assign 3600 cells to each thread or process.
Because all cell-cell interactions occur via Conduit-based messaging channels, logically-neighboring cells can interact fully whether or not they are located on the same thread or process (albeit with potential irregularities due to best-effort limitations).
An alternate approach to evolving large populations might be an island model, where Conduit-based messaging channels would be used solely to exchange genomes between otherwise independent populations \citep{bennett1999building}.
However, we chose to instead parallelize DISHTINY as a unified spatial realm in order to enable parent-offspring interaction and leave the door open for future work with multicells that exceed the scope of an individual thread or process.

\subsection{ Graph Coloring Benchmark } \label{sec:graph-coloring-benchmark}

The graph coloring benchmark employs a graph coloring algorithm designed for distributed WLAN channel selection \citep{leith2012wlan}.
In this algorithm, nodes begin by randomly choosing a color.
Each computational update, nodes test for any neighbor with the same color.
If and only if a conflicting neighbor is detected, nodes randomly select another color.
The probability of selecting each possible color is stored in array associated with each node.
Before selecting a new color, the stored probability of selecting the current (conflicting) color is decreased by a multiplicative factor $1 - b$.
We used $b=0.1$, as suggested by Leith et al.
Tandem adjustments increase the probability of selecting all other channels.
Regardless of whether their color changed, nodes always transmit their current color to their neighbor.

Our benchmarks focus on weak scalability, using a fixed problem size of \numprint{2048} graph nodes per thread or process.
These nodes were arranged in a two-dimensional grid topology where each node had three possible colors and four neighbors.
We implement the algorithm with a single Conduit communication layer carrying graph color as an unsigned integer.
We used Conduit's built-in pooling feature to consolidate color information into a single MPI message between pairs of communicating processes each update.
We performed five replicates, each with a five second simulation runtime.
Solution error was measured as the number of graph color conflicts remaining at the end of the benchmark.

\subsection{Asynchronicity Modes} \label{sec:asynchronicity_modes}

\begin{table}
  \begin{tabular}{ccl}
    \toprule
    Mode & Description\\
    \midrule
    0 & Barrier sync every update\\
    1 & Rolling barrier sync\\
    2 & Fixed barrier sync\\
    3 & No barrier sync\\
    4 & No inter-cpu communication\\
  \bottomrule
\end{tabular}
  \caption{Asynchronicity modes used for benchmarking experiments, arranged from most to least synchronized.}
  \label{tab:asynchronicity_modes}
\end{table}

For both benchmarks, we compared performance across a spectrum of synchronization settings, which we term ``asynchronicity modes'' (Table \ref{tab:asynchronicity_modes}).
Asynchronicity mode 0 represents traditional fully-synchronous methodology.
Under this treatment, full barrier synchronization was performed between each computational update.
Asynchronicity mode 3 represents fully asynchronous methodology.
Under this treatment, individual threads or processes performed computational updates freely, incorporating input from other threads or processes on a fully best-effort basis.

During early development of the library, we discovered episodes where unprocessed messages built up faster than they could be processed --- even if they were being skipped over to only get the latest message.
In some instances, this strongly degraded  quality of service or even caused runtime instability.
We switched to MPI communication primitives that could consume many backlogged messages per call and increased buffer size to address these issues, but remained interested in the possibility of partial synchronization to clear potential message backlogs.
So, we included two partially-synchronized treatments: asynchronicity modes 1 and 2.

In asynchronicity mode 1, threads and processes alternated between performing computational updates for a fixed-time duration and executing a global barrier synchronization.
For the graph coloring benchmark, work was performed in 10ms chunks.
For the digital evolution benchmark, which is more computationally intensive, work was performed in 100ms chunks.
In asynchronicity mode 2, threads and processes executed global barrier synchronizations at predetermined time points based on the UTC clock.

Finally, asynchronicity mode 4 disables all inter-thread and inter-process communication, including barrier synchronization.
We included this mode to isolate the impact on performance of communication between threads and processes from other factors potentially affecting performance, such as cache crowding.
In this run mode for the graph coloring benchmark, all calls send messages between processes or threads were skipped (except after the benchmark concluded, when assessing solution quality).
Because of its larger footprint, incorporating logic into the digital evolution simulation to disable all inter-thread and inter-process messaging was impractical.
Instead, we launched multiple instances of the simulation as fully-independent processes and measured performance of each.

\subsection{Quality of Service Metrics} \label{sec:quality-of-service-metrics}

\begin{figure*}
  \centering
  \begin{subfigure}[b]{0.5\textwidth}
    \centering
    \includegraphics[width=\linewidth]{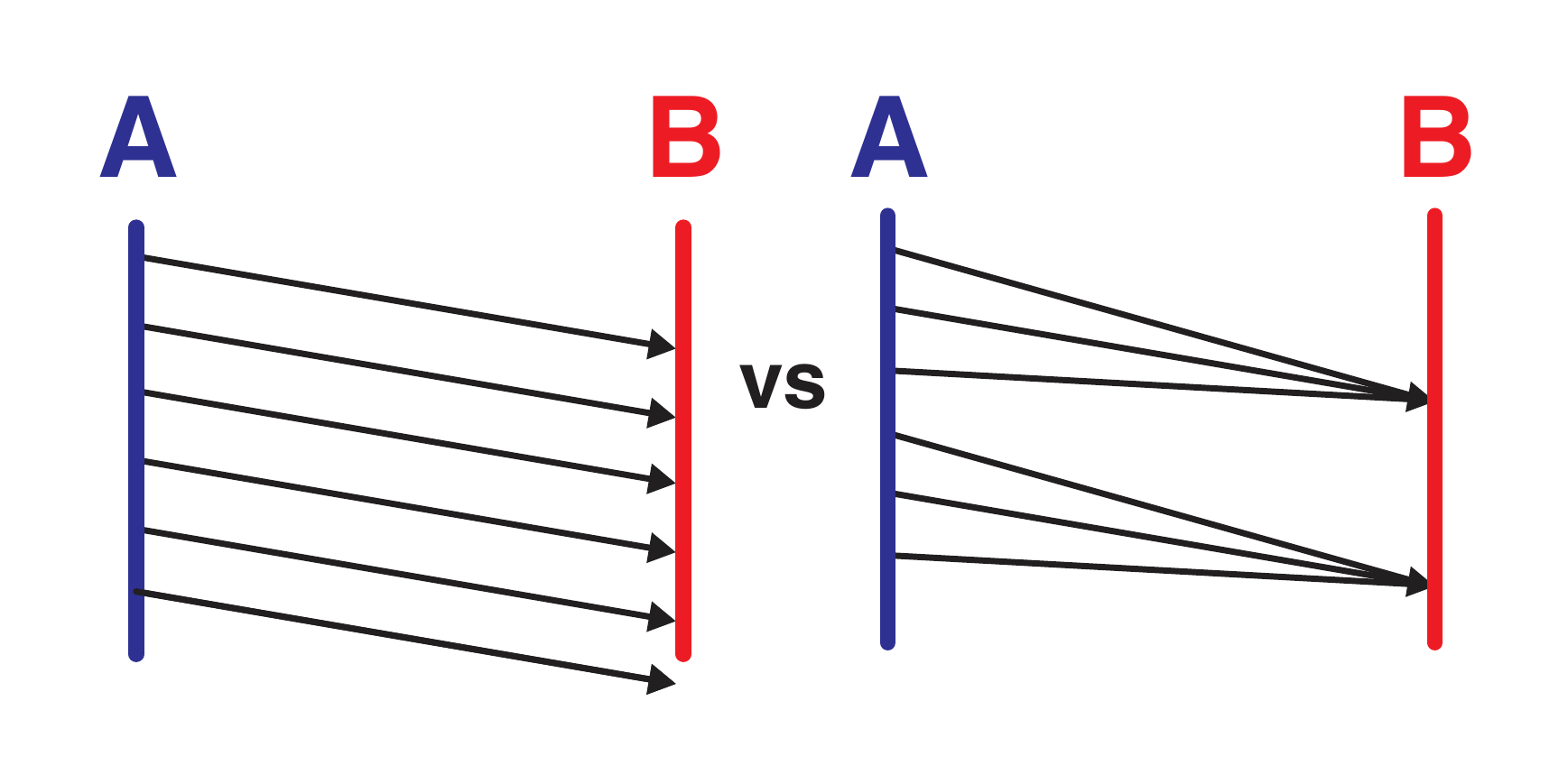}
    \caption{Clumpiness}
    \label{fig:quality-of-service-metric-definitions-clumpiness}
  \end{subfigure}%
  \begin{subfigure}[b]{0.5\textwidth}
    \centering
    \includegraphics[width=\linewidth]{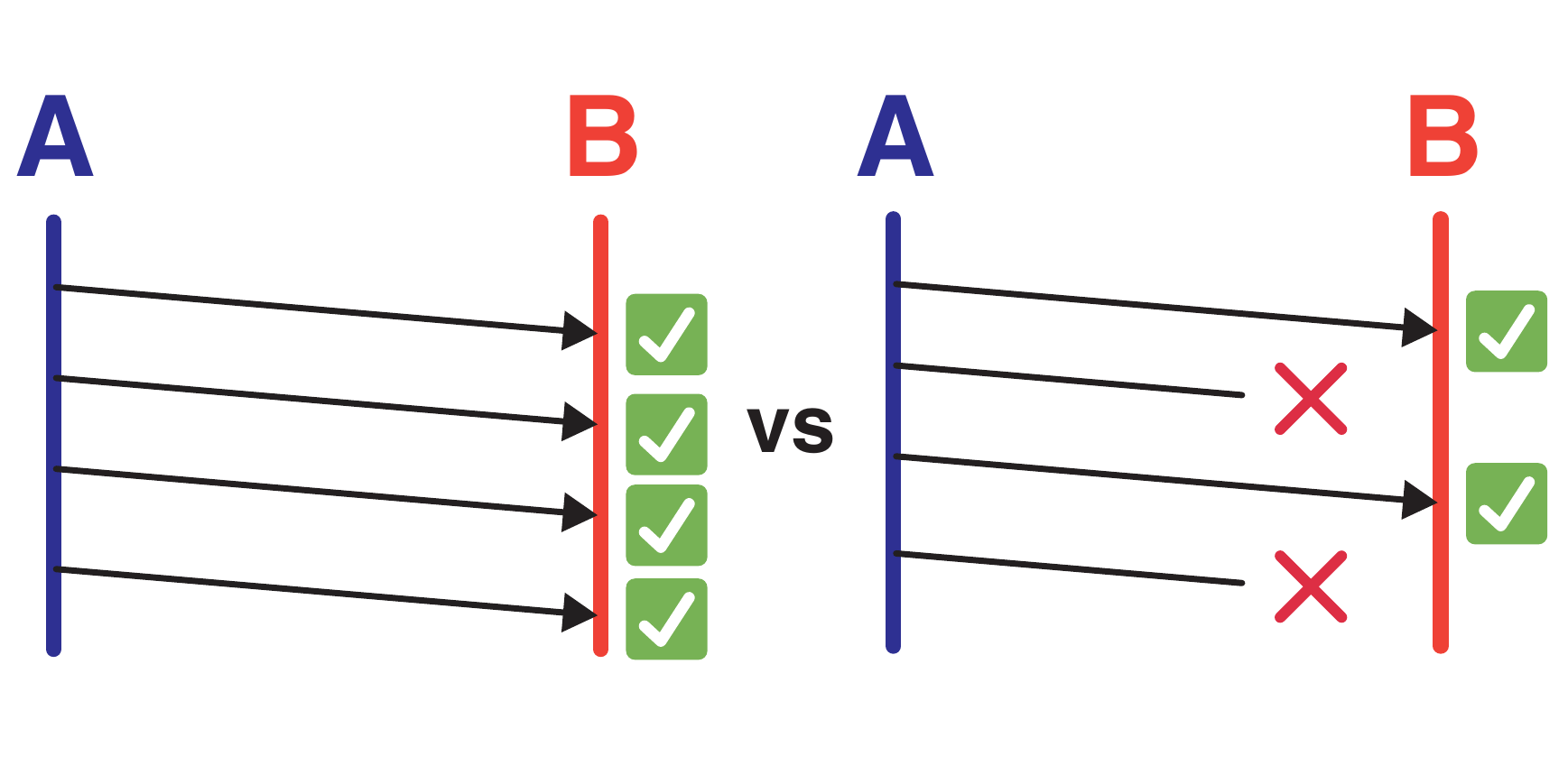}
    \caption{Delivery Failure Rate}
    \label{fig:quality-of-service-metric-definitions-delivery-failure-rate}
  \end{subfigure}
  \begin{subfigure}[b]{0.5\textwidth}
    \centering
    \includegraphics[width=\linewidth]{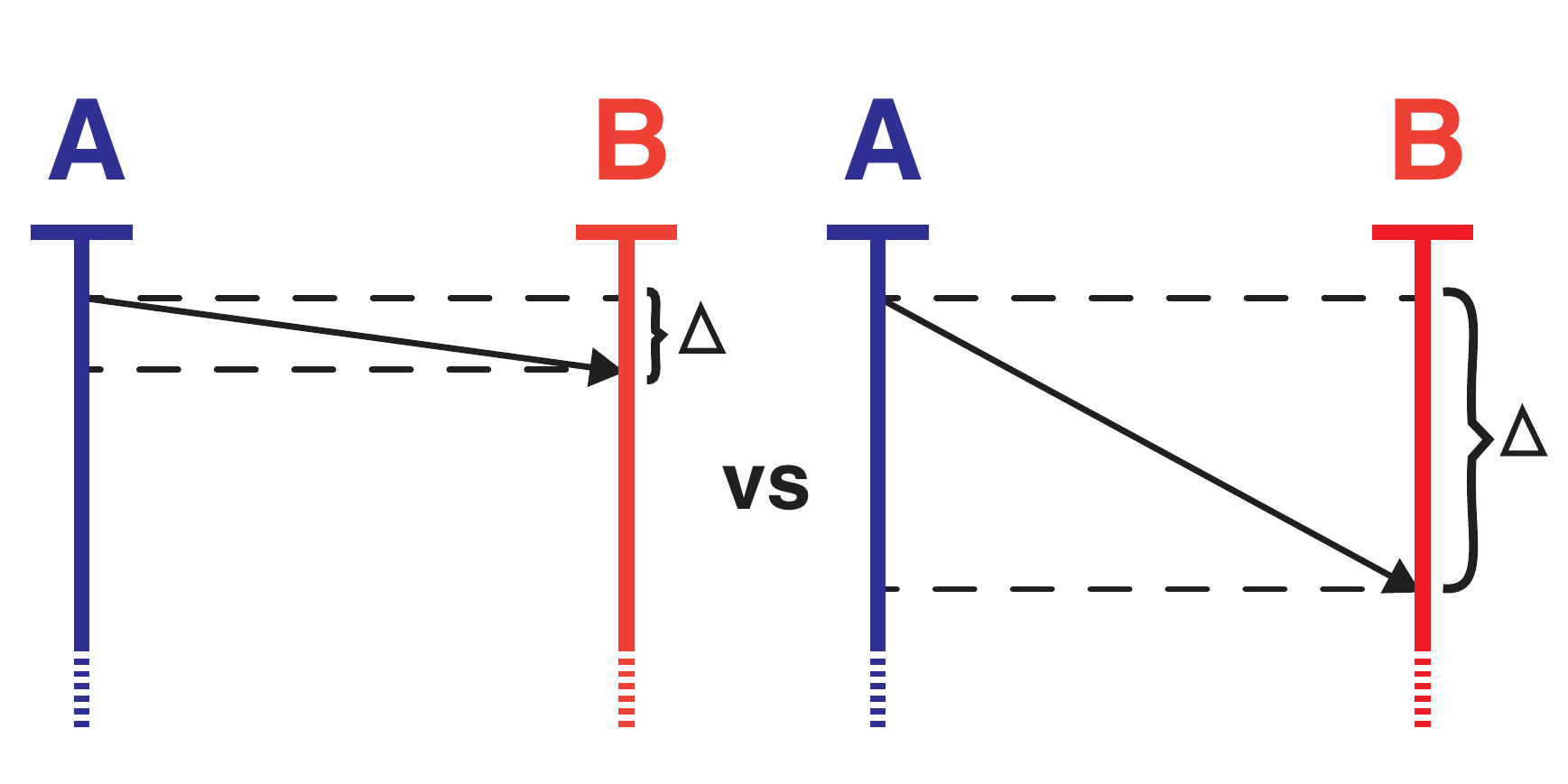}
    \caption{Latency}
    \label{fig:quality-of-service-metric-definitions-latency}
  \end{subfigure}%
  \begin{subfigure}[b]{0.5\textwidth}
    \centering
    \includegraphics[width=\linewidth]{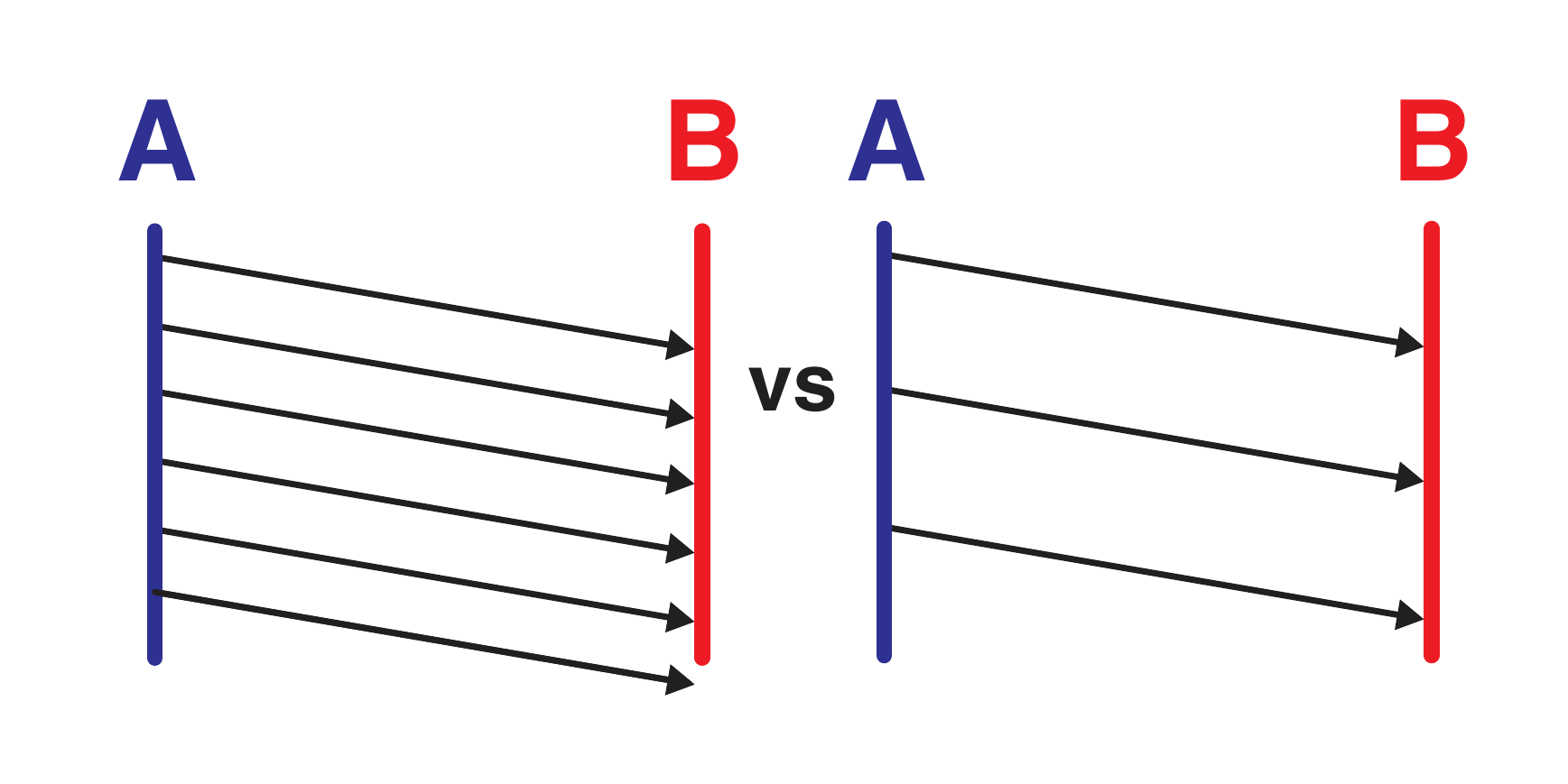}
    \caption{Simstep Period}
    \label{fig:quality-of-service-metric-definitions-simstep-period}
  \end{subfigure}%
  \caption{
  Quality of service metrics.
  Each illustration is a space-time diagram, with $A$ and $B$ representing independent processes.
  The vertical axis depicts the passage of time, from top to bottom.
  Solid black arrows represent message delivery.
  The left panel of each metric's diagram depicts a scenario with a lower (``better'') value for that metric compared to the right panel, which depicts a higher (``worse'') value for that metric.
  }
  \label{fig:quality-of-service-metric-definitions}
\end{figure*}

The best-effort communication model eschews effort to insulate computation from real-time message delivery dynamics.
Because these dynamics are difficult to predict \textit{a priori} and can bias computation, thorough, empirical runtime measurements are necessary to understand results of such computation.
To this end, we developed a suite of quality of service metrics.
Figure \ref{fig:quality-of-service-metric-definitions} provides space-time diagrams illustrating the metrics presented in this section.

For the purposes of these metrics, we assume that simulations proceed in an iterative fashion with alternating compute and communication phases.
For short, we refer to a single compute-communication cycle as a ``simstep.''
We derive formulas for metrics in terms of independent observations preceding and succeeding a ``snapshot'' window, during which the simulation and any associated best-effort communication proceeds unimpeded.
Snapshot observations are taken at one minute intervals over the course of each of our a replicate experiments.
The following section, \ref{sec:quality-of-service-experiments}, details the experimental apparatus used to generate quality of service metrics reported in this work.

\subsubsection{Simstep Period} \label{sec:simstep-period-metric}

We calculate the amount of wall-time elapsed per simulation update cycle (``Simstep Period'') during a snapshot window as
\begin{align*}
\frac{
  \mathrm{update\,count\,after} - \mathrm{update\,count\,before}
}{
  \mathrm{walltime\,after} - \mathrm{walltime\,before}
}.
\end{align*}
Figure \ref{fig:quality-of-service-metric-definitions-simstep-period} compares a scenario with low simstep period to a scenario with a higher simstep period.

\subsubsection{Simstep Latency} \label{sec:wall-time-latency-metric}

This metric reports the number of simulation iterations that elapse between message dispatch and message delivery.
Figure \ref{fig:quality-of-service-metric-definitions-latency} compares a scenario with low latency to a scenario with a higher latency.

To prevent interfering effects of imperfect clock synchronization, we estimate one-way wall-time latency from a round-trip measure.
As part of our instrumentation, each simulation element maintains an independent zero-initialized ``touch counter'' for each neighbor simulation element it communicates with.
Dispatched messages are bundled with the current value touch counter associated with the target element.
Every message received sets the corresponding touch counter to $1 + \mathrm{bundled\,touch\,count}$.
In this manner, the touch counter increments by two for each successful round trip completed.
(Because simulation elements are arranged as a toroidal mesh in all experiments, interaction between simulation elements is always reciprocal.)

We therefore calculate one-way latency during a snapshot window as,
\begin{align*}
  \frac{
    \mathrm{update\,count\,after} - \mathrm{update\,count\,before}
  }{
    \min\Big( \mathrm{ touch\,count\,after } - \mathrm{ touch\,count\,before }, 1 \Big)
  }.
\end{align*}
Note that, in the case where no touches elapse, we bound our latency measure to snapshot duration.

\subsubsection{Wall-time Latency} \label{sec:simulation-time-latency-metric}

Wall-time latency is closely related to simstep latency, simply denominating in elapsed real time instead of elapsed simulation steps.
To calculate wall-time latency we apply a conversion to simstep latency based on simstep period,
\begin{align*}
  \mathrm{simstep\,latency} \times \mathrm{simstep\,period}.
\end{align*}

This metric directly tells the real-time performance of message transmission.
Although it directly follows from the interaction between simstep period and wall-time latency, it complements simstep latency's convenient interpretation in terms of potential simulation mechanics (e.g., simulation elements tending to see data from two updates ago versus from ten).

In addition to simstep latency, Figure \ref{fig:quality-of-service-metric-definitions-latency} is also representative of wall-time latency --- simply interpreting the $y$ axis in terms of wall-time instead of elapsed simulation updates.

\subsubsection{Delivery Failure Rate} \label{sec:delivery-failure-rate-metric}

Delivery failure rate measures the fraction of messages sent that are dropped.
The only condition where messages are dropped is when a send buffer fills.
(Under the existing MPI-based implementation, messages that queue on the send buffer are guaranteed for delivery.)
So, we can calculate
\begin{align*}
  \frac{
    \mathrm{successful\,send\,count\,after} - \mathrm{successful\,send\,count\,before}
  }{
    \mathrm{attempted\,send\,count\,after} - \mathrm{attempted\,send\,count\,before}
  }.
\end{align*}

\subsubsection{Delivery Clumpiness} \label{sec:delivery-clumpiness-metric}

Delivery clumpiness quantifies the extent to which message arrival is consolidated to a subset of message pull attempts.
That is, the extent to which independently dispatched messages arrive in batches rather than as an even stream.

If messages all arrive in independent pull attempts, then clumpiness will be zero.
At the point where the pigeonhole principle applies (i.e., $\mathrm{num\,arriving\,messages} >= \mathrm{num\,pull\,attempts}$), clumpiness will also be zero so long as every pull attempt is laden.
If all messages arrive during a single pull attempt, then clumpiness will approach 1.

We formulate clumpiness as the compliment of steadiness.
(Reporting clumpiness provides a lower-is-better interpretation consistent with the rest of the quality of service metrics.)
Steadiness, in turn, stems from three component statistics,
\begin{align*}
\mathrm{num\,laden\,pulls\,elapsed} =& \mathrm{laden\,pull\,count\,after} \\
  &- \mathrm{laden\,pull\,count\,before} \\
\mathrm{num\,messages\,received} =& \mathrm{message\,count\,after} \\
  &- \mathrm{message\,count\,before} \\
\mathrm{num\,pulls\,attempted} =& \mathrm{pull\,attempt\,count\,after} \\
  &- \mathrm{pull\,attempt\,count\,before}
.
\end{align*}

Here, we refer to pull attempts that successfully retrieve a message as ``laden.''

We combine $\mathrm{num\,messages\,received}$ and $\mathrm{num\,pulls\,attempted}$ to derive,
\begin{align*}
  \mathrm{num\,opportunities\,for\,laden\,pulls} = \\
   \min\Big(\mathrm{num\,messages\,received}, \mathrm{num\,pulls\,attempted}\Big).
\end{align*}

Then, to calculate steadiness,
\begin{align*}
  \frac{
    \mathrm{num\,laden\,pulls\,elapsed}
  }{
    \mathrm{num\,opportunities\,for\,laden\,pulls}
  }.
\end{align*}

Delivery clumpiness follows as $1 - \mathrm{steadiness}$.
Figure \ref{fig:quality-of-service-metric-definitions-clumpiness} compares a scenario with low clumpiness to a scenario with higher clumpiness.

\subsection{Quality of Service Experiments} \label{sec:quality-of-service-experiments}

Quality of service experiments executed the graph coloring algorithm described in Section \ref{sec:graph-coloring-benchmark}.
In order to maximize communication intensity, only one graph vertex was assigned per CPU.
Ten experimental replicates were performed for each condition surveyed.
Slightly over five minutes of runtime was afforded to each replicate.
Over five minutes of runtime, snapshot observations were taken at one minute intervals.
The first snapshot observation was taken one minute after the beginning of runtime.

Snapshot observations lasted one second, with the graph coloring algorithm running fully unhampered during the entire snapshot.
This was accomplished by collecting and recording data via a separate thread.
That thread collected and recorded a first tranche of snapshot data, spin waited for one second, and then recorded a second tranche.
Because the underlying system runs in real-time while being observed, state changes can occur during data collection (somewhat akin to photographic motion blur).
Therefore, violations of intuitive invariants --- like strictly non-negative delivery failure rates --- occur in some cases.
However, the magnitude of such violations is generally minor.
Further, because data collection procedures were consistent across treatments, statistical comparisons between treatments nevertheless remain sound.

Snapshots were performed independently for each process at each timepoint.
So, for example, for two processes over the five minute window of a single replicate ten snapshots were collected.
For statistical tests comparing treatments, snapshots were aggregated by replicate by both mean and median.
For each quality of service statistic we estimate mean --- which captures effects of extreme-magnitude outliers --- and median --- which more better represents typicality --- across these window samples.
Statistical comparisons across treatment conditions are performed via regression.
We use ordinary least squares regression to analyze means \citep{geladi1986partial} and quantile regression to analyze medians \citep{koenker2001quantile}.
For comparisons between dichotomous, categorical treatment conditions, one condition is coded as 0 and the other as 1.
In the case of ordinary least squares regression, this boils down to an independent $t$-test.
Although quantile regression on categorical predictors is not precisely equivalent to a direct test on medians between two groups (i.e., Mood's median test), there is precedent for this approach \citep{petscher2014quantile, konstantopoulos2019using}.

Most statistics reported here can be calculated just as well in terms of incoming or outgoing messages.
That is, most statistics can be generated via data from instrumentation attached to message ``inlets'' or data from instrumentation attached to message ``outlets'' with no obvious reason to prefer one over the other.
As ``inlet-'' and ``outlet-''derived statistics are nearly identical in all cases, we simply report the mean over these two measurements.

\subsection{Code, Data, and Reproducibility}

\subsubsection{Benchmarking Experiments} \label{sec:methods-code-data-reproducibility-benchmarking-experiments}

Benchmarking experiments were performed on Michigan State University's High Performance Computing Center, a cluster of hundreds of heterogeneous x86 nodes linked with InfiniBand interconnects.
For multithread experiments, benchmarks for each thread count were collected from the same node.
For multiprocess experiments, each processes was assigned to a distinct node in order to ensure results were representative of performance in a distributed context.
All multiprocess benchmarks were recorded from the same collection of nodes.
Hostnames are recorded for each benchmark data point.
For an exact accounting of hardware architectures used, these hostnames can be crossreferenced with a table included with the data that summarizes the cluster's node configurations.

Code for the distributed graph coloring benchmark is available at \url{https://github.com/mmore500/conduit} under \\ \texttt{demos/channel\_selection}.
Code for the digital evolution simulation benchmark is available at \url{https://github.com/mmore500/dishtiny}.
Exact versions of software used are recorded with each benchmark data point.
Data is available via the Open Science Framework at \url{https://osf.io/7jkgp/} \citep{foster2017open}.
Notebook code for all reported statistics and data visualizations is available at \url{https://hopth.ru/ci}.

\subsubsection{Quality of Service Experiments}

Quality of service experiments were carried out on Michigan State University's High Performance Computing Center lac cluster, consisting of 28-core Intel(R) Xeon(R) CPU E5-2680 v4 \@ 2.40GHz nodes.
All statistical comparisons are performed between observations from the same job allocation.
(Except in the case where intranode and internode configurations were compared, where experiments were performed on separate allocations using comparable nodes on the same cluster.)

Benchmarking experiments described in Section \ref{sec:methods-code-data-reproducibility-benchmarking-experiments} used a send/receive buffer size of 2.
However, due to the high communication intensity of the graph coloring problem with just one simulation element per CPU, quality of service experiments required a larger buffer size of 64 to maintain runtime stability.
In early work developing the Conduit library, we discovered that real-time messaging channels can enter a destabilizing positive feedback spiral when incoming messages take longer to handle (e.g., skip past or read) than sending messages.
Under such conditions, when a process exchanging messages from a partner process experiences a delay it sends fewer messages to that partner process.
Due to fewer incoming messages, the partner the partner process can update more rapidly, increasing incoming message load on the delayed process.
This effect can snowball the partnership intended for even, two-way message exchange into effectively a unilateral producer-consumer relationship where (potentially unbounded) work piles up on the consumer.
To interrupt such a scenario, we use the bulk message pull call \verb|MPI_Testsome| to ensure fast message consumption under backlogged conditions.
So, receiver workload remains closer to constant under high traffic situations (instead of having to pull messages down one-by-one).
Larger receive buffer size, as configured for the quality of service experiments, increases the effectiveness of the bulk message consumption countermeasure.

Code for the distributed graph coloring benchmark is available at \url{https://github.com/mmore500/conduit} under \\ \texttt{demos/channel\_selection}.
Exact versions of software used are recorded with each benchmark data point.
Data is available via the Open Science Framework at \url{https://osf.io/72k5n/} \citep{foster2017open}.
Notebook code for all reported statistics is available at \url{https://hopth.ru/cj}.

\section{Results and Discussion}

Sections \ref{sec:multithread-benchmarks} and \ref{sec:multiprocess-benchmarks} compare execution performance under the best-effort communication versus the perfect communication models.
In particular, both sections investigate how the impact of best-effort communication on performance relates to CPU count scale.
Section \ref{sec:multithread-benchmarks} covers multithreading and Section \ref{sec:multiprocess-benchmarks} covers multiprocessing.

The following sections investigate how system configuration affects quality of service.
Specifically, these sections cover the impact of
\begin{itemize}
  \item increasing compute workload per simulation update step (Section \ref{sec:computation-vs-communication}),
  \item within-node versus between-node process placement (Section \ref{sec:intranode-vs-internode}), and
  \item multithreading versus multiprocessing (Section \ref{sec:multithreading-vs-multiprocessing}).
\end{itemize}

Section \ref{sec:weak-scaling} tests how quality of service changes with CPU count.
This analysis fleshes out the performance-centric picture of best-effort scalability established in Sections \ref{sec:multithread-benchmarks} and \ref{sec:multiprocess-benchmarks}.

Section \ref{sec:with-lac-417-vs-sans-lac-417} tests how inclusion of an apparently faulty node (i.e., that provided exceptionally poor quality of service) affects global quality of service.
This experiment provides insight into the robustness of best-effort approaches to single-point failure.

\subsection{Performance: Multithread Benchmarks}
\label{sec:multithread-benchmarks}

  \begin{figure}[thpb]
      \centering
    \begin{subfigure}[b]{0.49\textwidth}
      \centering
      \includegraphics[width=\linewidth]{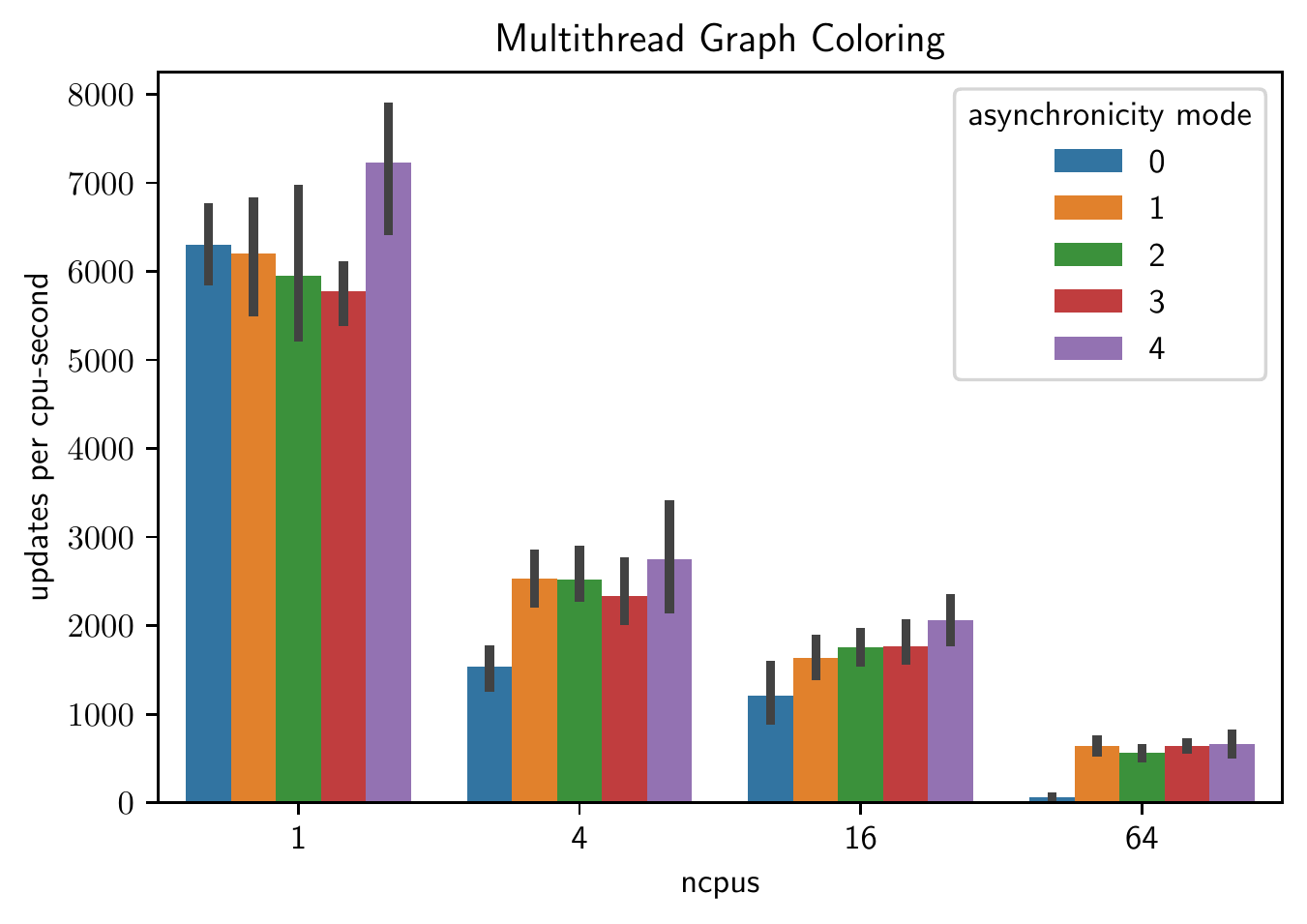}
     \caption{Graph coloring per-thread update rate. Higher is better.}
         \label{fig:multithread_graph_coloring_update_rate}
      \end{subfigure}

    \begin{subfigure}[b]{0.49\textwidth}
      \centering
      \includegraphics[width=\linewidth]{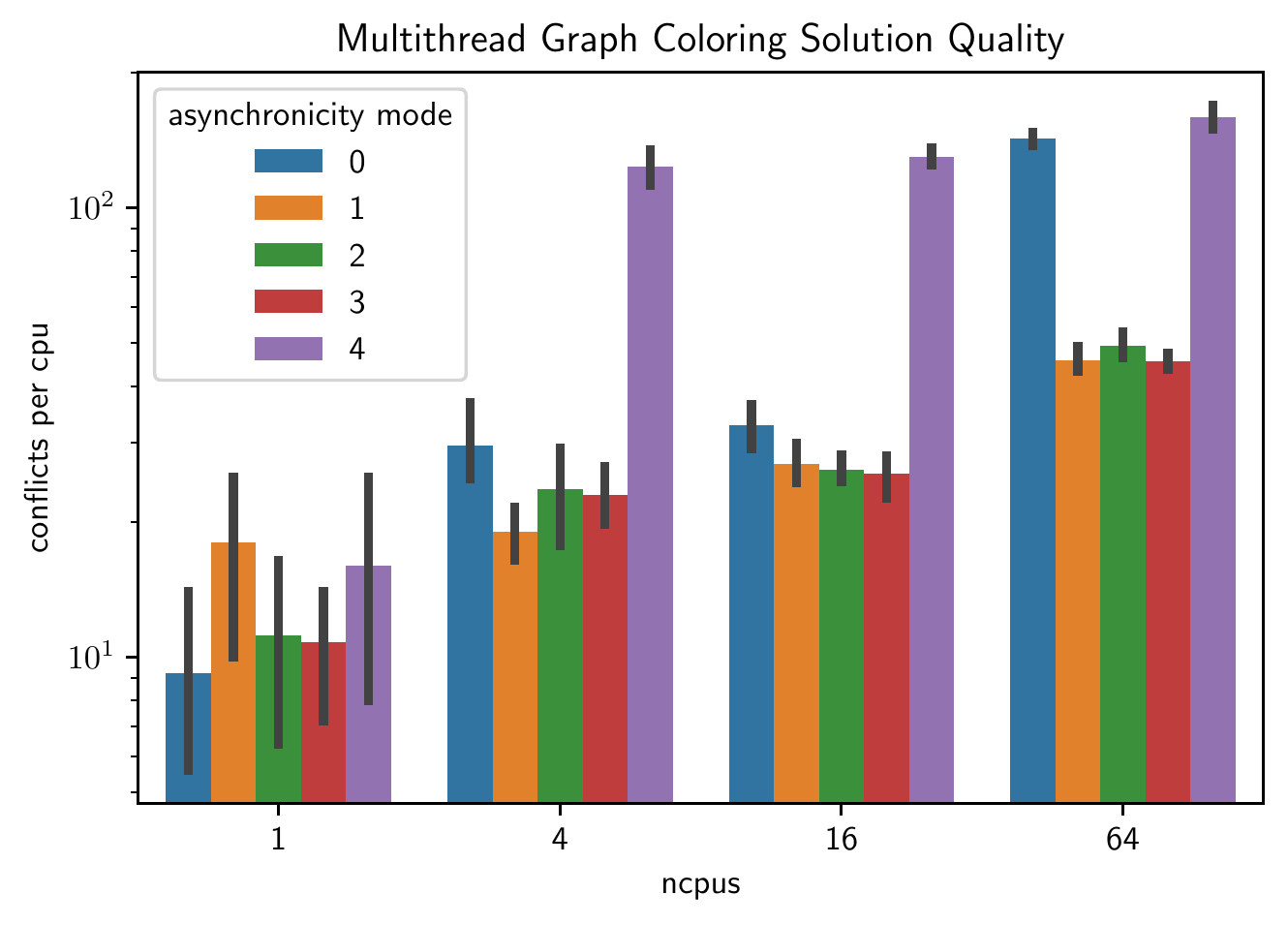}
      \caption{Graph coloring solution conflicts. Lower is better.}
         \label{fig:multithread_graph_coloring_solution_quality}
    \end{subfigure}

    \begin{subfigure}[b]{0.49\textwidth}
    \includegraphics[width=\linewidth]{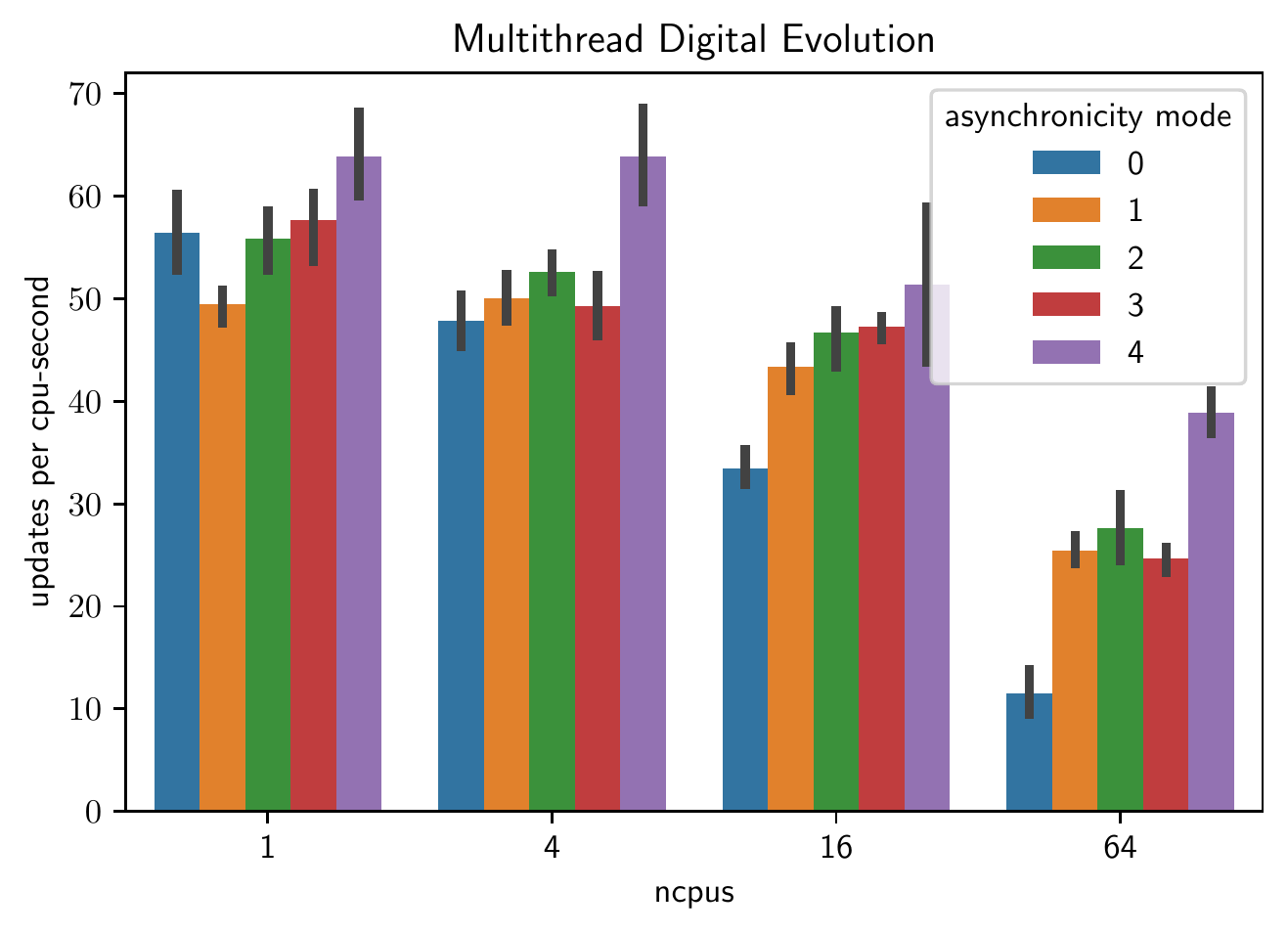}
    \caption{Digital evolution per-thread update rate. Higher is better.}
         \label{fig:multithread_digital_evolution_update_rate}
    \end{subfigure}

    \caption{Multithread benchmark results. Bars represent bootstrapped 95\% confidence intervals. }
      \label{fig:multithread_benchmarks}
  \end{figure}

We first tested how performance on the graph coloring and digital evolution benchmarks fared when increasing thread count on a single hardware node.

Figure \ref{fig:multithread_graph_coloring_update_rate} presents per-CPU algorithm update rate for the graph coloring benchmark at 1, 4, 16, and 64 threads.
Update rate performance decreased with increasing multithreading across all asynchronicity modes.
This performance degradation was rather severe --- per-CPU update rate decreased by 61\% between 1 and 4 threads and by about another 75\% between 4 and 64 threads.
Surprisingly, this issue appears largely unrelated to inter-thread communication, as it was also observed in asynchronicity mode 4, where all interthread communication is disabled.
Perhaps per-CPU update rate degradation under threading was induced by strain on a limited system resource like memory cache or access to the system clock (which was used to control run timing).

Nevertheless, we were able to observe significantly better performance of best-effort asynchronicity modes 1, 2, and 3 at high thread counts.
At 64 threads, these run modes significantly outperformed the fully-synchronized mode 0 ($p < 0.05$, non-overlapping 95\% confidence intervals).
Likewise, as shown in Figure \ref{fig:multithread_graph_coloring_solution_quality}, best-effort asynchronicity modes were able to deliver significantly better graph coloring solutions within the allotted compute time than the fully-synchronized mode 0 ($p < 0.05$, non-overlapping 95\% confidence intervals).

Figure \ref{fig:multithread_digital_evolution_update_rate} shows per-CPU algorithm update rate for the digital evolution benchmark at 1, 4, 16, and 64 threads.
Similarly to the graph coloring benchmark, update rate performance decreased with increasing multithreading across all asynchronicity modes --- including mode 4, which eschews inter-thread communication.
Even without communication between threads, with 64 threads each thread performed updates at only 61\% the rate of a lone thread.
At 64 threads, best-effort asynchronicity modes 1, 2, and 3 exhibit about 43\% the update-rate performance of a lone thread.
Although best-effort inter-thread communication only exhibits half the update-rate performance of completely decoupled execution at 64 threads, this update-rate performance is roughly $2.1\times$ that of the fully-synchronous mode 0.
Indeed, best-effort modes significantly outperform the fully-synchronous mode on the digital evolution benchmark at both 16 and 64 threads ($p < 0.05$, non-overlapping 95\% confidence intervals).

\subsection{Performance: Multiprocess Benchmarks}
\label{sec:multiprocess-benchmarks}

\begin{figure}[thpb]
      \centering

    \begin{subfigure}[b]{0.49\textwidth}
    \centering
    \includegraphics[width=\linewidth]{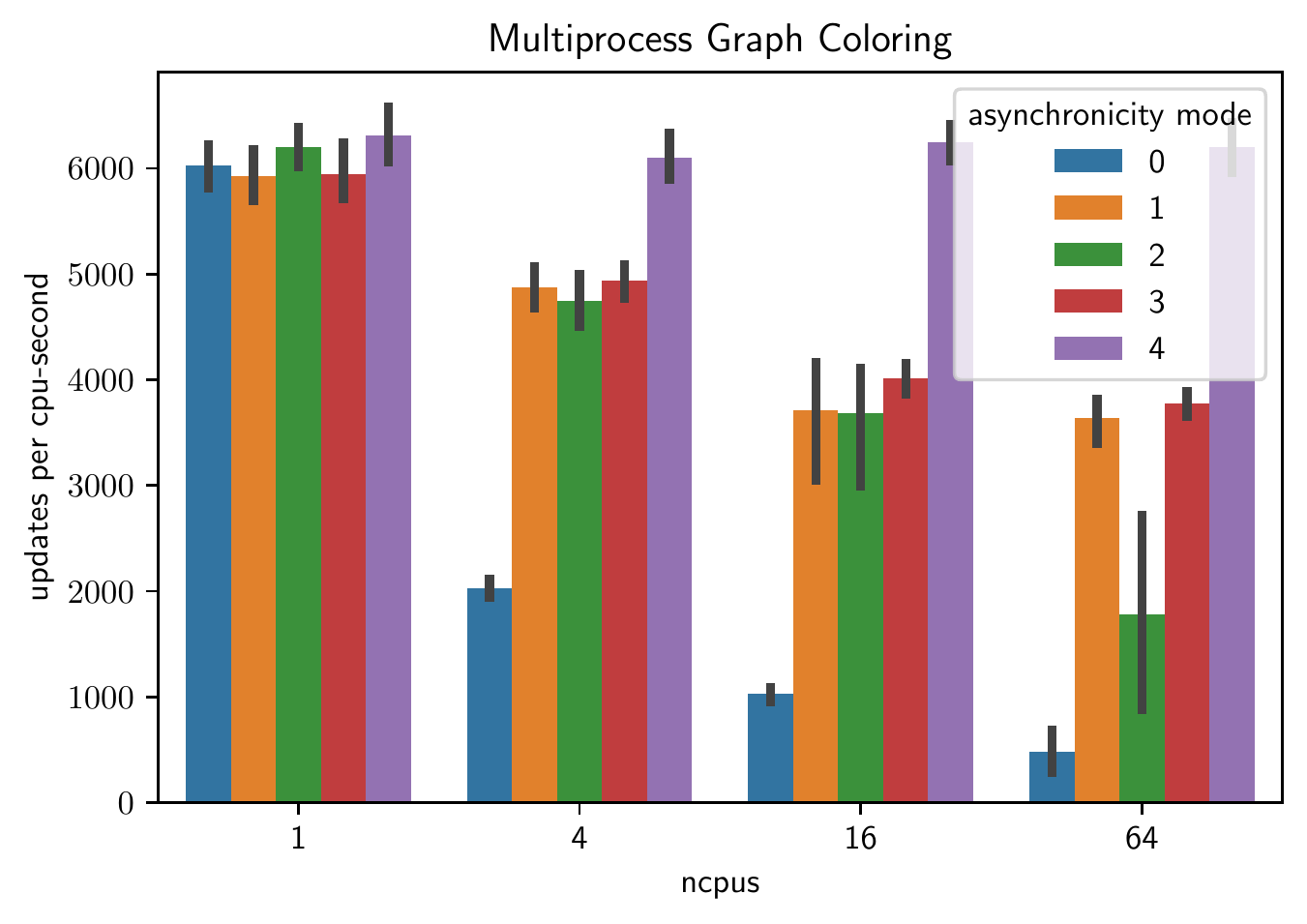}
    \caption{Graph coloring per-process update rate. Higher is better.}
    \label{fig:multiprocess_graph_coloring_update_rate}
    \end{subfigure}

    \begin{subfigure}[b]{0.49\textwidth}
      \centering
      \includegraphics[width=\linewidth]{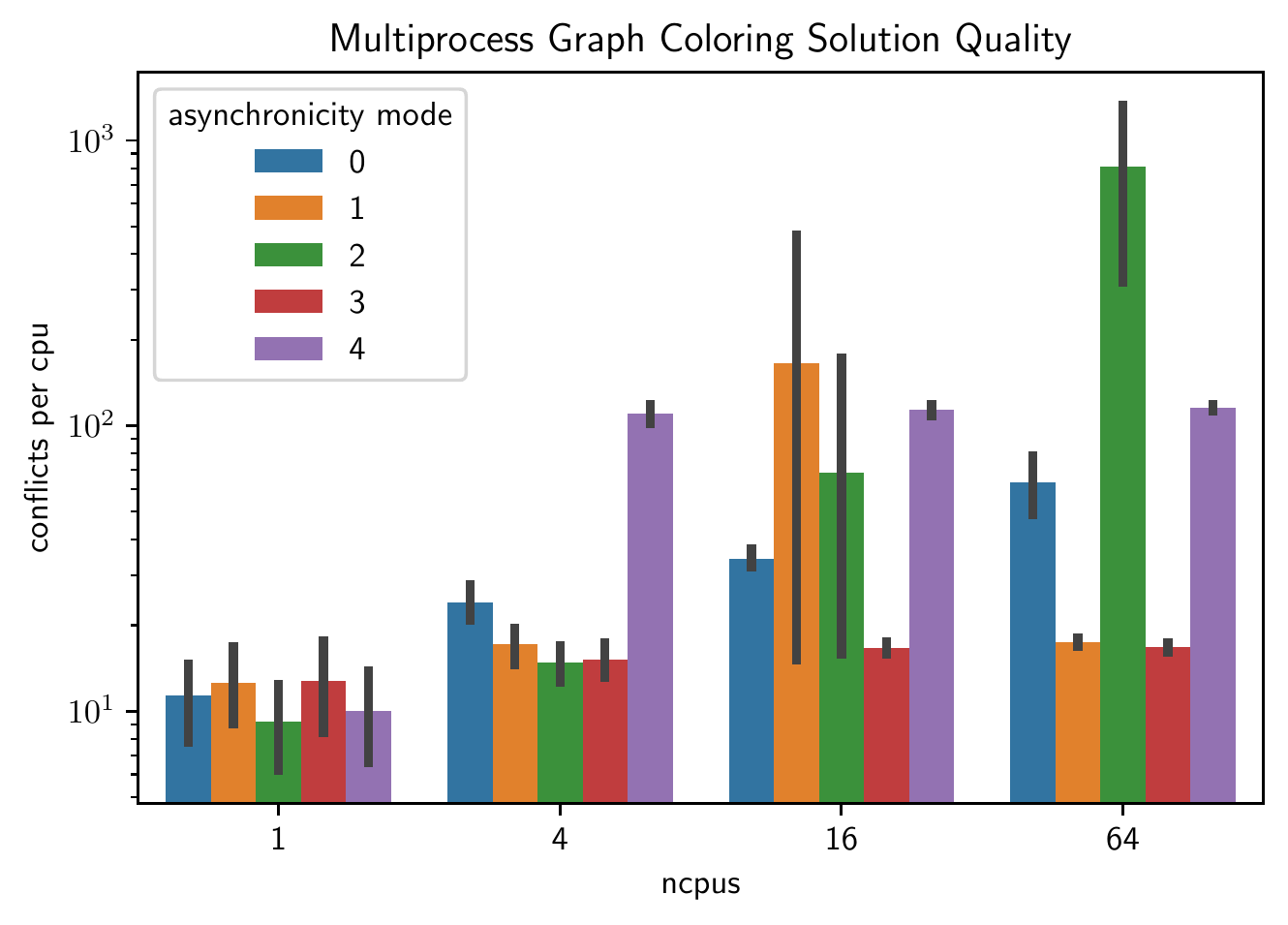}
      \caption{Graph coloring solution conflicts. Lower is better.}
      \label{fig:multiprocess_graph_coloring_solution_quality}
    \end{subfigure}

  \begin{subfigure}[b]{0.49\textwidth}
    \centering
  \includegraphics[width=\linewidth]{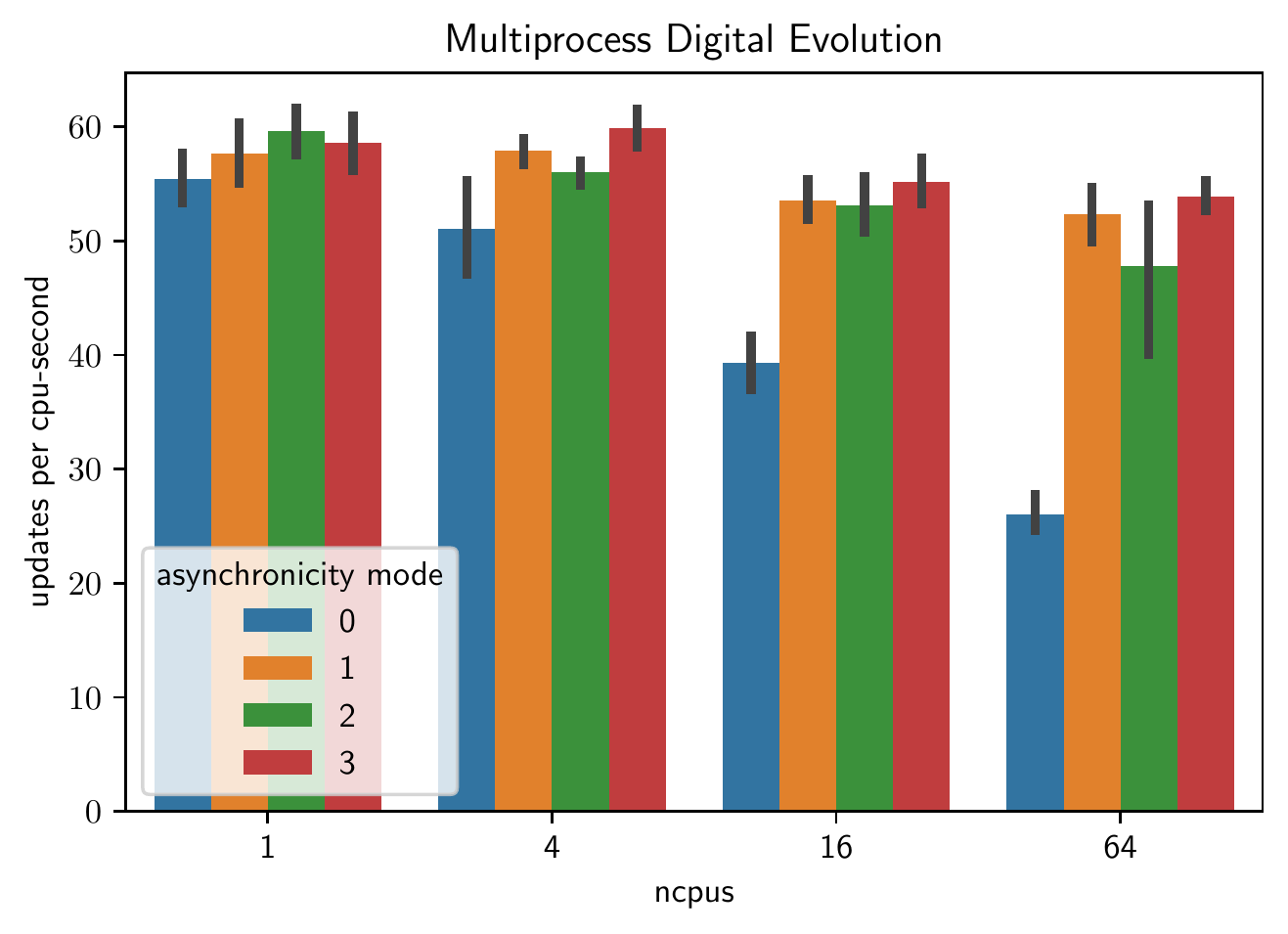}
  \caption{Digital evolution per-process update rate. Higher is better.}
  \label{fig:multiprocess_digital_evolution_update_rate}
  \end{subfigure}

  \caption{Multiprocess benchmark results. Bars represent bootstrapped 95\% confidence intervals. }
  \label{fig:multiprocess_benchmarks}
\end{figure}

Next, we tested how performance on the graph coloring and digital evolution benchmarks fared when scaling with fully independent processes located on different hardware nodes.

Figure \ref{fig:multiprocess_graph_coloring_update_rate} shows per-CPU algorithm update rate for the graph coloring benchmark at 1, 4, 16, and 64 processes.
Unlike the multithreaded benchmark, multiprocess graph coloring exhibits consistent update-rate performance across process counts under asynchronicity mode 4, where inter-CPU communication is entirely disabled.
This matches the unsurprising expectation that, indeed, with comparable hardware a single process should exhibit the same mean performance as any number of completely decoupled processes.

At 64 processes, best-effort asynchronicity mode 3 with the graph coloring benchmark exhibits about 63\% the update-rate performance of single-process execution.
This represents a $7.8\times$ speedup compared to fully-synchronous mode 0.
Indeed, best-effort mode 3 enables significantly better per-CPU update rates at 4, 16, and 64 processes ($p < 0.05$, non-overlapping 95\% confidence intervals).

Likewise, shown in Figure \ref{fig:multiprocess_graph_coloring_solution_quality}, best-effort asynchronicity mode 3 yields significantly better graph-coloring results within the allotted time at 4, 16, and 64 processes ($p < 0.05$, non-overlapping 95\% confidence intervals).
Interestingly, partial-synchronization modes 1 and 2 exhibited highly inconsistent solution quality performance at 16 and 64 process count benchmarks.
Fixed-timepoint barrier sync (mode 2) had particularly poor performance performance at 64 processes (note the log-scale axis).
We suspect this was caused by a race condition where workers would assign sync points to different fixed points different based on slightly different startup times (i.e., process 0 syncs at seconds 0, 1, 2... while process 1 syncs at seconds 1, 2, 3..).

Figure \ref{fig:multiprocess_digital_evolution_update_rate} presents per-CPU algorithm update rate for the digital evolution benchmark at 1, 4, 16, and 64 processes.
Relative performance fares well at high process counts under this relatively computation-heavy workload.
With 64 processes, fully best-effort simulation retains about 92\% the update rate performance of single-process simulation.
This represents a $2.1\times$ speedup compared to the fully-synchronous run mode 0.
Best-effort mode 3 significantly outperforms the per-CPU update rate of fully-synchronous mode 0 at process counts 16 and 64 ($p < 0.05$, non-overlapping 95\% confidence intervals).

\subsection{Quality of Service: Computation vs. Communication}
\label{sec:computation-vs-communication}

Having shown performance benefits of best-effort communication on the graph coloring and digital evolution benchmarks in Sections \ref{sec:multithread-benchmarks} and \ref{sec:multiprocess-benchmarks}, we next seek to more fully characterize the best-effort approach using a holistic suite of proposed quality of service metrics.

This section evaluates how a simulation's ratio of communication intensity to computational work affects these quality of service metrics.
The graph coloring benchmark serves as our experimental model.

For this experiment, arbitrary compute work (detached from the underlying algorithm) was added to the simulation update process.
We used a call to the \texttt{std::mt19937} random number engine as a unit of compute work.
In microbenchmarks, we found that one work unit consumed about 35ns of walltime and 21ns of compute time.
We performed 5 treatments, adding \numprint{0}, \numprint{64}, \numprint{4096}, \numprint{262144}, or \numprint{16777216} units of compute work to the update process.
For each treatment, measurements were made on a pair of processes split across different nodes.

\subsubsection{Simstep Period}

Unsurprisingly, we found a direct relationship between per-update computational workload and the walltime required per computational update.

Supplementary Figures \ref{fig:computation-vs-communication-distribution-simstep-period-inlet-ns} and \ref{fig:computation-vs-communication-distribution-simstep-period-outlet-ns} depict the distribution of walltime per computational update across snapshots.
Once added compute work supersedes the light compute work already associated with the graph coloring algorithm update step (at around 64 work units), simstep period scales in direct proportion with compute work.

Indeed, we found a significant positive relationship between both mean and median simstep period and added compute work (Supplementary Figures \ref{fig:computation-vs-communication-regression-simstep-period-inlet-ns} and \ref{fig:computation-vs-communication-regression-simstep-period-outlet-ns}).
At 0 units of added compute work, mean and median simstep period was 14.7 \SI{14.7}{\micro\second}.
At \numprint{16777216} units of added compute work, mean simstep period was 611ms and median simstep period was 507ms.
This largest workload brushes against instrumentation limitations, accounting for more than half the one second snapshot window.
Supplementary Tables \ref{tab:computation-vs-communication-ordinary-least-squares-regression} and \ref{tab:computation-vs-communication-quantile-regression} detail numerical results of these regressions.

\subsubsection{Simstep Latency}

Unsurprisingly, again, we observed a negative relationship between the number of simulation steps elapsed during message transit and added computational work.
Put simply, longer update steps provide more time for messages to transit.

Supplementary Figures \ref{fig:computation-vs-communication-distribution-latency-simsteps-inlet} and \ref{fig:computation-vs-communication-distribution-latency-simsteps-outlet} show the distribution of simstep latency across compute workloads.
With no added compute work, messages take between 20 and 100 simulation steps to transit (mean: 48.0 updates; median: 42.5 updates).
At maximum compute work per update, messages arrive at a median 1.00 update latency.

Regression analysis confirms a significant negative relationship between both mean and median log simstep latency and log added compute work (Supplementary Figures \ref{fig:computation-vs-communication-regression-latency-simsteps-inlet} and \ref{fig:computation-vs-communication-regression-latency-simsteps-outlet}).
Supplementary Tables \ref{tab:computation-vs-communication-ordinary-least-squares-regression} and \ref{tab:computation-vs-communication-quantile-regression} detail numerical results of these regressions.

\subsubsection{Walltime Latency}

Effects of compute work on walltime latency highlight an important caveat in interpretation of this metric: sensitivity in measuring walltime latency is limited by the simulation update pace.
If a message is dispatched while its recipient is busy, the soonest it can be received will be when that recipient completes the computational phase of its update.
Then, a full computational update will elapse before touch count propagates in a return message.
So, even approaching instantaneous message delivery, a lower floor of $\approx 1.5 \times$ simstep period should be expected for the walltime latency measure.
It should be noted, however, that distinctions in latency below the simulation update rate largely lack practical significance with respect to simullation outcomes.

For \numprint{0}, \numprint{64}, and \numprint{4096} work units, walltime latency measures $\approx 1$ ms (means: \SI{708}{\micro\second}, \SI{788}{\micro\second}, \SI{902}{\micro\second}; medians: \SI{622}{\micro\second}, \SI{640}{\micro\second}, \SI{738}{\micro\second}).
At higher computational loads, simstep period surpasses 1ms.
At \numprint{262144} work units, simulation update time reaches $\approx 8ms$ and, correspondingly, measured walltime latency registers slightly more than 50\% longer at $\approx 13ms$ (mean: 13.2ms, median 13.1ms).
So, true latency likely falls well below simulation update time.
At the highest computational workload, measured walltime latency stretches to the full one second snapshot window.

Supplementary Figures \ref{fig:computation-vs-communication-distribution-latency-walltime-inlet-ns} and \ref{fig:computation-vs-communication-distribution-latency-walltime-outlet-ns} show the distribution of walltime latency across computational workloads.
Supplementary Figures \ref{fig:computation-vs-communication-regression-latency-walltime-inlet-ns} and \ref{fig:computation-vs-communication-regression-latency-walltime-outlet-ns} summarize regression between walltime latency and added compute work.
Supplementary Tables \ref{tab:computation-vs-communication-ordinary-least-squares-regression} and \ref{tab:computation-vs-communication-quantile-regression} detail numerical results of those regressions.

\subsubsection{Delivery Clumpiness}

We observed a significant negative relationship between computation workload and delivery clumpiness.

At low computational intensity, we observed clumpiness greater than 0.95, meaning that fewer than 5\% of pull requests were laden with fresh messages (at 0 compute work mean: 0.96, median 0.96).
However, at high computational intensity clumpiness reached 0, indicating that messages arrived as a steady stream (at \numprint{16777216} compute work mean: 0.00, median 0.00).
Presumably, the reduction in clumpiness is due to increased real-time separation between dispatched messages.

Supplementary Figure \ref{fig:computation-vs-communication-distribution-delivery-clumpiness} shows the effect of computational workload on the distribution of observed clumpinesses.
We found a significant negative relationship between both mean and median clumpiness and computational intensity.
Supplementary Figure \ref{fig:computation-vs-communication-regression-delivery-clumpiness} visualizes these regressions and Supplementary Tables \ref{tab:computation-vs-communication-ordinary-least-squares-regression} and \ref{tab:computation-vs-communication-quantile-regression} provide numerical details.

\subsubsection{Delivery Failure Rate}

We did not observe any delivery failures across all replicates and all compute workloads.
So, compute workload had no observable effect on delivery reliability.

Supplementary Figure \ref{fig:computation-vs-communication-distribution-delivery-failure-rate} shows the distribution of delivery failure rates across computation workloads and Supplementary Figure \ref{fig:computation-vs-communication-regression-delivery-failure-rate} shows regressions of delivery failure rate against computational workload.
See Supplementary Tables \ref{tab:computation-vs-communication-ordinary-least-squares-regression} and \ref{tab:computation-vs-communication-quantile-regression} for numerical details.

\subsection{Quality of Service: Intranode vs. Internode}
\label{sec:intranode-vs-internode}

This section tests the effect of process assignment on best-effort quality of service, comparing multi-node and single-node assignments.
The graph coloring benchmark again serves as our experimental substrate.

For this experiment, processes were either assigned to the same node or were assigned to different nodes.
In both cases, we used two processes.

\subsubsection{Simstep Period}

Simstep period was significantly slower under internode conditions than under intranode conditions.

When processes shared the same node, simstep period was around \SI{9}{\micro\second} (mean: \SI{9.06}{\micro\second}; median: \SI{9.08}{\micro\second}).
Under internode conditions, simstep period was around \SI{14}{\micro\second} (mean: \SI{14.5}{\micro\second}; median: \SI{14.4}{\micro\second}).
Supplementary Figures \ref{fig:intranode-vs-internode-distribution-simstep-period-inlet-ns} and \ref{fig:intranode-vs-internode-distribution-simstep-period-outlet-ns} depict the distribution of walltime per computational update across intranode and internode conditions.

This result presumably attributes to an increased walltime cost for calls to the MPI implementation backing internode communication compared to the MPI implementation backing intranode communication.
Although this effect is clearly detectable, its magnitude is modest given the minimal computational intensity of the simulation update step --- only $\approx 56\%$ more expensive than intranode dispatch.

Both mean and median simstep period increased significantly under internode conditions.
(Supplementary Figures \ref{fig:intranode-vs-internode-regression-simstep-period-inlet-ns} and \ref{fig:intranode-vs-internode-regression-simstep-period-outlet-ns} visualize these regressions and Supplementary Tables \ref{tab:intranode-vs-internode-ordinary-least-squares-regression} and \ref{tab:intranode-vs-internode-quantile-regression} detail numerical results.)

\subsubsection{Simstep Latency}

Significantly more simulation updates transpired during message transmission under internode condtions compared to intranode conditions.

Supplementary Figures \ref{fig:intranode-vs-internode-distribution-latency-simsteps-inlet} and \ref{fig:intranode-vs-internode-distribution-latency-simsteps-outlet} compares the distributions of simstep latency across these conditions.
Simstep latency was around 1 update for intranode communication (mean: 1.00 updates; median 0.75 updates) and around 40 updates for internode communication (mean: 41.6 updates; median: 37.4 updates).

Regression analysis confirms the significant effect of process placement on simstep latency (Supplementary Figures \ref{fig:intranode-vs-internode-regression-latency-simsteps-inlet} and \ref{fig:intranode-vs-internode-regression-latency-simsteps-outlet}).
Supplementary Tables \ref{tab:intranode-vs-internode-ordinary-least-squares-regression} and \ref{tab:intranode-vs-internode-quantile-regression} detail numerical results of these regressions.

\subsubsection{Walltime Latency}

Significantly more walltime elapsed during message transmission under internode condtions compared to intranode conditions.

Walltime latency was less than \SI{10}{\micro\second} for intranode communication (mean: \SI{7.70}{\micro\second}; median: \SI{6.94}{\micro\second}).
Internode communication had approximately $50\times$ greater walltime latency, at around \SI{500}{\micro\second} (mean: \SI{600}{\micro\second}; median: \SI{551}{\micro\second}).

Supplementary Figures \ref{fig:intranode-vs-internode-distribution-latency-walltime-inlet-ns} and \ref{fig:intranode-vs-internode-distribution-latency-walltime-outlet-ns} show the distributions of walltime latency for intra- and inter-node communication.
Regression analysis confirmed a significant increase in walltime latency under inter-node communication (Supplementary Figures \ref{fig:intranode-vs-internode-regression-latency-walltime-inlet-ns}, \ref{fig:intranode-vs-internode-regression-latency-walltime-outlet-ns}; Supplementary Tables \ref{tab:intranode-vs-internode-ordinary-least-squares-regression} and \ref{tab:intranode-vs-internode-quantile-regression}).

\subsubsection{Delivery Clumpiness}

Delivery clumpiness was minimal under intranode communication and very high under internode communication.

Under intranode conditions, we observed a mean clumpiness value of 0.014 and a median of 0.002.
Under internode conditions, we observed mean and median clumpiness values of 0.96.
Supplementary Figures \ref{fig:intranode-vs-internode-distribution-delivery-clumpiness} and \ref{fig:intranode-vs-internode-distribution-delivery-clumpiness} show the distributions of clumpiness for intra- and inter-node communication.
Regression analysis confirmed a significant increase in clumpiness under inter-node communication (Supplementary Figures \ref{fig:intranode-vs-internode-regression-delivery-clumpiness}, \ref{fig:intranode-vs-internode-regression-delivery-clumpiness}; Supplementary Tables \ref{tab:intranode-vs-internode-ordinary-least-squares-regression} and \ref{tab:intranode-vs-internode-quantile-regression}).

\subsubsection{Delivery Failure Rate}

Somewhat counterintuitively, a significantly higher proportion of deliveries failed for intranode communication than for internode communication.

We observed a delivery failure rate of around 0.3 for intranode communication (mean: 0.33; median: 0.30) and no delivery failures for internode communication (mean: 0.00; median: 0.00).
In some intranode snapshot windows, we observed a delivery failure rate as high as 0.8.
Supplementary Figures \ref{fig:intranode-vs-internode-distribution-delivery-clumpiness} and \ref{fig:intranode-vs-internode-distribution-delivery-clumpiness} show the distributions of delivery failure rate for intra- and inter-node communication.

Because of Conduit's current MPI-based implementation, messages only drop when the underlying send buffer fills; queued messages are guaranteed for delivery.
Slower simstep period under internode allocation could improve stability of the send buffer due to more time, on average, between send attempts.
Underlying buffering or consolidation by the MPI backend for internode communication might also play a role by allowing data to be moved out of the userspace send buffer more promptly.

Regression analysis confirmed a significant increase in delivery failure under intra-node communication (Supplementary Figures \ref{fig:intranode-vs-internode-regression-delivery-clumpiness}, \ref{fig:intranode-vs-internode-regression-delivery-clumpiness}; Supplementary Tables \ref{tab:intranode-vs-internode-ordinary-least-squares-regression} and \ref{tab:intranode-vs-internode-quantile-regression}).

\subsection{Quality of Service: Multithreading vs. Multiprocessing}
\label{sec:multithreading-vs-multiprocessing}

This section compares best-effort quality of service under multithreading and multiprocessing schemes.
We hold hardware configuration constant by restricting multiprocessing to cores a single hardware node, as is the case for multithreading.
However, inter-process communication occurred via MPI calls while inter-thread communication occurred via shared memory access mediated by a C++ \texttt{std::mutex}.

Experiments again applied the graph coloring benchmark.
Both treatments used a single pair of CPUs.

\subsubsection{Simstep Period}

Multithreading enabled faster simulation update turnover than multiprocessing.

Under multithreading, simstep period was around \SI{5}{\micro\second} (mean: \SI{4.60}{\micro\second}; median: \SI{4.64}{\micro\second}).
Simstep period for multiprocessing was around \SI{9}{\micro\second} (mean: \SI{9.00}{\micro\second}; median: \SI{9.04}{\micro\second}).
Supplementary Figures \ref{fig:multithreading-vs-multiprocessing-distribution-simstep-period-inlet-ns} and \ref{fig:multithreading-vs-multiprocessing-distribution-simstep-period-outlet-ns} depict the distribution of walltime per computational update for both multiprocessing and multithreading.
This result falls in line with expectations that interaction via shared memory incurs lower overhead than via MPI calls.

Regression analysis showed that both mean and median simstep period were significantly slower under multiprocessing compared to multithreading.
(Supplementary Figures \ref{fig:multithreading-vs-multiprocessing-regression-simstep-period-inlet-ns} and \ref{fig:multithreading-vs-multiprocessing-regression-simstep-period-outlet-ns} visualize these regressions and Supplementary Tables \ref{tab:multithreading-vs-multiprocessing-ordinary-least-squares-regression} and \ref{tab:multithreading-vs-multiprocessing-quantile-regression} detail numerical results.)

\subsubsection{Walltime Latency}

No significant difference in walltime latency was detected between multiprocessing and multithreading.

In the median case, walltime latency was approximately \SI{5}{\micro\second} for multithreading and \SI{8}{\micro\second} for multiprocessing.
However, a pair of extreme outliers among snapshot windows --- with walltime latencies of approximately 12ms --- drove multithreading walltime latency much higher in the mean case, to \SI{451}{\micro\second}.
In the median case, multiprocessing walltime latency was \SI{8.56}{\micro\second}.

Cache invalidation or mutex contention provide possible explanations for the observed episodes of extreme multithreading latency, although magnitude on the order of milliseconds for such effects is surprising.
Multithreading appears to provide marginally lower latency service in the median case, but at the cost of vulnerability to extreme high-latency disruptions.

Supplementary Figures \ref{fig:multithreading-vs-multiprocessing-distribution-latency-walltime-inlet-ns} and \ref{fig:multithreading-vs-multiprocessing-distribution-latency-walltime-outlet-ns} show the distributions of walltime latency for multithread and multiprocess runs.
Regression analysis did not detect any significant difference in walltime latency between multithreading and multiprocessing (Supplementary Figures \ref{fig:multithreading-vs-multiprocessing-regression-latency-walltime-inlet-ns}, \ref{fig:multithreading-vs-multiprocessing-regression-latency-walltime-outlet-ns}; Supplementary Tables \ref{tab:multithreading-vs-multiprocessing-ordinary-least-squares-regression} and \ref{tab:multithreading-vs-multiprocessing-quantile-regression}).

\subsubsection{Simstep Latency}

No significant difference in simstep latency was detected between multiprocessing and multithreading.

In the median case, multiprocessing offered marginally lower simstep latency than multithreading.
Median simstep latency was 0.84 updates under multiprocessing and 1.10 updates under multithreading.
However, just as for walltime latency, extreme magintude outliers ($\approx$ \numprint{2000} simsteps) boosted mean simstep latency for multithreading.
Mean simstep latency was 0.94 updates under multiprocessing and 78.0 updates under multithreading.
Supplementary Figures \ref{fig:multithreading-vs-multiprocessing-distribution-latency-simsteps-inlet} and \ref{fig:multithreading-vs-multiprocessing-distribution-latency-simsteps-outlet} compare the distributions of simstep latency across these conditions.

Direct measurements of simstep period and walltime latency suggest that faster simstep period, rather than slower walltime latency, explain the marginally higher simstep latency under multithreading.

Regression analysis detected no significant effect of threading versus processing on simstep latency in both the mean and median cases (Supplementary Figures \ref{fig:multithreading-vs-multiprocessing-regression-latency-simsteps-inlet} and \ref{fig:multithreading-vs-multiprocessing-regression-latency-simsteps-outlet}).
Supplementary Tables \ref{tab:multithreading-vs-multiprocessing-ordinary-least-squares-regression} and \ref{tab:multithreading-vs-multiprocessing-quantile-regression} detail numerical results of these regressions.

\subsubsection{Delivery Clumpiness}

Multithreading exhibited higher median clumpiness and greater variance in clumpiness than multiprocessing.

Under multithreading, clumpiness was nearly 1 within some snapshot windows and less than 0.1 within others.
Under multiprocessing, clumpiness was consistently less than 0.1.
Supplementary Figures \ref{fig:multithreading-vs-multiprocessing-distribution-delivery-clumpiness} and \ref{fig:multithreading-vs-multiprocessing-distribution-delivery-clumpiness} show the distributions of clumpiness under both multiprocessing and multithreading.
Multithreading median clumpiness was 0.54.
Multiprocessing median clumpiness was 0.03.
Multithreading and multiprocessing mean clumpinesses were 0.56 and 0.03, respectively.

Regression analysis confirmed a significantly greater clumpiness under both multithreading compared to multiprocessing (Supplementary Figures \ref{fig:multithreading-vs-multiprocessing-regression-delivery-clumpiness}, \ref{fig:multithreading-vs-multiprocessing-regression-delivery-clumpiness}; Supplementary Tables \ref{tab:multithreading-vs-multiprocessing-ordinary-least-squares-regression} and \ref{tab:multithreading-vs-multiprocessing-quantile-regression}).

\subsubsection{Delivery Failure Rate}

We observed a higher proportion of deliveries fail for multiprocessing than for multithreading.
(This is as expected; the multithread implementation directly wrote updates to a piece of shared memory, so there was no send buffer to backlog and induce message drops.)

Multiprocessing exhibited both mean and median delivery failure rate of 0.38.
In individual multiprocessing snapshot windows, we observed a delivery failure rates ranging from less than 0.1 to as high as 0.7.
We observed no multithreaded delivery failures.
Supplementary Figures \ref{fig:multithreading-vs-multiprocessing-distribution-delivery-clumpiness} and \ref{fig:multithreading-vs-multiprocessing-distribution-delivery-clumpiness} show the distributions of delivery failure rate for multithreading and multiprocessing.

Regression analysis confirmed a significant increase in delivery failure under multiprocessing (Supplementary Figures \ref{fig:multithreading-vs-multiprocessing-regression-delivery-clumpiness}, \ref{fig:multithreading-vs-multiprocessing-regression-delivery-clumpiness}; Supplementary Tables \ref{tab:multithreading-vs-multiprocessing-ordinary-least-squares-regression} and \ref{tab:multithreading-vs-multiprocessing-quantile-regression}).

\subsection{Quality of Service: Weak Scaling}
\label{sec:weak-scaling}

Sections \ref{sec:multiprocess-benchmarks} and \ref{sec:multithread-benchmarks} showed how best-effort communication could improve application performance, particularly when scaling up processor count.
Multiprocess performance scales well under the best-effort approach, with overlapping performance estimate intervals for 16 and 64 processor counts on both surveyed benchmark problems.

This section aims to flesh out a more holistic picture of the effects of increasing processor count on best-effort computation by considering a comprehensive suite of quality of service metrics.
Our particular interest is in which, if any, aspects of quality of service degrade under larger processing pools.

To address these questions, we performed weak scaling experiments on 16, 64, and 256 processes using the graph coloring benchmark.
To broaden the survey, we tested scaling with different numbers of processors allocated per node and different numbers of simulation elements assigned per processor.

For the first variable, we tested scaling on allocations with each processor hosted on an independent node and allocations where each node hosted an average of four processors.
This allowed us to examine how quality of service fared in homogeneous network conditions, where all communication between processes was inter-node, compared to heteregeneous conditions, where some inter-process communication was inter-node and some was intra-node.

For the second variable, we tested with \numprint{2048} simulation elements (``simels'') per processor (consistent with the benchmarking experiments performed in Sections \ref{sec:multiprocess-benchmarks} and \ref{sec:multithread-benchmarks}) and just one simulation element per processor.
This allowed us to vary the amount of computational work performed per process.

\subsubsection{Simstep Period}

Supplementary Figures \ref{fig:weak-scaling-distribution-simstep-period-inlet-ns} and \ref{fig:weak-scaling-distribution-simstep-period-outlet-ns} survey the distributions of simstep periods observed within snapshot windows.
Across process counts, simstep period registers around \SI{80}{\micro\second} with one simel per CPU and around \SI{200}{\micro\second} with \numprint{2048} simels per CPU.
However, on heterogeneous allocations (4 CPUs per node) this metric is more variable between snapshots.
Outlier observations range up to around 10ms with \numprint{2048} simels per CPU.
With 1 simel per CPU, outlying values fell upwards of 100ms.

\providecommand{\figweakscalingregressionolsmacro}[2]{
\def\metric{#1}\def\metricslug{#2}\begin{figure*}[h]
  \centering

  \begin{subfigure}[b]{0.49\textwidth}
    \centering
    \includegraphics[width=\textwidth]{submodule/conduit-binder/binder/date=2021+project=72k5n/a=weak-scaling/teeplots/\metricslug/col=cpus-per-node+marker=+regression=ordinary-least-squares-regression+row=num-simels-per-cpu+transform=snapshot_diffs-groupby_exec_instance-mean+viz=facet-unsplit-regression+x=log-num-processes+y=\metricslug+ext=}
    \caption{
      Complete ordinary least squares regression plot.
      Observations are means per replicate.
    }
    \label{fig:weak-scaling-regression-ols-\metricslug-complete-regression}
  \end{subfigure}%
  \hfill%
  \begin{subfigure}[b]{0.49\textwidth}
    \centering
    \includegraphics[width=\textwidth]{submodule/conduit-binder/binder/date=2021+project=72k5n/a=weak-scaling/teeplots/\metricslug/col=cpus-per-node+estimated-statistic=\metricslug-mean+num-processes=16-64-256+row=num-simels-per-cpu+transform=fit_regression+viz=lambda+y=absolute-effect-size+ext=}
    \caption{Estimated regression coefficient for complete regression. Zero corresponds to no effect.}
    \label{fig:weak-scaling-regression-ols-\metricslug-complete-effect-size}
  \end{subfigure}

  \begin{subfigure}[b]{0.49\textwidth}
    \centering
    \includegraphics[width=\textwidth]{submodule/conduit-binder/binder/date=2021+project=72k5n/a=weak-scaling/teeplots/\metricslug/col=cpus-per-node+marker=+regression=ordinary-least-squares-regression+row=num-simels-per-cpu+transform=snapshot_diffs-groupby_exec_instance-mean+viz=facet-split-regression+x=log-num-processes+y=\metricslug+ext=}
    \caption{
      Piecewise ordinary least squares regression plot.
      Observations are means per replicate.
    }
    \label{fig:weak-scaling-regression-ols-\metricslug-partial-regression}
  \end{subfigure}%
  \hfill%
  \begin{subfigure}[b]{0.49\textwidth}
    \centering
    \includegraphics[width=\textwidth]{submodule/conduit-binder/binder/date=2021+project=72k5n/a=weak-scaling/teeplots/\metricslug/col=cpus-per-node+estimated-statistic=\metricslug-mean+num-processes=64-256+row=num-simels-per-cpu+transform=fit_regression+viz=lambda+y=absolute-effect-size+ext=}
    \caption{Estimated regression coefficient for rightmost partial regression. Zero corresponds to no effect.}
    \label{fig:weak-scaling-regression-ols-\metricslug-partial-effect-size}
  \end{subfigure}
  \caption{
  Ordinary least squares regressions of \metric{} against log processor count for weak scaling experiment (Section \ref{sec:weak-scaling}).
  Lower is better.
  Top row shows complete regression and bottom row shows piecewise regression.
  Ordinary least squares regression estimates relationship between independent variable and mean of response variable.
  Error bands and bars are 95\% confidence intervals.
  Note that log is base 4, so processor counts correspond to 16, 64, and 256.
  }
  \label{fig:weak-scaling-regression-ols-\metricslug}
\end{figure*}

}

\def\metric{Simstep Period Inlet (ns)}\def\metricslug{simstep-period-inlet-ns}\begin{figure*}[h]
  \centering

  \begin{subfigure}[b]{0.49\textwidth}
    \centering
    \includegraphics[width=\textwidth]{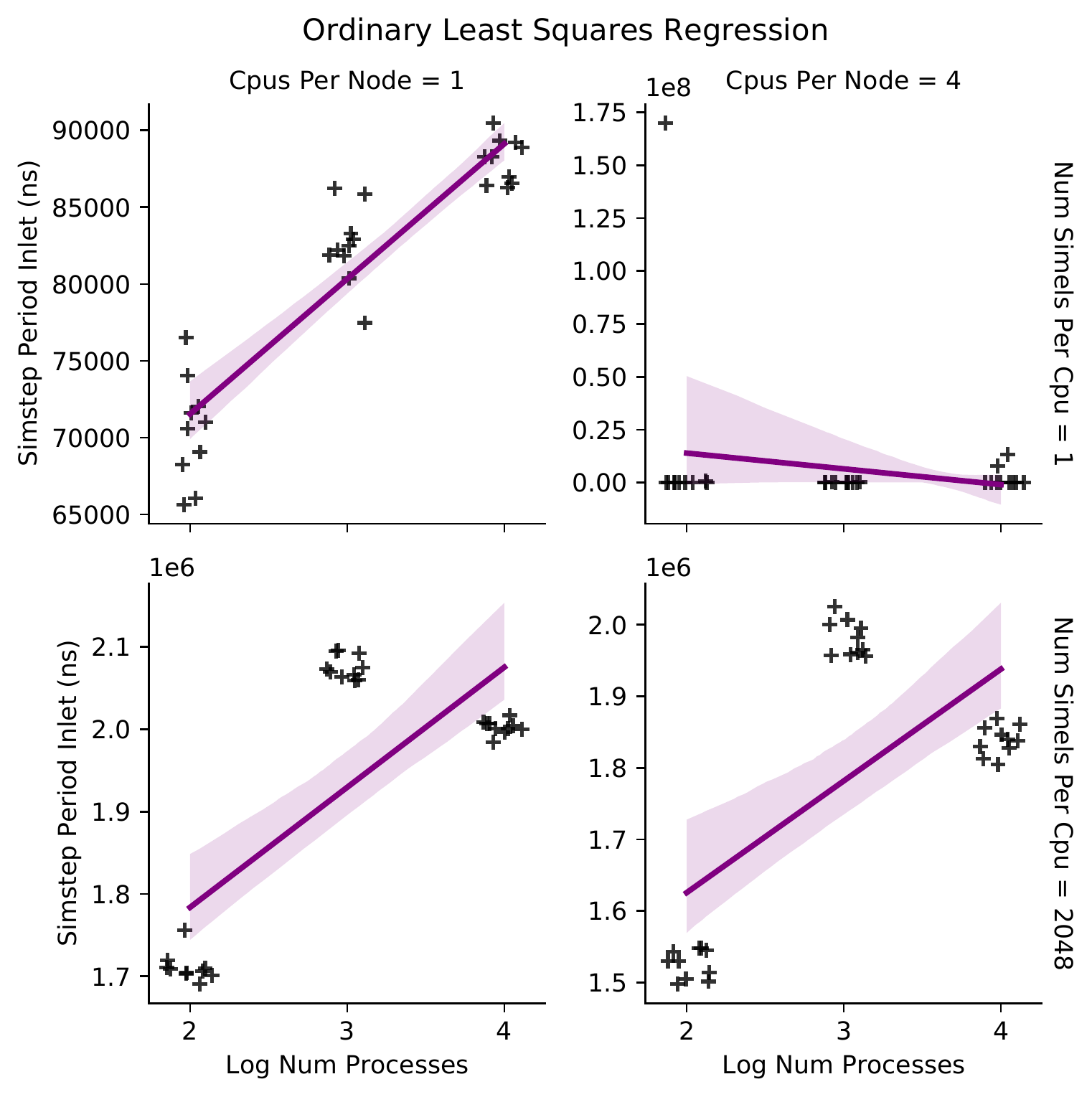}
    \caption{
      Complete ordinary least squares regression plot.
      Observations are means per replicate.
    }
    \label{fig:weak-scaling-regression-ols-\metricslug-complete-regression}
  \end{subfigure}%
  \hfill%
  \begin{subfigure}[b]{0.49\textwidth}
    \centering
    \includegraphics[width=\textwidth]{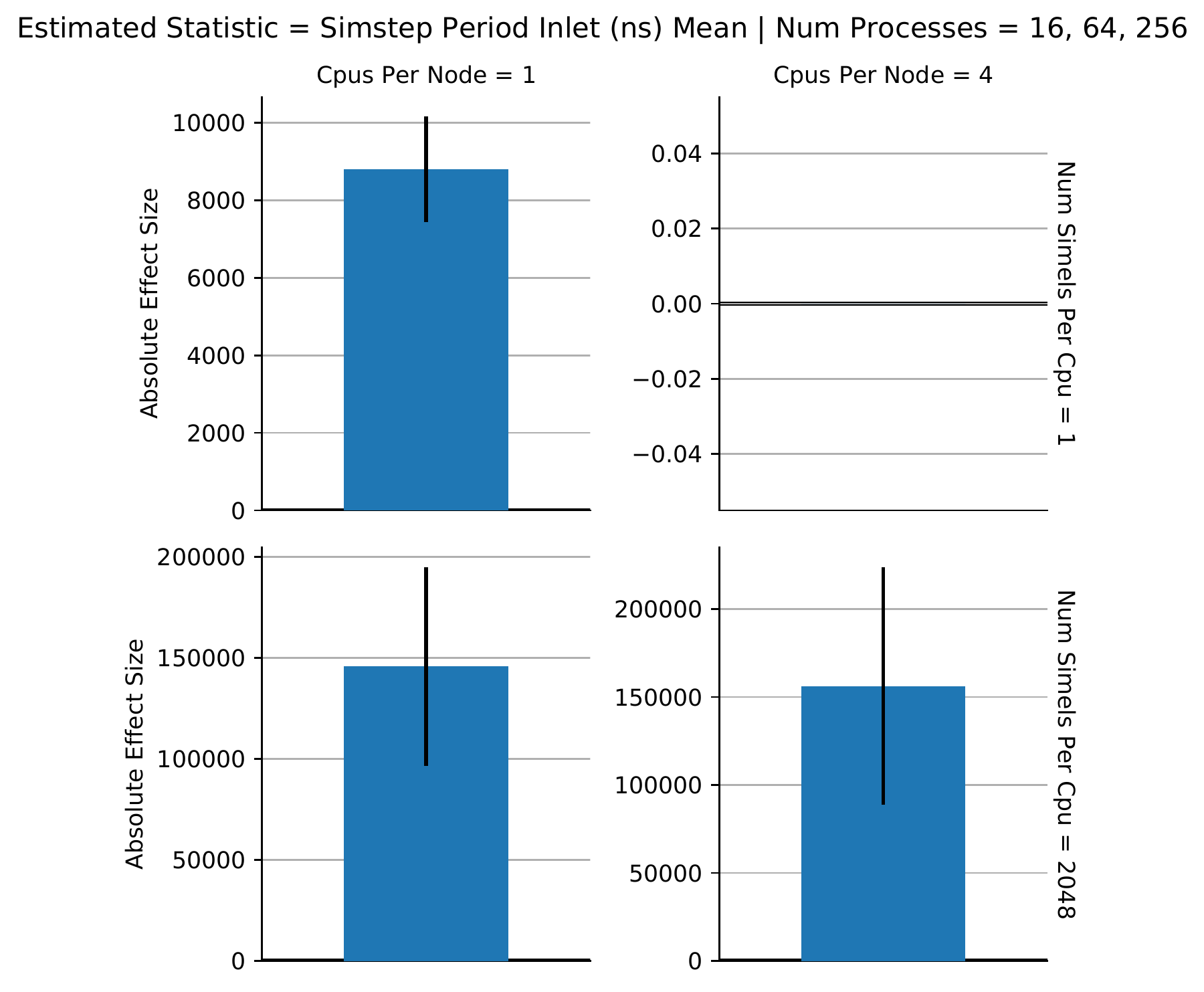}
    \caption{Estimated regression coefficient for complete regression. Zero corresponds to no effect.}
    \label{fig:weak-scaling-regression-ols-\metricslug-complete-effect-size}
  \end{subfigure}

  \begin{subfigure}[b]{0.49\textwidth}
    \centering
    \includegraphics[width=\textwidth]{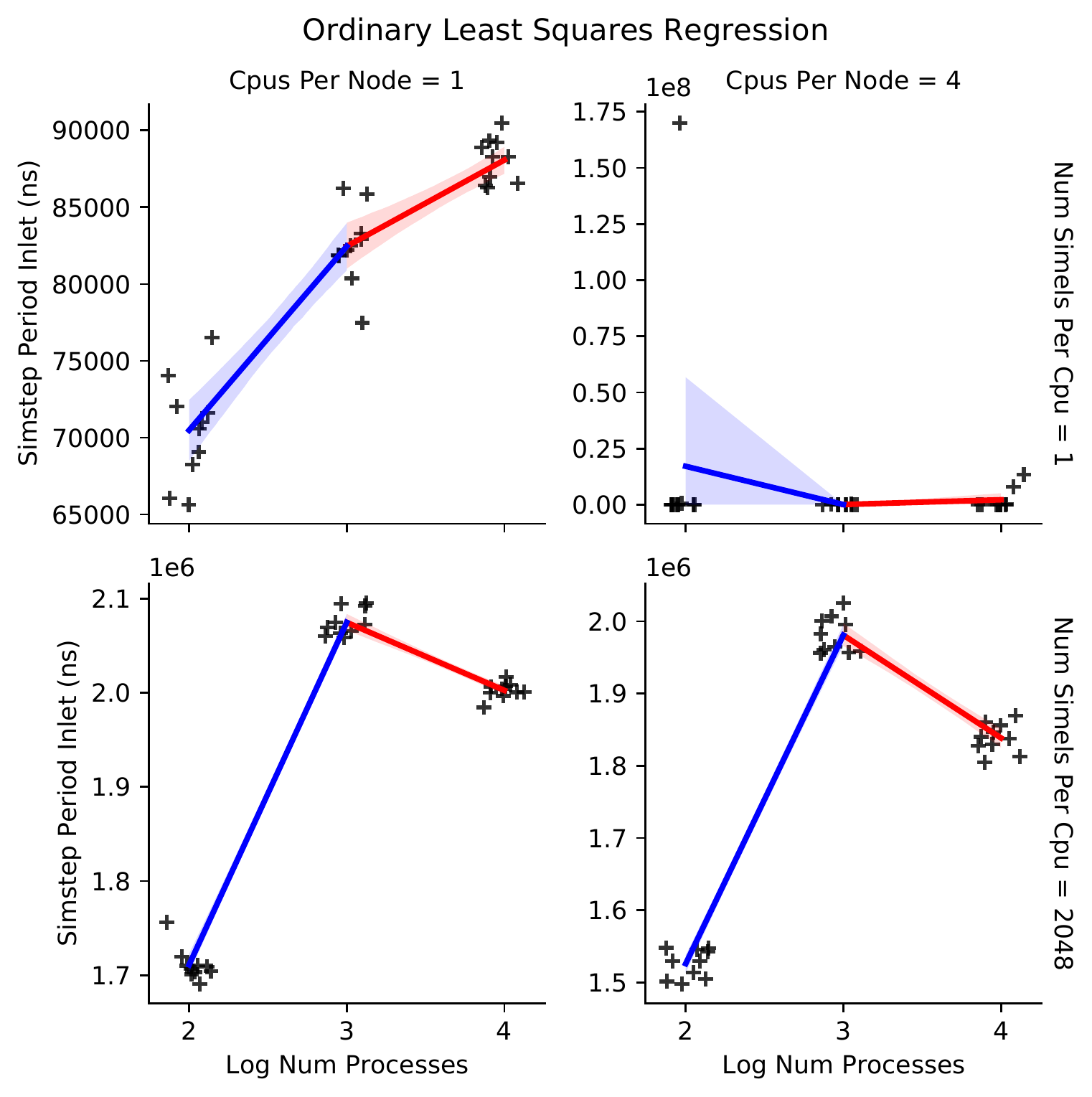}
    \caption{
      Piecewise ordinary least squares regression plot.
      Observations are means per replicate.
    }
    \label{fig:weak-scaling-regression-ols-\metricslug-partial-regression}
  \end{subfigure}%
  \hfill%
  \begin{subfigure}[b]{0.49\textwidth}
    \centering
    \includegraphics[width=\textwidth]{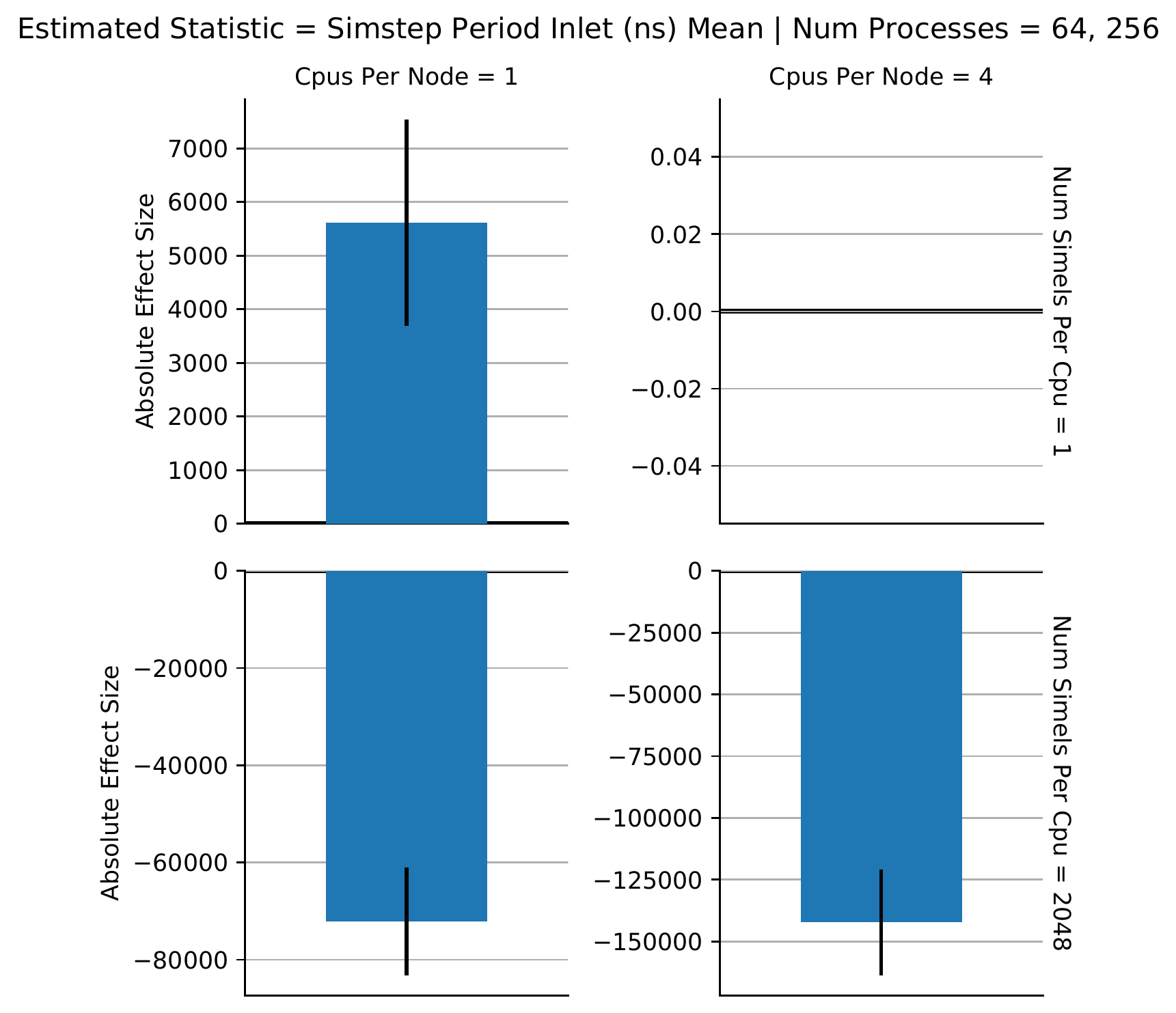}
    \caption{Estimated regression coefficient for rightmost partial regression. Zero corresponds to no effect.}
    \label{fig:weak-scaling-regression-ols-\metricslug-partial-effect-size}
  \end{subfigure}
  \caption{
  Ordinary least squares regressions of \metric{} against log processor count for weak scaling experiment (Section \ref{sec:weak-scaling}).
  Lower is better.
  Top row shows complete regression and bottom row shows piecewise regression.
  Ordinary least squares regression estimates relationship between independent variable and mean of response variable.
  Error bands and bars are 95\% confidence intervals.
  Note that log is base 4, so processor counts correspond to 16, 64, and 256.
  }
  \label{fig:weak-scaling-regression-ols-\metricslug}
\end{figure*}

We performed an ordinary least squares (OLS) regression to test how mean simstep period changed with processor count.
In all cases except one simel per CPU with four CPUs per node, mean simstep period increased significantly with processor count from 16 to 64 to 256 CPUs.
However, from 64 to 256 processors mean simstep period only increased significantly with one simel per CPU per node.
Between 64 and 256 processes, mean simstep period actually decreased significantly for runs with \numprint{2048} simels per CPU.
Figure \ref{fig:weak-scaling-regression-ols-simstep-period-inlet-ns} and Supplementary Figure \ref{fig:weak-scaling-regression-ols-simstep-period-outlet-ns} visualize reported OLS regressions.
Supplementary Tables \ref{tab:weak-scaling-simstep-period-inlet-ns-regression-ols} and \ref{tab:weak-scaling-simstep-period-outlet-ns-regression-ols} provide numerical details on reported OLS regressions.

Median simstep period exhibited the same relationships with processor count, tested with quartile regression.
Supplementary Figures \ref{fig:weak-scaling-regression-quantile-simstep-period-inlet-ns} and \ref{fig:weak-scaling-regression-quantile-simstep-period-outlet-ns} visualize corresponding quartile regressions.
Supplementary Tables \ref{tab:weak-scaling-simstep-period-inlet-ns-regression-quantile} and \ref{tab:weak-scaling-simstep-period-outlet-ns-regression-quantile} report numerical details on those quartile regressions.

Except for the extreme case of one simel per CPU and one CPU per node, simstep period quality of service is stable in scaling from 64 to 256 processes.

\subsubsection{Walltime Latency}

Walltime latency sits at around \SI{500}{\micro\second} for one-simel runs and around 2ms for \numprint{2048}-simel runs.
However, variability is greater for heterogeneous (four CPUs per node) allocations.
Extreme outliers of up to almost 100ms inlet/2s outlet occur in four CPUs per node, one-simel runs.
In 256 process, \numprint{2048}-simel, one CPU per node runs, outliers of more than 10s occur.
Supplementary Figures \ref{fig:weak-scaling-distribution-latency-walltime-inlet-ns} and \ref{fig:weak-scaling-distribution-latency-walltime-outlet-ns} show the distribution of walltime latencies observed across run conditions.

We performed OLS regressions to test how mean walltime latency changed with processor count.
Over 16, 64, and 256 processes, mean walltime latency increased significantly with processor count only with \numprint{2048} simels per CPU.
Between 64 and 256 processes, mean walltime latency increased significantly with processor count only for one CPU per node with \numprint{2048} simels per CPU.
Supplementary Figures \ref{fig:weak-scaling-regression-ols-latency-walltime-inlet-ns} and Figure \ref{fig:weak-scaling-regression-ols-latency-walltime-outlet-ns} show these regressions.
Supplementary Tables \ref{tab:weak-scaling-latency-walltime-inlet-ns-regression-ols} and \ref{tab:weak-scaling-latency-walltime-outlet-ns-regression-ols} provide numerical details.

\providecommand{\figweakscalingregressionquantilemacro}[2]{
\def\metric{#1}\def\metricslug{#2}\begin{figure*}[h]
  \centering

  \begin{subfigure}[b]{0.49\textwidth}
    \centering
    \includegraphics[width=\textwidth]{submodule/conduit-binder/binder/date=2021+project=72k5n/a=weak-scaling/teeplots/\metricslug/col=cpus-per-node+marker=+regression=quantile-regression+row=num-simels-per-cpu+transform=snapshot_diffs-groupby_exec_instance-median+viz=facet-unsplit-regression+x=log-num-processes+y=\metricslug+ext=}
    \caption{
      Complete quantile regression plot.
      Observations are medians per replicate.
    }
    \label{fig:weak-scaling-regression-quantile-\metricslug-complete-regression}
  \end{subfigure}%
  \hfill%
  \begin{subfigure}[b]{0.49\textwidth}
    \centering
    \includegraphics[width=\textwidth]{submodule/conduit-binder/binder/date=2021+project=72k5n/a=weak-scaling/teeplots/\metricslug/col=cpus-per-node+estimated-statistic=\metricslug-median+num-processes=16-64-256+row=num-simels-per-cpu+transform=fit_regression+viz=lambda+y=absolute-effect-size+ext=}
    \caption{Estimated regression coefficient for ordinary least squares regression. Zero corresponds to no effect.}
    \label{fig:weak-scaling-regression-quantile-\metricslug-complete-effect-size}
  \end{subfigure}

  \begin{subfigure}[b]{0.49\textwidth}
    \centering
    \includegraphics[width=\textwidth]{submodule/conduit-binder/binder/date=2021+project=72k5n/a=weak-scaling/teeplots/\metricslug/col=cpus-per-node+marker=+regression=quantile-regression+row=num-simels-per-cpu+transform=snapshot_diffs-groupby_exec_instance-median+viz=facet-split-regression+x=log-num-processes+y=\metricslug+ext=}
    \caption{
      Piecewise quantile regression plot.
      Observations are medians per replicate.
    }
    \label{fig:weak-scaling-regression-quantile-\metricslug-partial-regression}
  \end{subfigure}%
  \hfill%
  \begin{subfigure}[b]{0.49\textwidth}
    \centering
    \includegraphics[width=\textwidth]{submodule/conduit-binder/binder/date=2021+project=72k5n/a=weak-scaling/teeplots/\metricslug/col=cpus-per-node+estimated-statistic=\metricslug-median+num-processes=64-256+row=num-simels-per-cpu+transform=fit_regression+viz=lambda+y=absolute-effect-size+ext=}
    \caption{Estimated regression coefficient for rightmost partial regression. Zero corresponds to no effect.}
    \label{fig:weak-scaling-regression-quantile-\metricslug-partial-effect-size}
  \end{subfigure}
  \caption{
  Quantile Regressions of \metric{} against log processor count for weak scaling experiment (Section \ref{sec:weak-scaling}).
  Lower is better.
  Top row shows complete regression and bottom row shows piecewise regression.
  Quantile regression estimates relationship between independent variable and median of response variable.
  Note that log is base 4, so processor counts correspond to 16, 64, and 256.
  }
  \label{fig:weak-scaling-regression-quantile-\metricslug}
\end{figure*}

}

\def\metric{Latency Walltime Inlet (ns)}\def\metricslug{latency-walltime-inlet-ns}\begin{figure*}[h]
  \centering

  \begin{subfigure}[b]{0.49\textwidth}
    \centering
    \includegraphics[width=\textwidth]{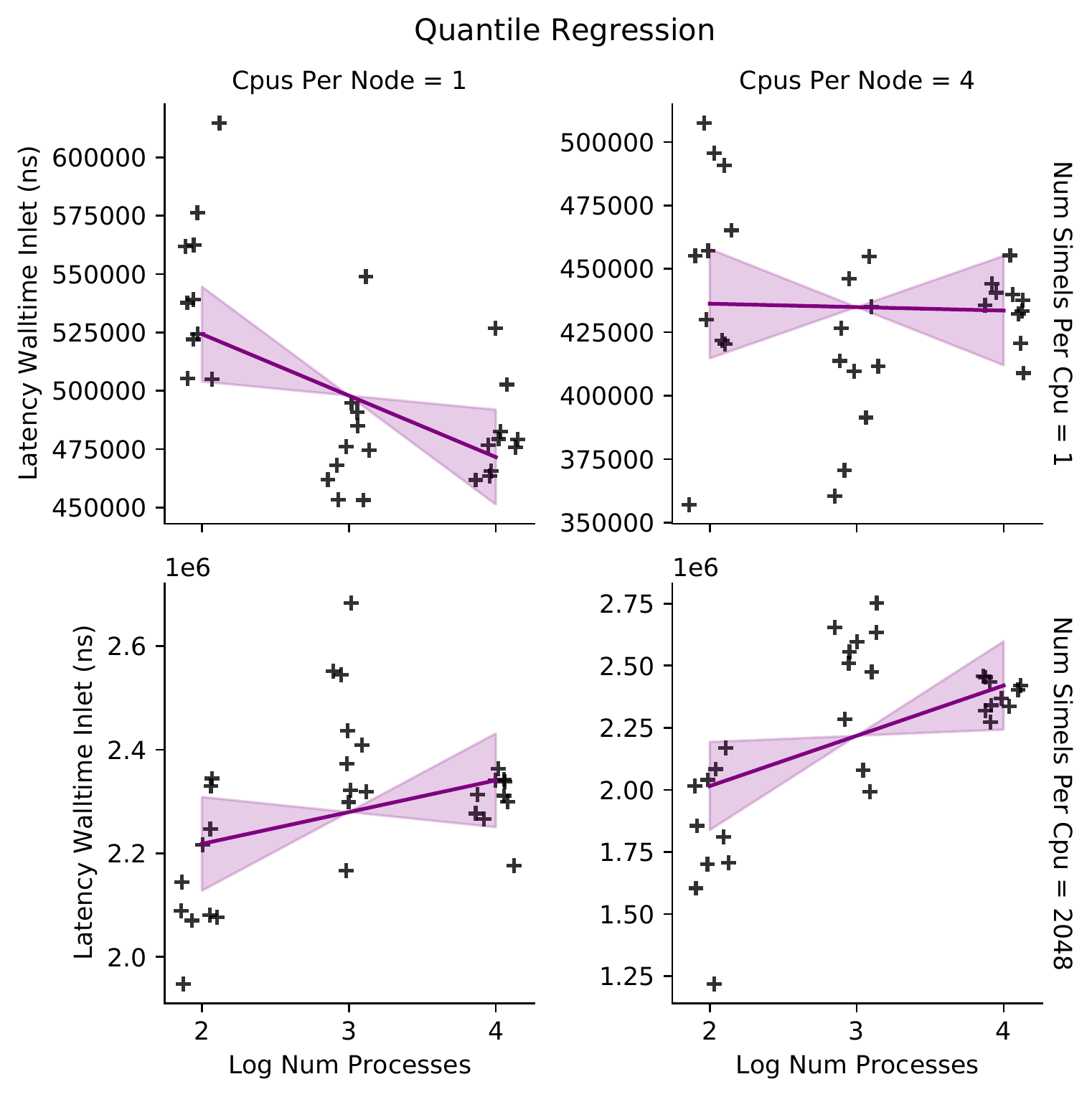}
    \caption{
      Complete quantile regression plot.
      Observations are medians per replicate.
    }
    \label{fig:weak-scaling-regression-quantile-\metricslug-complete-regression}
  \end{subfigure}%
  \hfill%
  \begin{subfigure}[b]{0.49\textwidth}
    \centering
    \includegraphics[width=\textwidth]{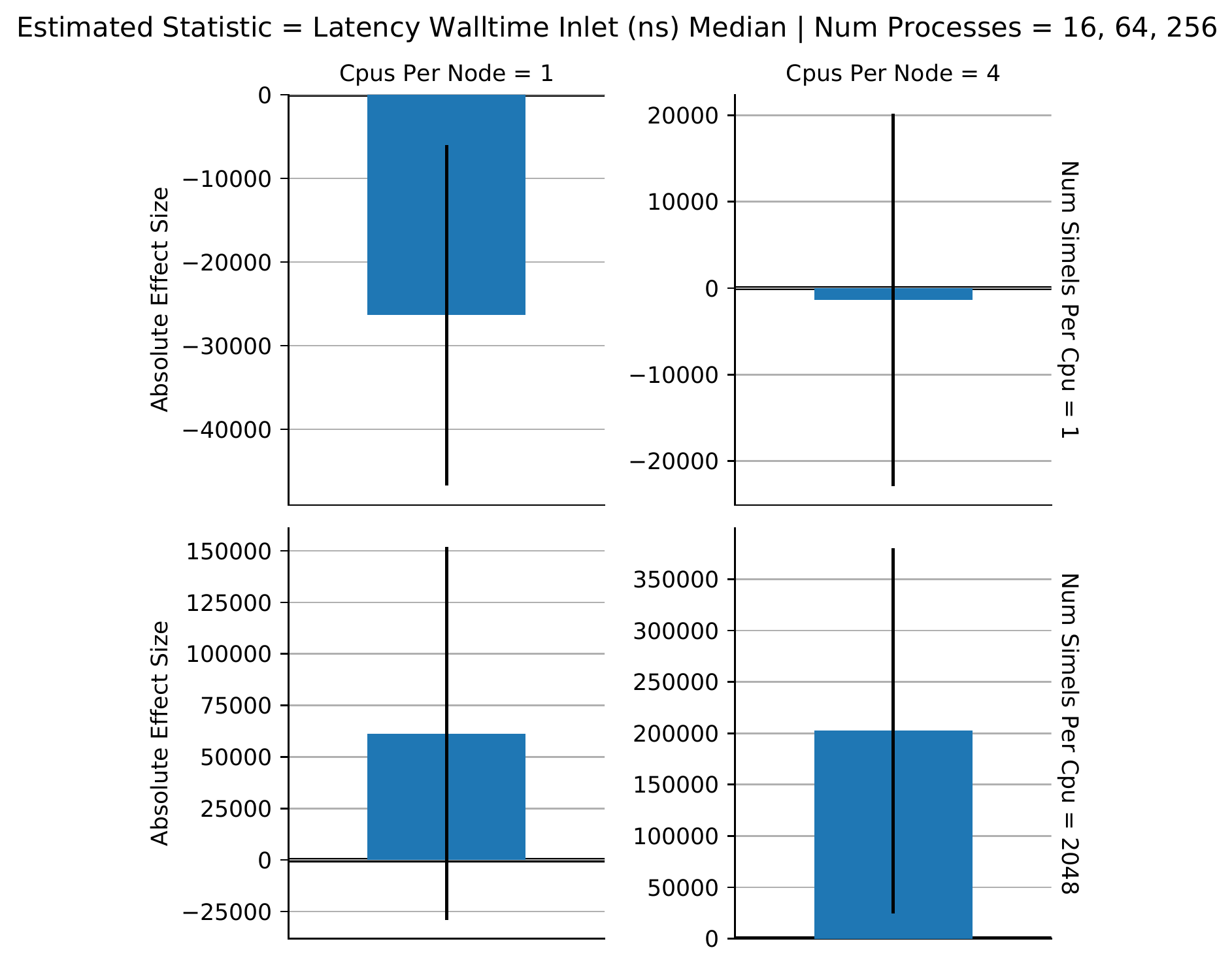}
    \caption{Estimated regression coefficient for ordinary least squares regression. Zero corresponds to no effect.}
    \label{fig:weak-scaling-regression-quantile-\metricslug-complete-effect-size}
  \end{subfigure}

  \begin{subfigure}[b]{0.49\textwidth}
    \centering
    \includegraphics[width=\textwidth]{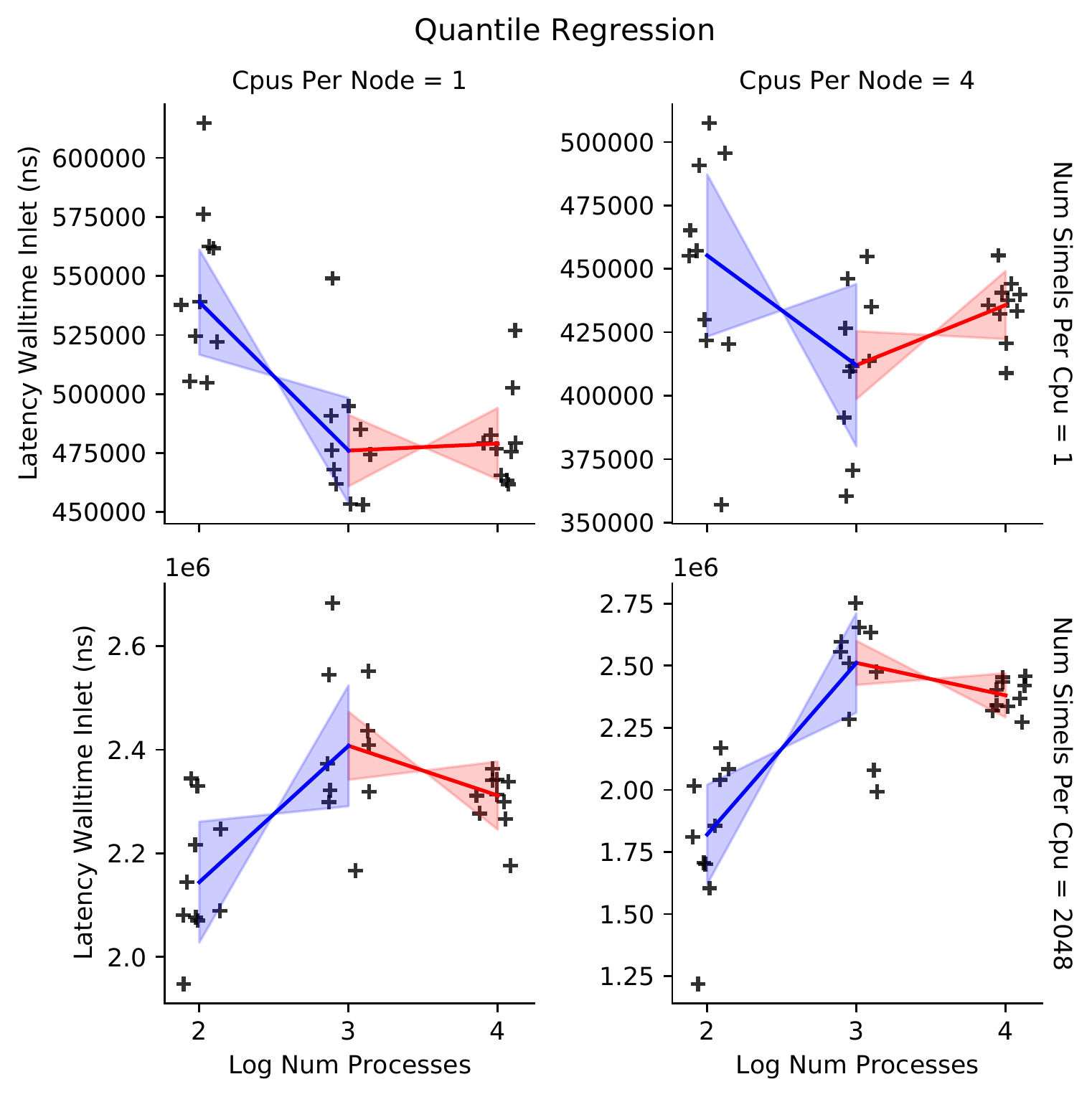}
    \caption{
      Piecewise quantile regression plot.
      Observations are medians per replicate.
    }
    \label{fig:weak-scaling-regression-quantile-\metricslug-partial-regression}
  \end{subfigure}%
  \hfill%
  \begin{subfigure}[b]{0.49\textwidth}
    \centering
    \includegraphics[width=\textwidth]{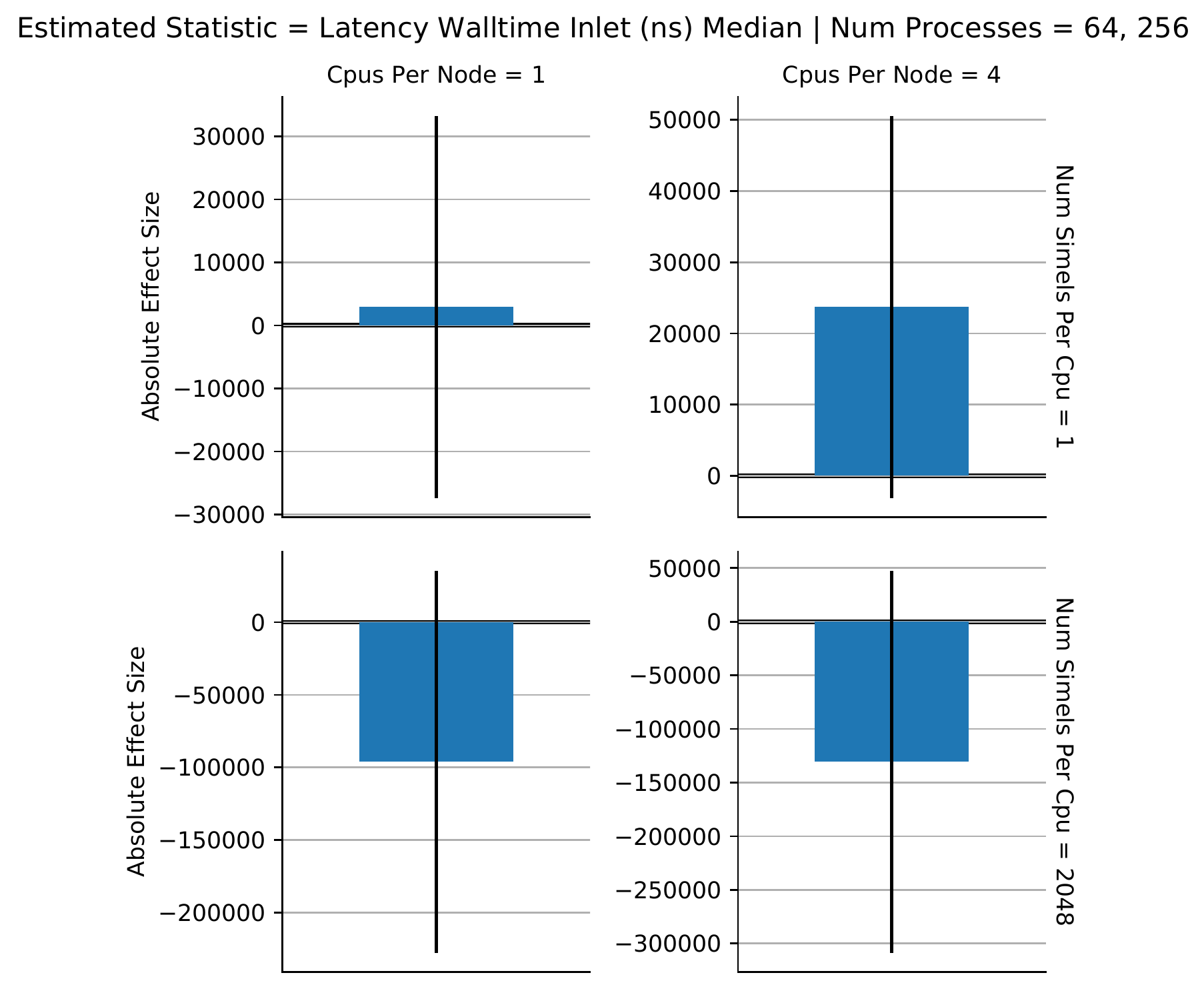}
    \caption{Estimated regression coefficient for rightmost partial regression. Zero corresponds to no effect.}
    \label{fig:weak-scaling-regression-quantile-\metricslug-partial-effect-size}
  \end{subfigure}
  \caption{
  Quantile Regressions of \metric{} against log processor count for weak scaling experiment (Section \ref{sec:weak-scaling}).
  Lower is better.
  Top row shows complete regression and bottom row shows piecewise regression.
  Quantile regression estimates relationship between independent variable and median of response variable.
  Note that log is base 4, so processor counts correspond to 16, 64, and 256.
  }
  \label{fig:weak-scaling-regression-quantile-\metricslug}
\end{figure*}

Next, we performed quantile regressions to test how processor count affected median walltime latency.
Over 16, 64, and 256 processes, median walltime latency increased significantly only with 4 CPUs per node and \numprint{2048} simels per CPU.
Over 64 and 256 processes, there was no significant relationship between processor count and median walltime latency under any condition.
Figure \ref{fig:weak-scaling-regression-quantile-latency-walltime-inlet-ns} and Supplementary Figure \ref{fig:weak-scaling-regression-quantile-latency-walltime-outlet-ns} show regression results.
Supplementary Tables \ref{tab:weak-scaling-latency-walltime-inlet-ns-regression-quantile} and \ref{tab:weak-scaling-latency-walltime-outlet-ns-regression-quantile} provide numerical details.

\subsubsection{Simstep Latency}

Simstep latency sits around 7 updates for runs with one simel per CPU and around 1.2 updates for runs with \numprint{2048} simels per CPU.
For runs with one simel per CPU, outlier snapshot windows reach up to 50 updates under homogeneous allocations and up to almost 100 updates under heterogeneous allocations.
The \numprint{2048} simels per CPU, one CPU per node, 256 process condition exhibited outliers of up to almost \numprint{8000} update simstep latency.
Supplementary Figures \ref{fig:weak-scaling-distribution-latency-simsteps-inlet} and \ref{fig:weak-scaling-distribution-latency-simsteps-outlet}  show the distribution of simstep latencies observed across run conditions.

Over 16, 64, and 256 processes, mean simstep latency increased with process count only under 1 CPU per node, \numprint{2048} simel per CPU conditions.
The same was true over just 64 to 256 processes.
Supplementary Figures \ref{fig:weak-scaling-regression-ols-latency-simsteps-inlet} and \ref{fig:weak-scaling-regression-ols-latency-simsteps-outlet} show the OLS regressions performed, with
Supplementary Tables \ref{tab:weak-scaling-latency-simsteps-inlet-regression-ols} and \ref{tab:weak-scaling-latency-simsteps-outlet-regression-ols} providing numerical details.

\def\metric{Latency Simsteps Inlet}\def\metricslug{latency-simsteps-inlet}

For median simstep latency, however, there was no condition where latency increased significantly with process count.
Figure \ref{fig:weak-scaling-regression-quantile-latency-simsteps-inlet} and Supplementary Figure \ref{fig:weak-scaling-regression-quantile-latency-simsteps-outlet} show the quantile regressions performed, with Supplementary Tables \ref{tab:weak-scaling-latency-simsteps-inlet-regression-quantile} and \ref{tab:weak-scaling-latency-simsteps-outlet-regression-quantile} providing numerical details.

\subsubsection{Delivery Clumpiness}

For one-simel-per-CPU runs, median delivery clumpiness registered between 0.8 and 0.6.
On \numprint{2048}-simel-per-CPU runs, median delivery clumpiness was lower at around 0.4.
Supplementary Figure \ref{fig:weak-scaling-distribution-delivery-clumpiness}
shows the distribution of delivery clumpiness values observed across run conditions.

\def\metric{Delivery Clumpiness}\def\metricslug{delivery-clumpiness}

Using OLS regression, we found no evidence of mean clumpiness worsening with increased process count.
In fact, over 16, 64, and 256 processes clumpiness significantly decreased with process count in all conditions except four CPUs per node with \numprint{2048} simels per CPU.
Figure \ref{fig:weak-scaling-regression-ols-delivery-clumpiness} and Supplementary Table \ref{tab:weak-scaling-delivery-clumpiness-regression-ols} detail regressions performed to test the relationship between mean clumpiness and process count.

Median delivery clumpiness exhibited the same relationships with processor count, tested with quartile regression.
Supplementary Figure \ref{fig:weak-scaling-regression-quantile-delivery-clumpiness} and Supplementary Table \ref{tab:weak-scaling-delivery-clumpiness-regression-quantile} detail regressions between median clumpiness and process count.

\subsubsection{Delivery Failure Rate}

Typical delivery failure rate was near zero, except with one simel per CPU and four CPUs per node where median delivery failure rate was approximately 0.1.
However, outlier delivery failure rates of up to 0.7 were observed with 1 CPU per node, \numprint{2048} simels per CPU, and 256 processes.
Outlier delivery failure rates of up to 0.2 were observed with 4 CPUs per node, \numprint{2048} simels per CPU, and 256 processes.
Supplementary Figure \ref{fig:weak-scaling-distribution-delivery-failure-rate} shows the distribution of delivery failure rates observed across run conditions.

Mean delivery failure rate increased significantly between 64 and 256 processes with 1 CPU per node and \numprint{2048} simels per CPU as well as with 4 CPUs per node an 1 simel per CPU.
However, the median delivery failure rate only increased significantly with processor count with 4 CPUs per node and 1 simel per CPU.

\def\metric{Delivery Failure Rate}\def\metricslug{delivery-failure-rate}

Supplementary Figure \ref{fig:weak-scaling-regression-ols-delivery-failure-rate} and Supplementary Table \ref{tab:weak-scaling-delivery-failure-rate-regression-ols} detail the OLS regression testing mean delivery failure rate against processor count.
Figure \ref{fig:weak-scaling-regression-quantile-delivery-failure-rate} and Supplementary Table \ref{tab:weak-scaling-delivery-failure-rate-regression-quantile} detail the quantile regression testing median delivery failure rate against processor count.

\subsection{Quality of Service: Faulty Hardware}
\label{sec:with-lac-417-vs-sans-lac-417}

The extreme magnitude of outliers for metrics reported in Section \ref{sec:weak-scaling} prompted further investigation of the conditions under which these outliers arose.
Closer inspection revealed that the most extreme outliers were all associated with snapshots on a single node: lac-417.

So, we acquired two separate 256 process allocations on the lac cluster: one including lac-417 and one excluding lac-417.

Supplementary Figures \ref{fig:with-lac-417-vs-sans-lac-417-distribution-simstep-period-inlet-ns}, \ref{fig:with-lac-417-vs-sans-lac-417-distribution-simstep-period-outlet-ns}, \ref{fig:with-lac-417-vs-sans-lac-417-distribution-latency-simsteps-inlet}, \ref{fig:with-lac-417-vs-sans-lac-417-distribution-latency-simsteps-outlet}, \ref{fig:with-lac-417-vs-sans-lac-417-distribution-latency-walltime-inlet-ns}, \ref{fig:with-lac-417-vs-sans-lac-417-distribution-latency-walltime-outlet-ns}, \ref{fig:with-lac-417-vs-sans-lac-417-distribution-delivery-clumpiness}, and \ref{fig:with-lac-417-vs-sans-lac-417-distribution-delivery-failure-rate} compare the distributions of quality of service metrics between allocations with and without lac-417.
Extreme outliers are present exclusively in the lac-417 allocation for walltime latency, simstep latency, and delivery failure rate.
Otherwise, the metrics' distributions across snapshots are very similar between allocations.

Supplementary Figures \ref{fig:with-lac-417-vs-sans-lac-417-regression-simstep-period-inlet-ns}, \ref{fig:with-lac-417-vs-sans-lac-417-regression-simstep-period-outlet-ns}, \ref{fig:with-lac-417-vs-sans-lac-417-regression-latency-simsteps-inlet}, \ref{fig:with-lac-417-vs-sans-lac-417-regression-latency-simsteps-outlet}, \ref{fig:with-lac-417-vs-sans-lac-417-regression-latency-walltime-inlet-ns}, \ref{fig:with-lac-417-vs-sans-lac-417-regression-latency-walltime-outlet-ns}, \ref{fig:with-lac-417-vs-sans-lac-417-regression-delivery-clumpiness}, \ref{fig:with-lac-417-vs-sans-lac-417-regression-delivery-clumpiness}, and
\ref{fig:with-lac-417-vs-sans-lac-417-regression-delivery-failure-rate} chart OLS and quantile regressions of quality of service metrics on job composition.
Mean walltime latency, simstep latency, and delivery failure rate are all significantly greater with lac-417.
Surprisingly, mean simstep period is significantly longer without lac-417.

However, there is no significant difference in median value for any quality of service metric between allocations including or excluding lac-417.
This stability of metric medians within allocations containing lac-417 --- which have significantly different means due to outlier values induced by the presence of lac-417 --- demonstrates how the best-effort system maintains overall quality of service stability despite defective or degraded components.

Supplementary Tables \ref{tab:with-lac-417-vs-sans-lac-417-ordinary-least-squares-regression} and \ref{tab:with-lac-417-vs-sans-lac-417-quantile-regression} provide numerical details on regressions reported above.

\section{Conclusion}

The fundamental motivation for best-effort communication is efficient scalability.
Our results confirm that best-effort communication can fulfill on this goal.

We found that the best-effort approach significantly increases performance at high CPU count.
This finding was consistent across the communication-intensive graph coloring benchmark and the computation-intensive digital evolution benchmark.
The computation-heavy digital evolution benchmark yielded the strongest scaling efficiency, achieving at 64 processes 92\% the update-rate of single-process execution.
We observed the greatest relative speedup under distributed communication-heavy workloads --- about $7.8\times$ on the graph coloring benchmark.
In the case of the graph coloring benchmark, we found that best-effort communication can help achieve tangibly better solution quality within a fixed time constraint.

Because real-time volatility affects the outcome of computation under the best-effort model, raw execution speed performance does not suffice to fully understand the consequences of the best-effort communication model.
In order to characterize the real-time dynamics under the best-effort model, we designed and measured a suite of quality of service metrics: simstep period, simstep latency, wall-time latency, delivery failure rate, and delivery clumpiness.

We performed several experiments to validate and characterize these metrics.
Comparing quality of service between multithreading and multiprocessing, we found that multithreading had lower runtime overhead cost but that multiprocessing reduced delivery erraticity, curbing especially extreme poor quality of service outlier events.
We found better quality of service, especially with respect to latency, for processes occupying the same node.
Finally, varying the ratio of computational work to communication, we found lower communication intensity associated with less volatile quality of service.

In order for best-effort communication to succeed in facilitating scale-up, median quality of service must stabilize with increasing CPU count.
Put another way, best-effort communication cannot succeed at scale if communication quality tends toward complete degradation.
In Section \ref{sec:weak-scaling}, we used weak scaling experiments to test the effect of scale-up on quality of service at 8, 64, and 256 processes.
Under a lower communication-intensivity task parameterization, we found that all median quality of service metrics were stable when scaling from 64 to 256 process.
Under maximal communication intensivity, we found in one case that median simstep period degraded from around \SI{80}{\micro\second} to around \SI{85}{\micro\second}.
In another case, median message delivery failure rate increased from around 7\% to around 9\%.
Such minor --- and, in most cases, nil --- degradation in median quality of service despite maximal communication intensivity bodes well for the viability of best-effort communication at scale.

Resilience is a second major motivating factor for best-effort computing.
In another promising result, we found that the presence of an apparently faulty compute node did not degrade median performance or quality of service.
Despite extreme quality of service degradation measured among that node and its clique, collective performance and quality of service remained steady.
In effect, the best-effort approach successfully decoupled global performance from the worst performer.
Such so-called ``straggler effects'' plague traditional approaches to large-scale high-performance computing \citep{aktacs2019straggler}, so avoiding them is a major boon.

Development of the Conduit library stemmed from a practical need for an abstract, prepackaged best-effort commmunication interface to support our digital evolution research.
Because real-time effects are fundamentally application-dependent and arise without any explicit in-program specification (and therefore may be unanticipated) it is important to be able to perform such quality of service profiling case-by-case in applications of best-effort communication.
The instrumentation used in these experiments is written as wrappers around the library's \texttt{Inlet} and \texttt{Outlet} classes that may be enabled via compile-time configuration switch.
This makes data generation for quality of service analysis trivial to perform in any system built with the Conduit library.
We hope that making this library and quality of service metrics available to the community can reduce domain expertise and programmability barriers to taking advantage of the best-effort communication model to efficiently leverage burgeoning parallel and distributed computing power.

In future work, it may be of interest to design systems that monitor and proactively react to real-time quality of service conditions.
For example, imposing a variable cost for cell-cell messaging to agents based on traffic levels or increasing per-update resource generation rates for agents on slow-running nodes.
We are eager to investigate how Conduit's best-effort communication model scales on much larger process counts on the order of thousands of cores.

\section*{Acknowledgment}

Thanks to Santiago Rodriguez Papa for contributing graphics illustrating quality of service metrics.
This research was supported in part by NSF grants DEB-1655715 and DBI-0939454 as well as by Michigan State University through the computational resources provided by the Institute for Cyber-Enabled Research.
This material is based upon work supported by the National Science Foundation Graduate Research Fellowship under Grant No. DGE-1424871.
Any opinions, findings, and conclusions or recommendations expressed in this material are those of the author(s) and do not necessarily reflect the views of the National Science Foundation.

\bibliographystyle{apalike}
\bibliography{bibl}

\clearpage


\providecommand{\dissertationexclude}[1]{%
\ifdefined\DISSERTATION
\else
#1
\fi
}

\providecommand{\dissertationonly}[1]{%
\ifdefined\DISSERTATION%
#1%
\else%
\fi
}

\dissertationexclude{\section{Supplemental Materials}}

\dissertationonly{
\newenvironment{levelup}
  {\let\subparagraph\paragraph%
   \let\paragraph\subsubsection%
   \let\subsubsection\subsection%
   \let\subsection\section%
  }{}
  \begin{levelup}
}

\subsection{Weak Scaling}

This section provides full results from weak scaling experiments discussed in \ref{sec:weak-scaling}.

\providecommand{\figweakscalingdistributionmacro}[2]{
\def\metric{#1}\def\metricslug{#2}\begin{figure*}[h]
  \centering
  \begin{subfigure}[b]{0.49\textwidth}
    \centering
    \includegraphics[width=\textwidth]{submodule/conduit-binder/binder/date=2021+project=72k5n/a=weak-scaling/teeplots/\metricslug/col=cpus-per-node+row=num-simels-per-cpu+transform=snapshot_diffs+viz=facet-boxplot-nofliers+x=num-processes+y=\metricslug+ext=}
    \caption{Distribution of \metric{} for each snapshot, without outliers.}
    \label{fig:weak-scaling-distribution-\metricslug-nofliers}
  \end{subfigure}%
  \hfill%
  \begin{subfigure}[b]{0.49\textwidth}
    \centering
    \includegraphics[width=\textwidth]{submodule/conduit-binder/binder/date=2021+project=72k5n/a=weak-scaling/teeplots/\metricslug/col=cpus-per-node+row=num-simels-per-cpu+transform=snapshot_diffs+viz=facet-boxplot-withfliers+x=num-processes+y=\metricslug+ext=}
    \caption{Distribution of \metric{} for each snapshot, with outliers.}
    \label{fig:weak-scaling-distribution-\metricslug-withfliers}
  \end{subfigure}
  \caption{Distribution of \metric{} for individual snapshot measurements for weak scaling experiment (Section \ref{sec:weak-scaling}). Lower is better.}
  \label{fig:weak-scaling-distribution-\metricslug}
\end{figure*}

}

\def\metric{Latency Walltime Inlet (ns)}\def\metricslug{latency-walltime-inlet-ns}\begin{figure*}[h]
  \centering
  \begin{subfigure}[b]{0.49\textwidth}
    \centering
    \includegraphics[width=\textwidth]{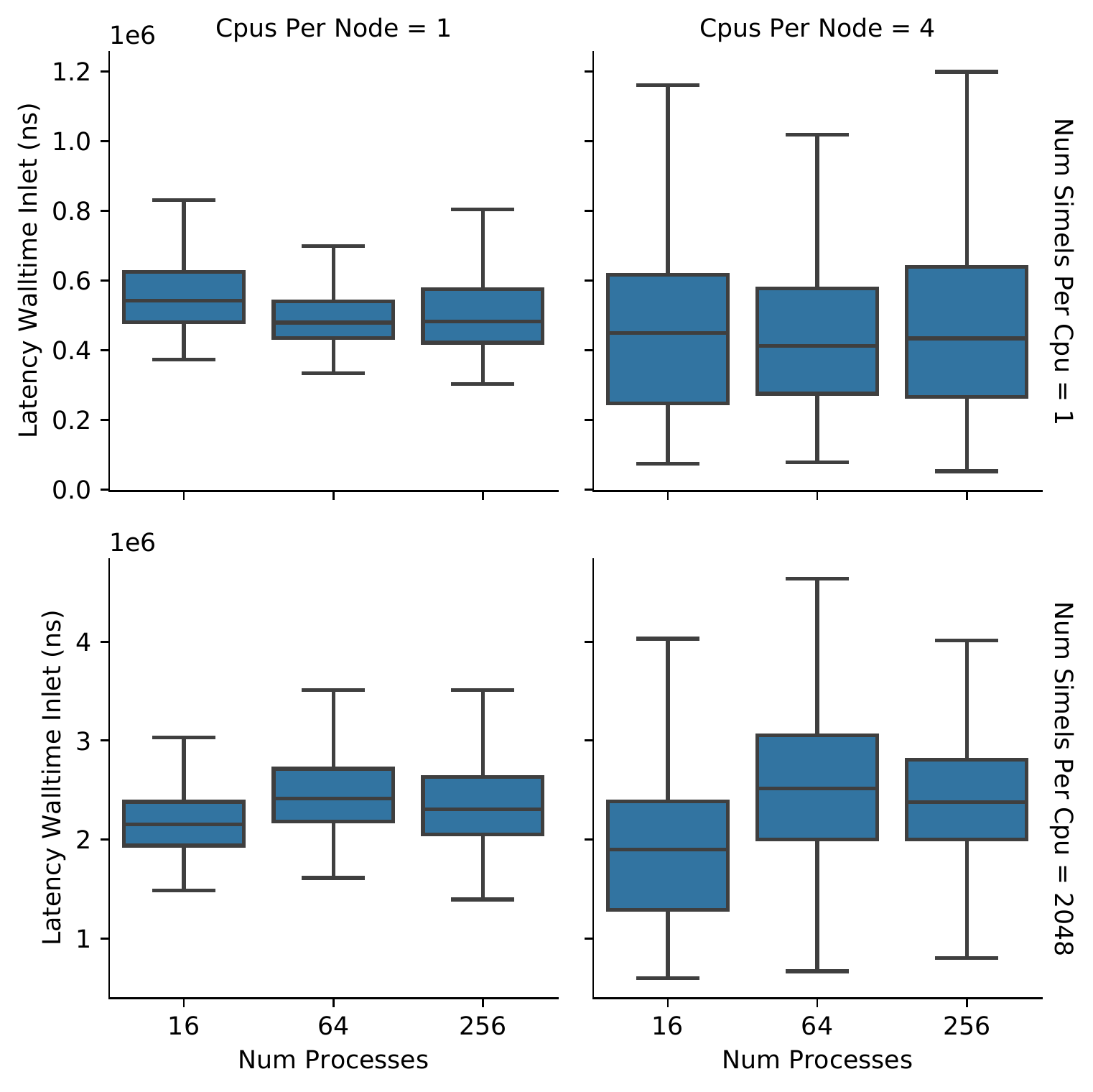}
    \caption{Distribution of \metric{} for each snapshot, without outliers.}
    \label{fig:weak-scaling-distribution-\metricslug-nofliers}
  \end{subfigure}%
  \hfill%
  \begin{subfigure}[b]{0.49\textwidth}
    \centering
    \includegraphics[width=\textwidth]{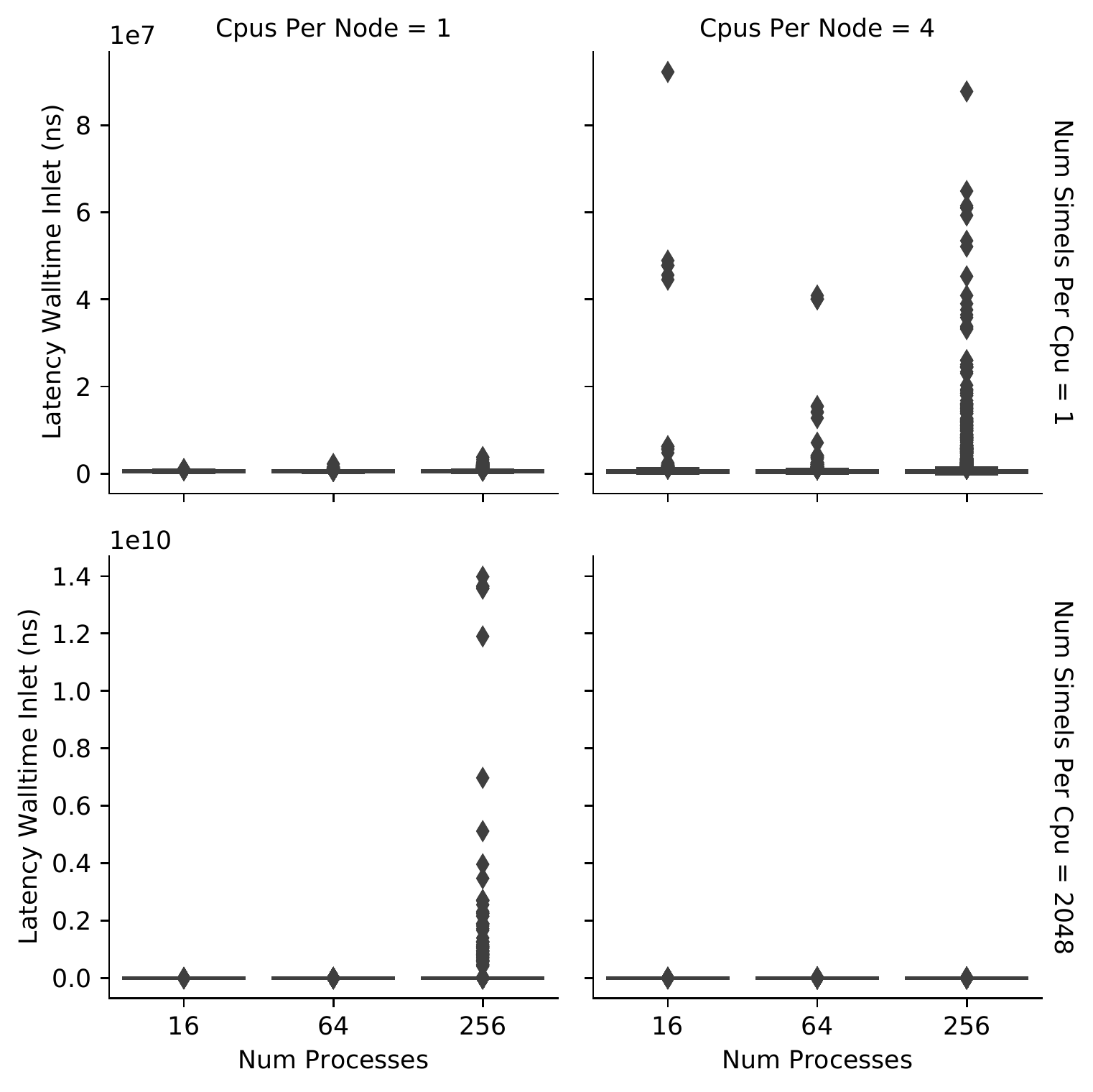}
    \caption{Distribution of \metric{} for each snapshot, with outliers.}
    \label{fig:weak-scaling-distribution-\metricslug-withfliers}
  \end{subfigure}
  \caption{Distribution of \metric{} for individual snapshot measurements for weak scaling experiment (Section \ref{sec:weak-scaling}). Lower is better.}
  \label{fig:weak-scaling-distribution-\metricslug}
\end{figure*}

\def\metric{Latency Simsteps Outlet}\def\metricslug{latency-simsteps-outlet}\begin{figure*}[h]
  \centering
  \begin{subfigure}[b]{0.49\textwidth}
    \centering
    \includegraphics[width=\textwidth]{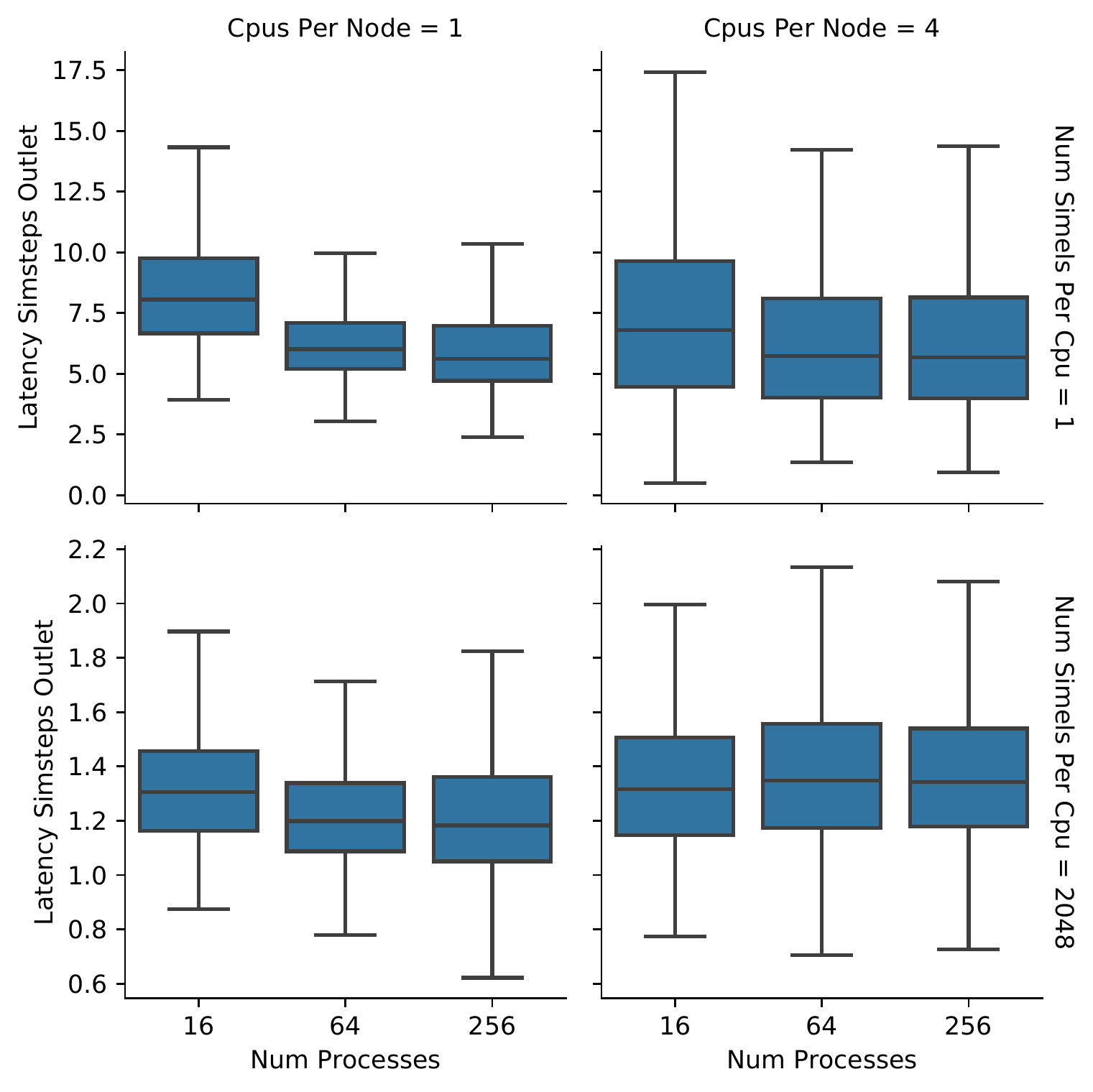}
    \caption{Distribution of \metric{} for each snapshot, without outliers.}
    \label{fig:weak-scaling-distribution-\metricslug-nofliers}
  \end{subfigure}%
  \hfill%
  \begin{subfigure}[b]{0.49\textwidth}
    \centering
    \includegraphics[width=\textwidth]{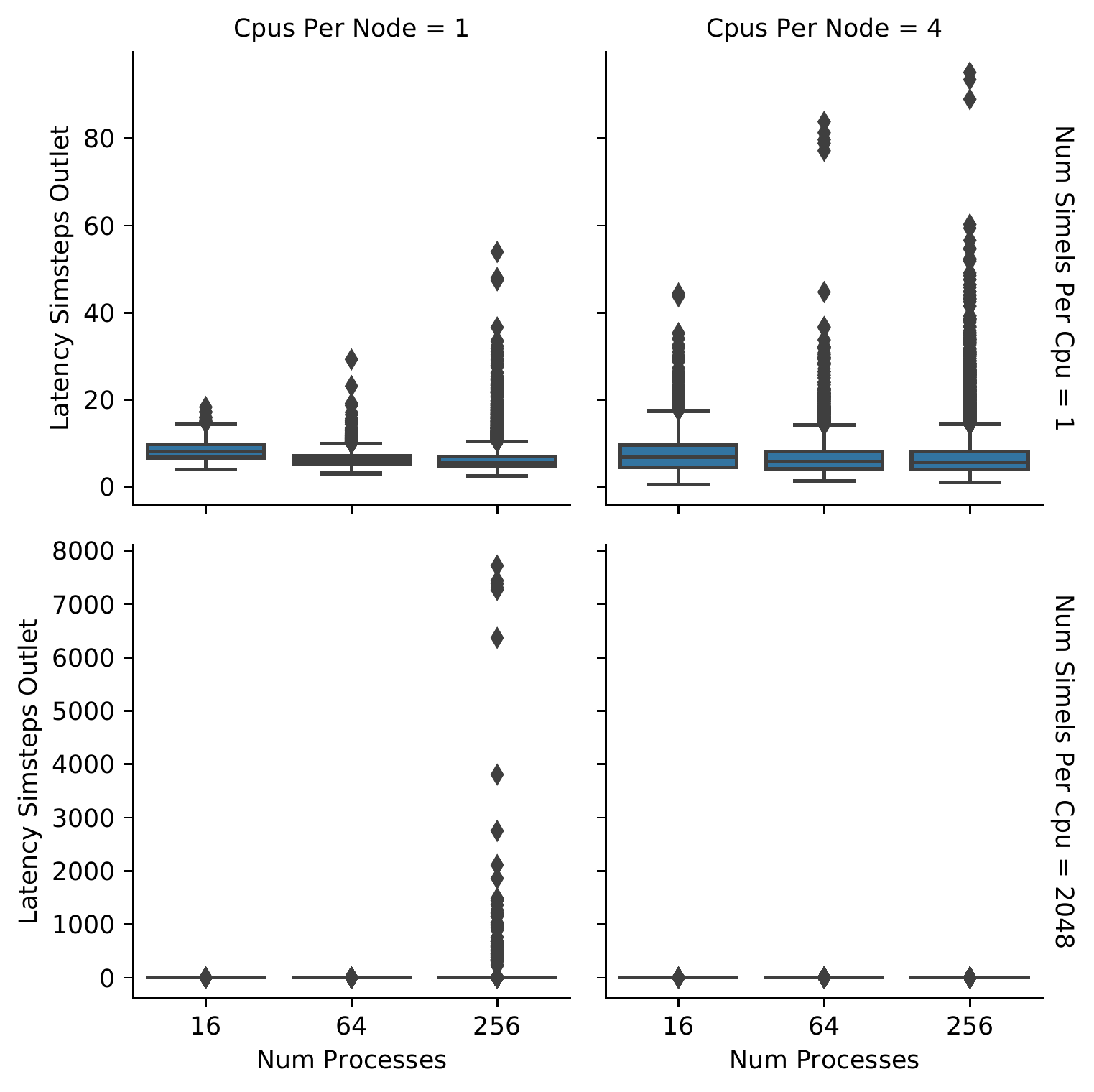}
    \caption{Distribution of \metric{} for each snapshot, with outliers.}
    \label{fig:weak-scaling-distribution-\metricslug-withfliers}
  \end{subfigure}
  \caption{Distribution of \metric{} for individual snapshot measurements for weak scaling experiment (Section \ref{sec:weak-scaling}). Lower is better.}
  \label{fig:weak-scaling-distribution-\metricslug}
\end{figure*}

\def\metric{Latency Walltime Outlet (ns)}\def\metricslug{latency-walltime-outlet-ns}\begin{figure*}[h]
  \centering
  \begin{subfigure}[b]{0.49\textwidth}
    \centering
    \includegraphics[width=\textwidth]{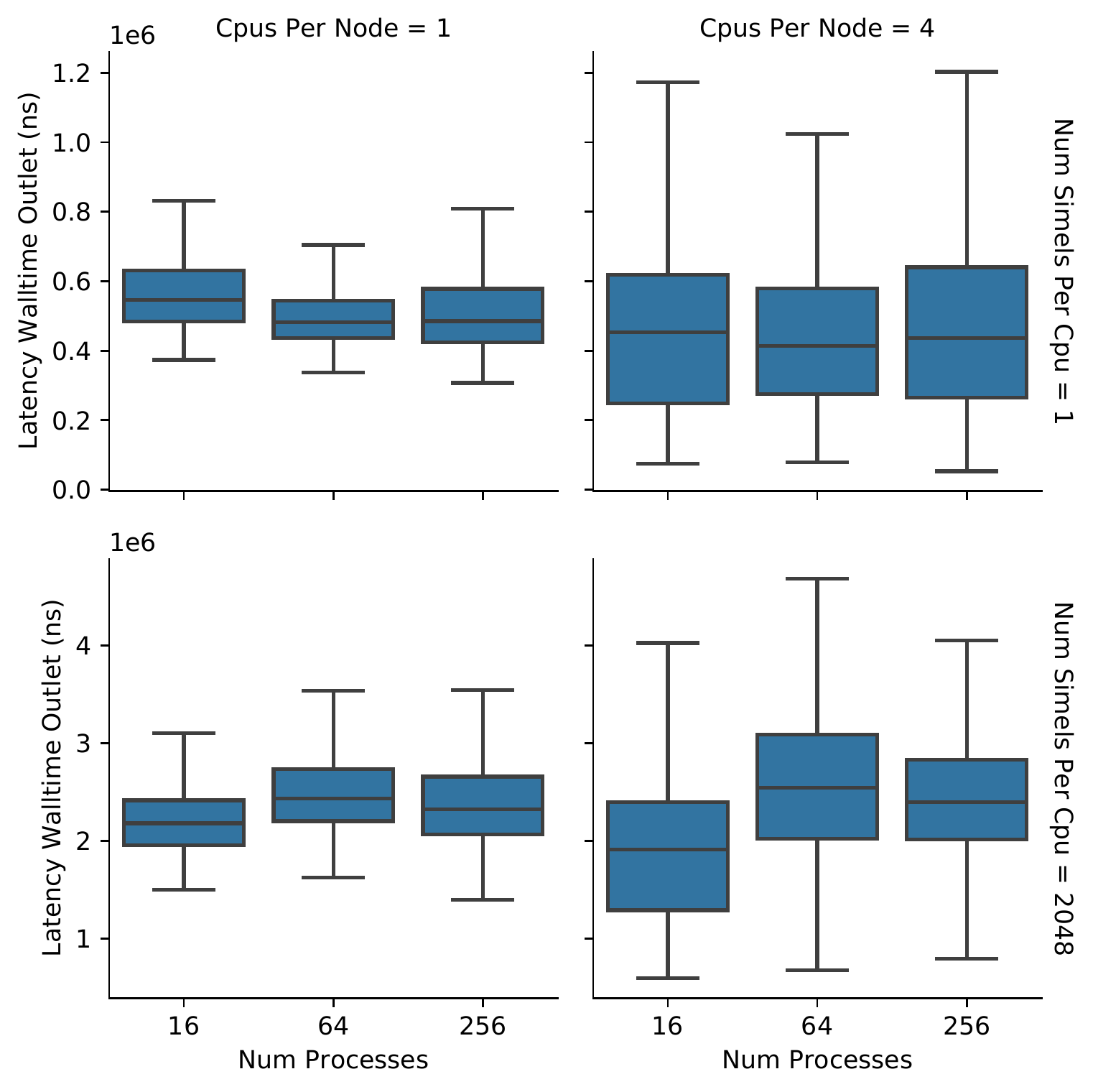}
    \caption{Distribution of \metric{} for each snapshot, without outliers.}
    \label{fig:weak-scaling-distribution-\metricslug-nofliers}
  \end{subfigure}%
  \hfill%
  \begin{subfigure}[b]{0.49\textwidth}
    \centering
    \includegraphics[width=\textwidth]{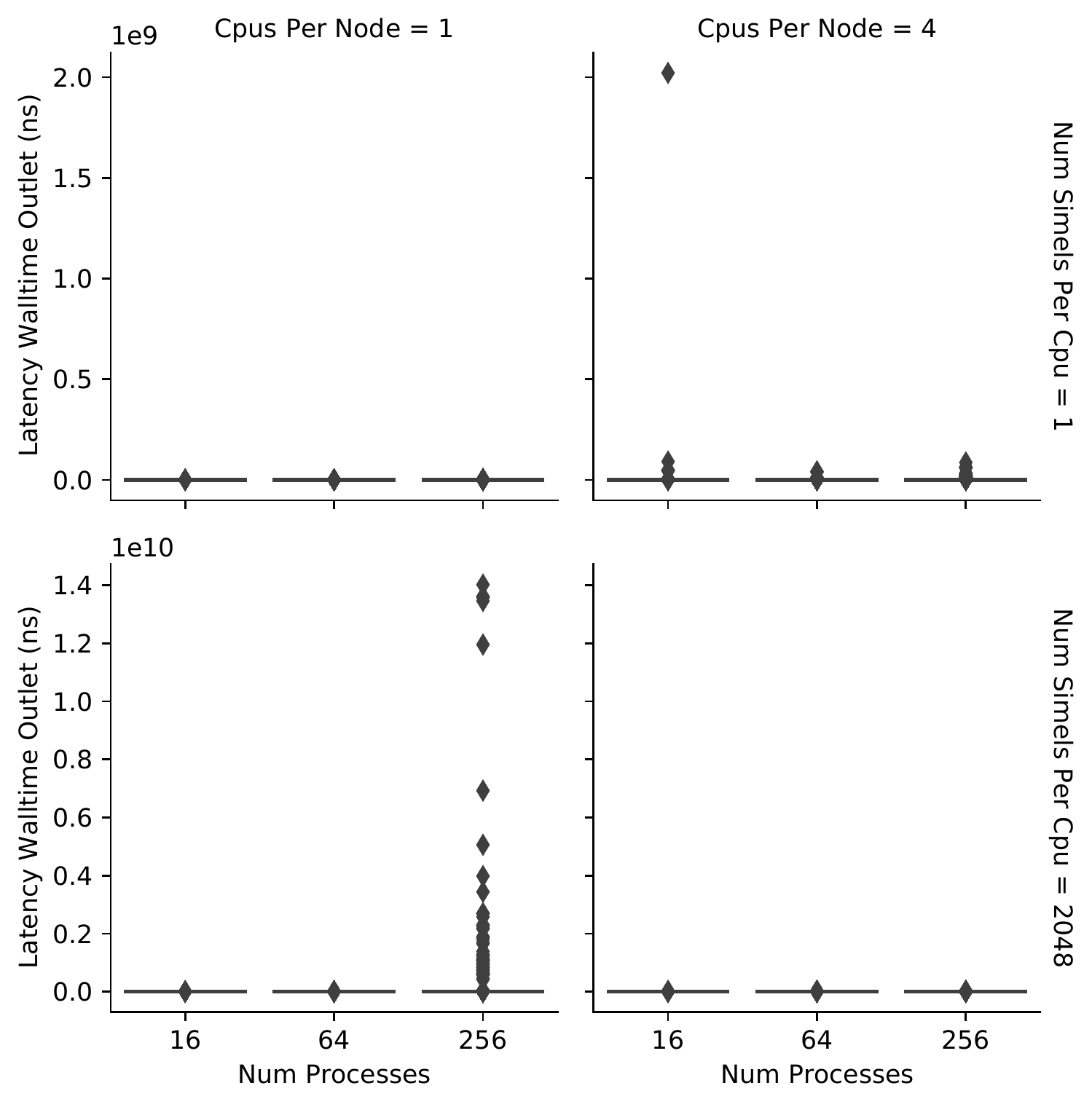}
    \caption{Distribution of \metric{} for each snapshot, with outliers.}
    \label{fig:weak-scaling-distribution-\metricslug-withfliers}
  \end{subfigure}
  \caption{Distribution of \metric{} for individual snapshot measurements for weak scaling experiment (Section \ref{sec:weak-scaling}). Lower is better.}
  \label{fig:weak-scaling-distribution-\metricslug}
\end{figure*}

\def\metric{Delivery Clumpiness}\def\metricslug{delivery-clumpiness}\begin{figure*}[h]
  \centering
  \begin{subfigure}[b]{0.49\textwidth}
    \centering
    \includegraphics[width=\textwidth]{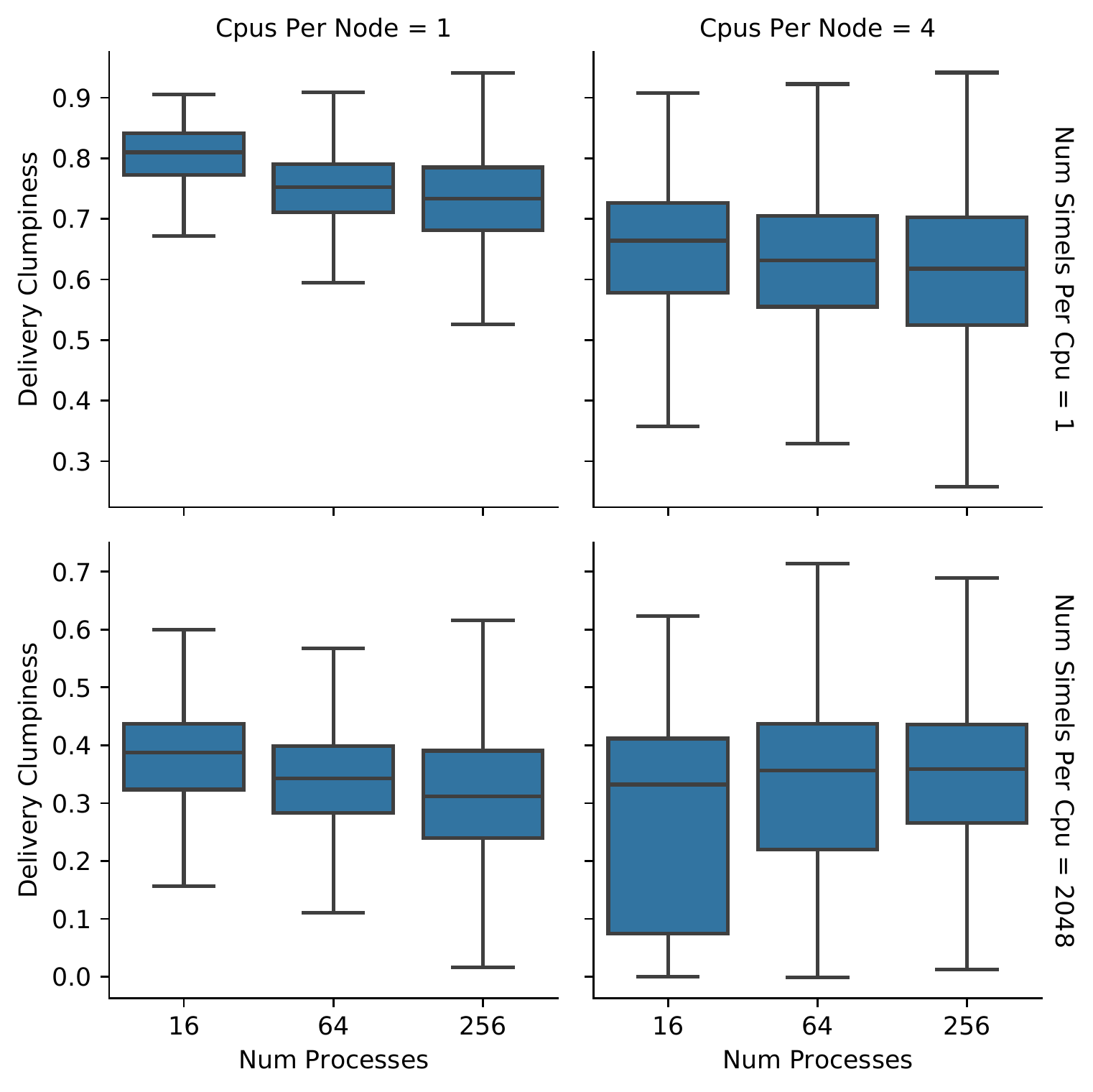}
    \caption{Distribution of \metric{} for each snapshot, without outliers.}
    \label{fig:weak-scaling-distribution-\metricslug-nofliers}
  \end{subfigure}%
  \hfill%
  \begin{subfigure}[b]{0.49\textwidth}
    \centering
    \includegraphics[width=\textwidth]{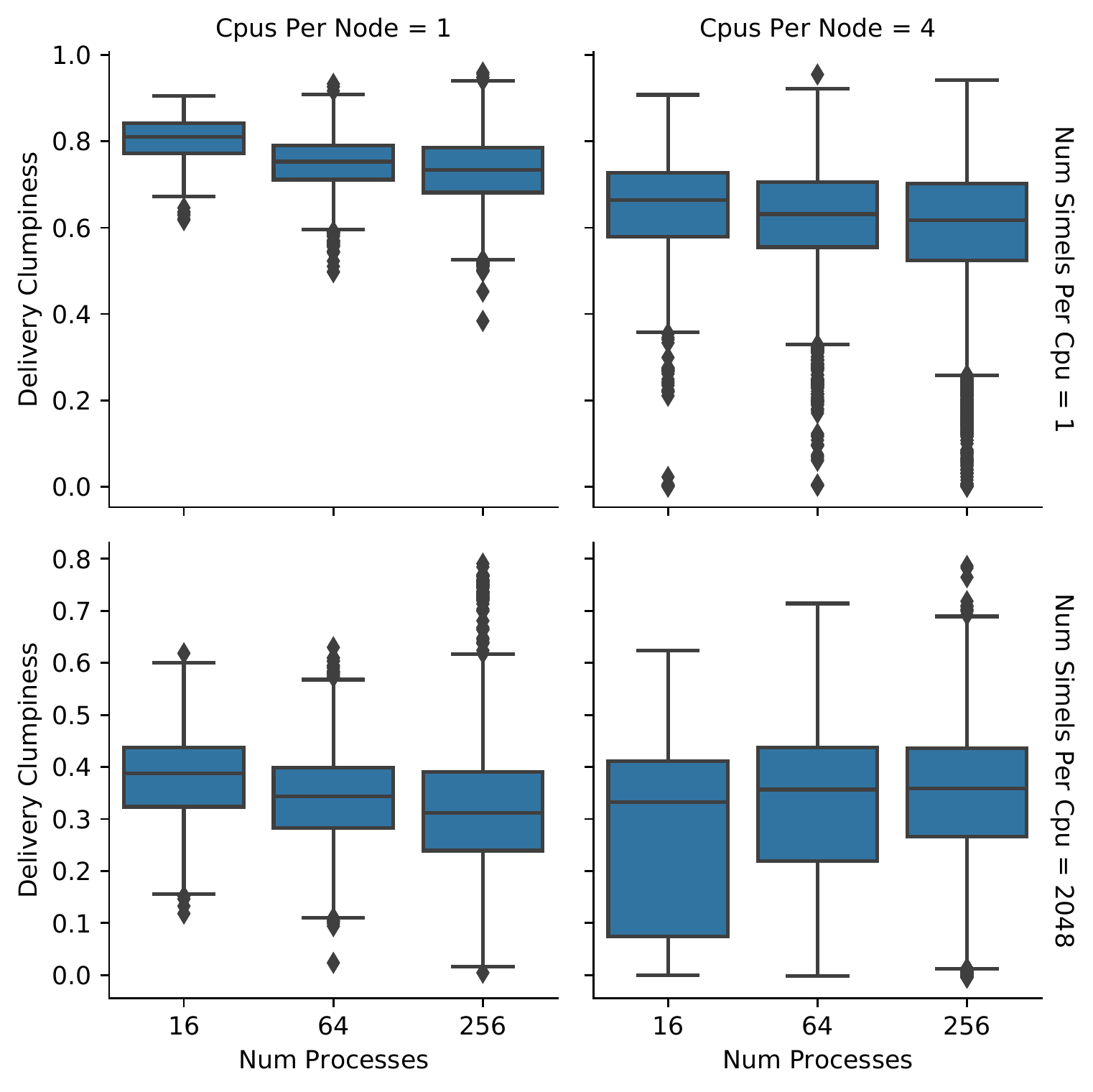}
    \caption{Distribution of \metric{} for each snapshot, with outliers.}
    \label{fig:weak-scaling-distribution-\metricslug-withfliers}
  \end{subfigure}
  \caption{Distribution of \metric{} for individual snapshot measurements for weak scaling experiment (Section \ref{sec:weak-scaling}). Lower is better.}
  \label{fig:weak-scaling-distribution-\metricslug}
\end{figure*}

\def\metric{Simstep Period Inlet (ns)}\def\metricslug{simstep-period-inlet-ns}\begin{figure*}[h]
  \centering
  \begin{subfigure}[b]{0.49\textwidth}
    \centering
    \includegraphics[width=\textwidth]{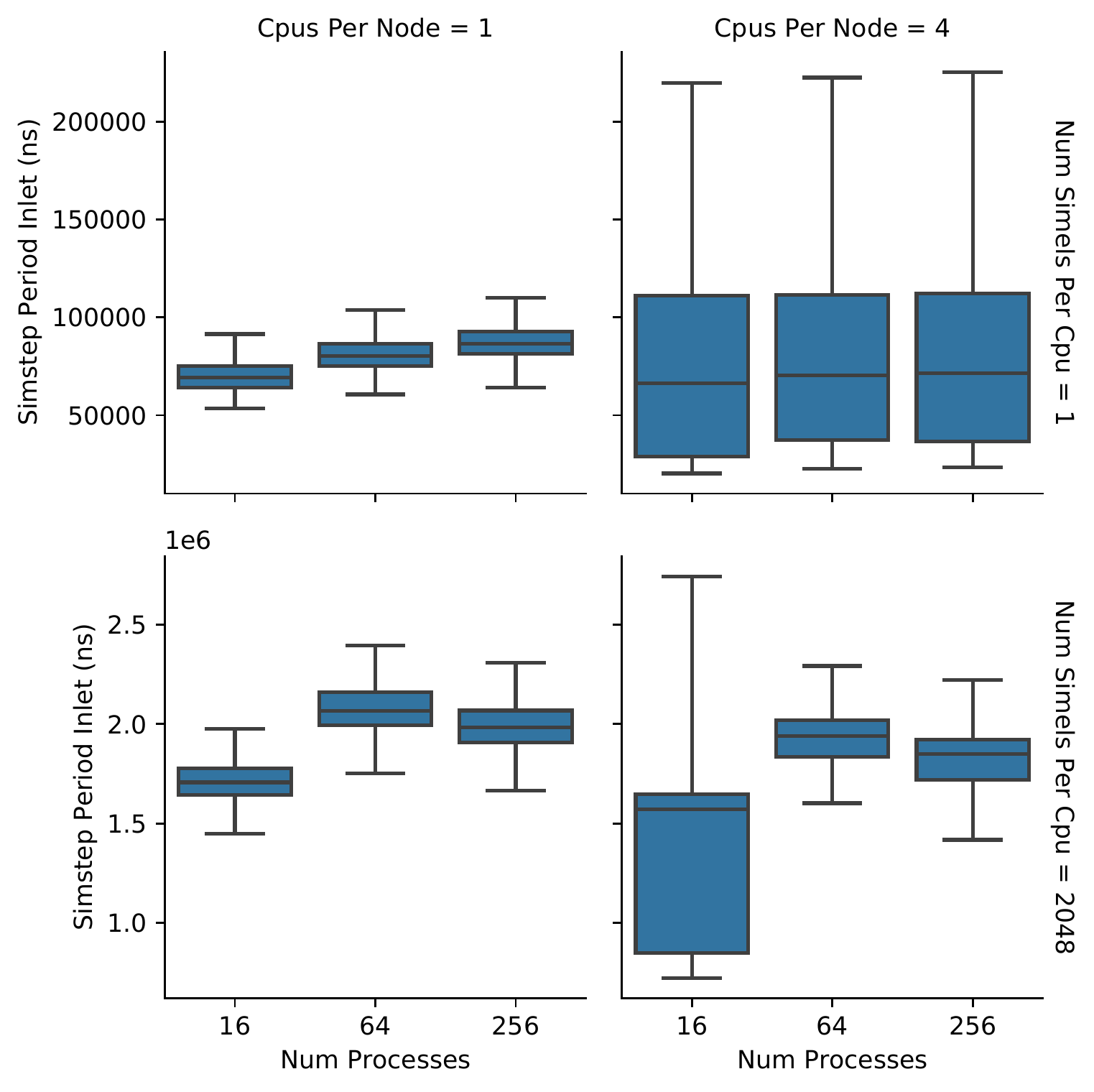}
    \caption{Distribution of \metric{} for each snapshot, without outliers.}
    \label{fig:weak-scaling-distribution-\metricslug-nofliers}
  \end{subfigure}%
  \hfill%
  \begin{subfigure}[b]{0.49\textwidth}
    \centering
    \includegraphics[width=\textwidth]{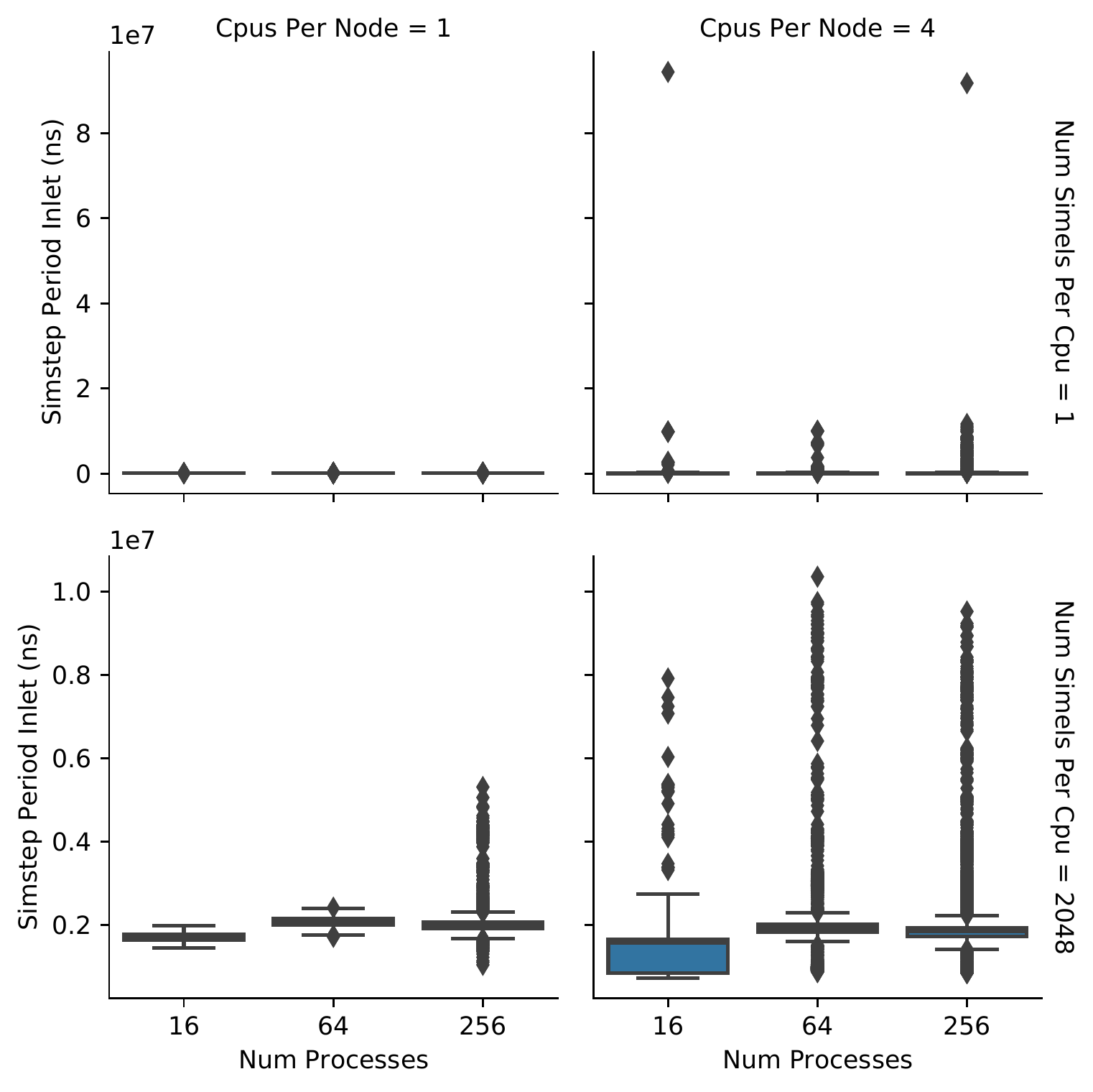}
    \caption{Distribution of \metric{} for each snapshot, with outliers.}
    \label{fig:weak-scaling-distribution-\metricslug-withfliers}
  \end{subfigure}
  \caption{Distribution of \metric{} for individual snapshot measurements for weak scaling experiment (Section \ref{sec:weak-scaling}). Lower is better.}
  \label{fig:weak-scaling-distribution-\metricslug}
\end{figure*}

\def\metric{Latency Simsteps Inlet}\def\metricslug{latency-simsteps-inlet}\begin{figure*}[h]
  \centering
  \begin{subfigure}[b]{0.49\textwidth}
    \centering
    \includegraphics[width=\textwidth]{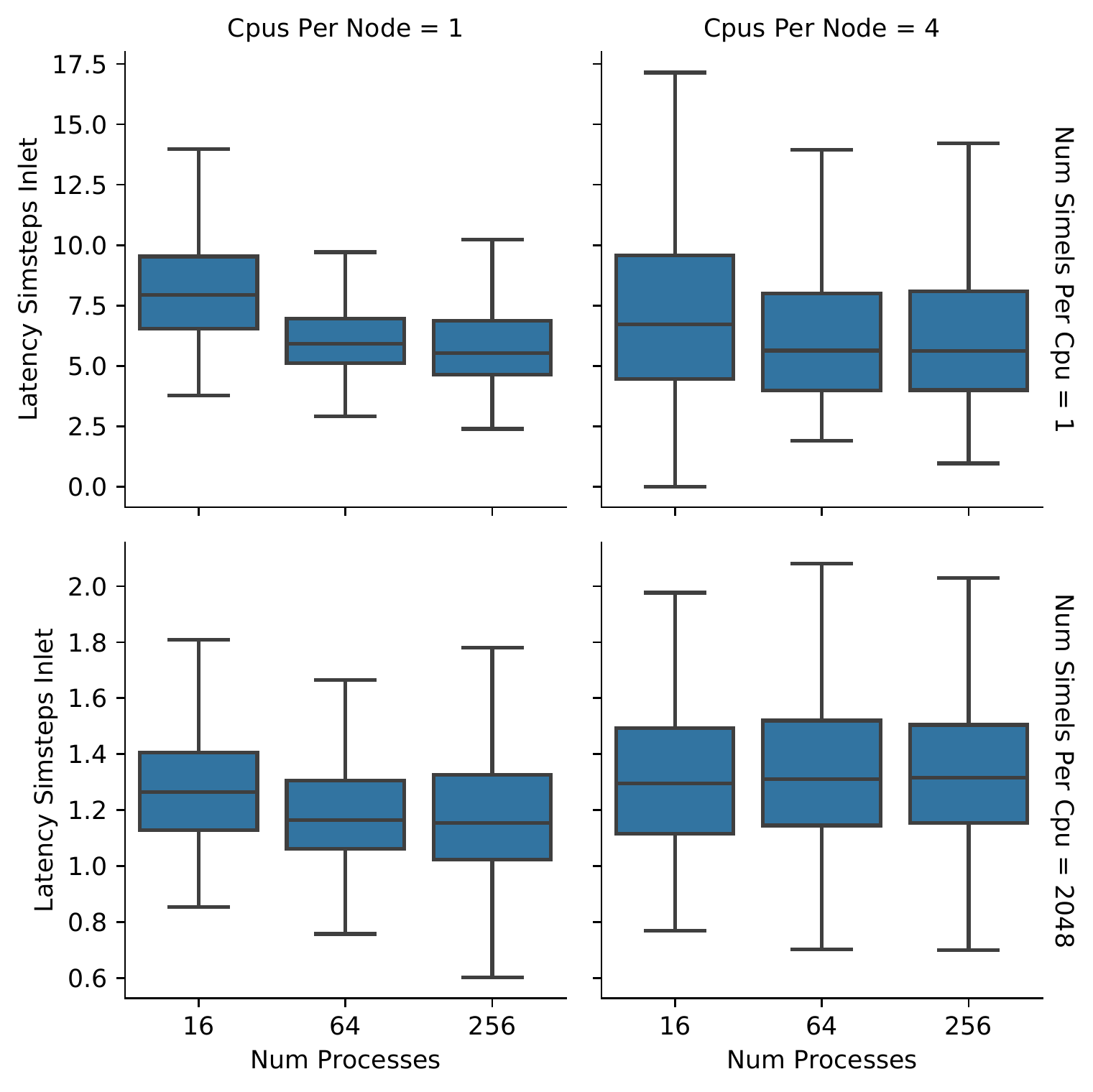}
    \caption{Distribution of \metric{} for each snapshot, without outliers.}
    \label{fig:weak-scaling-distribution-\metricslug-nofliers}
  \end{subfigure}%
  \hfill%
  \begin{subfigure}[b]{0.49\textwidth}
    \centering
    \includegraphics[width=\textwidth]{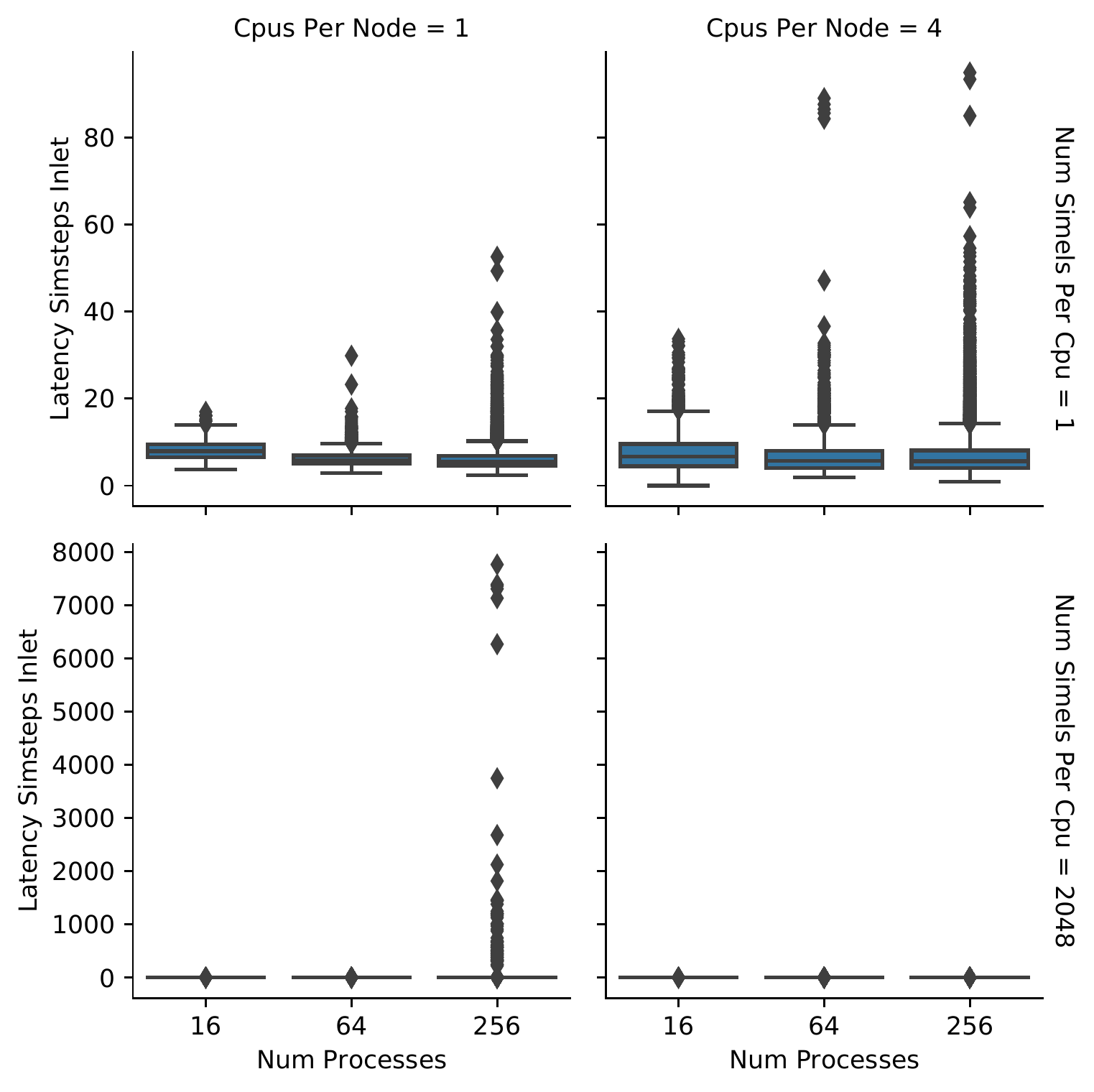}
    \caption{Distribution of \metric{} for each snapshot, with outliers.}
    \label{fig:weak-scaling-distribution-\metricslug-withfliers}
  \end{subfigure}
  \caption{Distribution of \metric{} for individual snapshot measurements for weak scaling experiment (Section \ref{sec:weak-scaling}). Lower is better.}
  \label{fig:weak-scaling-distribution-\metricslug}
\end{figure*}

\def\metric{Simstep Period Outlet (ns)}\def\metricslug{simstep-period-outlet-ns}\begin{figure*}[h]
  \centering
  \begin{subfigure}[b]{0.49\textwidth}
    \centering
    \includegraphics[width=\textwidth]{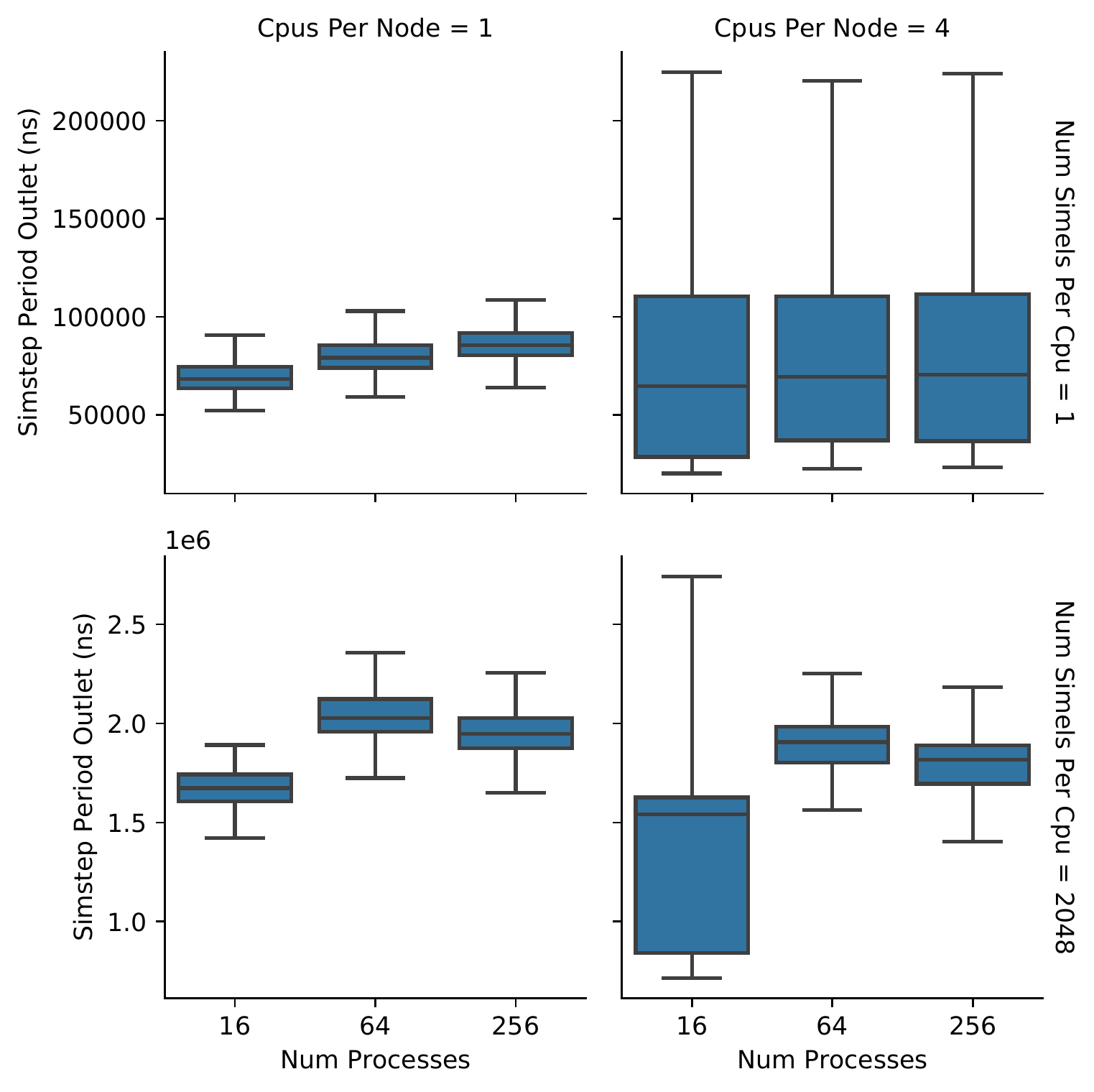}
    \caption{Distribution of \metric{} for each snapshot, without outliers.}
    \label{fig:weak-scaling-distribution-\metricslug-nofliers}
  \end{subfigure}%
  \hfill%
  \begin{subfigure}[b]{0.49\textwidth}
    \centering
    \includegraphics[width=\textwidth]{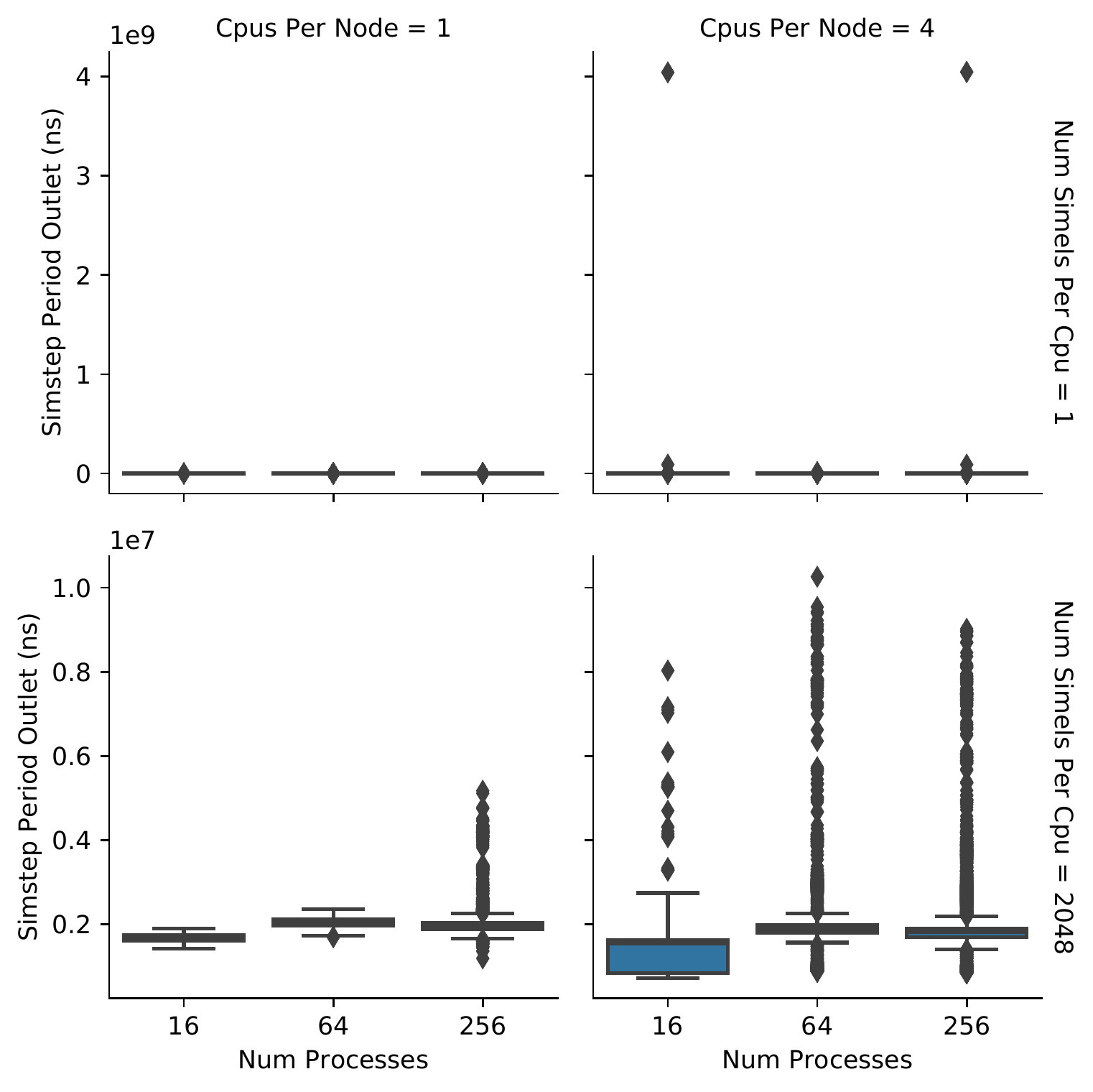}
    \caption{Distribution of \metric{} for each snapshot, with outliers.}
    \label{fig:weak-scaling-distribution-\metricslug-withfliers}
  \end{subfigure}
  \caption{Distribution of \metric{} for individual snapshot measurements for weak scaling experiment (Section \ref{sec:weak-scaling}). Lower is better.}
  \label{fig:weak-scaling-distribution-\metricslug}
\end{figure*}

\def\metric{Delivery Failure Rate}\def\metricslug{delivery-failure-rate}\begin{figure*}[h]
  \centering
  \begin{subfigure}[b]{0.49\textwidth}
    \centering
    \includegraphics[width=\textwidth]{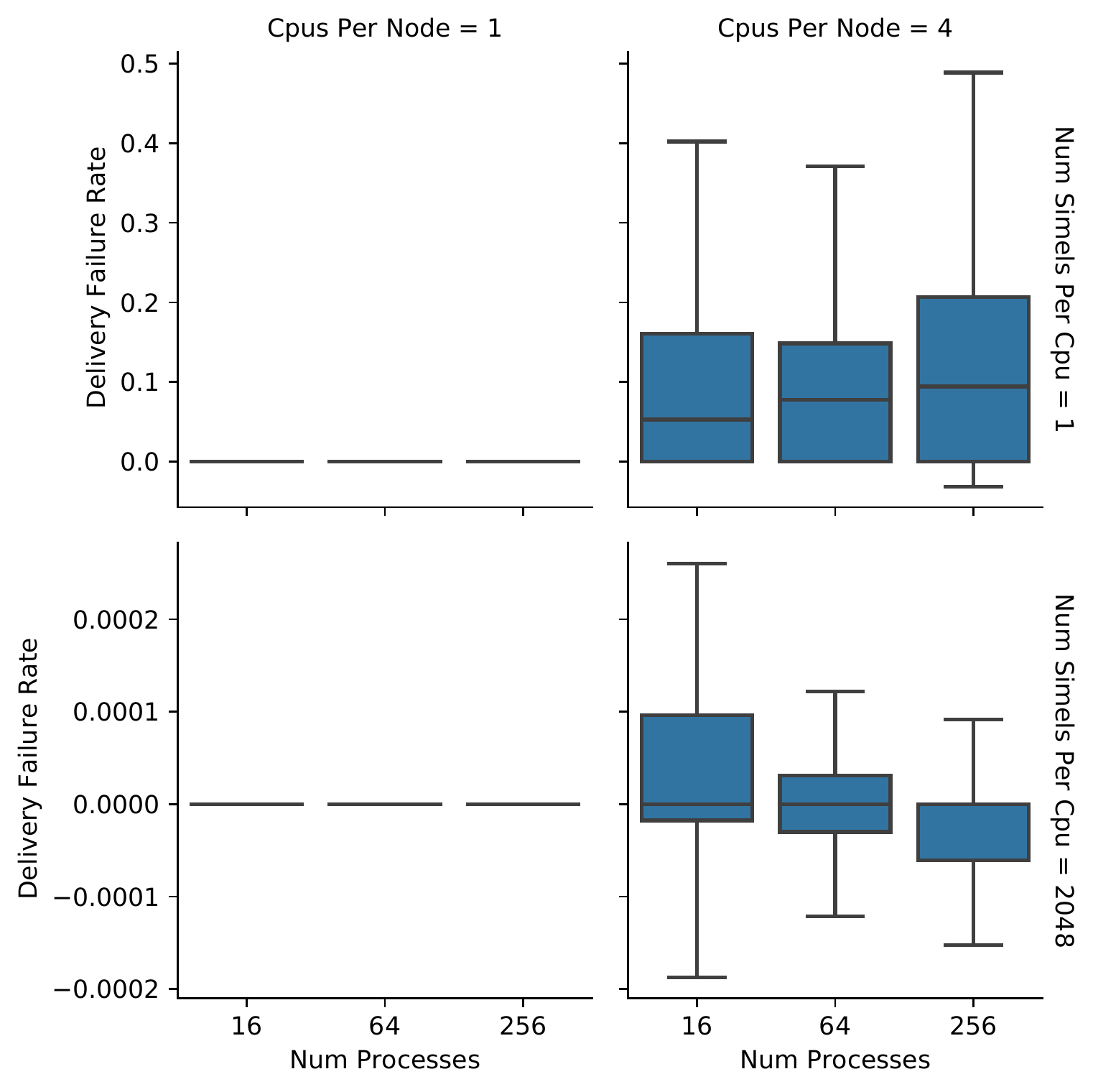}
    \caption{Distribution of \metric{} for each snapshot, without outliers.}
    \label{fig:weak-scaling-distribution-\metricslug-nofliers}
  \end{subfigure}%
  \hfill%
  \begin{subfigure}[b]{0.49\textwidth}
    \centering
    \includegraphics[width=\textwidth]{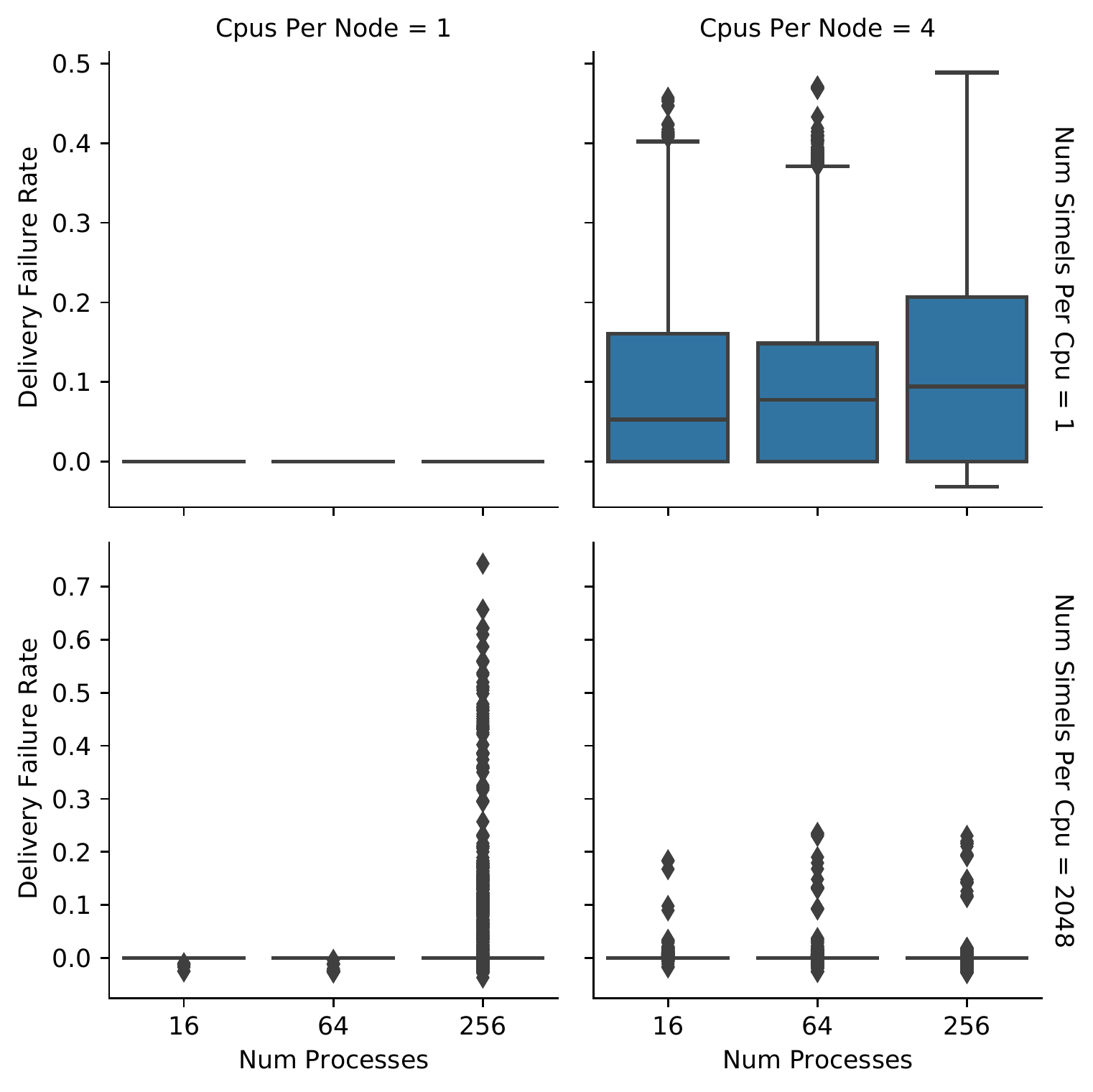}
    \caption{Distribution of \metric{} for each snapshot, with outliers.}
    \label{fig:weak-scaling-distribution-\metricslug-withfliers}
  \end{subfigure}
  \caption{Distribution of \metric{} for individual snapshot measurements for weak scaling experiment (Section \ref{sec:weak-scaling}). Lower is better.}
  \label{fig:weak-scaling-distribution-\metricslug}
\end{figure*}

\def\metric{Latency Walltime Inlet (ns)}\def\metricslug{latency-walltime-inlet-ns}

\def\metric{Latency Simsteps Outlet}\def\metricslug{latency-simsteps-outlet}

\def\metric{Latency Walltime Outlet (ns)}\def\metricslug{latency-walltime-outlet-ns}

\def\metric{Latency Simsteps Inlet}\def\metricslug{latency-simsteps-inlet}

\def\metric{Simstep Period Outlet (ns)}\def\metricslug{simstep-period-outlet-ns}

\def\metric{Delivery Failure Rate}\def\metricslug{delivery-failure-rate}

\def\metric{Latency Simsteps Outlet}\def\metricslug{latency-simsteps-outlet}

\def\metric{Latency Walltime Outlet (ns)}\def\metricslug{latency-walltime-outlet-ns}

\def\metric{Delivery Clumpiness}\def\metricslug{delivery-clumpiness}

\def\metric{Simstep Period Inlet (ns)}\def\metricslug{simstep-period-inlet-ns}

\def\metric{Simstep Period Outlet (ns)}\def\metricslug{simstep-period-outlet-ns}

\providecommand{\tabweakscalingregressionmacro}[5]{
\def\regression{#1}\def\regressionslug{#2}\def\regressionlongslug{#3}\def\metric{#4}\def\metricslug{#5}

\providecommand{\dissertationelse}[2]{%
\ifdefined\DISSERTATION
#1
\else
#2
\fi
}

\makeatletter
  \providecommand\importpath{\import@path}
\makeatother

\pragmaonce

\def\replacesinglequote#1{%
  \IfSubStr{#1}{'}{%
    \StrSubstitute{#1}{'}{\,}}{#1}}

\pragmaonce

\providecommand\rotcol[1]{ \multicolumn{1}{@{}@{}}{} \adjustbox{angle=60,lap=\width}{\rule{0.4\textwidth}{0.8pt}} \hfill \makebox[3ex]{\adjustbox{angle=60,lap=\width}{\bfseries \hspace{1ex} #1}} \hfill }


\clearpage
{
\onecolumn

\begin{landscape}
\dissertationelse{\fontsize{8}{9}\selectfont}{\footnotesize}
\csvstyle{myTableStyle}{
  longtable=@{\hspace{-.5\arrayrulewidth}}c@{\hspace{-.5\arrayrulewidth}}|l|l|l|l|l|l|l|l|l|l|l|l|l|l|,
  table head=\caption{
    Full \regression{} results of \metric{} against against log processor count (Section \ref{sec:weak-scaling}).
    Significance level $p < 0.05$ used.
    Inf or NaN values may occur due to multicollinearity or due to inf or NaN observations.
  }\label{tab:weak-scaling-\metricslug-\regressionslug} \\
    \toprule%
    \multicolumn{1}{@{\hspace{-.5\arrayrulewidth}}l@{\hspace{-.5\arrayrulewidth}}}{} &%
    \rotcol{ Metric } &%
    \rotcol{ Statistic } &%
    \rotcol{ Significant Effect Sign } &%
    \rotcol{ Cpus Per Node } &%
    \rotcol{ Num Simels Per Cpu} &%
    \rotcol{ Num Processes} &%
    \rotcol{ Absolute Effect Size} &%
    \rotcol{ Absolute Effect Size 95\% CI Lower Bound} &%
    \rotcol{ Absolute Effect Size 95\% CI Upper Bound} &%
    \rotcol{ Relative Effect Size} &%
    \rotcol{ Relative Effect Size 95\% CI Lower Bound} &%
    \rotcol{ Relative Effect Size 95\% CI Upper Bound} &%
    \rotcol{$\mathbf{n}$} & %
    \rotcol{$\mathbf{p}$} %
    \vspace{-1ex} \\ \midrule\endfirsthead
    \caption*{\tablename{} \thetable{} (cont'd)} \\
    \toprule%
    \multicolumn{1}{@{\hspace{-.5\arrayrulewidth}}l@{\hspace{-.5\arrayrulewidth}}}{} &%
    \rotcol{ Metric } &%
    \rotcol{ Statistic } &%
    \rotcol{ Significant Effect Sign } &%
    \rotcol{ Cpus Per Node } &%
    \rotcol{ Num Simels Per Cpu} &%
    \rotcol{ Num Processes} &%
    \rotcol{ Absolute Effect Size} &%
    \rotcol{ Absolute Effect Size 95\% CI Lower Bound} &%
    \rotcol{ Absolute Effect Size 95\% CI Upper Bound} &%
    \rotcol{ Relative Effect Size} &%
    \rotcol{ Relative Effect Size 95\% CI Lower Bound} &%
    \rotcol{ Relative Effect Size 95\% CI Upper Bound} &%
    \rotcol{$\mathbf{n}$} & %
    \rotcol{$\mathbf{p}$} %
    \vspace{-1ex} \\ \midrule\endhead
    \bottomrule\endfoot,
  late after line=\\\hline,
  late after last line=\\\bottomrule,
  head to column names,
  filter=\equal{\csuse{DependentVariableSlug}}{\metricslug} \and \equal{\csuse{RegressionModelSlug}}{\regressionlongslug},
}

\catcode`\%=12
\providecommand\filename{}
\renewcommand\filename{\detokenize{submodule/conduit-binder/binder/date=2021+project=72k5n/a=weak-scaling/outplots/a=weak_scaling_regression_results+async_mode=3+comp_work=0+impl=proc+nproc=16_64_256+nthread=1+proc=0-255+readability=latexcsvreader+replicate=0-9+slurm_nnodes=64_256+slurm_ntasks=256+subject=inlet_outlet+view=summary+ext=.csv}}
\catcode`\%=14

\csvreader[myTableStyle]{\importpath\filename}{}%
{ %
  &%
  \csuse{DependentVariable} &%
  \csuse{Statistic} &%
  \ifthenelse{\equal{\csuse{SignificantEffectSign}}{-}}{\cellcolor{red!25}}{}%
  \ifthenelse{\equal{\csuse{SignificantEffectSign}}{+}}{\cellcolor{green!25}}{}%
  \ifthenelse{\equal{\csuse{SignificantEffectSign}}{0}}{\cellcolor{gray!25}}{}%
  \csuse{SignificantEffectSign} &%
  \replacesinglequote{\csuse{CpusPerNode}} &%
  \replacesinglequote{\csuse{NumSimelsPerCpu}} &%
  \replacesinglequote{\csuse{NumProcessesPrettyprint}} &%
  \replacesinglequote{\csuse{AbsoluteEffectSize}} &%
  \replacesinglequote{\csuse{AbsoluteEffectSize95CILowerBound}} &%
  \replacesinglequote{\csuse{AbsoluteEffectSize95CIUpperBound}} &%
  \replacesinglequote{\csuse{RelativeEffectSize}} &%
  \replacesinglequote{\csuse{RelativeEffectSize95CILowerBound}} &%
  \replacesinglequote{\csuse{RelativeEffectSize95CIUpperBound}} &%
  \replacesinglequote{\csuse{n}} &%
  \replacesinglequote{\csuse{p}} %
}

\end{landscape}

}

\clearpage

}

\def\regression{%
Ordinary Least Squares Regression%
}\def\regressionslug{%
regression-ols%
}\def\regressionlongslug{%
ordinary-least-squares-regression%
}\def\metric{%
Latency Walltime Inlet (ns)%
}\def\metricslug{%
latency-walltime-inlet-ns%
}

\providecommand{\dissertationelse}[2]{%
\ifdefined\DISSERTATION
#1
\else
#2
\fi
}

\makeatletter
  \providecommand\importpath{\import@path}
\makeatother

\pragmaonce

\def\replacesinglequote#1{%
  \IfSubStr{#1}{'}{%
    \StrSubstitute{#1}{'}{\,}}{#1}}

\pragmaonce

\providecommand\rotcol[1]{ \multicolumn{1}{@{}@{}}{} \adjustbox{angle=60,lap=\width}{\rule{0.4\textwidth}{0.8pt}} \hfill \makebox[3ex]{\adjustbox{angle=60,lap=\width}{\bfseries \hspace{1ex} #1}} \hfill }


\clearpage
{
\onecolumn

\begin{landscape}
\dissertationelse{\fontsize{8}{9}\selectfont}{\footnotesize}
\csvstyle{myTableStyle}{
  longtable=@{\hspace{-.5\arrayrulewidth}}c@{\hspace{-.5\arrayrulewidth}}|l|l|l|l|l|l|l|l|l|l|l|l|l|l|,
  table head=\caption{
    Full \regression{} results of \metric{} against against log processor count (Section \ref{sec:weak-scaling}).
    Significance level $p < 0.05$ used.
    Inf or NaN values may occur due to multicollinearity or due to inf or NaN observations.
  }\label{tab:weak-scaling-\metricslug-\regressionslug} \\
    \toprule%
    \multicolumn{1}{@{\hspace{-.5\arrayrulewidth}}l@{\hspace{-.5\arrayrulewidth}}}{} &%
    \rotcol{ Metric } &%
    \rotcol{ Statistic } &%
    \rotcol{ Significant Effect Sign } &%
    \rotcol{ Cpus Per Node } &%
    \rotcol{ Num Simels Per Cpu} &%
    \rotcol{ Num Processes} &%
    \rotcol{ Absolute Effect Size} &%
    \rotcol{ Absolute Effect Size 95\% CI Lower Bound} &%
    \rotcol{ Absolute Effect Size 95\% CI Upper Bound} &%
    \rotcol{ Relative Effect Size} &%
    \rotcol{ Relative Effect Size 95\% CI Lower Bound} &%
    \rotcol{ Relative Effect Size 95\% CI Upper Bound} &%
    \rotcol{$\mathbf{n}$} & %
    \rotcol{$\mathbf{p}$} %
    \vspace{-1ex} \\ \midrule\endfirsthead
    \caption*{\tablename{} \thetable{} (cont'd)} \\
    \toprule%
    \multicolumn{1}{@{\hspace{-.5\arrayrulewidth}}l@{\hspace{-.5\arrayrulewidth}}}{} &%
    \rotcol{ Metric } &%
    \rotcol{ Statistic } &%
    \rotcol{ Significant Effect Sign } &%
    \rotcol{ Cpus Per Node } &%
    \rotcol{ Num Simels Per Cpu} &%
    \rotcol{ Num Processes} &%
    \rotcol{ Absolute Effect Size} &%
    \rotcol{ Absolute Effect Size 95\% CI Lower Bound} &%
    \rotcol{ Absolute Effect Size 95\% CI Upper Bound} &%
    \rotcol{ Relative Effect Size} &%
    \rotcol{ Relative Effect Size 95\% CI Lower Bound} &%
    \rotcol{ Relative Effect Size 95\% CI Upper Bound} &%
    \rotcol{$\mathbf{n}$} & %
    \rotcol{$\mathbf{p}$} %
    \vspace{-1ex} \\ \midrule\endhead
    \bottomrule\endfoot,
  late after line=\\\hline,
  late after last line=\\\bottomrule,
  head to column names,
  filter=\equal{\csuse{DependentVariableSlug}}{\metricslug} \and \equal{\csuse{RegressionModelSlug}}{\regressionlongslug},
}

\catcode`\%=12
\providecommand\filename{}
\renewcommand\filename{\detokenize{submodule/conduit-binder/binder/date=2021+project=72k5n/a=weak-scaling/outplots/a=weak_scaling_regression_results+async_mode=3+comp_work=0+impl=proc+nproc=16_64_256+nthread=1+proc=0-255+readability=latexcsvreader+replicate=0-9+slurm_nnodes=64_256+slurm_ntasks=256+subject=inlet_outlet+view=summary+ext=.csv}}
\catcode`\%=14

\csvreader[myTableStyle]{\importpath\filename}{}%
{ %
  &%
  \csuse{DependentVariable} &%
  \csuse{Statistic} &%
  \ifthenelse{\equal{\csuse{SignificantEffectSign}}{-}}{\cellcolor{red!25}}{}%
  \ifthenelse{\equal{\csuse{SignificantEffectSign}}{+}}{\cellcolor{green!25}}{}%
  \ifthenelse{\equal{\csuse{SignificantEffectSign}}{0}}{\cellcolor{gray!25}}{}%
  \csuse{SignificantEffectSign} &%
  \replacesinglequote{\csuse{CpusPerNode}} &%
  \replacesinglequote{\csuse{NumSimelsPerCpu}} &%
  \replacesinglequote{\csuse{NumProcessesPrettyprint}} &%
  \replacesinglequote{\csuse{AbsoluteEffectSize}} &%
  \replacesinglequote{\csuse{AbsoluteEffectSize95CILowerBound}} &%
  \replacesinglequote{\csuse{AbsoluteEffectSize95CIUpperBound}} &%
  \replacesinglequote{\csuse{RelativeEffectSize}} &%
  \replacesinglequote{\csuse{RelativeEffectSize95CILowerBound}} &%
  \replacesinglequote{\csuse{RelativeEffectSize95CIUpperBound}} &%
  \replacesinglequote{\csuse{n}} &%
  \replacesinglequote{\csuse{p}} %
}

\end{landscape}

}

\clearpage

\def\regression{%
Ordinary Least Squares Regression%
}\def\regressionslug{%
regression-ols%
}\def\regressionlongslug{%
ordinary-least-squares-regression%
}\def\metric{%
Latency Simsteps Outlet%
}\def\metricslug{%
latency-simsteps-outlet%
}

\def\regression{%
Ordinary Least Squares Regression%
}\def\regressionslug{%
regression-ols%
}\def\regressionlongslug{%
ordinary-least-squares-regression%
}\def\metric{%
Latency Walltime Outlet (ns)%
}\def\metricslug{%
latency-walltime-outlet-ns%
}

\def\regression{%
Ordinary Least Squares Regression%
}\def\regressionslug{%
regression-ols%
}\def\regressionlongslug{%
ordinary-least-squares-regression%
}\def\metric{%
Delivery Clumpiness%
}\def\metricslug{%
delivery-clumpiness%
}

\def\regression{%
Ordinary Least Squares Regression%
}\def\regressionslug{%
regression-ols%
}\def\regressionlongslug{%
ordinary-least-squares-regression%
}\def\metric{%
Simstep Period Inlet (ns)%
}\def\metricslug{%
simstep-period-inlet-ns%
}

\def\regression{%
Ordinary Least Squares Regression%
}\def\regressionslug{%
regression-ols%
}\def\regressionlongslug{%
ordinary-least-squares-regression%
}\def\metric{%
Latency Simsteps Inlet%
}\def\metricslug{%
latency-simsteps-inlet%
}

\def\regression{%
Ordinary Least Squares Regression%
}\def\regressionslug{%
regression-ols%
}\def\regressionlongslug{%
ordinary-least-squares-regression%
}\def\metric{%
Simstep Period Outlet (ns)%
}\def\metricslug{%
simstep-period-outlet-ns%
}

\def\regression{%
Ordinary Least Squares Regression%
}\def\regressionslug{%
regression-ols%
}\def\regressionlongslug{%
ordinary-least-squares-regression%
}\def\metric{%
Delivery Failure Rate%
}\def\metricslug{%
delivery-failure-rate%
}

\def\regression{%
Quantile Regression%
}\def\regressionslug{%
regression-quantile%
}\def\regressionlongslug{%
quantile-regression%
}\def\metric{%
Latency Walltime Inlet (ns)%
}\def\metricslug{%
latency-walltime-inlet-ns%
}

\def\regression{%
Quantile Regression%
}\def\regressionslug{%
regression-quantile%
}\def\regressionlongslug{%
quantile-regression%
}\def\metric{%
Latency Simsteps Outlet%
}\def\metricslug{%
latency-simsteps-outlet%
}

\def\regression{%
Quantile Regression%
}\def\regressionslug{%
regression-quantile%
}\def\regressionlongslug{%
quantile-regression%
}\def\metric{%
Latency Walltime Outlet (ns)%
}\def\metricslug{%
latency-walltime-outlet-ns%
}

\def\regression{%
Quantile Regression%
}\def\regressionslug{%
regression-quantile%
}\def\regressionlongslug{%
quantile-regression%
}\def\metric{%
Delivery Clumpiness%
}\def\metricslug{%
delivery-clumpiness%
}

\def\regression{%
Quantile Regression%
}\def\regressionslug{%
regression-quantile%
}\def\regressionlongslug{%
quantile-regression%
}\def\metric{%
Simsteps Period Inlet (ns)%
}\def\metricslug{%
simstep-period-inlet-ns%
}

\def\regression{%
Quantile Regression%
}\def\regressionslug{%
regression-quantile%
}\def\regressionlongslug{%
quantile-regression%
}\def\metric{%
Latency Simsteps Inlet%
}\def\metricslug{%
latency-simsteps-inlet%
}

\def\regression{%
Quantile Regression%
}\def\regressionslug{%
regression-quantile%
}\def\regressionlongslug{%
quantile-regression%
}\def\metric{%
Simstep Period Outlet (ns)%
}\def\metricslug{%
simstep-period-outlet-ns%
}

\def\regression{%
Quantile Regression%
}\def\regressionslug{%
regression-quantile%
}\def\regressionlongslug{%
quantile-regression%
}\def\metric{%
Delivery Failure Rate%
}\def\metricslug{%
delivery-failure-rate%
}

\clearpage

\subsection{Computation vs. Communication}

This section provides full results from computation vs. communication experiments discussed in Section \ref{sec:computation-vs-communication}.

\providecommand{\figcomputationvscommunicationdistributionmacro}[2]{
\def\metric{#1}\def\metricslug{#2}\begin{figure*}[h]
  \centering
  \begin{subfigure}[b]{0.5\textwidth}
    \centering
    \includegraphics[width=\textwidth]{submodule/conduit-binder/binder/date=2021+project=72k5n/a=computation-vs-communication/teeplots/\metricslug/col=cpus-per-node+row=num-simels-per-cpu+transform=snapshot_diffs+viz=facet-boxplot-nofliers+x=log-compute-work+y=\metricslug+ext=}
    \caption{Distribution of \metric{} for each snapshot, without outliers.}
    \label{fig:computation-vs-communication-distribution-\metricslug-nofliers}
  \end{subfigure}%
  \begin{subfigure}[b]{0.5\textwidth}
    \centering
    \includegraphics[width=\textwidth]{submodule/conduit-binder/binder/date=2021+project=72k5n/a=computation-vs-communication/teeplots/\metricslug/col=cpus-per-node+row=num-simels-per-cpu+transform=snapshot_diffs+viz=facet-boxplot-withfliers+x=log-compute-work+y=\metricslug+ext=}
    \caption{Distribution of \metric{} for each snapshot, with outliers.}
    \label{fig:computation-vs-communication-distribution-\metricslug-withfliers}
  \end{subfigure}
  \caption{Distribution of \metric{} for individual snapshot measurements for computation vs. communication experiment (Section \ref{sec:computation-vs-communication}). Lower is better.}
  \label{fig:computation-vs-communication-distribution-\metricslug}
\end{figure*}

}

\def\metric{Latency Walltime Inlet (ns)}\def\metricslug{latency-walltime-inlet-ns}\begin{figure*}[h]
  \centering
  \begin{subfigure}[b]{0.5\textwidth}
    \centering
    \includegraphics[width=\textwidth]{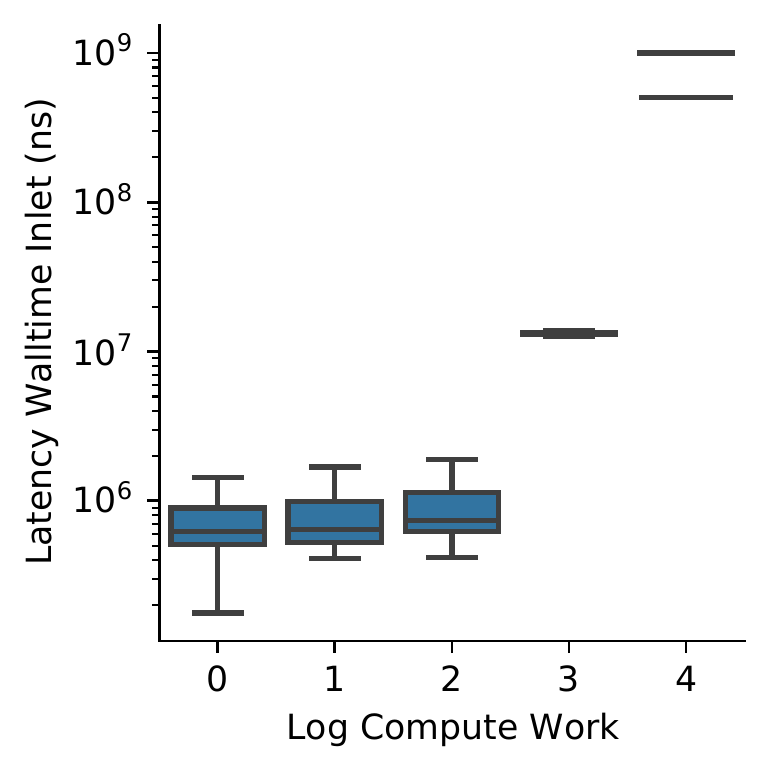}
    \caption{Distribution of \metric{} for each snapshot, without outliers.}
    \label{fig:computation-vs-communication-distribution-\metricslug-nofliers}
  \end{subfigure}%
  \begin{subfigure}[b]{0.5\textwidth}
    \centering
    \includegraphics[width=\textwidth]{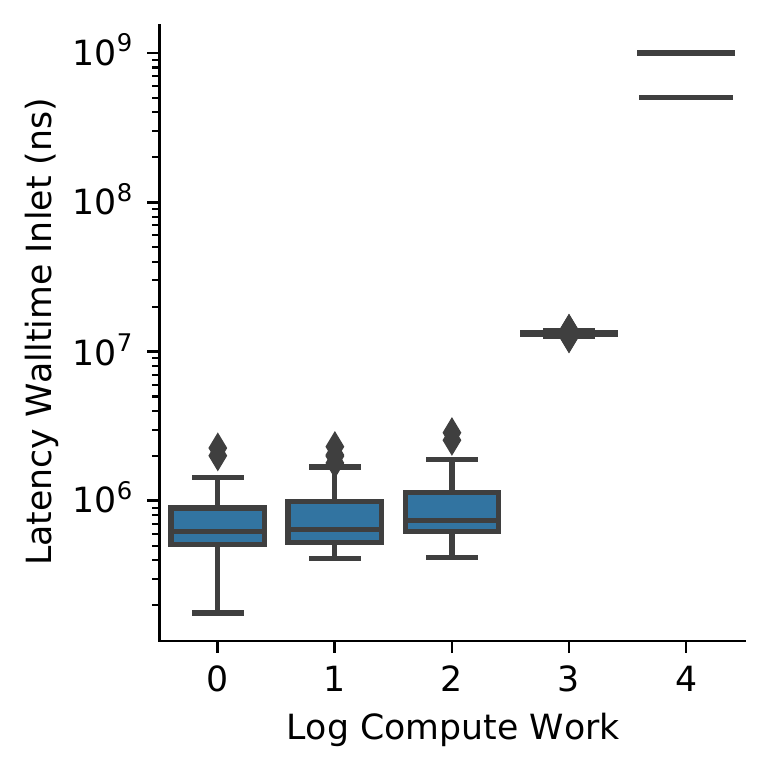}
    \caption{Distribution of \metric{} for each snapshot, with outliers.}
    \label{fig:computation-vs-communication-distribution-\metricslug-withfliers}
  \end{subfigure}
  \caption{Distribution of \metric{} for individual snapshot measurements for computation vs. communication experiment (Section \ref{sec:computation-vs-communication}). Lower is better.}
  \label{fig:computation-vs-communication-distribution-\metricslug}
\end{figure*}

\def\metric{Latency Simsteps Outlet}\def\metricslug{latency-simsteps-outlet}\begin{figure*}[h]
  \centering
  \begin{subfigure}[b]{0.5\textwidth}
    \centering
    \includegraphics[width=\textwidth]{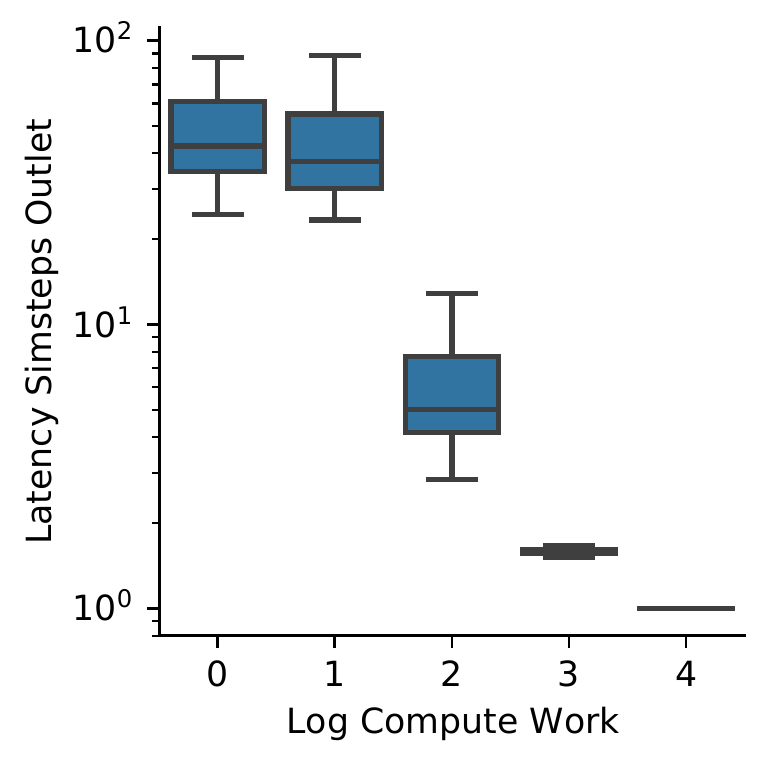}
    \caption{Distribution of \metric{} for each snapshot, without outliers.}
    \label{fig:computation-vs-communication-distribution-\metricslug-nofliers}
  \end{subfigure}%
  \begin{subfigure}[b]{0.5\textwidth}
    \centering
    \includegraphics[width=\textwidth]{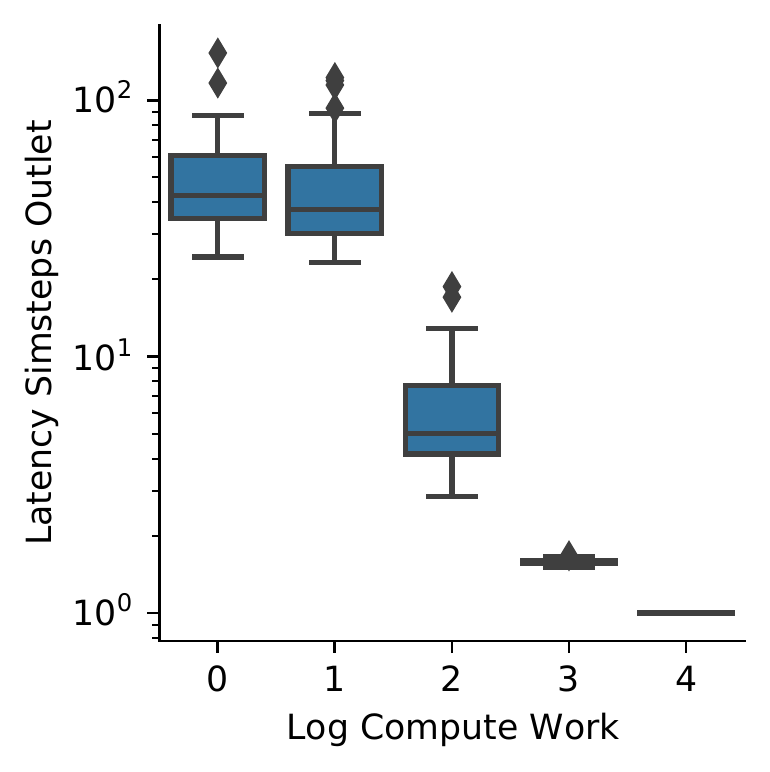}
    \caption{Distribution of \metric{} for each snapshot, with outliers.}
    \label{fig:computation-vs-communication-distribution-\metricslug-withfliers}
  \end{subfigure}
  \caption{Distribution of \metric{} for individual snapshot measurements for computation vs. communication experiment (Section \ref{sec:computation-vs-communication}). Lower is better.}
  \label{fig:computation-vs-communication-distribution-\metricslug}
\end{figure*}

\def\metric{Latency Walltime Outlet (ns)}\def\metricslug{latency-walltime-outlet-ns}\begin{figure*}[h]
  \centering
  \begin{subfigure}[b]{0.5\textwidth}
    \centering
    \includegraphics[width=\textwidth]{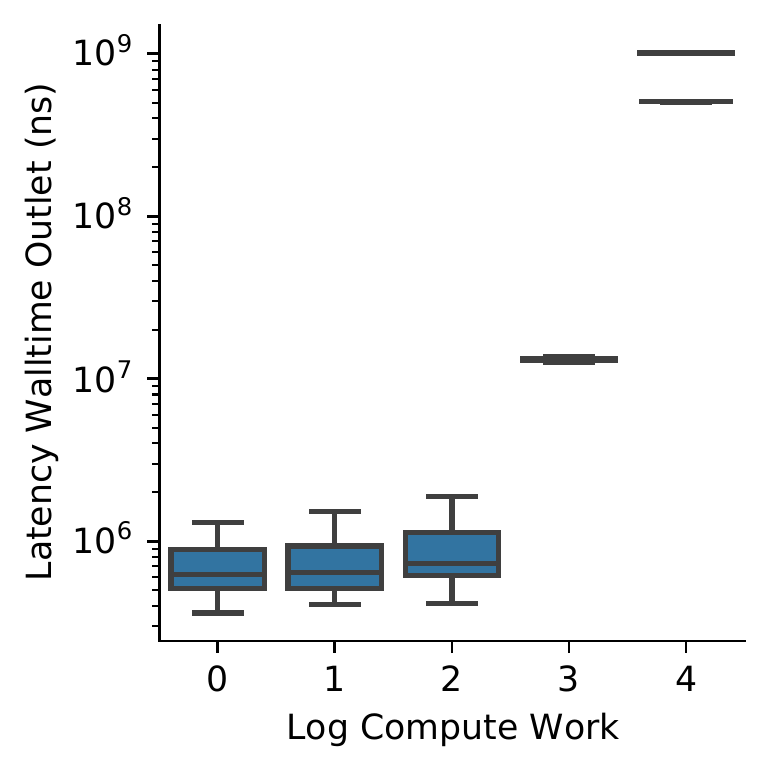}
    \caption{Distribution of \metric{} for each snapshot, without outliers.}
    \label{fig:computation-vs-communication-distribution-\metricslug-nofliers}
  \end{subfigure}%
  \begin{subfigure}[b]{0.5\textwidth}
    \centering
    \includegraphics[width=\textwidth]{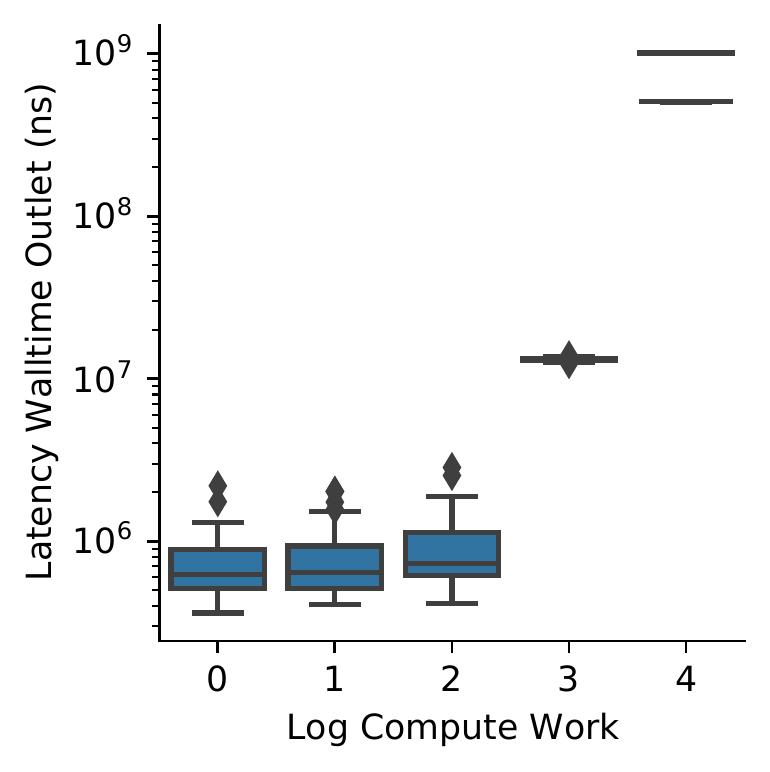}
    \caption{Distribution of \metric{} for each snapshot, with outliers.}
    \label{fig:computation-vs-communication-distribution-\metricslug-withfliers}
  \end{subfigure}
  \caption{Distribution of \metric{} for individual snapshot measurements for computation vs. communication experiment (Section \ref{sec:computation-vs-communication}). Lower is better.}
  \label{fig:computation-vs-communication-distribution-\metricslug}
\end{figure*}

\def\metric{Delivery Clumpiness}\def\metricslug{delivery-clumpiness}\begin{figure*}[h]
  \centering
  \begin{subfigure}[b]{0.5\textwidth}
    \centering
    \includegraphics[width=\textwidth]{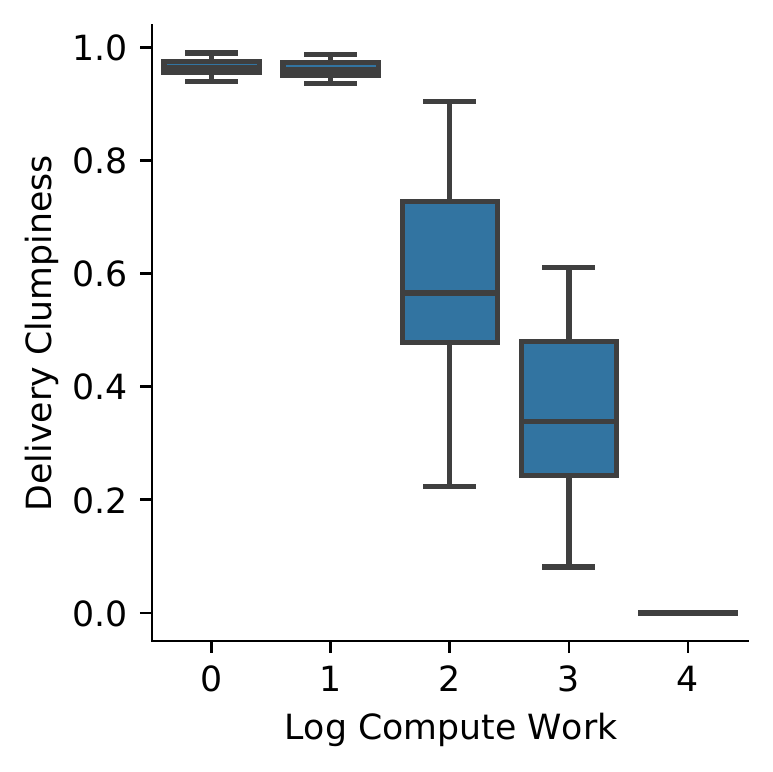}
    \caption{Distribution of \metric{} for each snapshot, without outliers.}
    \label{fig:computation-vs-communication-distribution-\metricslug-nofliers}
  \end{subfigure}%
  \begin{subfigure}[b]{0.5\textwidth}
    \centering
    \includegraphics[width=\textwidth]{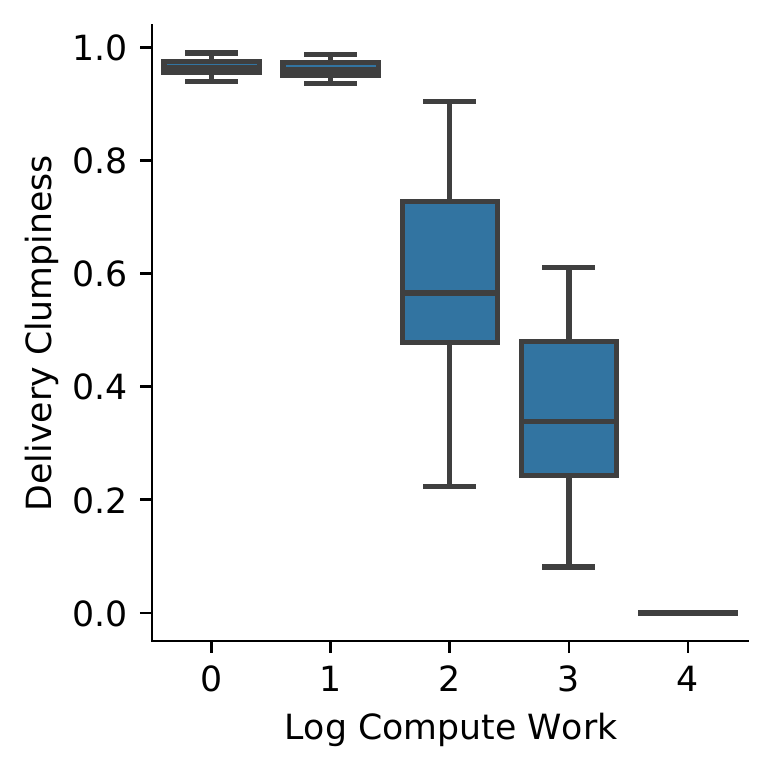}
    \caption{Distribution of \metric{} for each snapshot, with outliers.}
    \label{fig:computation-vs-communication-distribution-\metricslug-withfliers}
  \end{subfigure}
  \caption{Distribution of \metric{} for individual snapshot measurements for computation vs. communication experiment (Section \ref{sec:computation-vs-communication}). Lower is better.}
  \label{fig:computation-vs-communication-distribution-\metricslug}
\end{figure*}

\def\metric{Simstep Period Inlet (ns)}\def\metricslug{simstep-period-inlet-ns}\begin{figure*}[h]
  \centering
  \begin{subfigure}[b]{0.5\textwidth}
    \centering
    \includegraphics[width=\textwidth]{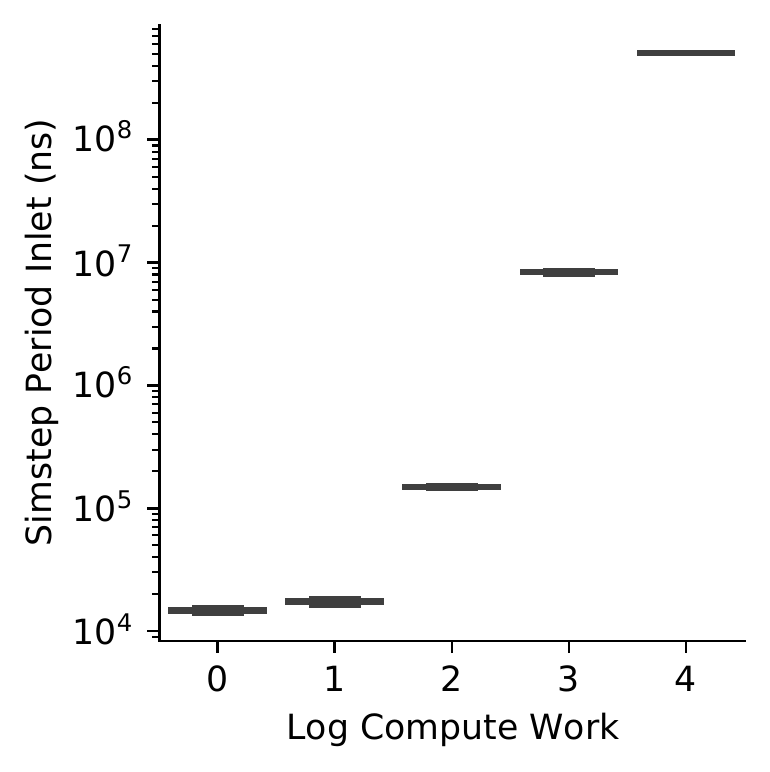}
    \caption{Distribution of \metric{} for each snapshot, without outliers.}
    \label{fig:computation-vs-communication-distribution-\metricslug-nofliers}
  \end{subfigure}%
  \begin{subfigure}[b]{0.5\textwidth}
    \centering
    \includegraphics[width=\textwidth]{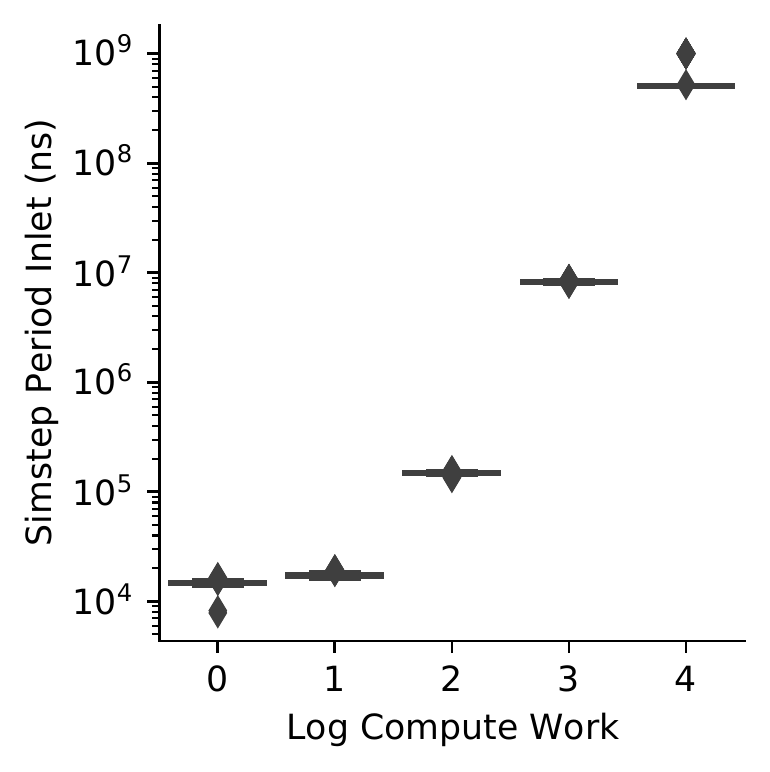}
    \caption{Distribution of \metric{} for each snapshot, with outliers.}
    \label{fig:computation-vs-communication-distribution-\metricslug-withfliers}
  \end{subfigure}
  \caption{Distribution of \metric{} for individual snapshot measurements for computation vs. communication experiment (Section \ref{sec:computation-vs-communication}). Lower is better.}
  \label{fig:computation-vs-communication-distribution-\metricslug}
\end{figure*}

\def\metric{Latency Simsteps Inlet}\def\metricslug{latency-simsteps-inlet}\begin{figure*}[h]
  \centering
  \begin{subfigure}[b]{0.5\textwidth}
    \centering
    \includegraphics[width=\textwidth]{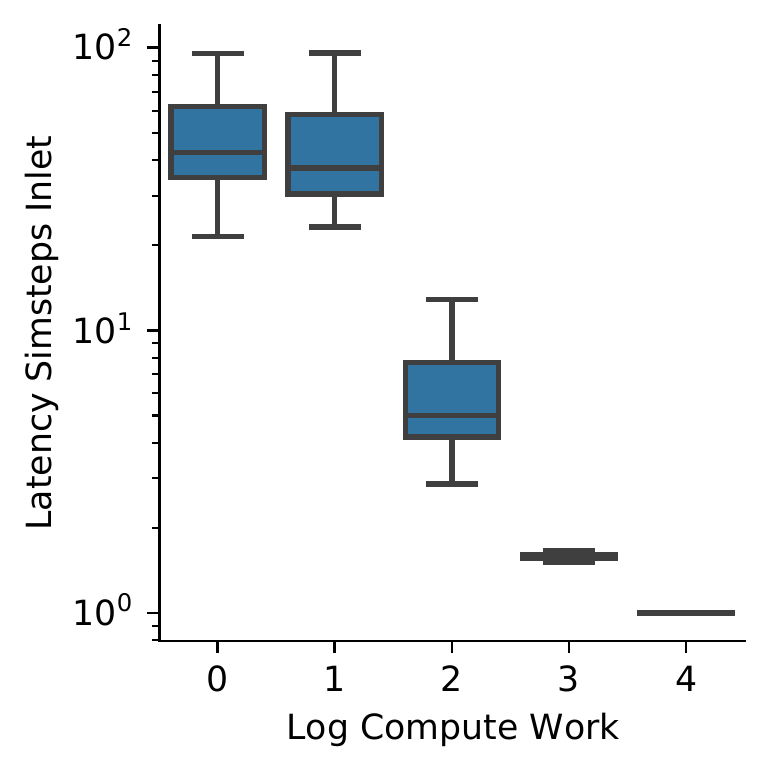}
    \caption{Distribution of \metric{} for each snapshot, without outliers.}
    \label{fig:computation-vs-communication-distribution-\metricslug-nofliers}
  \end{subfigure}%
  \begin{subfigure}[b]{0.5\textwidth}
    \centering
    \includegraphics[width=\textwidth]{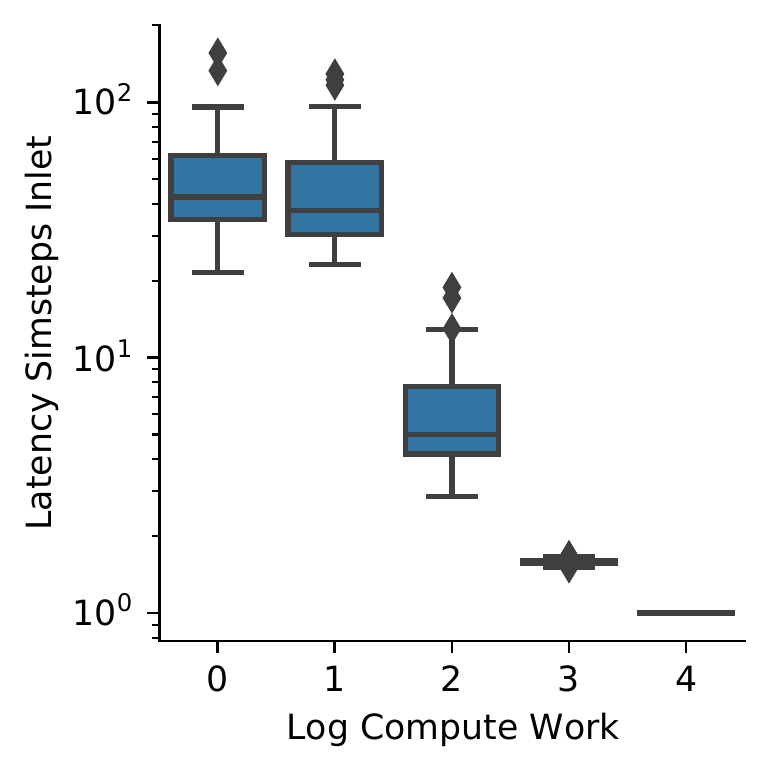}
    \caption{Distribution of \metric{} for each snapshot, with outliers.}
    \label{fig:computation-vs-communication-distribution-\metricslug-withfliers}
  \end{subfigure}
  \caption{Distribution of \metric{} for individual snapshot measurements for computation vs. communication experiment (Section \ref{sec:computation-vs-communication}). Lower is better.}
  \label{fig:computation-vs-communication-distribution-\metricslug}
\end{figure*}

\def\metric{Simstep Period Outlet (ns)}\def\metricslug{simstep-period-outlet-ns}\begin{figure*}[h]
  \centering
  \begin{subfigure}[b]{0.5\textwidth}
    \centering
    \includegraphics[width=\textwidth]{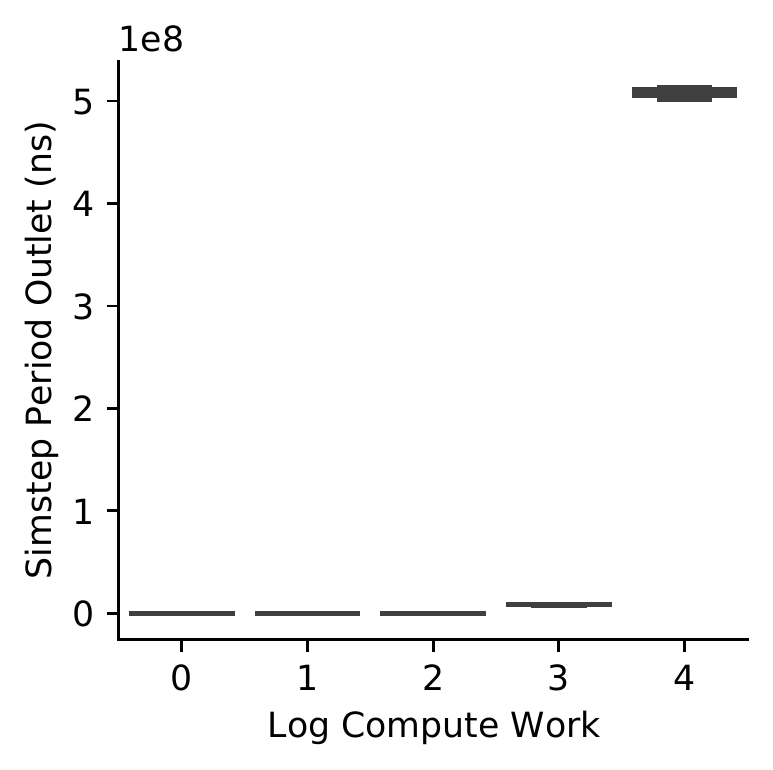}
    \caption{Distribution of \metric{} for each snapshot, without outliers.}
    \label{fig:computation-vs-communication-distribution-\metricslug-nofliers}
  \end{subfigure}%
  \begin{subfigure}[b]{0.5\textwidth}
    \centering
    \includegraphics[width=\textwidth]{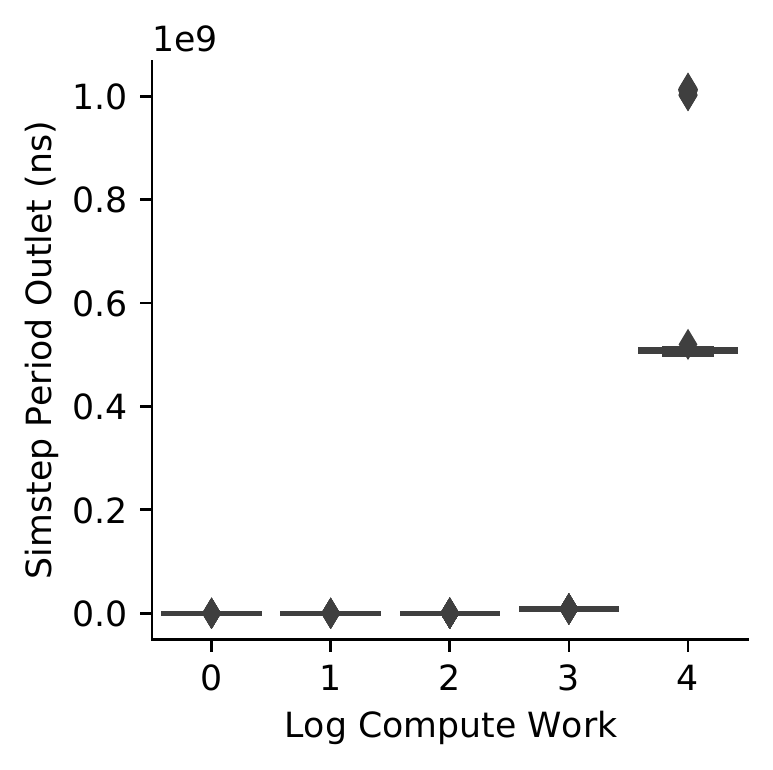}
    \caption{Distribution of \metric{} for each snapshot, with outliers.}
    \label{fig:computation-vs-communication-distribution-\metricslug-withfliers}
  \end{subfigure}
  \caption{Distribution of \metric{} for individual snapshot measurements for computation vs. communication experiment (Section \ref{sec:computation-vs-communication}). Lower is better.}
  \label{fig:computation-vs-communication-distribution-\metricslug}
\end{figure*}

\def\metric{Delivery Failure Rate}\def\metricslug{delivery-failure-rate}\begin{figure*}[h]
  \centering
  \begin{subfigure}[b]{0.5\textwidth}
    \centering
    \includegraphics[width=\textwidth]{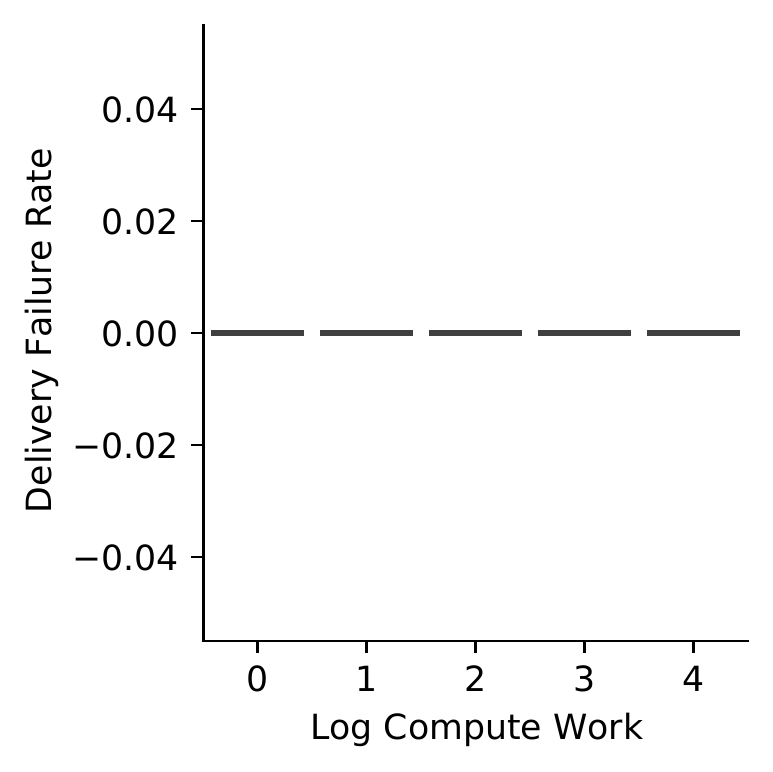}
    \caption{Distribution of \metric{} for each snapshot, without outliers.}
    \label{fig:computation-vs-communication-distribution-\metricslug-nofliers}
  \end{subfigure}%
  \begin{subfigure}[b]{0.5\textwidth}
    \centering
    \includegraphics[width=\textwidth]{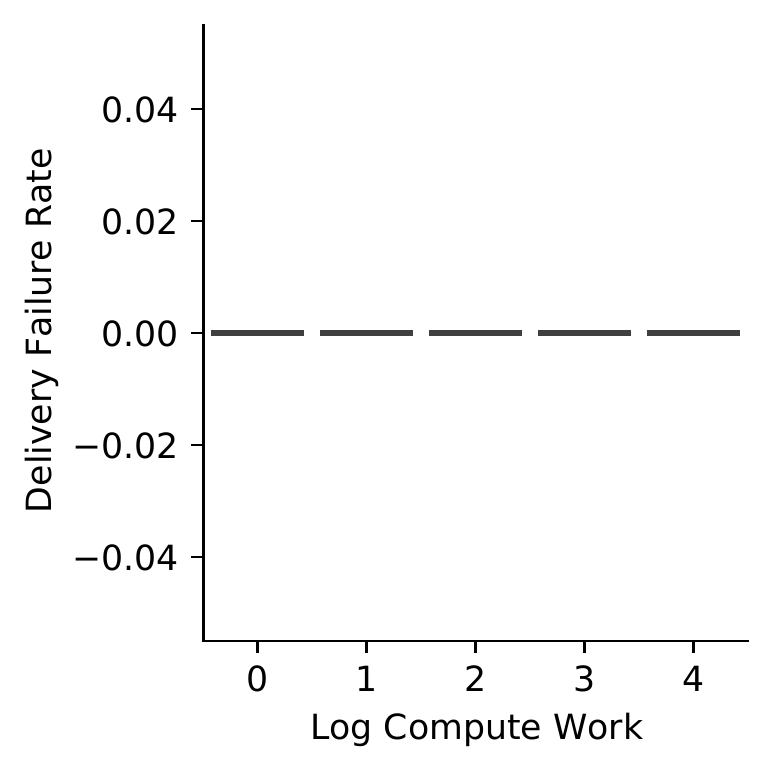}
    \caption{Distribution of \metric{} for each snapshot, with outliers.}
    \label{fig:computation-vs-communication-distribution-\metricslug-withfliers}
  \end{subfigure}
  \caption{Distribution of \metric{} for individual snapshot measurements for computation vs. communication experiment (Section \ref{sec:computation-vs-communication}). Lower is better.}
  \label{fig:computation-vs-communication-distribution-\metricslug}
\end{figure*}

\providecommand{\figcomputationvscommunicationregressionmacro}[4]{
\def\metric{#1}\def\metricslug{#2}\def\xvarslug{#3}\def\yvarprefix{#4}\begin{figure*}[h]
  \centering

  \begin{subfigure}[b]{0.5\textwidth}
    \centering
    \includegraphics[width=\textwidth]{submodule/conduit-binder/binder/date=2021+project=72k5n/a=computation-vs-communication/teeplots/\metricslug/col=cpus-per-node+marker=+regression=ordinary-least-squares-regression+row=num-simels-per-cpu+transform=snapshot_diffs-groupby_exec_instance-mean+viz=facet-unsplit-regression+x=\xvarslug+y=\yvarprefix\metricslug+ext=}
    \caption{
    Ordinary least squares regression plot.
    Observations are means per replicate.
    }
    \label{fig:computation-vs-communication-regression-ols-\metricslug-complete-regression}
  \end{subfigure}%
  \begin{subfigure}[b]{0.5\textwidth}
    \centering
    \includegraphics[width=\textwidth]{submodule/conduit-binder/binder/date=2021+project=72k5n/a=computation-vs-communication/teeplots/\metricslug/col=cpus-per-node+estimated-statistic=\metricslug-mean+row=num-simels-per-cpu+transform=fit_regression+viz=lambda+y=absolute-effect-size+ext=}
    \caption{Estimated regression coefficient for ordinary least squares regression. Zero corresponds to no effect.}
    \label{fig:computation-vs-communication-regression-ols-\metricslug-complete-effect-size}
  \end{subfigure}

  \begin{subfigure}[b]{0.5\textwidth}
    \centering
    \includegraphics[width=\textwidth]{submodule/conduit-binder/binder/date=2021+project=72k5n/a=computation-vs-communication/teeplots/\metricslug/col=cpus-per-node+marker=+regression=quantile-regression+row=num-simels-per-cpu+transform=snapshot_diffs-groupby_exec_instance-median+viz=facet-unsplit-regression+x=\xvarslug+y=\yvarprefix\metricslug+ext=}
    \caption{
      Quantile regression plot.
      Observations are medians per replicate.
    }
    \label{fig:computation-vs-communication-regression-quantile-\metricslug-complete-regression}
  \end{subfigure}%
  \begin{subfigure}[b]{0.5\textwidth}
    \centering
    \includegraphics[width=\textwidth]{submodule/conduit-binder/binder/date=2021+project=72k5n/a=computation-vs-communication/teeplots/\metricslug/col=cpus-per-node+estimated-statistic=\metricslug-median+row=num-simels-per-cpu+transform=fit_regression+viz=lambda+y=absolute-effect-size+ext=}
    \caption{Estimated regression coefficient for quantile regression. Zero corresponds to no effect.}
    \label{fig:computation-vs-communication-regression-quantile-\metricslug-complete-effect-size}
  \end{subfigure}

  \caption{
  Regressions of \metric{} against log computational intensity for computation vs. communication experiment (Section \ref{sec:computation-vs-communication}).
  Lower is better.
  Ordinary least squares regression (top row) estimates relationship between dependent variable and mean of response variable.
  Quantile regression (bottom row) estimates relationship between independent variable and median of response variable.
  Error bands and bars are 95\% confidence intervals.
  }
  \label{fig:computation-vs-communication-regression-\metricslug}
\end{figure*}

}

\def\metric{Latency Walltime Inlet (ns)}\def\metricslug{latency-walltime-inlet-ns}\def\xvarslug{log-compute-work}\def\yvarprefix{log-}\begin{figure*}[h]
  \centering

  \begin{subfigure}[b]{0.5\textwidth}
    \centering
    \includegraphics[width=\textwidth]{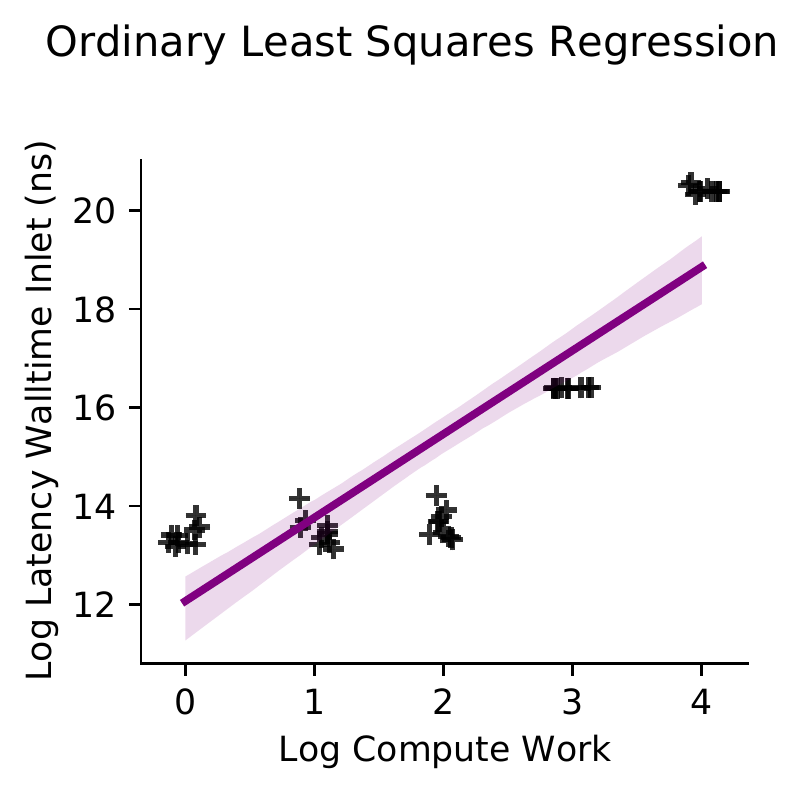}
    \caption{
    Ordinary least squares regression plot.
    Observations are means per replicate.
    }
    \label{fig:computation-vs-communication-regression-ols-\metricslug-complete-regression}
  \end{subfigure}%
  \begin{subfigure}[b]{0.5\textwidth}
    \centering
    \includegraphics[width=\textwidth]{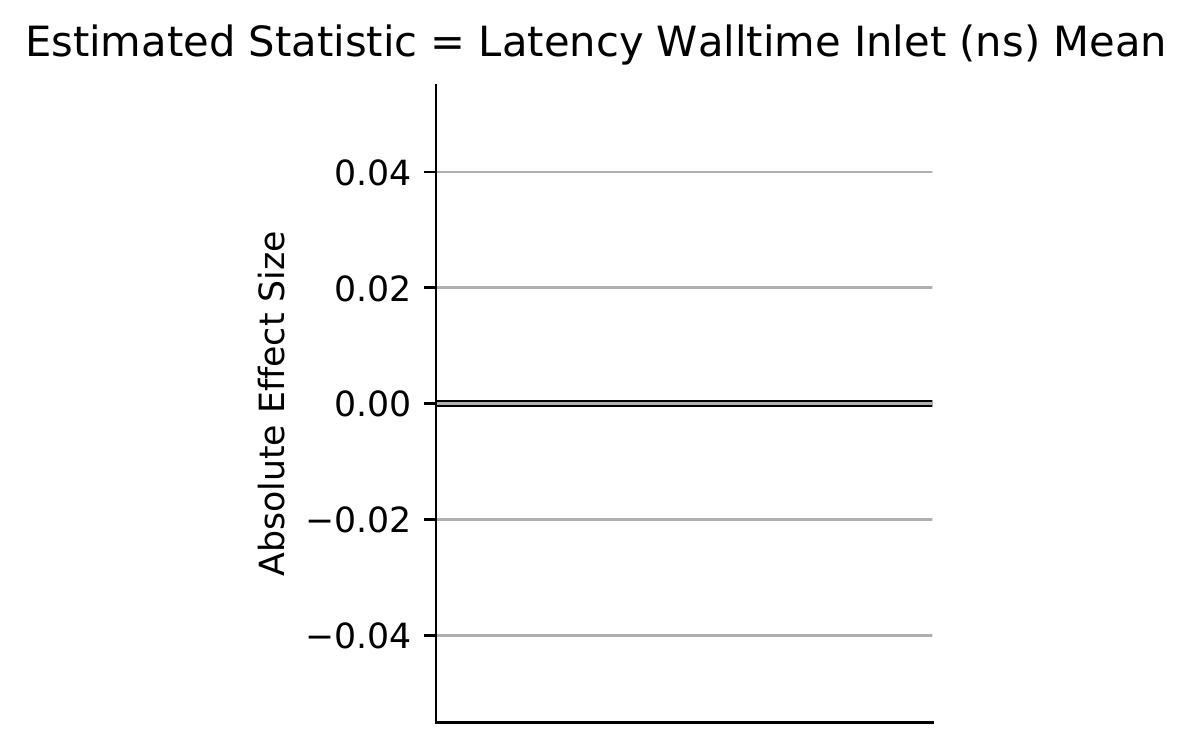}
    \caption{Estimated regression coefficient for ordinary least squares regression. Zero corresponds to no effect.}
    \label{fig:computation-vs-communication-regression-ols-\metricslug-complete-effect-size}
  \end{subfigure}

  \begin{subfigure}[b]{0.5\textwidth}
    \centering
    \includegraphics[width=\textwidth]{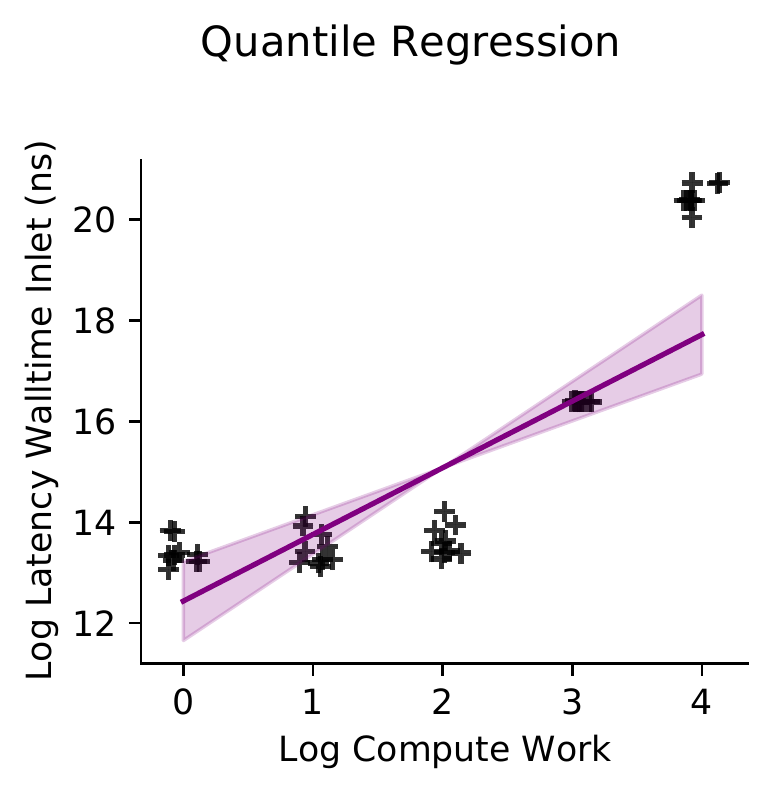}
    \caption{
      Quantile regression plot.
      Observations are medians per replicate.
    }
    \label{fig:computation-vs-communication-regression-quantile-\metricslug-complete-regression}
  \end{subfigure}%
  \begin{subfigure}[b]{0.5\textwidth}
    \centering
    \includegraphics[width=\textwidth]{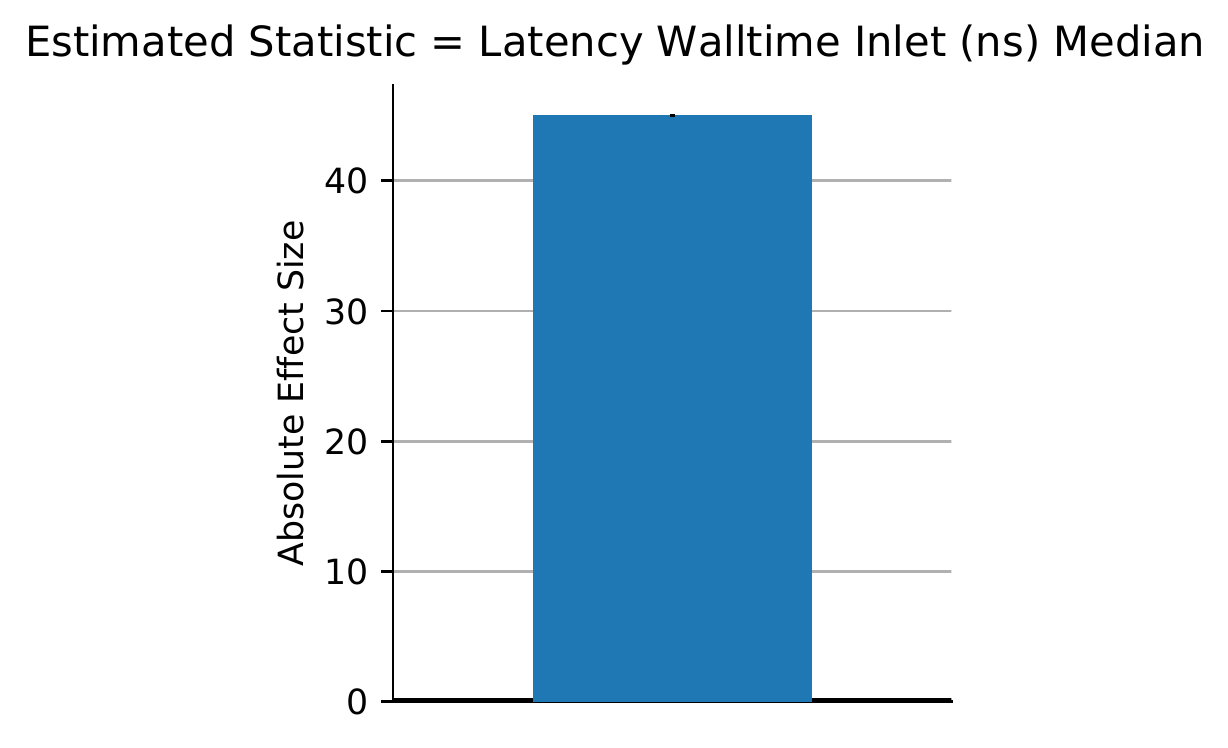}
    \caption{Estimated regression coefficient for quantile regression. Zero corresponds to no effect.}
    \label{fig:computation-vs-communication-regression-quantile-\metricslug-complete-effect-size}
  \end{subfigure}

  \caption{
  Regressions of \metric{} against log computational intensity for computation vs. communication experiment (Section \ref{sec:computation-vs-communication}).
  Lower is better.
  Ordinary least squares regression (top row) estimates relationship between dependent variable and mean of response variable.
  Quantile regression (bottom row) estimates relationship between independent variable and median of response variable.
  Error bands and bars are 95\% confidence intervals.
  }
  \label{fig:computation-vs-communication-regression-\metricslug}
\end{figure*}

\def\metric{Latency Simsteps Outlet}\def\metricslug{latency-simsteps-outlet}\def\xvarslug{log-compute-work}\def\yvarprefix{log-}

\def\metric{Latency Walltime Outlet (ns)}\def\metricslug{latency-walltime-outlet-ns}\def\xvarslug{log-compute-work}\def\yvarprefix{log-}

\def\metric{Delivery Clumpiness}\def\metricslug{delivery-clumpiness}\def\xvarslug{log-compute-work}\def\yvarprefix{}

\def\metric{Simstep Period Inlet (ns)}\def\metricslug{simstep-period-inlet-ns}\def\xvarslug{log-compute-work}\def\yvarprefix{log-}

\def\metric{Latency Simsteps Inlet}\def\metricslug{latency-simsteps-inlet}\def\xvarslug{log-compute-work}\def\yvarprefix{log-}

\def\metric{Simstep Period Outlet (ns)}\def\metricslug{simstep-period-outlet-ns}\def\xvarslug{log-compute-work}\def\yvarprefix{log-}

\def\metric{Delivery Failure Rate}\def\metricslug{delivery-failure-rate}\def\xvarslug{log-compute-work}\def\yvarprefix{}

\providecommand{\tabcomputationvscommunicationregressionmacro}[2]{
\def\regression{#1}\def\regressionslug{#2}

\providecommand{\dissertationelse}[2]{%
\ifdefined\DISSERTATION
#1
\else
#2
\fi
}

\makeatletter
  \providecommand\importpath{\import@path}
\makeatother

\pragmaonce

\def\replacesinglequote#1{%
  \IfSubStr{#1}{'}{%
    \StrSubstitute{#1}{'}{\,}}{#1}}

\pragmaonce

\providecommand\rotcol[1]{ \multicolumn{1}{@{}@{}}{} \adjustbox{angle=60,lap=\width}{\rule{0.4\textwidth}{0.8pt}} \hfill \makebox[3ex]{\adjustbox{angle=60,lap=\width}{\bfseries \hspace{1ex} #1}} \hfill }


\clearpage
{
\onecolumn

\begin{landscape}
\dissertationelse{\fontsize{8}{9}\selectfont}{\footnotesize}
\csvstyle{myTableStyle}{
  longtable=@{\hspace{-.5\arrayrulewidth}}c@{\hspace{-.5\arrayrulewidth}}|l|l|l|l|l|l|l|l|l|l|l|l|l|l|,
  table head=\caption{
    Full \regression{} results of quality of service metrics against log computational intensity for computation vs. communication experiment (Section \ref{sec:computation-vs-communication}).
    Significance level $p < 0.05$ used.
    Inf or NaN values may occur due to multicollinearity or due to inf or NaN observations.
    }\label{tab:computation-vs-communication-\regressionslug} \\
    \toprule%
    \multicolumn{1}{@{\hspace{-.5\arrayrulewidth}}l@{\hspace{-.5\arrayrulewidth}}}{} &%
    \rotcol{ Metric } &%
    \rotcol{ Statistic } &%
    \rotcol{ Significant Effect Sign } &%
    \rotcol{ Cpus Per Node } &%
    \rotcol{ Num Simels Per Cpu} &%
    \rotcol{ Num Processes} &%
    \rotcol{ Absolute Effect Size} &%
    \rotcol{ Absolute Effect Size 95\% CI Lower Bound} &%
    \rotcol{ Absolute Effect Size 95\% CI Upper Bound} &%
    \rotcol{ Relative Effect Size} &%
    \rotcol{ Relative Effect Size 95\% CI Lower Bound} &%
    \rotcol{ Relative Effect Size 95\% CI Upper Bound} &%
    \rotcol{$\mathbf{n}$} & %
    \rotcol{$\mathbf{p}$} %
    \vspace{-1ex} \\ \midrule\endfirsthead
    \caption*{\tablename{} \thetable{} (cont'd)} \\
    \toprule%
    \multicolumn{1}{@{\hspace{-.5\arrayrulewidth}}l@{\hspace{-.5\arrayrulewidth}}}{} &%
    \rotcol{ Metric } &%
    \rotcol{ Statistic } &%
    \rotcol{ Significant Effect Sign } &%
    \rotcol{ Cpus Per Node } &%
    \rotcol{ Num Simels Per Cpu} &%
    \rotcol{ Num Processes} &%
    \rotcol{ Absolute Effect Size} &%
    \rotcol{ Absolute Effect Size 95\% CI Lower Bound} &%
    \rotcol{ Absolute Effect Size 95\% CI Upper Bound} &%
    \rotcol{ Relative Effect Size} &%
    \rotcol{ Relative Effect Size 95\% CI Lower Bound} &%
    \rotcol{ Relative Effect Size 95\% CI Upper Bound} &%
    \rotcol{$\mathbf{n}$} & %
    \rotcol{$\mathbf{p}$} %
    \vspace{-1ex} \\ \midrule\endhead
    \bottomrule\endfoot,
  late after line=\\\hline,
  late after last line=\\\bottomrule,
  head to column names,
  filter=\equal{\csuse{RegressionModelSlug}}{\regressionslug},
}

\catcode`\%=12
\providecommand\filename{}
\renewcommand\filename{\detokenize{submodule/conduit-binder/binder/date=2021+project=72k5n/a=computation-vs-communication/outplots/a=computation_vs_communication_regression_results+async_mode=3+comp_work=0-64_64_4096_262144_16777216+impl=proc+nproc=2+nthread=1+proc=0-1+readability=latexcsvreader+replicate=0-9+slurm_nnodes=2+slurm_ntasks=2+subject=inlet_outlet+view=summary+ext=.csv}}
\catcode`\%=14

\csvreader[myTableStyle]{\importpath\filename}{}%
{ %
  &%
  \csuse{DependentVariable} &%
  \csuse{Statistic} &%
  \ifthenelse{\equal{\csuse{SignificantEffectSign}}{-}}{\cellcolor{red!25}}{}%
  \ifthenelse{\equal{\csuse{SignificantEffectSign}}{+}}{\cellcolor{green!25}}{}%
  \ifthenelse{\equal{\csuse{SignificantEffectSign}}{0}}{\cellcolor{gray!25}}{}%
  \csuse{SignificantEffectSign} &%
  \replacesinglequote{\csuse{CpusPerNode}} &%
  \replacesinglequote{\csuse{NumSimelsPerCpu}} &%
  \replacesinglequote{\csuse{NumProcessesPrettyprint}} &%
  \replacesinglequote{\csuse{AbsoluteEffectSize}} &%
  \replacesinglequote{\csuse{AbsoluteEffectSize95CILowerBound}} &%
  \replacesinglequote{\csuse{AbsoluteEffectSize95CIUpperBound}} &%
  \replacesinglequote{\csuse{RelativeEffectSize}} &%
  \replacesinglequote{\csuse{RelativeEffectSize95CILowerBound}} &%
  \replacesinglequote{\csuse{RelativeEffectSize95CIUpperBound}} &%
  \replacesinglequote{\csuse{n}} &%
  \replacesinglequote{\csuse{p}} %
}

\end{landscape}

\clearpage
}

}

\def\regression{Ordinary Least Squares Regression}\def\regressionslug{ordinary-least-squares-regression}

\providecommand{\dissertationelse}[2]{%
\ifdefined\DISSERTATION
#1
\else
#2
\fi
}

\makeatletter
  \providecommand\importpath{\import@path}
\makeatother

\pragmaonce

\def\replacesinglequote#1{%
  \IfSubStr{#1}{'}{%
    \StrSubstitute{#1}{'}{\,}}{#1}}

\pragmaonce

\providecommand\rotcol[1]{ \multicolumn{1}{@{}@{}}{} \adjustbox{angle=60,lap=\width}{\rule{0.4\textwidth}{0.8pt}} \hfill \makebox[3ex]{\adjustbox{angle=60,lap=\width}{\bfseries \hspace{1ex} #1}} \hfill }


\clearpage
{
\onecolumn

\begin{landscape}
\dissertationelse{\fontsize{8}{9}\selectfont}{\footnotesize}
\csvstyle{myTableStyle}{
  longtable=@{\hspace{-.5\arrayrulewidth}}c@{\hspace{-.5\arrayrulewidth}}|l|l|l|l|l|l|l|l|l|l|l|l|l|l|,
  table head=\caption{
    Full \regression{} results of quality of service metrics against log computational intensity for computation vs. communication experiment (Section \ref{sec:computation-vs-communication}).
    Significance level $p < 0.05$ used.
    Inf or NaN values may occur due to multicollinearity or due to inf or NaN observations.
    }\label{tab:computation-vs-communication-\regressionslug} \\
    \toprule%
    \multicolumn{1}{@{\hspace{-.5\arrayrulewidth}}l@{\hspace{-.5\arrayrulewidth}}}{} &%
    \rotcol{ Metric } &%
    \rotcol{ Statistic } &%
    \rotcol{ Significant Effect Sign } &%
    \rotcol{ Cpus Per Node } &%
    \rotcol{ Num Simels Per Cpu} &%
    \rotcol{ Num Processes} &%
    \rotcol{ Absolute Effect Size} &%
    \rotcol{ Absolute Effect Size 95\% CI Lower Bound} &%
    \rotcol{ Absolute Effect Size 95\% CI Upper Bound} &%
    \rotcol{ Relative Effect Size} &%
    \rotcol{ Relative Effect Size 95\% CI Lower Bound} &%
    \rotcol{ Relative Effect Size 95\% CI Upper Bound} &%
    \rotcol{$\mathbf{n}$} & %
    \rotcol{$\mathbf{p}$} %
    \vspace{-1ex} \\ \midrule\endfirsthead
    \caption*{\tablename{} \thetable{} (cont'd)} \\
    \toprule%
    \multicolumn{1}{@{\hspace{-.5\arrayrulewidth}}l@{\hspace{-.5\arrayrulewidth}}}{} &%
    \rotcol{ Metric } &%
    \rotcol{ Statistic } &%
    \rotcol{ Significant Effect Sign } &%
    \rotcol{ Cpus Per Node } &%
    \rotcol{ Num Simels Per Cpu} &%
    \rotcol{ Num Processes} &%
    \rotcol{ Absolute Effect Size} &%
    \rotcol{ Absolute Effect Size 95\% CI Lower Bound} &%
    \rotcol{ Absolute Effect Size 95\% CI Upper Bound} &%
    \rotcol{ Relative Effect Size} &%
    \rotcol{ Relative Effect Size 95\% CI Lower Bound} &%
    \rotcol{ Relative Effect Size 95\% CI Upper Bound} &%
    \rotcol{$\mathbf{n}$} & %
    \rotcol{$\mathbf{p}$} %
    \vspace{-1ex} \\ \midrule\endhead
    \bottomrule\endfoot,
  late after line=\\\hline,
  late after last line=\\\bottomrule,
  head to column names,
  filter=\equal{\csuse{RegressionModelSlug}}{\regressionslug},
}

\catcode`\%=12
\providecommand\filename{}
\renewcommand\filename{\detokenize{submodule/conduit-binder/binder/date=2021+project=72k5n/a=computation-vs-communication/outplots/a=computation_vs_communication_regression_results+async_mode=3+comp_work=0-64_64_4096_262144_16777216+impl=proc+nproc=2+nthread=1+proc=0-1+readability=latexcsvreader+replicate=0-9+slurm_nnodes=2+slurm_ntasks=2+subject=inlet_outlet+view=summary+ext=.csv}}
\catcode`\%=14

\csvreader[myTableStyle]{\importpath\filename}{}%
{ %
  &%
  \csuse{DependentVariable} &%
  \csuse{Statistic} &%
  \ifthenelse{\equal{\csuse{SignificantEffectSign}}{-}}{\cellcolor{red!25}}{}%
  \ifthenelse{\equal{\csuse{SignificantEffectSign}}{+}}{\cellcolor{green!25}}{}%
  \ifthenelse{\equal{\csuse{SignificantEffectSign}}{0}}{\cellcolor{gray!25}}{}%
  \csuse{SignificantEffectSign} &%
  \replacesinglequote{\csuse{CpusPerNode}} &%
  \replacesinglequote{\csuse{NumSimelsPerCpu}} &%
  \replacesinglequote{\csuse{NumProcessesPrettyprint}} &%
  \replacesinglequote{\csuse{AbsoluteEffectSize}} &%
  \replacesinglequote{\csuse{AbsoluteEffectSize95CILowerBound}} &%
  \replacesinglequote{\csuse{AbsoluteEffectSize95CIUpperBound}} &%
  \replacesinglequote{\csuse{RelativeEffectSize}} &%
  \replacesinglequote{\csuse{RelativeEffectSize95CILowerBound}} &%
  \replacesinglequote{\csuse{RelativeEffectSize95CIUpperBound}} &%
  \replacesinglequote{\csuse{n}} &%
  \replacesinglequote{\csuse{p}} %
}

\end{landscape}

\clearpage
}

\def\regression{Quantile Regression}\def\regressionslug{quantile-regression}

\clearpage

\section{Intranode vs Internode}

This section provides full results from intranode vs. internode experiments discussed in Section \ref{sec:intranode-vs-internode}.

\providecommand{\figintranodevsinternodedistributionmacro}[2]{
\def\metric{#1}\def\metricslug{#2}\begin{figure*}[h]
  \centering
  \begin{subfigure}[b]{0.5\textwidth}
    \centering
    \includegraphics[width=\textwidth]{submodule/conduit-binder/binder/date=2021+project=72k5n/a=intranode-vs-internode/teeplots/\metricslug/row=num-simels-per-cpu+transform=snapshot_diffs+viz=facet-boxplot-nofliers+x=0-intranode-1-internode+y=\metricslug+ext=}
    \caption{Distribution of \metric{} for each snapshot, without outliers.}
    \label{fig:intranode-vs-internode-distribution-\metricslug-nofliers}
  \end{subfigure}%
  \begin{subfigure}[b]{0.5\textwidth}
    \centering
    \includegraphics[width=\textwidth]{submodule/conduit-binder/binder/date=2021+project=72k5n/a=intranode-vs-internode/teeplots/\metricslug/row=num-simels-per-cpu+transform=snapshot_diffs+viz=facet-boxplot-withfliers+x=0-intranode-1-internode+y=\metricslug+ext=}
    \caption{Distribution of \metric{} for each snapshot, with outliers.}
    \label{fig:intranode-vs-internode-distribution-\metricslug-withfliers}
  \end{subfigure}
  \caption{Distribution of \metric{} for individual snapshot measurements for intranode vs. internode experiment (Section \ref{sec:intranode-vs-internode}). Lower is better.}
  \label{fig:intranode-vs-internode-distribution-\metricslug}
\end{figure*}

}

\def\metric{Latency Walltime Inlet (ns)}\def\metricslug{latency-walltime-inlet-ns}\begin{figure*}[h]
  \centering
  \begin{subfigure}[b]{0.5\textwidth}
    \centering
    \includegraphics[width=\textwidth]{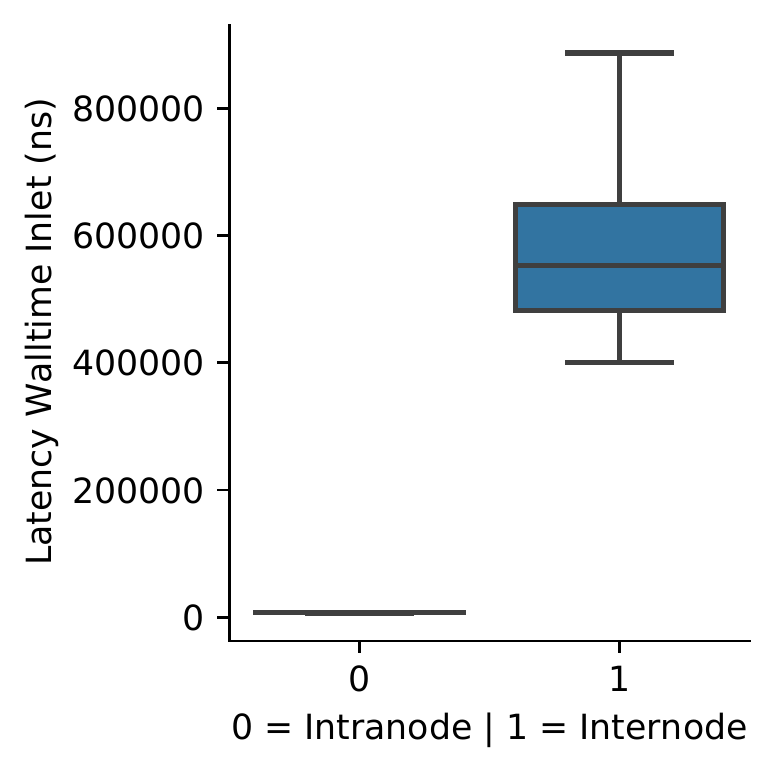}
    \caption{Distribution of \metric{} for each snapshot, without outliers.}
    \label{fig:intranode-vs-internode-distribution-\metricslug-nofliers}
  \end{subfigure}%
  \begin{subfigure}[b]{0.5\textwidth}
    \centering
    \includegraphics[width=\textwidth]{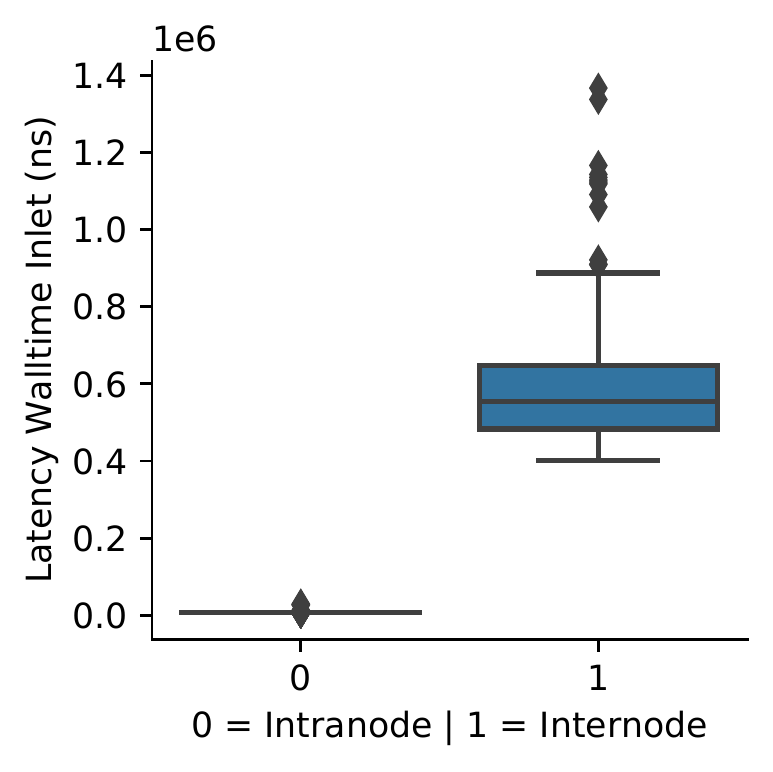}
    \caption{Distribution of \metric{} for each snapshot, with outliers.}
    \label{fig:intranode-vs-internode-distribution-\metricslug-withfliers}
  \end{subfigure}
  \caption{Distribution of \metric{} for individual snapshot measurements for intranode vs. internode experiment (Section \ref{sec:intranode-vs-internode}). Lower is better.}
  \label{fig:intranode-vs-internode-distribution-\metricslug}
\end{figure*}

\def\metric{Latency Simsteps Outlet}\def\metricslug{latency-simsteps-outlet}\begin{figure*}[h]
  \centering
  \begin{subfigure}[b]{0.5\textwidth}
    \centering
    \includegraphics[width=\textwidth]{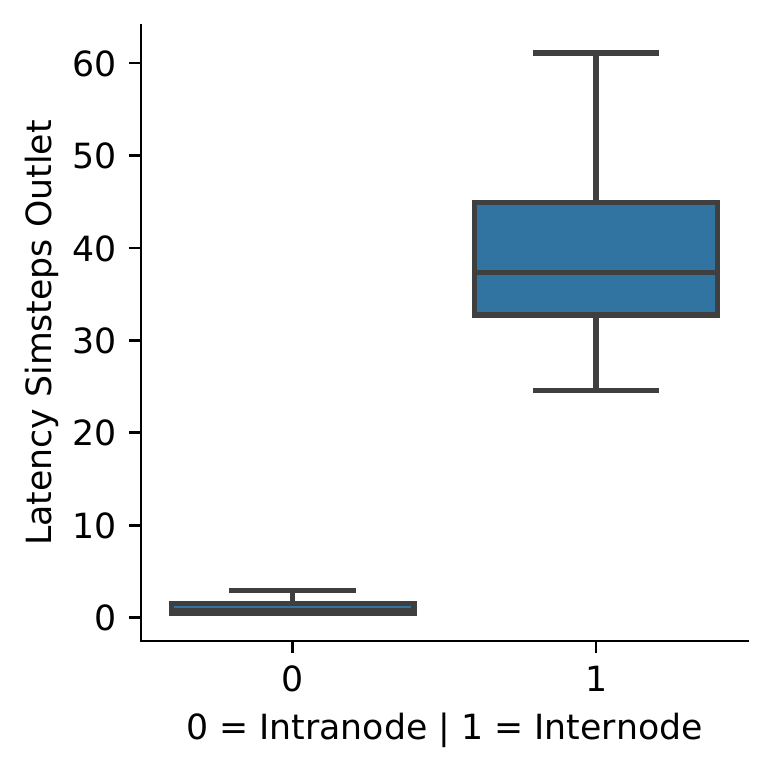}
    \caption{Distribution of \metric{} for each snapshot, without outliers.}
    \label{fig:intranode-vs-internode-distribution-\metricslug-nofliers}
  \end{subfigure}%
  \begin{subfigure}[b]{0.5\textwidth}
    \centering
    \includegraphics[width=\textwidth]{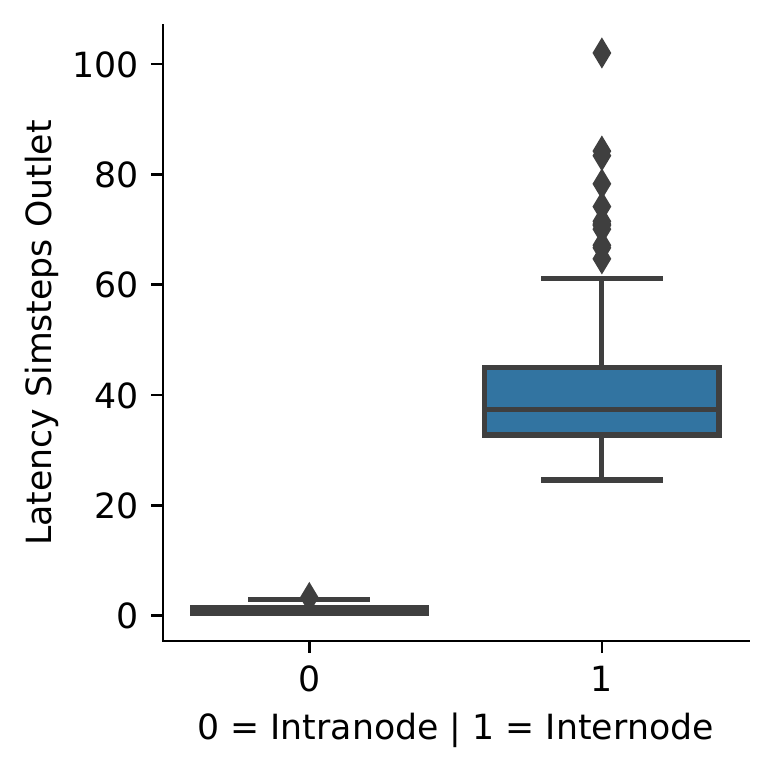}
    \caption{Distribution of \metric{} for each snapshot, with outliers.}
    \label{fig:intranode-vs-internode-distribution-\metricslug-withfliers}
  \end{subfigure}
  \caption{Distribution of \metric{} for individual snapshot measurements for intranode vs. internode experiment (Section \ref{sec:intranode-vs-internode}). Lower is better.}
  \label{fig:intranode-vs-internode-distribution-\metricslug}
\end{figure*}

\def\metric{Latency Walltime Outlet (ns)}\def\metricslug{latency-walltime-outlet-ns}\begin{figure*}[h]
  \centering
  \begin{subfigure}[b]{0.5\textwidth}
    \centering
    \includegraphics[width=\textwidth]{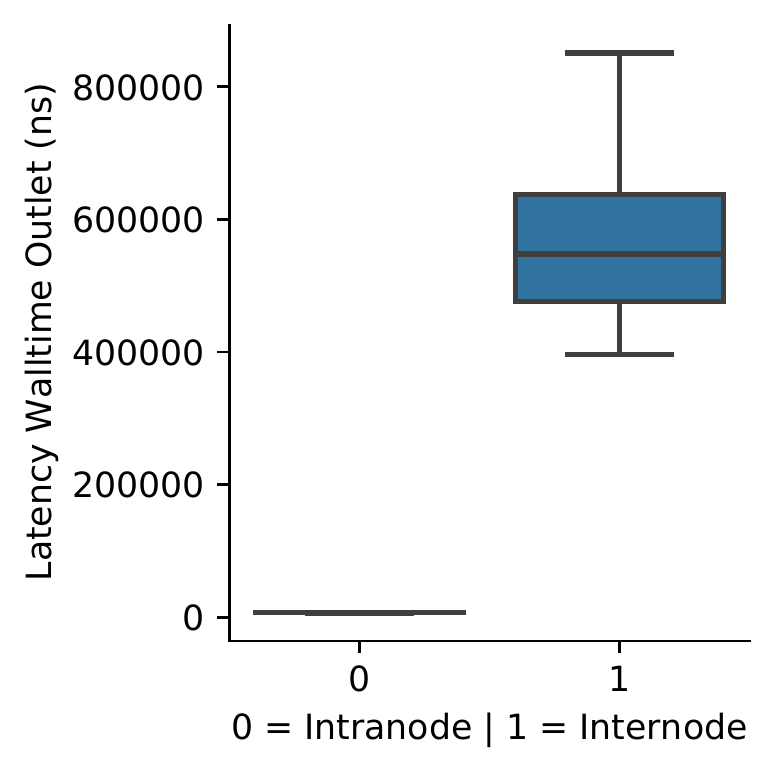}
    \caption{Distribution of \metric{} for each snapshot, without outliers.}
    \label{fig:intranode-vs-internode-distribution-\metricslug-nofliers}
  \end{subfigure}%
  \begin{subfigure}[b]{0.5\textwidth}
    \centering
    \includegraphics[width=\textwidth]{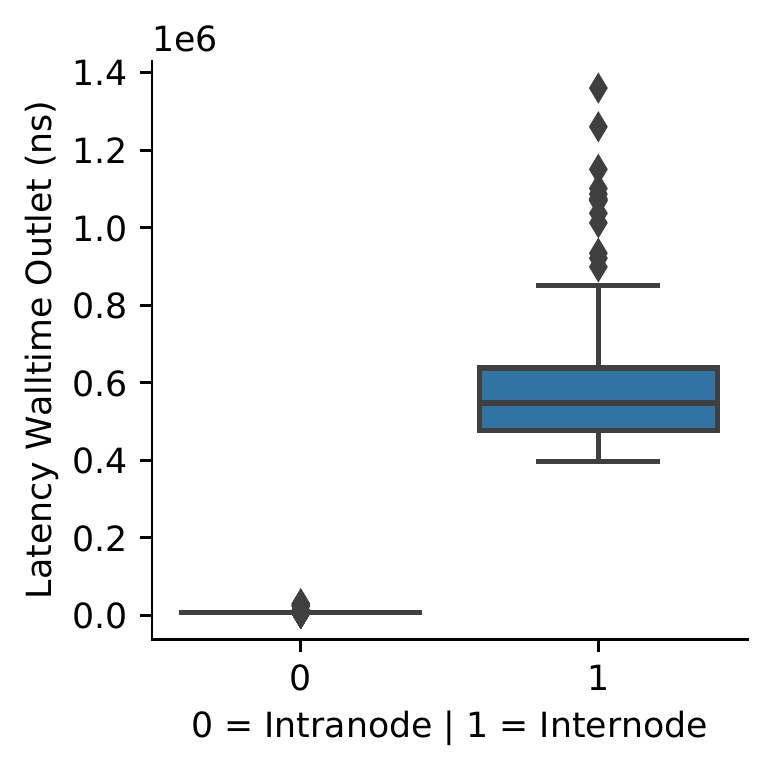}
    \caption{Distribution of \metric{} for each snapshot, with outliers.}
    \label{fig:intranode-vs-internode-distribution-\metricslug-withfliers}
  \end{subfigure}
  \caption{Distribution of \metric{} for individual snapshot measurements for intranode vs. internode experiment (Section \ref{sec:intranode-vs-internode}). Lower is better.}
  \label{fig:intranode-vs-internode-distribution-\metricslug}
\end{figure*}

\def\metric{Delivery Clumpiness}\def\metricslug{delivery-clumpiness}\begin{figure*}[h]
  \centering
  \begin{subfigure}[b]{0.5\textwidth}
    \centering
    \includegraphics[width=\textwidth]{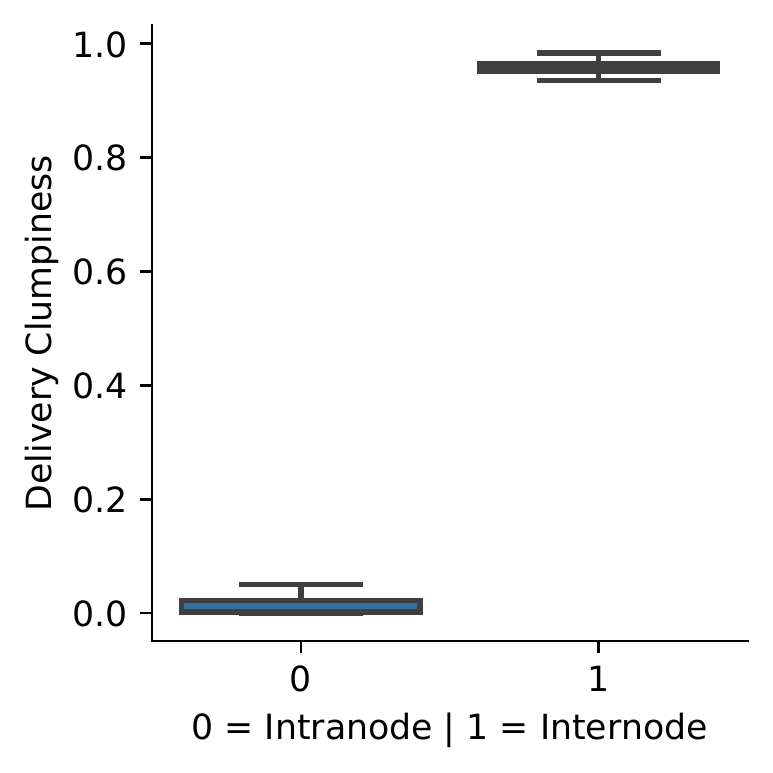}
    \caption{Distribution of \metric{} for each snapshot, without outliers.}
    \label{fig:intranode-vs-internode-distribution-\metricslug-nofliers}
  \end{subfigure}%
  \begin{subfigure}[b]{0.5\textwidth}
    \centering
    \includegraphics[width=\textwidth]{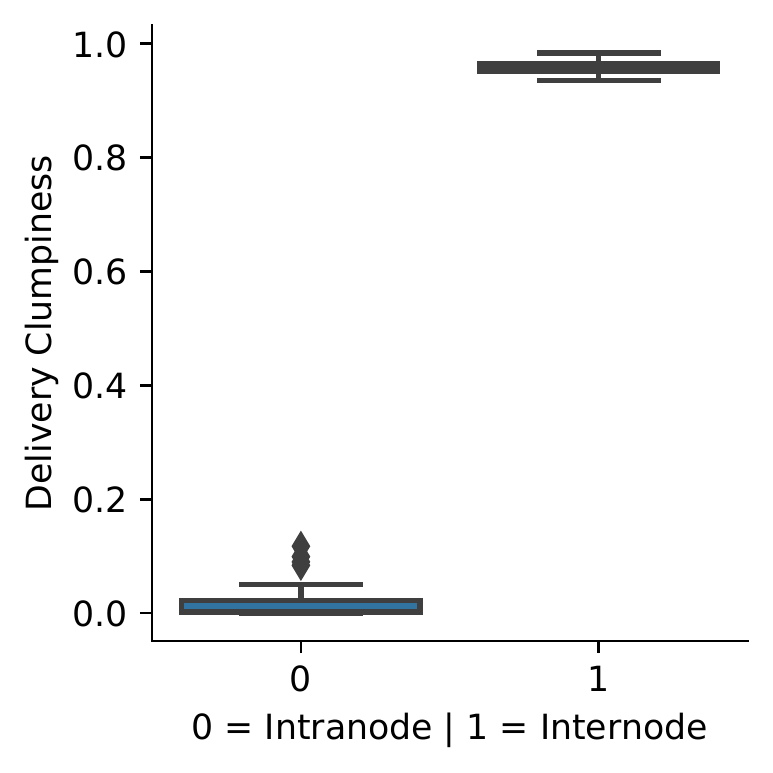}
    \caption{Distribution of \metric{} for each snapshot, with outliers.}
    \label{fig:intranode-vs-internode-distribution-\metricslug-withfliers}
  \end{subfigure}
  \caption{Distribution of \metric{} for individual snapshot measurements for intranode vs. internode experiment (Section \ref{sec:intranode-vs-internode}). Lower is better.}
  \label{fig:intranode-vs-internode-distribution-\metricslug}
\end{figure*}

\def\metric{Simstep Period Inlet (ns)}\def\metricslug{simstep-period-inlet-ns}\begin{figure*}[h]
  \centering
  \begin{subfigure}[b]{0.5\textwidth}
    \centering
    \includegraphics[width=\textwidth]{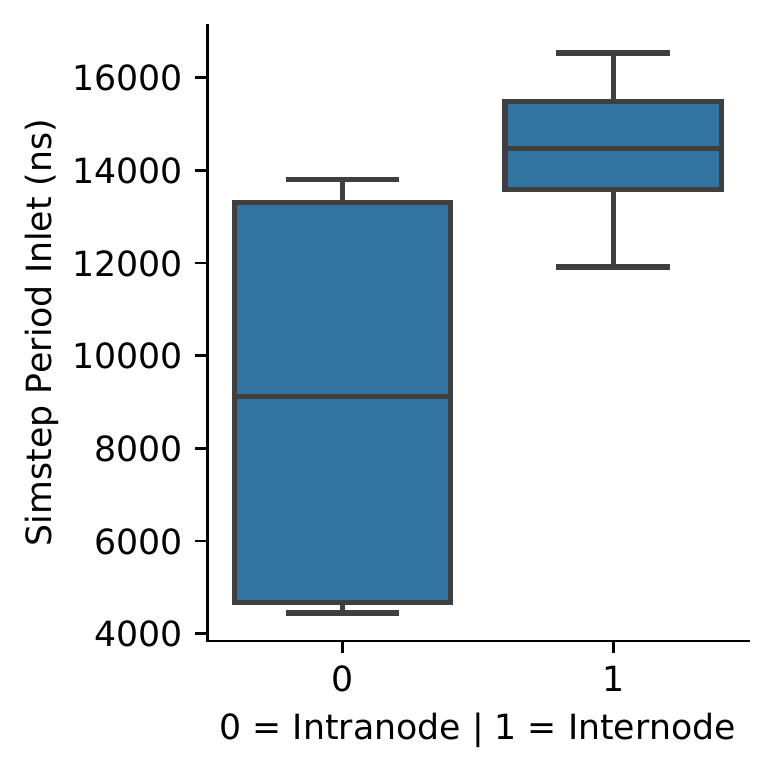}
    \caption{Distribution of \metric{} for each snapshot, without outliers.}
    \label{fig:intranode-vs-internode-distribution-\metricslug-nofliers}
  \end{subfigure}%
  \begin{subfigure}[b]{0.5\textwidth}
    \centering
    \includegraphics[width=\textwidth]{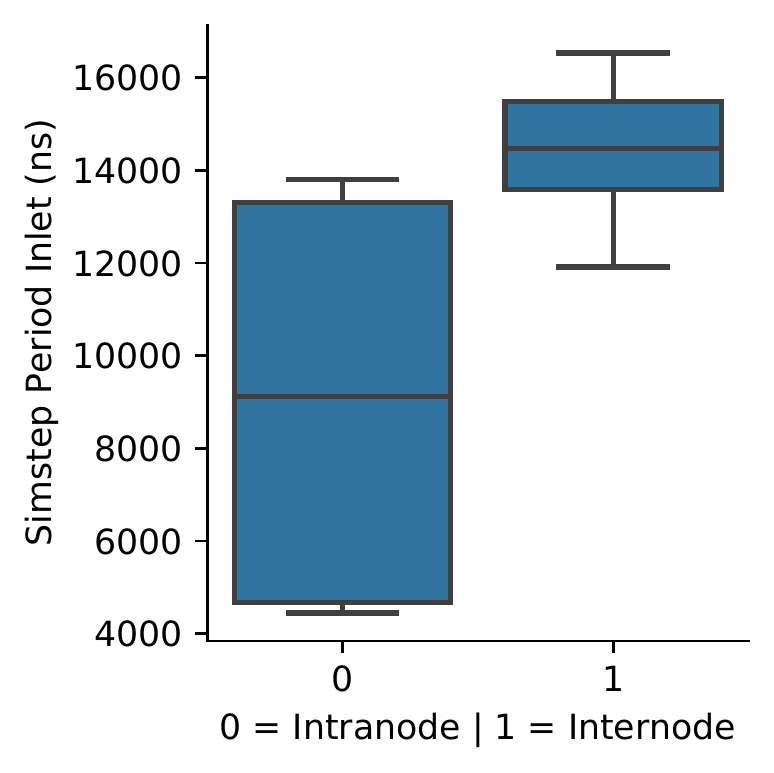}
    \caption{Distribution of \metric{} for each snapshot, with outliers.}
    \label{fig:intranode-vs-internode-distribution-\metricslug-withfliers}
  \end{subfigure}
  \caption{Distribution of \metric{} for individual snapshot measurements for intranode vs. internode experiment (Section \ref{sec:intranode-vs-internode}). Lower is better.}
  \label{fig:intranode-vs-internode-distribution-\metricslug}
\end{figure*}

\def\metric{Latency Simsteps Inlet}\def\metricslug{latency-simsteps-inlet}\begin{figure*}[h]
  \centering
  \begin{subfigure}[b]{0.5\textwidth}
    \centering
    \includegraphics[width=\textwidth]{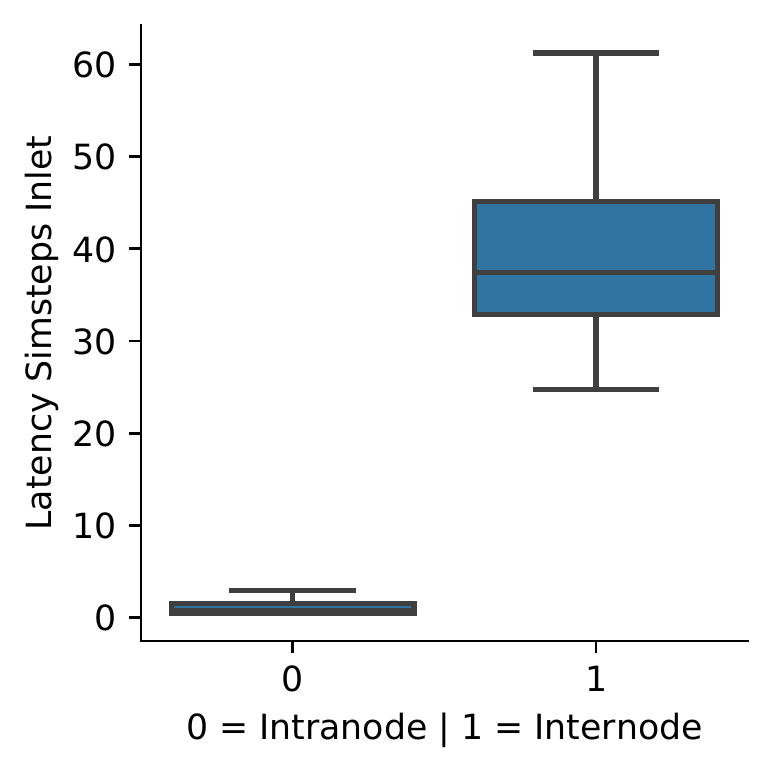}
    \caption{Distribution of \metric{} for each snapshot, without outliers.}
    \label{fig:intranode-vs-internode-distribution-\metricslug-nofliers}
  \end{subfigure}%
  \begin{subfigure}[b]{0.5\textwidth}
    \centering
    \includegraphics[width=\textwidth]{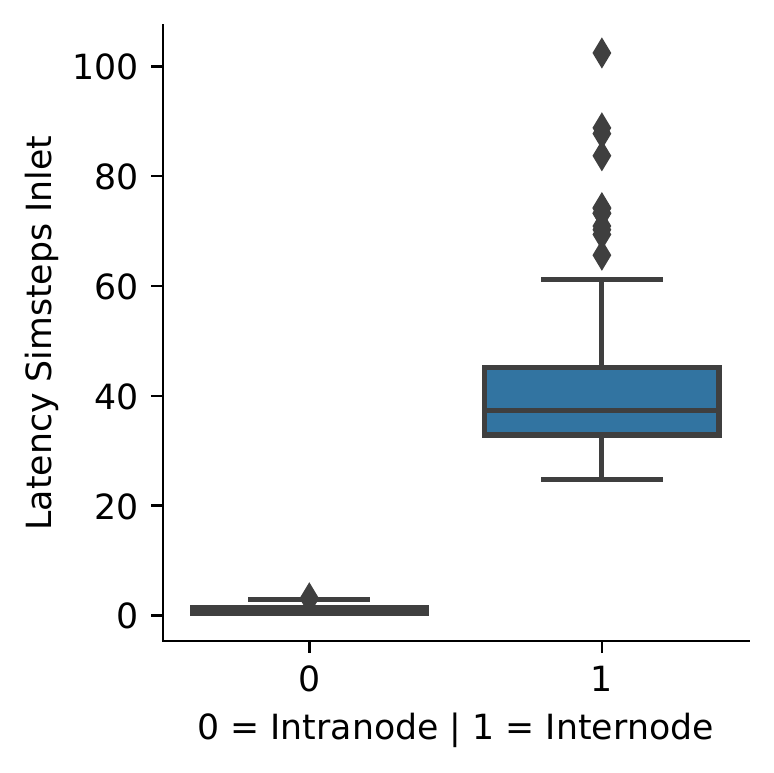}
    \caption{Distribution of \metric{} for each snapshot, with outliers.}
    \label{fig:intranode-vs-internode-distribution-\metricslug-withfliers}
  \end{subfigure}
  \caption{Distribution of \metric{} for individual snapshot measurements for intranode vs. internode experiment (Section \ref{sec:intranode-vs-internode}). Lower is better.}
  \label{fig:intranode-vs-internode-distribution-\metricslug}
\end{figure*}

\def\metric{Simstep Period Outlet (ns)}\def\metricslug{simstep-period-outlet-ns}\begin{figure*}[h]
  \centering
  \begin{subfigure}[b]{0.5\textwidth}
    \centering
    \includegraphics[width=\textwidth]{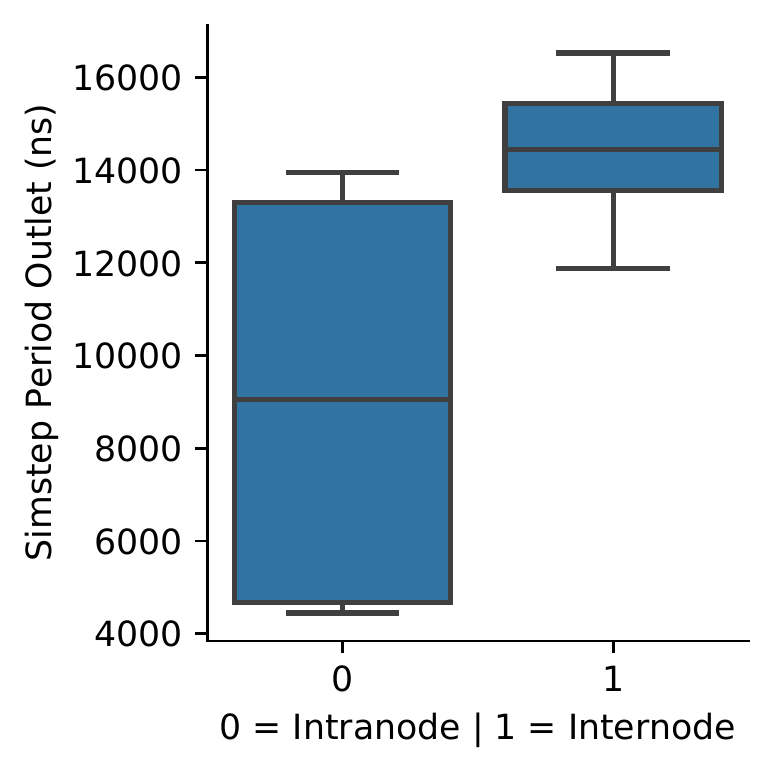}
    \caption{Distribution of \metric{} for each snapshot, without outliers.}
    \label{fig:intranode-vs-internode-distribution-\metricslug-nofliers}
  \end{subfigure}%
  \begin{subfigure}[b]{0.5\textwidth}
    \centering
    \includegraphics[width=\textwidth]{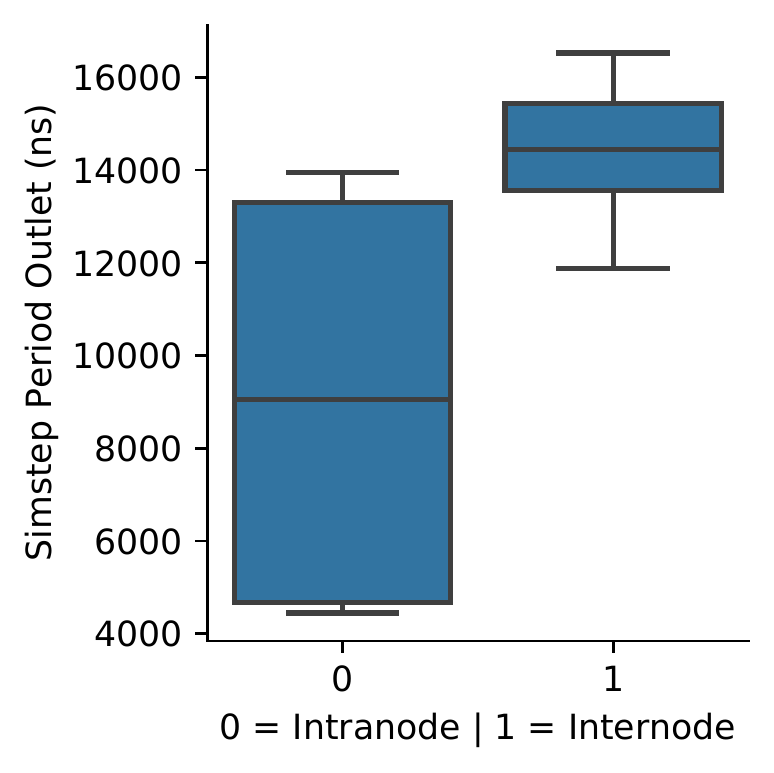}
    \caption{Distribution of \metric{} for each snapshot, with outliers.}
    \label{fig:intranode-vs-internode-distribution-\metricslug-withfliers}
  \end{subfigure}
  \caption{Distribution of \metric{} for individual snapshot measurements for intranode vs. internode experiment (Section \ref{sec:intranode-vs-internode}). Lower is better.}
  \label{fig:intranode-vs-internode-distribution-\metricslug}
\end{figure*}

\def\metric{Delivery Failure Rate}\def\metricslug{delivery-failure-rate}\begin{figure*}[h]
  \centering
  \begin{subfigure}[b]{0.5\textwidth}
    \centering
    \includegraphics[width=\textwidth]{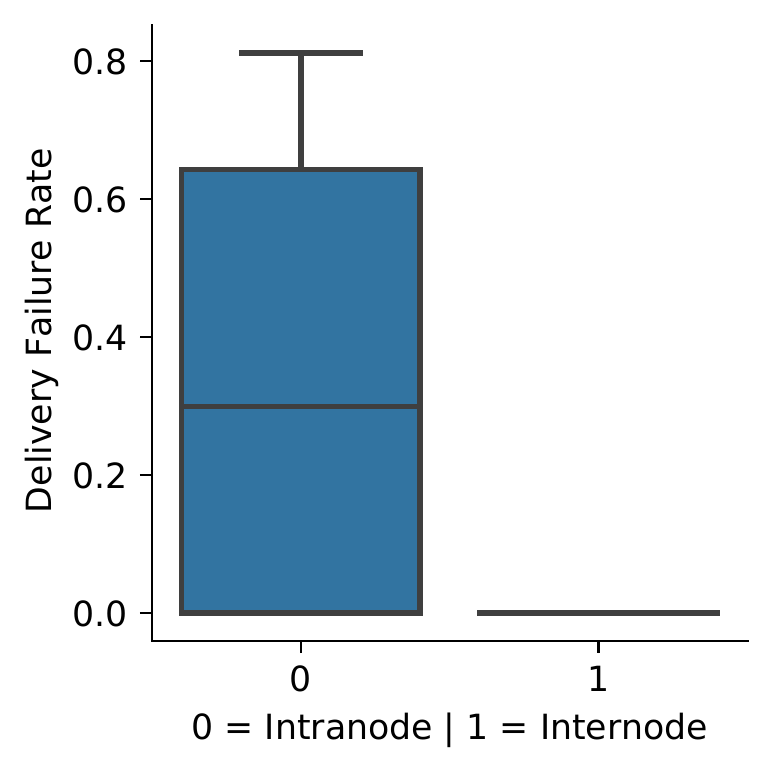}
    \caption{Distribution of \metric{} for each snapshot, without outliers.}
    \label{fig:intranode-vs-internode-distribution-\metricslug-nofliers}
  \end{subfigure}%
  \begin{subfigure}[b]{0.5\textwidth}
    \centering
    \includegraphics[width=\textwidth]{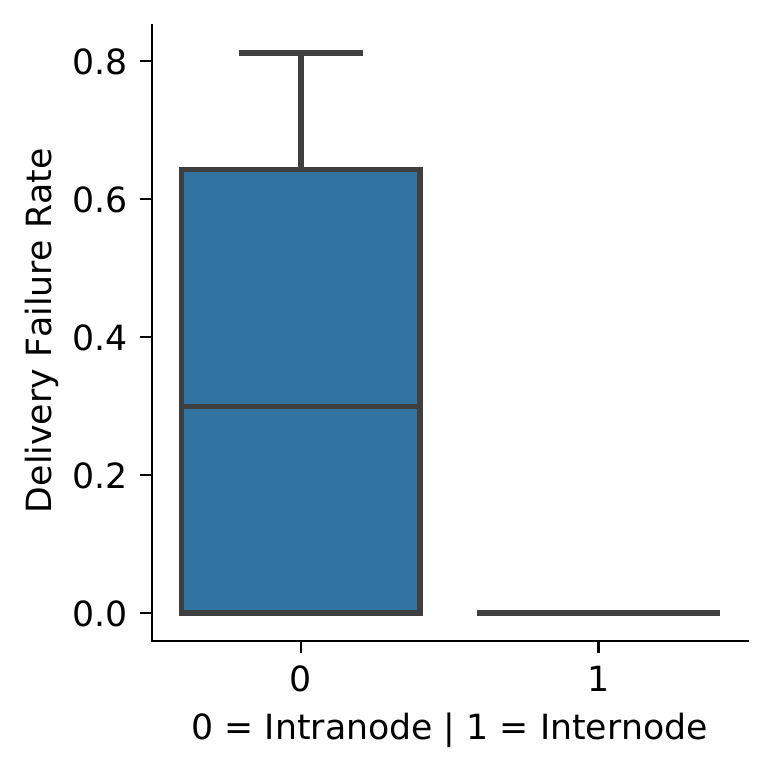}
    \caption{Distribution of \metric{} for each snapshot, with outliers.}
    \label{fig:intranode-vs-internode-distribution-\metricslug-withfliers}
  \end{subfigure}
  \caption{Distribution of \metric{} for individual snapshot measurements for intranode vs. internode experiment (Section \ref{sec:intranode-vs-internode}). Lower is better.}
  \label{fig:intranode-vs-internode-distribution-\metricslug}
\end{figure*}

\providecommand{\figintranodevsinternoderegressionmacro}[4]{
\def\metric{#1}\def\metricslug{#2}\def\xvarslug{#3}\def\yvarprefix{#4}\begin{figure*}[h]
  \centering

  \begin{subfigure}[b]{0.5\textwidth}
    \centering
    \includegraphics[width=\textwidth]{submodule/conduit-binder/binder/date=2021+project=72k5n/a=intranode-vs-internode/teeplots/\metricslug/marker=+regression=ordinary-least-squares-regression+row=num-simels-per-cpu+transform=snapshot_diffs-groupby_exec_instance-mean+viz=facet-unsplit-regression+x=\xvarslug+y=\yvarprefix\metricslug+ext=}
    \caption{
      Ordinary least squares regression plot.
      Observations are means per replicate.
    }
    \label{fig:intranode-vs-internode-regression-ols-\metricslug-complete-regression}
  \end{subfigure}%
  \begin{subfigure}[b]{0.5\textwidth}
    \centering
    \includegraphics[width=\textwidth]{submodule/conduit-binder/binder/date=2021+project=72k5n/a=intranode-vs-internode/teeplots/\metricslug/estimated-statistic=\metricslug-mean+row=num-simels-per-cpu+transform=fit_regression+viz=lambda+y=absolute-effect-size+ext=}
    \caption{Estimated regression coefficient for ordinary least squares regression. Zero corresponds to no effect.}
    \label{fig:intranode-vs-internode-regression-ols-\metricslug-complete-effect-size}
  \end{subfigure}

  \begin{subfigure}[b]{0.5\textwidth}
    \centering
    \includegraphics[width=\textwidth]{submodule/conduit-binder/binder/date=2021+project=72k5n/a=intranode-vs-internode/teeplots/\metricslug/marker=+regression=quantile-regression+row=num-simels-per-cpu+transform=snapshot_diffs-groupby_exec_instance-median+viz=facet-unsplit-regression+x=\xvarslug+y=\yvarprefix\metricslug+ext=}
    \caption{
      Quantile regression plot.
      Observations are medians per replicate.
    }
    \label{fig:intranode-vs-internode-regression-quantile-\metricslug-complete-regression}
  \end{subfigure}%
  \begin{subfigure}[b]{0.5\textwidth}
    \centering
    \includegraphics[width=\textwidth]{submodule/conduit-binder/binder/date=2021+project=72k5n/a=intranode-vs-internode/teeplots/\metricslug/estimated-statistic=\metricslug-median+row=num-simels-per-cpu+transform=fit_regression+viz=lambda+y=absolute-effect-size+ext=}
    \caption{Estimated regression coefficient for quantile regression. Zero corresponds to no effect.}
    \label{fig:intranode-vs-internode-regression-quantile-\metricslug-complete-effect-size}
  \end{subfigure}

  \caption{
  Regressions of \metric{} against categorically coded treatment for intranode vs. internode experiment (Section \ref{sec:intranode-vs-internode}).
  Lower is better.
  Ordinary least squares regression (top row) estimates relationship between categorical dependent variable and mean of response variable.
  Quantile regression (bottom row) estimates relationship between categorical independent variable and median of response variable.
  Error bands and bars are 95\% confidence intervals.
  }
  \label{fig:intranode-vs-internode-regression-\metricslug}
\end{figure*}

}

\def\metric{Latency Walltime Inlet (ns)}\def\metricslug{latency-walltime-inlet-ns}\def\xvarslug{0-intranode-1-internode}\def\yvarprefix{}\begin{figure*}[h]
  \centering

  \begin{subfigure}[b]{0.5\textwidth}
    \centering
    \includegraphics[width=\textwidth]{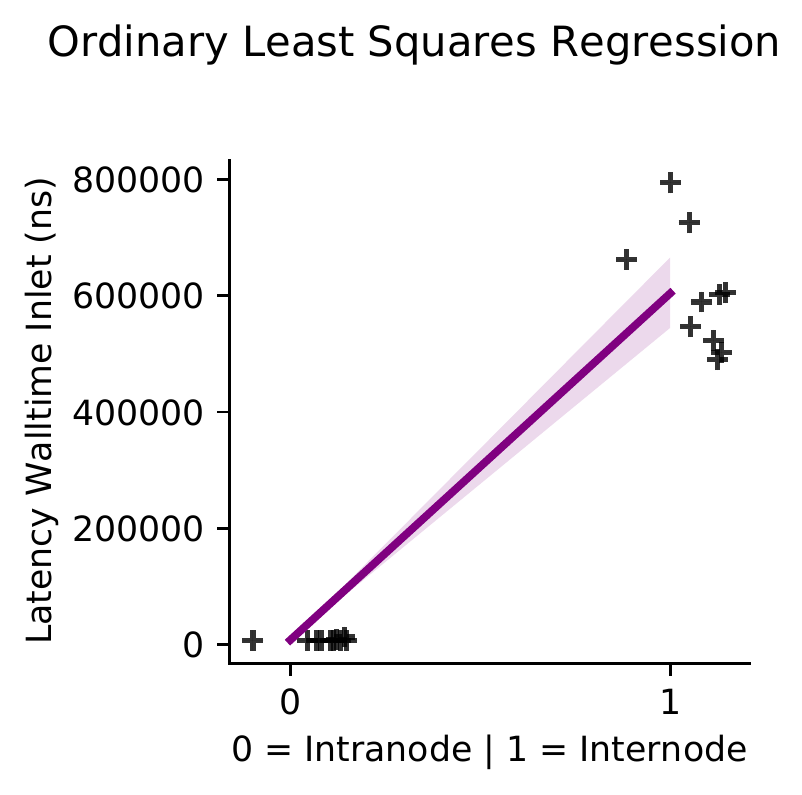}
    \caption{
      Ordinary least squares regression plot.
      Observations are means per replicate.
    }
    \label{fig:intranode-vs-internode-regression-ols-\metricslug-complete-regression}
  \end{subfigure}%
  \begin{subfigure}[b]{0.5\textwidth}
    \centering
    \includegraphics[width=\textwidth]{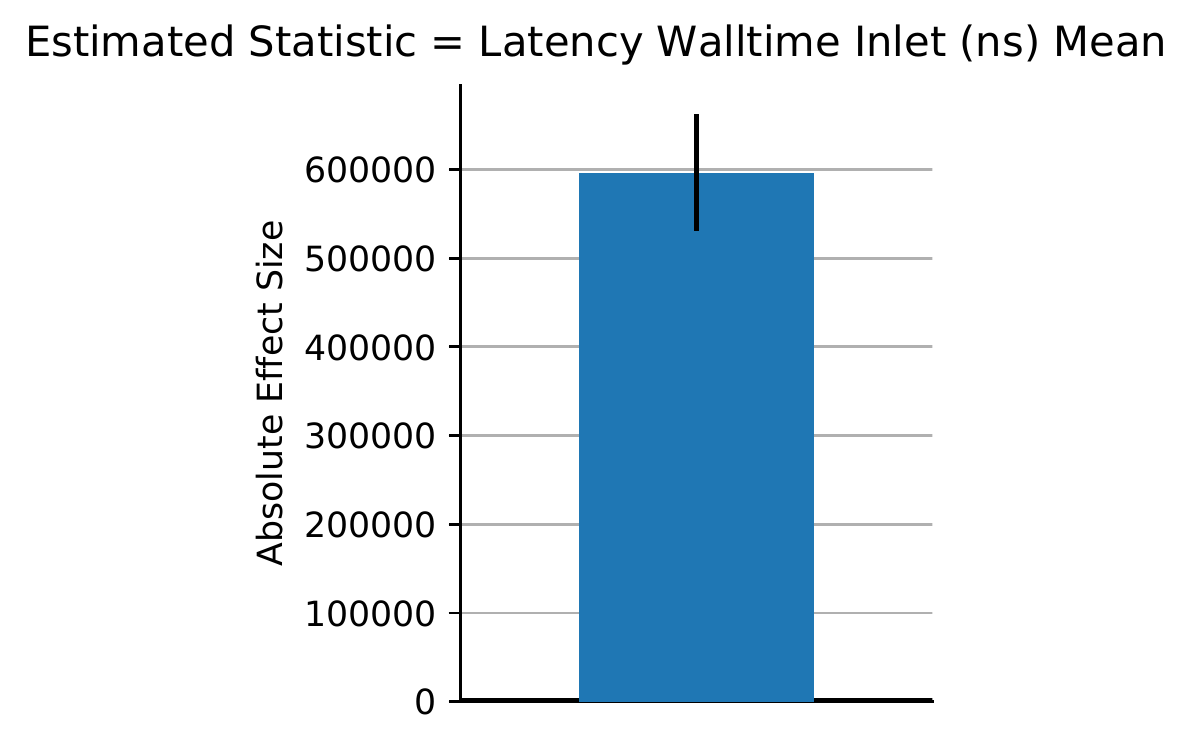}
    \caption{Estimated regression coefficient for ordinary least squares regression. Zero corresponds to no effect.}
    \label{fig:intranode-vs-internode-regression-ols-\metricslug-complete-effect-size}
  \end{subfigure}

  \begin{subfigure}[b]{0.5\textwidth}
    \centering
    \includegraphics[width=\textwidth]{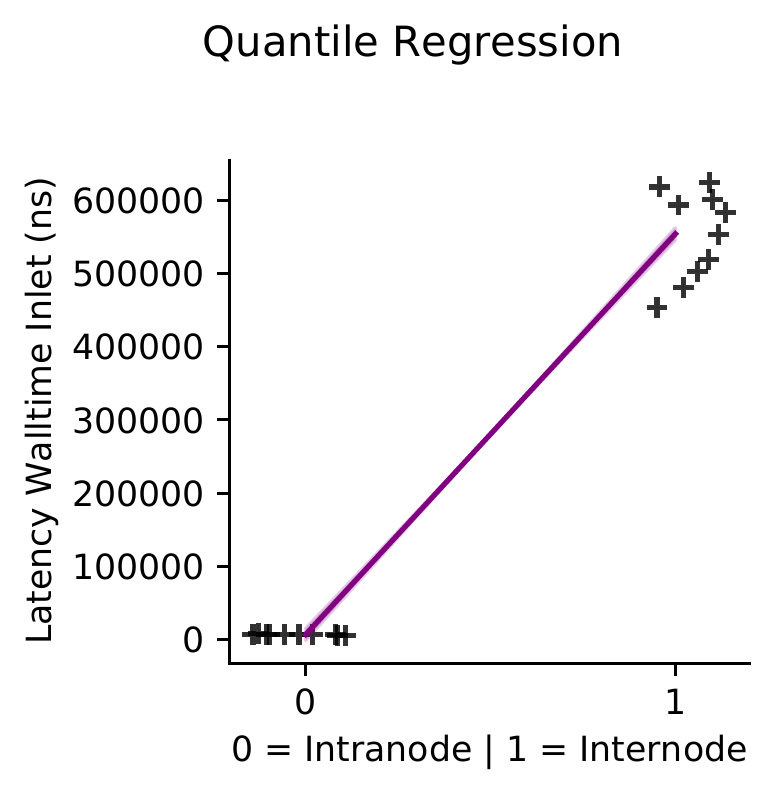}
    \caption{
      Quantile regression plot.
      Observations are medians per replicate.
    }
    \label{fig:intranode-vs-internode-regression-quantile-\metricslug-complete-regression}
  \end{subfigure}%
  \begin{subfigure}[b]{0.5\textwidth}
    \centering
    \includegraphics[width=\textwidth]{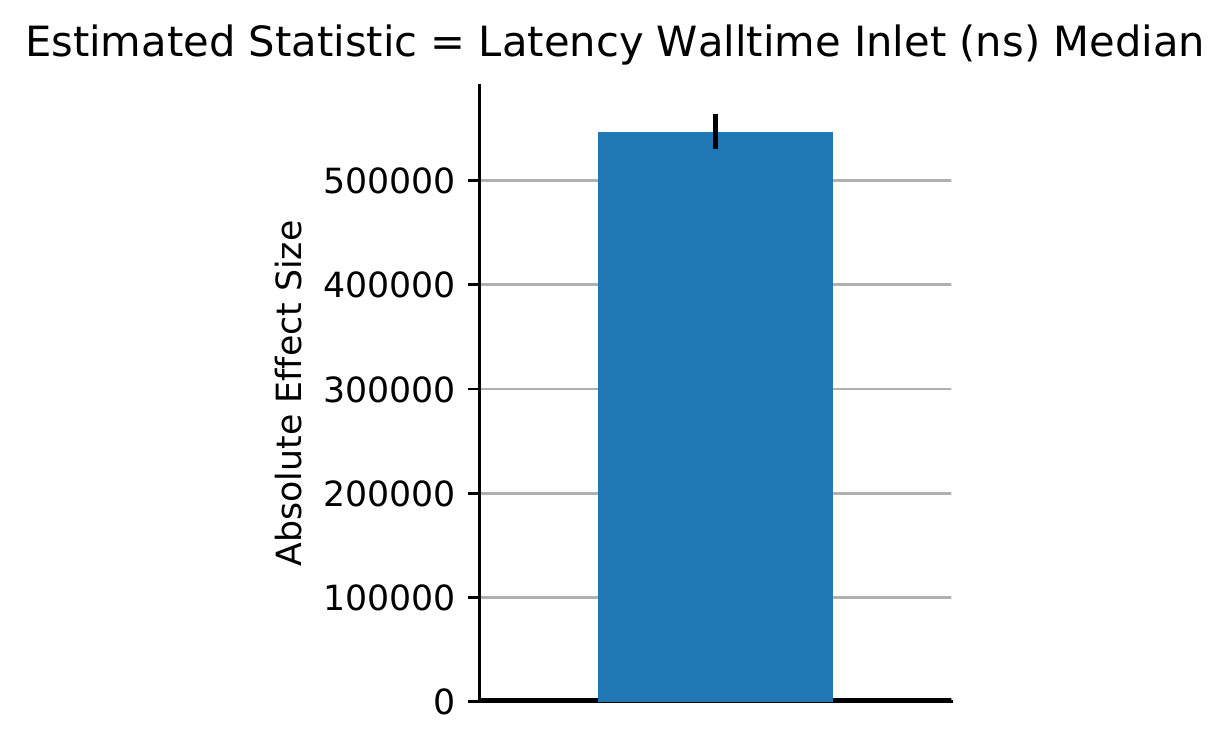}
    \caption{Estimated regression coefficient for quantile regression. Zero corresponds to no effect.}
    \label{fig:intranode-vs-internode-regression-quantile-\metricslug-complete-effect-size}
  \end{subfigure}

  \caption{
  Regressions of \metric{} against categorically coded treatment for intranode vs. internode experiment (Section \ref{sec:intranode-vs-internode}).
  Lower is better.
  Ordinary least squares regression (top row) estimates relationship between categorical dependent variable and mean of response variable.
  Quantile regression (bottom row) estimates relationship between categorical independent variable and median of response variable.
  Error bands and bars are 95\% confidence intervals.
  }
  \label{fig:intranode-vs-internode-regression-\metricslug}
\end{figure*}

\def\metric{Latency Simsteps Outlet}\def\metricslug{latency-simsteps-outlet}\def\xvarslug{0-intranode-1-internode}\def\yvarprefix{}

\def\metric{Latency Walltime Outlet (ns)}\def\metricslug{latency-walltime-outlet-ns}\def\xvarslug{0-intranode-1-internode}\def\yvarprefix{}

\def\metric{Delivery Clumpiness}\def\metricslug{delivery-clumpiness}\def\xvarslug{0-intranode-1-internode}\def\yvarprefix{}

\def\metric{Simstep Period Inlet (ns)}\def\metricslug{simstep-period-inlet-ns}\def\xvarslug{0-intranode-1-internode}\def\yvarprefix{}

\def\metric{Latency Simsteps Inlet}\def\metricslug{latency-simsteps-inlet}\def\xvarslug{0-intranode-1-internode}\def\yvarprefix{}

\def\metric{Simstep Period Outlet (ns)}\def\metricslug{simstep-period-outlet-ns}\def\xvarslug{0-intranode-1-internode}\def\yvarprefix{}

\def\metric{Delivery Failure Rate}\def\metricslug{delivery-failure-rate}\def\xvarslug{0-intranode-1-internode}\def\yvarprefix{}

\providecommand{\tabintranodevsinternoderegressionmacro}[2]{
\def\regression{#1}\def\regressionslug{#2}

\providecommand{\dissertationelse}[2]{%
\ifdefined\DISSERTATION
#1
\else
#2
\fi
}

\makeatletter
  \providecommand\importpath{\import@path}
\makeatother

\pragmaonce

\def\replacesinglequote#1{%
  \IfSubStr{#1}{'}{%
    \StrSubstitute{#1}{'}{\,}}{#1}}

\pragmaonce

\providecommand\rotcol[1]{ \multicolumn{1}{@{}@{}}{} \adjustbox{angle=60,lap=\width}{\rule{0.4\textwidth}{0.8pt}} \hfill \makebox[3ex]{\adjustbox{angle=60,lap=\width}{\bfseries \hspace{1ex} #1}} \hfill }


\clearpage
{
\onecolumn

\begin{landscape}
\dissertationelse{\fontsize{8}{9}\selectfont}{\footnotesize}
\csvstyle{myTableStyle}{
  longtable=@{\hspace{-.5\arrayrulewidth}}c@{\hspace{-.5\arrayrulewidth}}|l|l|l|l|l|l|l|l|l|l|l|l|l|l|,
  table head=\caption{
    Full \regression{} results of quality of service metrics against log processor count for weak scaling experiment (Section \ref{sec:weak-scaling}).
    Listed results include both piecewise and complete regression.
    Ordinary least squares regression estimates relationship between independent variable and mean of response variable.
    Quantile regression estimates relationship between independent variable and median of response variable.
    Significance level $p < 0.05$ used.
    Inf or NaN values may occur due to multicollinearity or due to inf or NaN observations.
  }\label{tab:intranode-vs-internode-\regressionslug} \\
    \toprule%
    \multicolumn{1}{@{\hspace{-.5\arrayrulewidth}}l@{\hspace{-.5\arrayrulewidth}}}{} &%
    \rotcol{ Metric } &%
    \rotcol{ Statistic } &%
    \rotcol{ Significant Effect Sign } &%
    \rotcol{ Cpus Per Node } &%
    \rotcol{ Num Simels Per Cpu} &%
    \rotcol{ Num Processes} &%
    \rotcol{ Absolute Effect Size} &%
    \rotcol{ Absolute Effect Size 95\% CI Lower Bound} &%
    \rotcol{ Absolute Effect Size 95\% CI Upper Bound} &%
    \rotcol{ Relative Effect Size} &%
    \rotcol{ Relative Effect Size 95\% CI Lower Bound} &%
    \rotcol{ Relative Effect Size 95\% CI Upper Bound} &%
    \rotcol{$\mathbf{n}$} & %
    \rotcol{$\mathbf{p}$} %
    \vspace{-1ex} \\ \midrule\endfirsthead
    \caption*{\tablename{} \thetable{} (cont'd)} \\
    \toprule%
    \multicolumn{1}{@{\hspace{-.5\arrayrulewidth}}l@{\hspace{-.5\arrayrulewidth}}}{} &%
    \rotcol{ Metric } &%
    \rotcol{ Statistic } &%
    \rotcol{ Significant Effect Sign } &%
    \rotcol{ Cpus Per Node } &%
    \rotcol{ Num Simels Per Cpu} &%
    \rotcol{ Num Processes} &%
    \rotcol{ Absolute Effect Size} &%
    \rotcol{ Absolute Effect Size 95\% CI Lower Bound} &%
    \rotcol{ Absolute Effect Size 95\% CI Upper Bound} &%
    \rotcol{ Relative Effect Size} &%
    \rotcol{ Relative Effect Size 95\% CI Lower Bound} &%
    \rotcol{ Relative Effect Size 95\% CI Upper Bound} &%
    \rotcol{$\mathbf{n}$} & %
    \rotcol{$\mathbf{p}$} %
    \vspace{-1ex} \\ \midrule\endhead
    \bottomrule\endfoot,
  late after line=\\\hline,
  late after last line=\\\bottomrule,
  head to column names,
  filter=\equal{\csuse{RegressionModelSlug}}{\regressionslug},
}

\catcode`\%=12
\providecommand\filename{}
\renewcommand\filename{\detokenize{submodule/conduit-binder/binder/date=2021+project=72k5n/a=intranode-vs-internode/outplots/a=intranode_vs_internode_regression_results+async_mode=3+impl=proc+nproc=2+nthread=1+proc=0-1+readability=latexcsvreader+replicate=0-9+slurm_nnodes=1-2+slurm_ntasks=2+subject=inlet_outlet+view=summary+ext=.csv}}
\catcode`\%=14

\csvreader[myTableStyle]{\importpath\filename}{}%
{ %
  &%
  \csuse{DependentVariable} &%
  \csuse{Statistic} &%
  \ifthenelse{\equal{\csuse{SignificantEffectSign}}{-}}{\cellcolor{red!25}}{}%
  \ifthenelse{\equal{\csuse{SignificantEffectSign}}{+}}{\cellcolor{green!25}}{}%
  \ifthenelse{\equal{\csuse{SignificantEffectSign}}{0}}{\cellcolor{gray!25}}{}%
  \csuse{SignificantEffectSign} &%
  \replacesinglequote{\csuse{CpusPerNode}} &%
  \replacesinglequote{\csuse{NumSimelsPerCpu}} &%
  \replacesinglequote{\csuse{NumProcessesPrettyprint}} &%
  \replacesinglequote{\csuse{AbsoluteEffectSize}} &%
  \replacesinglequote{\csuse{AbsoluteEffectSize95CILowerBound}} &%
  \replacesinglequote{\csuse{AbsoluteEffectSize95CIUpperBound}} &%
  \replacesinglequote{\csuse{RelativeEffectSize}} &%
  \replacesinglequote{\csuse{RelativeEffectSize95CILowerBound}} &%
  \replacesinglequote{\csuse{RelativeEffectSize95CIUpperBound}} &%
  \replacesinglequote{\csuse{n}} &%
  \replacesinglequote{\csuse{p}} %
}

\end{landscape}

}

\clearpage

}

\def\regression{Ordinary Least Squares Regression}\def\regressionslug{ordinary-least-squares-regression}

\providecommand{\dissertationelse}[2]{%
\ifdefined\DISSERTATION
#1
\else
#2
\fi
}

\makeatletter
  \providecommand\importpath{\import@path}
\makeatother

\pragmaonce

\def\replacesinglequote#1{%
  \IfSubStr{#1}{'}{%
    \StrSubstitute{#1}{'}{\,}}{#1}}

\pragmaonce

\providecommand\rotcol[1]{ \multicolumn{1}{@{}@{}}{} \adjustbox{angle=60,lap=\width}{\rule{0.4\textwidth}{0.8pt}} \hfill \makebox[3ex]{\adjustbox{angle=60,lap=\width}{\bfseries \hspace{1ex} #1}} \hfill }


\clearpage
{
\onecolumn

\begin{landscape}
\dissertationelse{\fontsize{8}{9}\selectfont}{\footnotesize}
\csvstyle{myTableStyle}{
  longtable=@{\hspace{-.5\arrayrulewidth}}c@{\hspace{-.5\arrayrulewidth}}|l|l|l|l|l|l|l|l|l|l|l|l|l|l|,
  table head=\caption{
    Full \regression{} results of quality of service metrics against log processor count for weak scaling experiment (Section \ref{sec:weak-scaling}).
    Listed results include both piecewise and complete regression.
    Ordinary least squares regression estimates relationship between independent variable and mean of response variable.
    Quantile regression estimates relationship between independent variable and median of response variable.
    Significance level $p < 0.05$ used.
    Inf or NaN values may occur due to multicollinearity or due to inf or NaN observations.
  }\label{tab:intranode-vs-internode-\regressionslug} \\
    \toprule%
    \multicolumn{1}{@{\hspace{-.5\arrayrulewidth}}l@{\hspace{-.5\arrayrulewidth}}}{} &%
    \rotcol{ Metric } &%
    \rotcol{ Statistic } &%
    \rotcol{ Significant Effect Sign } &%
    \rotcol{ Cpus Per Node } &%
    \rotcol{ Num Simels Per Cpu} &%
    \rotcol{ Num Processes} &%
    \rotcol{ Absolute Effect Size} &%
    \rotcol{ Absolute Effect Size 95\% CI Lower Bound} &%
    \rotcol{ Absolute Effect Size 95\% CI Upper Bound} &%
    \rotcol{ Relative Effect Size} &%
    \rotcol{ Relative Effect Size 95\% CI Lower Bound} &%
    \rotcol{ Relative Effect Size 95\% CI Upper Bound} &%
    \rotcol{$\mathbf{n}$} & %
    \rotcol{$\mathbf{p}$} %
    \vspace{-1ex} \\ \midrule\endfirsthead
    \caption*{\tablename{} \thetable{} (cont'd)} \\
    \toprule%
    \multicolumn{1}{@{\hspace{-.5\arrayrulewidth}}l@{\hspace{-.5\arrayrulewidth}}}{} &%
    \rotcol{ Metric } &%
    \rotcol{ Statistic } &%
    \rotcol{ Significant Effect Sign } &%
    \rotcol{ Cpus Per Node } &%
    \rotcol{ Num Simels Per Cpu} &%
    \rotcol{ Num Processes} &%
    \rotcol{ Absolute Effect Size} &%
    \rotcol{ Absolute Effect Size 95\% CI Lower Bound} &%
    \rotcol{ Absolute Effect Size 95\% CI Upper Bound} &%
    \rotcol{ Relative Effect Size} &%
    \rotcol{ Relative Effect Size 95\% CI Lower Bound} &%
    \rotcol{ Relative Effect Size 95\% CI Upper Bound} &%
    \rotcol{$\mathbf{n}$} & %
    \rotcol{$\mathbf{p}$} %
    \vspace{-1ex} \\ \midrule\endhead
    \bottomrule\endfoot,
  late after line=\\\hline,
  late after last line=\\\bottomrule,
  head to column names,
  filter=\equal{\csuse{RegressionModelSlug}}{\regressionslug},
}

\catcode`\%=12
\providecommand\filename{}
\renewcommand\filename{\detokenize{submodule/conduit-binder/binder/date=2021+project=72k5n/a=intranode-vs-internode/outplots/a=intranode_vs_internode_regression_results+async_mode=3+impl=proc+nproc=2+nthread=1+proc=0-1+readability=latexcsvreader+replicate=0-9+slurm_nnodes=1-2+slurm_ntasks=2+subject=inlet_outlet+view=summary+ext=.csv}}
\catcode`\%=14

\csvreader[myTableStyle]{\importpath\filename}{}%
{ %
  &%
  \csuse{DependentVariable} &%
  \csuse{Statistic} &%
  \ifthenelse{\equal{\csuse{SignificantEffectSign}}{-}}{\cellcolor{red!25}}{}%
  \ifthenelse{\equal{\csuse{SignificantEffectSign}}{+}}{\cellcolor{green!25}}{}%
  \ifthenelse{\equal{\csuse{SignificantEffectSign}}{0}}{\cellcolor{gray!25}}{}%
  \csuse{SignificantEffectSign} &%
  \replacesinglequote{\csuse{CpusPerNode}} &%
  \replacesinglequote{\csuse{NumSimelsPerCpu}} &%
  \replacesinglequote{\csuse{NumProcessesPrettyprint}} &%
  \replacesinglequote{\csuse{AbsoluteEffectSize}} &%
  \replacesinglequote{\csuse{AbsoluteEffectSize95CILowerBound}} &%
  \replacesinglequote{\csuse{AbsoluteEffectSize95CIUpperBound}} &%
  \replacesinglequote{\csuse{RelativeEffectSize}} &%
  \replacesinglequote{\csuse{RelativeEffectSize95CILowerBound}} &%
  \replacesinglequote{\csuse{RelativeEffectSize95CIUpperBound}} &%
  \replacesinglequote{\csuse{n}} &%
  \replacesinglequote{\csuse{p}} %
}

\end{landscape}

}

\clearpage

\def\regression{Quantile Regression}\def\regressionslug{quantile-regression}

\clearpage

\section{Multithreading vs Multiprocessing}

This section provides full results from multithreading vs. multiprocessing experiments discussed in Section \ref{sec:multithreading-vs-multiprocessing}.

\providecommand{\figmultithreadingvsmultiprocessingdistributionmacro}[2]{
\def\metric{#1}\def\metricslug{#2}\begin{figure*}[h]
  \centering
  \begin{subfigure}[b]{0.5\textwidth}
    \centering
    \includegraphics[width=\textwidth]{submodule/conduit-binder/binder/date=2021+project=72k5n/a=multithreading-vs-multiprocessing/teeplots/\metricslug/col=cpus-per-node+row=num-simels-per-cpu+transform=snapshot_diffs+viz=facet-boxplot-nofliers+x=0-multithreading-1-multiprocessing+y=\metricslug+ext=}
    \caption{Distribution of \metric{} for each snapshot, without outliers.}
    \label{fig:multithreading-vs-multiprocessing-distribution-\metricslug-nofliers}
  \end{subfigure}%
  \begin{subfigure}[b]{0.5\textwidth}
    \centering
    \includegraphics[width=\textwidth]{submodule/conduit-binder/binder/date=2021+project=72k5n/a=multithreading-vs-multiprocessing/teeplots/\metricslug/col=cpus-per-node+row=num-simels-per-cpu+transform=snapshot_diffs+viz=facet-boxplot-withfliers+x=0-multithreading-1-multiprocessing+y=\metricslug+ext=}
    \caption{Distribution of \metric{} for each snapshot, with outliers.}
    \label{fig:multithreading-vs-multiprocessing-distribution-\metricslug-withfliers}
  \end{subfigure}
  \caption{Distribution of \metric{} for individual snapshot measurements for multithreading vs. multiprocessing experiment (Section \ref{sec:multithreading-vs-multiprocessing}). Lower is better.}
  \label{fig:multithreading-vs-multiprocessing-distribution-\metricslug}
\end{figure*}

}

\def\metric{Latency Walltime Inlet (ns)}\def\metricslug{latency-walltime-inlet-ns}\begin{figure*}[h]
  \centering
  \begin{subfigure}[b]{0.5\textwidth}
    \centering
    \includegraphics[width=\textwidth]{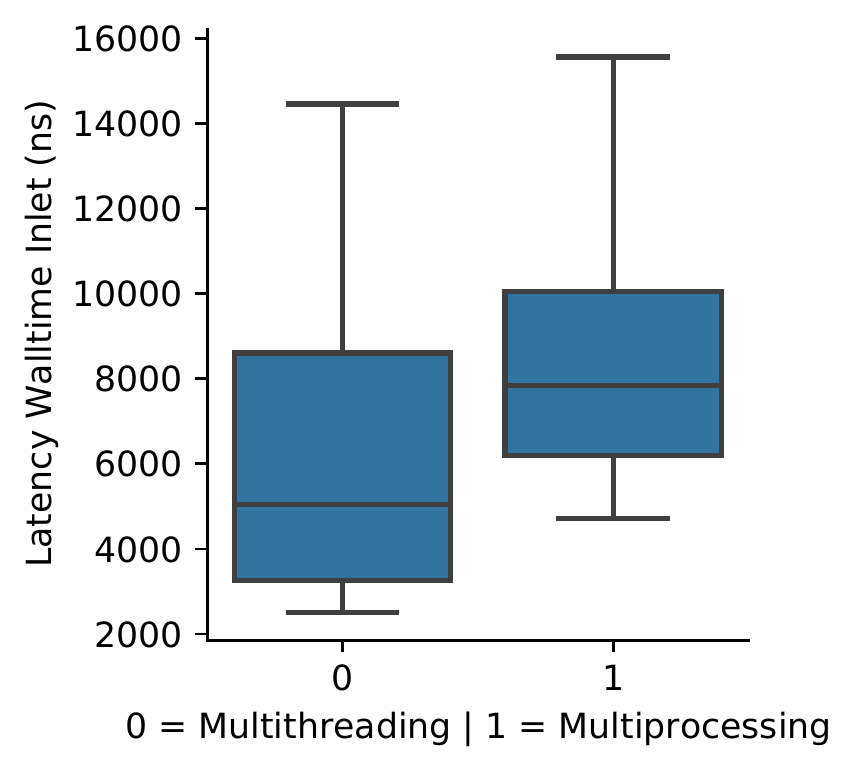}
    \caption{Distribution of \metric{} for each snapshot, without outliers.}
    \label{fig:multithreading-vs-multiprocessing-distribution-\metricslug-nofliers}
  \end{subfigure}%
  \begin{subfigure}[b]{0.5\textwidth}
    \centering
    \includegraphics[width=\textwidth]{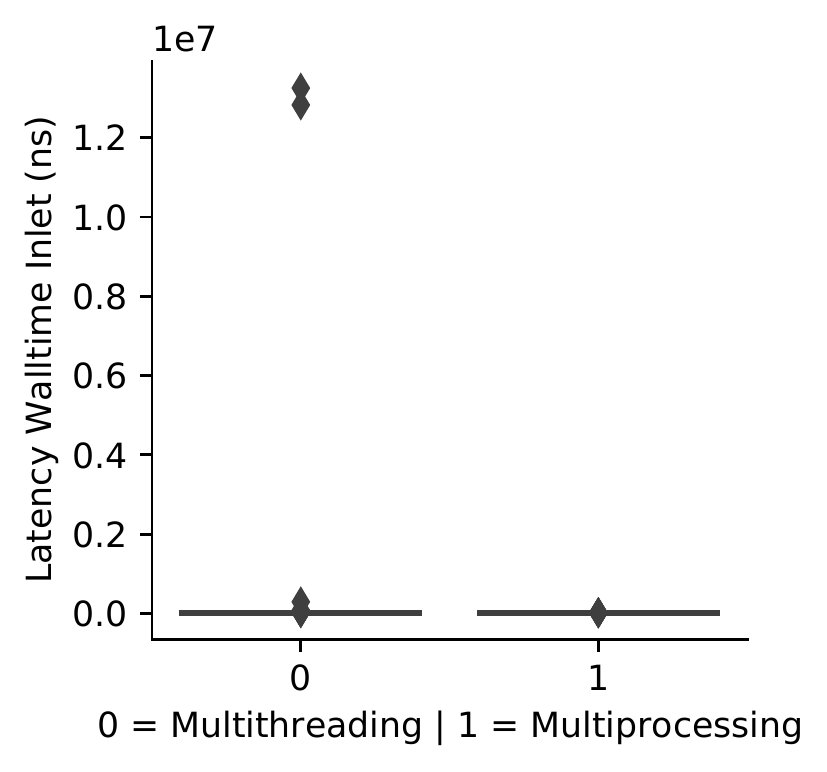}
    \caption{Distribution of \metric{} for each snapshot, with outliers.}
    \label{fig:multithreading-vs-multiprocessing-distribution-\metricslug-withfliers}
  \end{subfigure}
  \caption{Distribution of \metric{} for individual snapshot measurements for multithreading vs. multiprocessing experiment (Section \ref{sec:multithreading-vs-multiprocessing}). Lower is better.}
  \label{fig:multithreading-vs-multiprocessing-distribution-\metricslug}
\end{figure*}

\def\metric{Latency Simsteps Outlet}\def\metricslug{latency-simsteps-outlet}\begin{figure*}[h]
  \centering
  \begin{subfigure}[b]{0.5\textwidth}
    \centering
    \includegraphics[width=\textwidth]{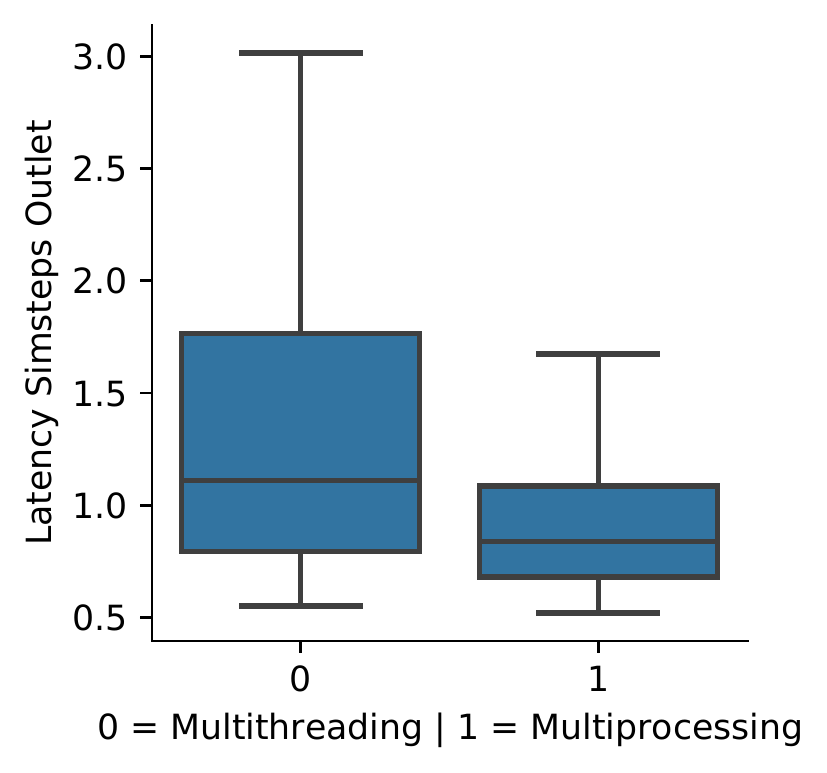}
    \caption{Distribution of \metric{} for each snapshot, without outliers.}
    \label{fig:multithreading-vs-multiprocessing-distribution-\metricslug-nofliers}
  \end{subfigure}%
  \begin{subfigure}[b]{0.5\textwidth}
    \centering
    \includegraphics[width=\textwidth]{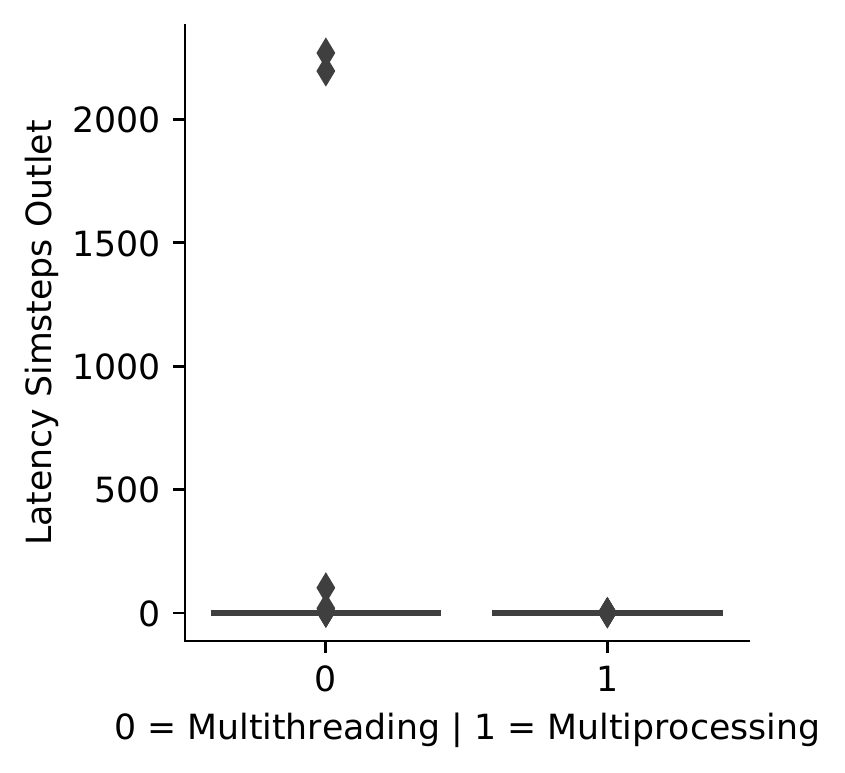}
    \caption{Distribution of \metric{} for each snapshot, with outliers.}
    \label{fig:multithreading-vs-multiprocessing-distribution-\metricslug-withfliers}
  \end{subfigure}
  \caption{Distribution of \metric{} for individual snapshot measurements for multithreading vs. multiprocessing experiment (Section \ref{sec:multithreading-vs-multiprocessing}). Lower is better.}
  \label{fig:multithreading-vs-multiprocessing-distribution-\metricslug}
\end{figure*}

\def\metric{Latency Walltime Outlet (ns)}\def\metricslug{latency-walltime-outlet-ns}\begin{figure*}[h]
  \centering
  \begin{subfigure}[b]{0.5\textwidth}
    \centering
    \includegraphics[width=\textwidth]{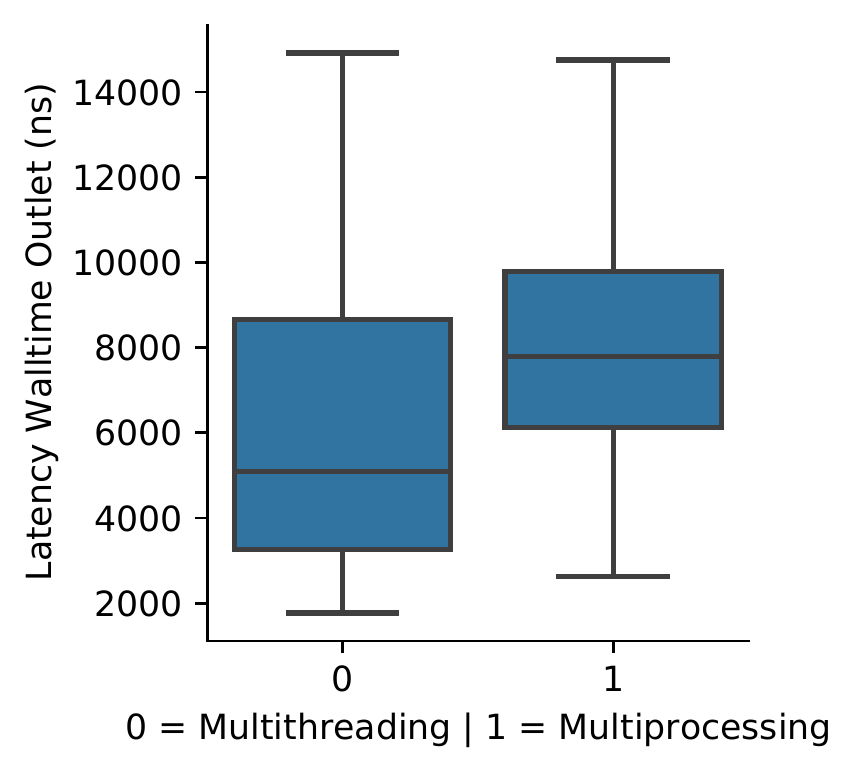}
    \caption{Distribution of \metric{} for each snapshot, without outliers.}
    \label{fig:multithreading-vs-multiprocessing-distribution-\metricslug-nofliers}
  \end{subfigure}%
  \begin{subfigure}[b]{0.5\textwidth}
    \centering
    \includegraphics[width=\textwidth]{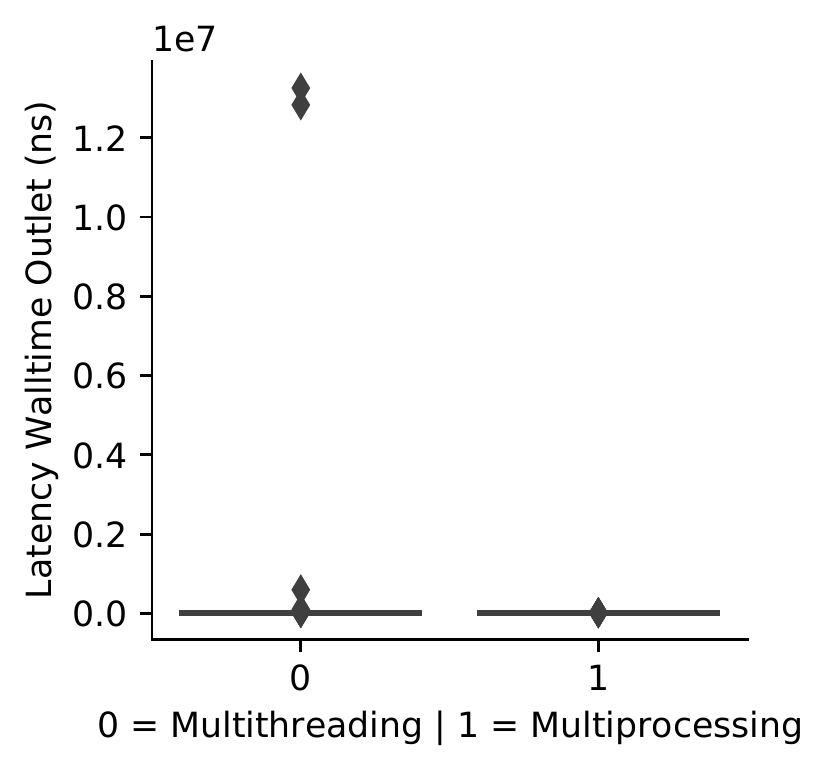}
    \caption{Distribution of \metric{} for each snapshot, with outliers.}
    \label{fig:multithreading-vs-multiprocessing-distribution-\metricslug-withfliers}
  \end{subfigure}
  \caption{Distribution of \metric{} for individual snapshot measurements for multithreading vs. multiprocessing experiment (Section \ref{sec:multithreading-vs-multiprocessing}). Lower is better.}
  \label{fig:multithreading-vs-multiprocessing-distribution-\metricslug}
\end{figure*}

\def\metric{Delivery Clumpiness}\def\metricslug{delivery-clumpiness}\begin{figure*}[h]
  \centering
  \begin{subfigure}[b]{0.5\textwidth}
    \centering
    \includegraphics[width=\textwidth]{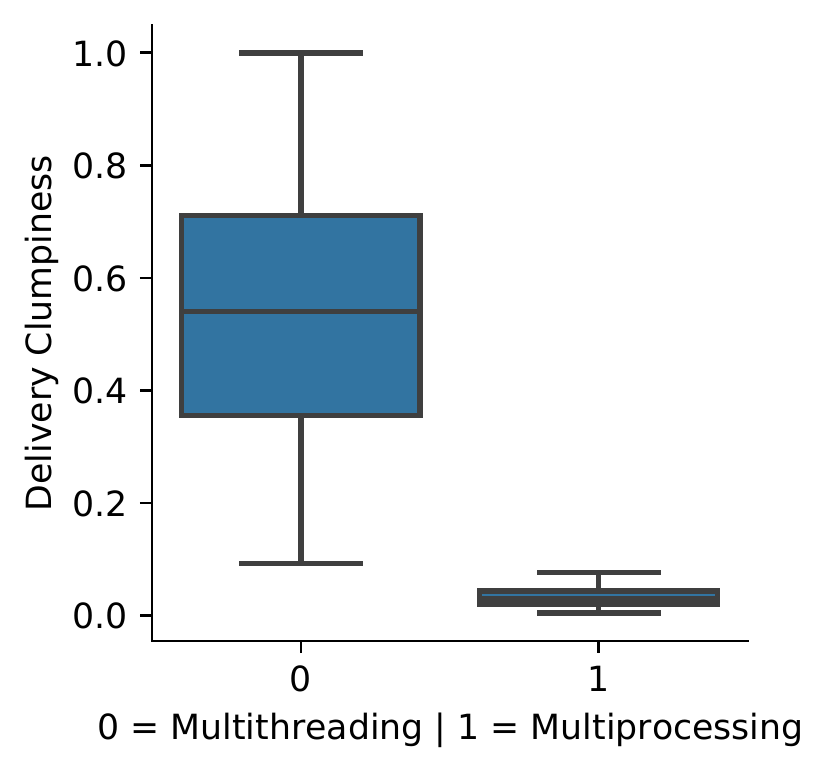}
    \caption{Distribution of \metric{} for each snapshot, without outliers.}
    \label{fig:multithreading-vs-multiprocessing-distribution-\metricslug-nofliers}
  \end{subfigure}%
  \begin{subfigure}[b]{0.5\textwidth}
    \centering
    \includegraphics[width=\textwidth]{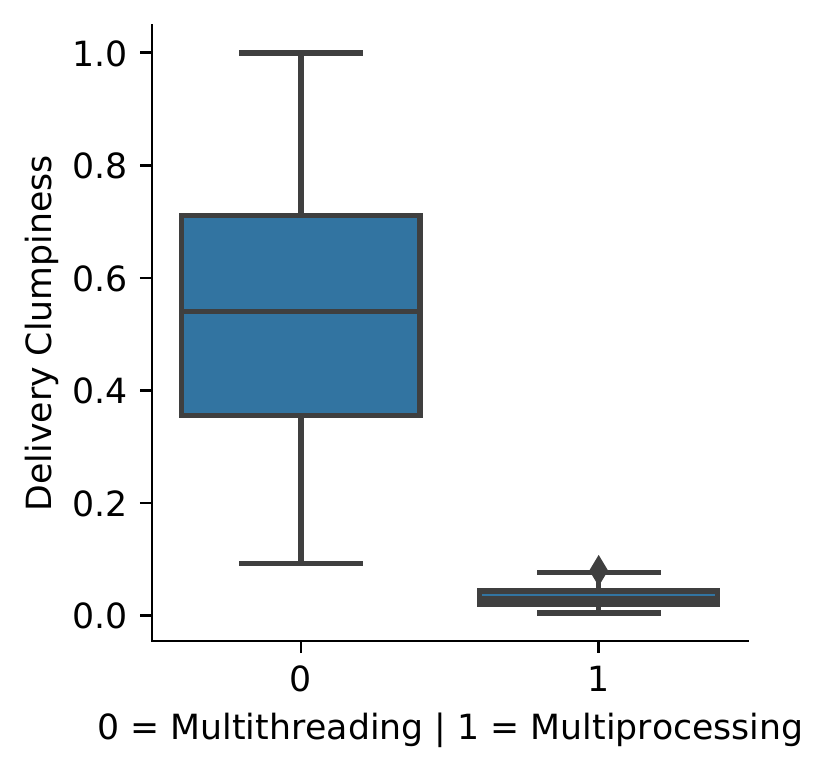}
    \caption{Distribution of \metric{} for each snapshot, with outliers.}
    \label{fig:multithreading-vs-multiprocessing-distribution-\metricslug-withfliers}
  \end{subfigure}
  \caption{Distribution of \metric{} for individual snapshot measurements for multithreading vs. multiprocessing experiment (Section \ref{sec:multithreading-vs-multiprocessing}). Lower is better.}
  \label{fig:multithreading-vs-multiprocessing-distribution-\metricslug}
\end{figure*}

\def\metric{Simstep Period Inlet (ns)}\def\metricslug{simstep-period-inlet-ns}\begin{figure*}[h]
  \centering
  \begin{subfigure}[b]{0.5\textwidth}
    \centering
    \includegraphics[width=\textwidth]{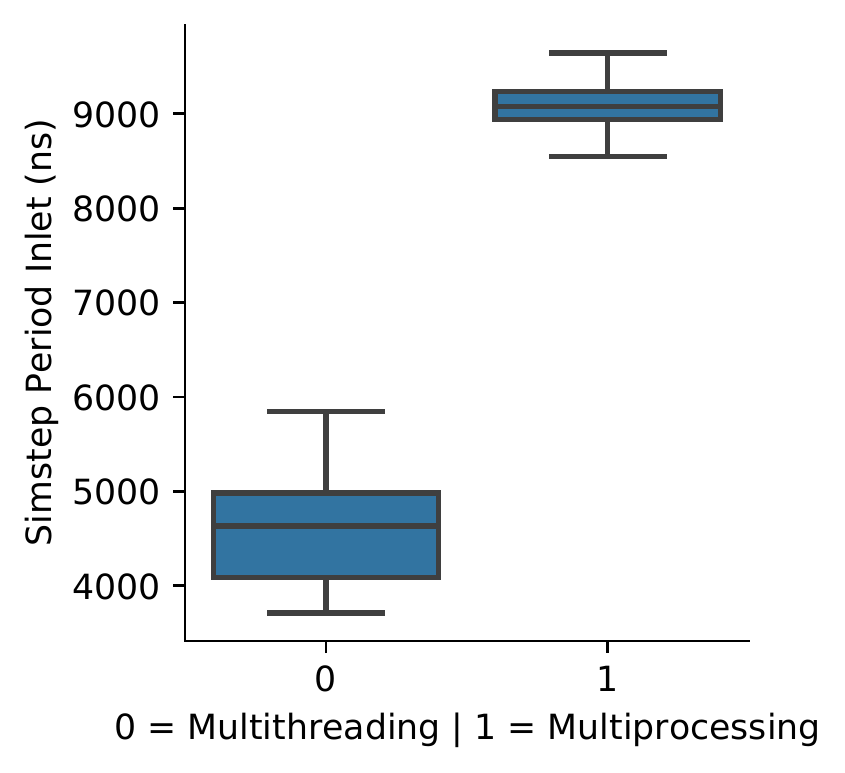}
    \caption{Distribution of \metric{} for each snapshot, without outliers.}
    \label{fig:multithreading-vs-multiprocessing-distribution-\metricslug-nofliers}
  \end{subfigure}%
  \begin{subfigure}[b]{0.5\textwidth}
    \centering
    \includegraphics[width=\textwidth]{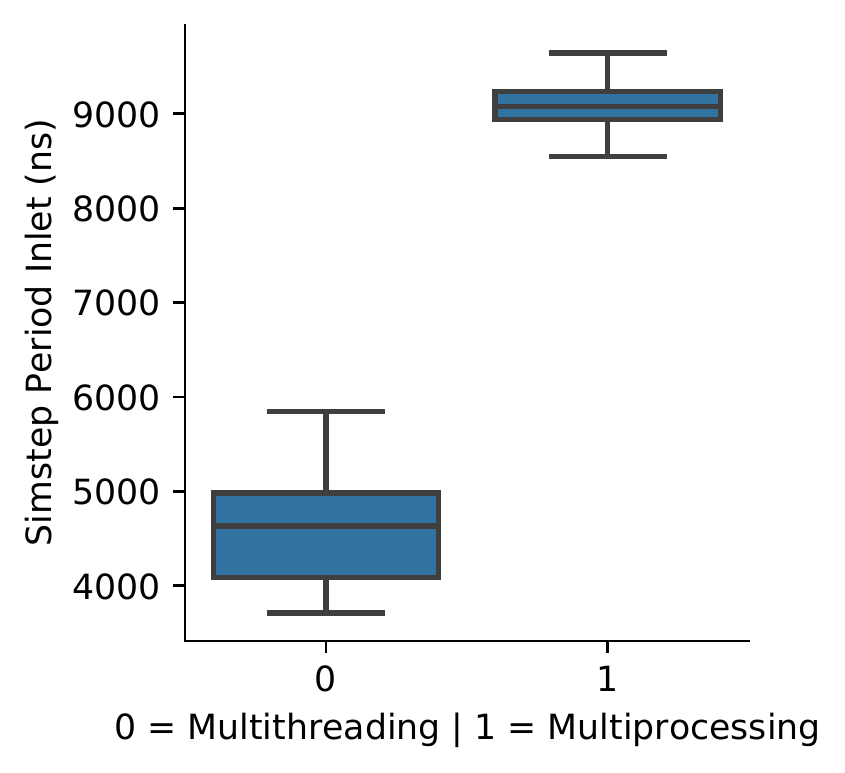}
    \caption{Distribution of \metric{} for each snapshot, with outliers.}
    \label{fig:multithreading-vs-multiprocessing-distribution-\metricslug-withfliers}
  \end{subfigure}
  \caption{Distribution of \metric{} for individual snapshot measurements for multithreading vs. multiprocessing experiment (Section \ref{sec:multithreading-vs-multiprocessing}). Lower is better.}
  \label{fig:multithreading-vs-multiprocessing-distribution-\metricslug}
\end{figure*}

\def\metric{Latency Simsteps Inlet}\def\metricslug{latency-simsteps-inlet}\begin{figure*}[h]
  \centering
  \begin{subfigure}[b]{0.5\textwidth}
    \centering
    \includegraphics[width=\textwidth]{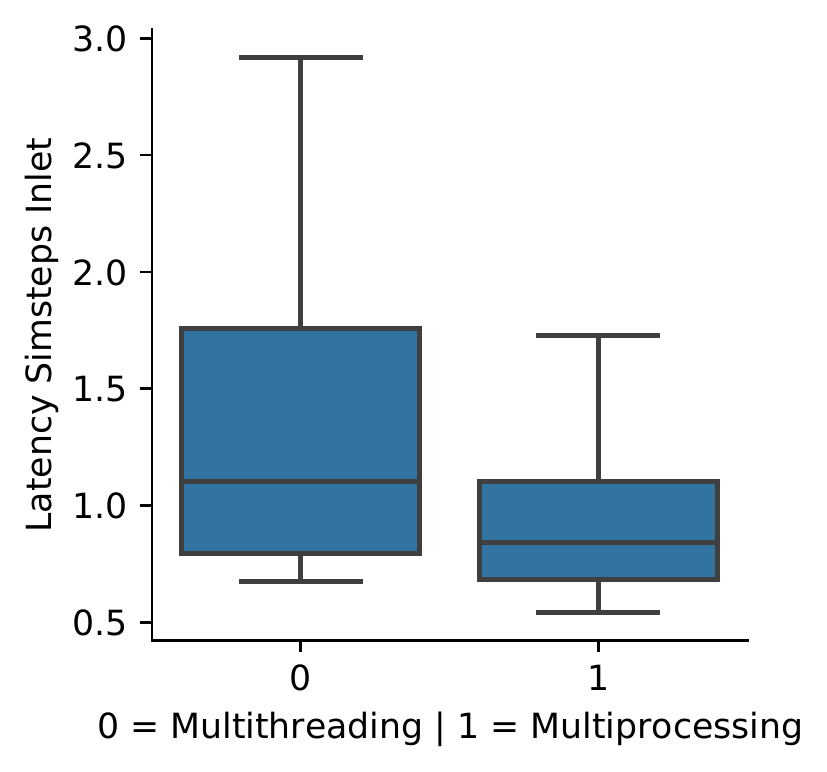}
    \caption{Distribution of \metric{} for each snapshot, without outliers.}
    \label{fig:multithreading-vs-multiprocessing-distribution-\metricslug-nofliers}
  \end{subfigure}%
  \begin{subfigure}[b]{0.5\textwidth}
    \centering
    \includegraphics[width=\textwidth]{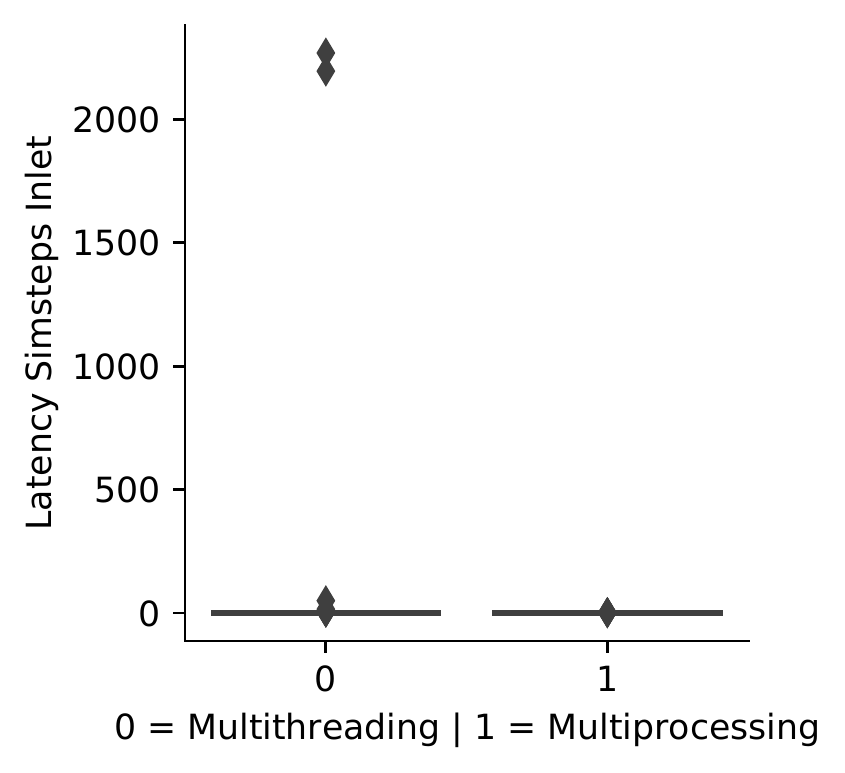}
    \caption{Distribution of \metric{} for each snapshot, with outliers.}
    \label{fig:multithreading-vs-multiprocessing-distribution-\metricslug-withfliers}
  \end{subfigure}
  \caption{Distribution of \metric{} for individual snapshot measurements for multithreading vs. multiprocessing experiment (Section \ref{sec:multithreading-vs-multiprocessing}). Lower is better.}
  \label{fig:multithreading-vs-multiprocessing-distribution-\metricslug}
\end{figure*}

\def\metric{Simstep Period Outlet (ns)}\def\metricslug{simstep-period-outlet-ns}\begin{figure*}[h]
  \centering
  \begin{subfigure}[b]{0.5\textwidth}
    \centering
    \includegraphics[width=\textwidth]{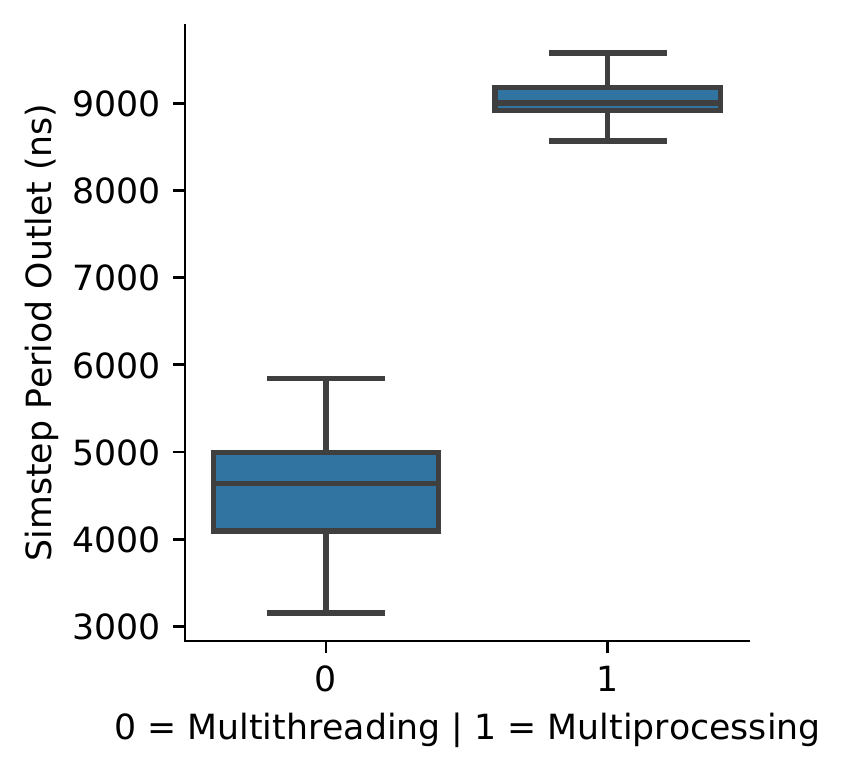}
    \caption{Distribution of \metric{} for each snapshot, without outliers.}
    \label{fig:multithreading-vs-multiprocessing-distribution-\metricslug-nofliers}
  \end{subfigure}%
  \begin{subfigure}[b]{0.5\textwidth}
    \centering
    \includegraphics[width=\textwidth]{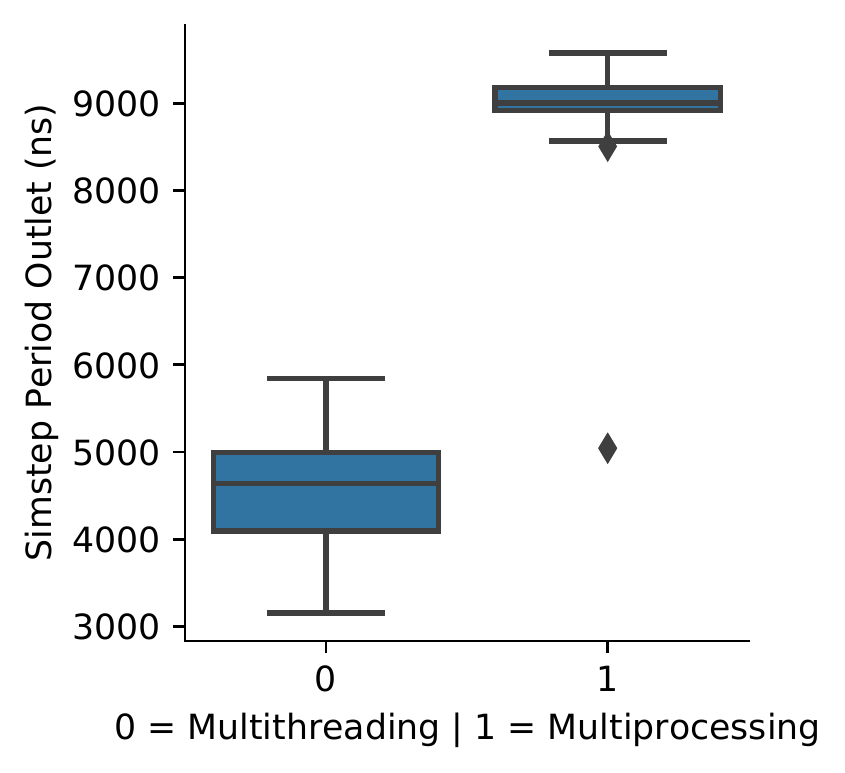}
    \caption{Distribution of \metric{} for each snapshot, with outliers.}
    \label{fig:multithreading-vs-multiprocessing-distribution-\metricslug-withfliers}
  \end{subfigure}
  \caption{Distribution of \metric{} for individual snapshot measurements for multithreading vs. multiprocessing experiment (Section \ref{sec:multithreading-vs-multiprocessing}). Lower is better.}
  \label{fig:multithreading-vs-multiprocessing-distribution-\metricslug}
\end{figure*}

\def\metric{Delivery Failure Rate}\def\metricslug{delivery-failure-rate}\begin{figure*}[h]
  \centering
  \begin{subfigure}[b]{0.5\textwidth}
    \centering
    \includegraphics[width=\textwidth]{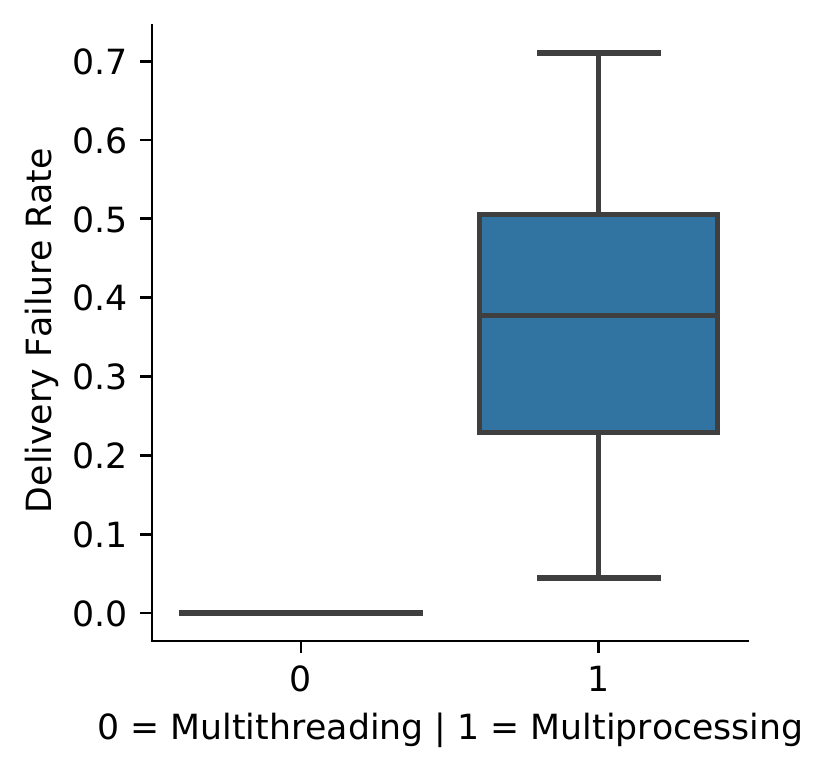}
    \caption{Distribution of \metric{} for each snapshot, without outliers.}
    \label{fig:multithreading-vs-multiprocessing-distribution-\metricslug-nofliers}
  \end{subfigure}%
  \begin{subfigure}[b]{0.5\textwidth}
    \centering
    \includegraphics[width=\textwidth]{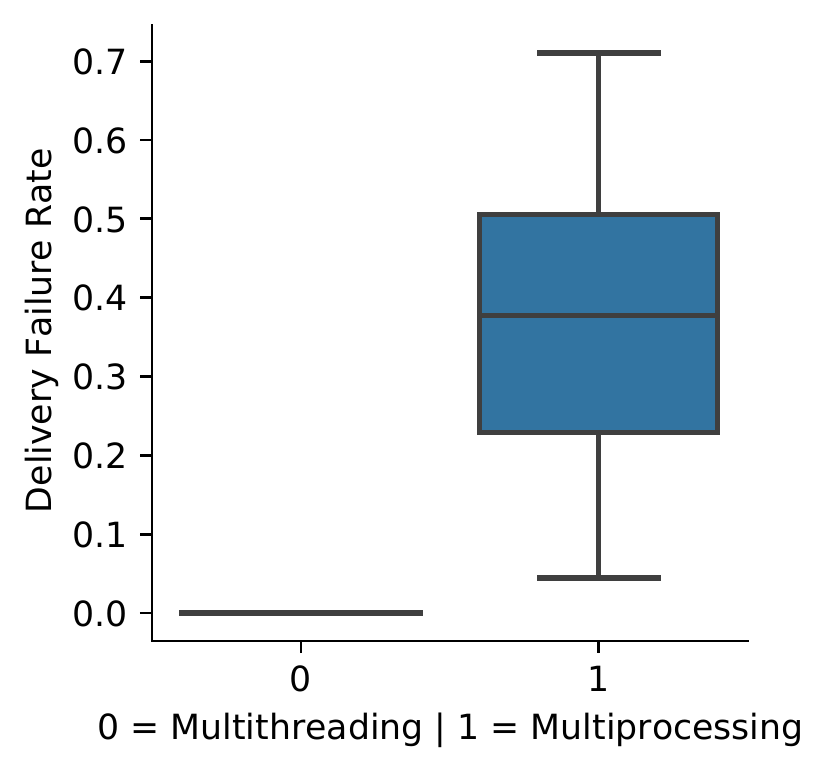}
    \caption{Distribution of \metric{} for each snapshot, with outliers.}
    \label{fig:multithreading-vs-multiprocessing-distribution-\metricslug-withfliers}
  \end{subfigure}
  \caption{Distribution of \metric{} for individual snapshot measurements for multithreading vs. multiprocessing experiment (Section \ref{sec:multithreading-vs-multiprocessing}). Lower is better.}
  \label{fig:multithreading-vs-multiprocessing-distribution-\metricslug}
\end{figure*}

\providecommand{\figmultithreadingvsmultiprocessingregressionmacro}[4]{
\def\metric{#1}\def\metricslug{#2}\def\xvarslug{#3}\def\yvarprefix{#4}\begin{figure*}[h]
  \centering

  \begin{subfigure}[b]{0.5\textwidth}
    \centering
    \includegraphics[width=\textwidth]{submodule/conduit-binder/binder/date=2021+project=72k5n/a=multithreading-vs-multiprocessing/teeplots/\metricslug/col=cpus-per-node+marker=+regression=ordinary-least-squares-regression+row=num-simels-per-cpu+transform=snapshot_diffs-groupby_exec_instance-mean+viz=facet-unsplit-regression+x=\xvarslug+y=\yvarprefix\metricslug+ext=}
    \caption{
      Ordinary least squares regression plot.
      Observations are means per replicate.
    }
    \label{fig:multithreading-vs-multiprocessing-regression-ols-\metricslug-complete-regression}
  \end{subfigure}%
  \begin{subfigure}[b]{0.5\textwidth}
    \centering
    \includegraphics[width=\textwidth]{submodule/conduit-binder/binder/date=2021+project=72k5n/a=multithreading-vs-multiprocessing/teeplots/\metricslug/col=cpus-per-node+estimated-statistic=\metricslug-mean+row=num-simels-per-cpu+transform=fit_regression+viz=lambda+y=absolute-effect-size+ext=}
    \caption{Estimated regression coefficient for ordinary least squares regression. Zero corresponds to no effect.}
    \label{fig:multithreading-vs-multiprocessing-regression-ols-\metricslug-complete-effect-size}
  \end{subfigure}

  \begin{subfigure}[b]{0.5\textwidth}
    \centering
    \includegraphics[width=\textwidth]{submodule/conduit-binder/binder/date=2021+project=72k5n/a=multithreading-vs-multiprocessing/teeplots/\metricslug/col=cpus-per-node+marker=+regression=quantile-regression+row=num-simels-per-cpu+transform=snapshot_diffs-groupby_exec_instance-median+viz=facet-unsplit-regression+x=\xvarslug+y=\yvarprefix\metricslug+ext=}
    \caption{
      Quantile regression plot.
      Observations are medians per replicate.
    }
    \label{fig:multithreading-vs-multiprocessing-regression-quantile-\metricslug-complete-regression}
  \end{subfigure}%
  \begin{subfigure}[b]{0.5\textwidth}
    \centering
    \includegraphics[width=\textwidth]{submodule/conduit-binder/binder/date=2021+project=72k5n/a=multithreading-vs-multiprocessing/teeplots/\metricslug/col=cpus-per-node+estimated-statistic=\metricslug-median+row=num-simels-per-cpu+transform=fit_regression+viz=lambda+y=absolute-effect-size+ext=}
    \caption{Estimated regression coefficient for quantile regression. Zero corresponds to no effect.}
    \label{fig:multithreading-vs-multiprocessing-regression-quantile-\metricslug-complete-effect-size}
  \end{subfigure}

  \caption{
  Regressions of \metric{} against categorically coded treatment for multithreading vs. multiprocessing experiment (Section \ref{sec:multithreading-vs-multiprocessing}).
  Lower is better.
  Ordinary least squares regression (top row) estimates relationship between categorical dependent variable and mean of response variable.
  Quantile regression (bottom row) estimates relationship between categorical independent variable and median of response variable.
  Error bands and bars are 95\% confidence intervals.
  }
  \label{fig:multithreading-vs-multiprocessing-regression-\metricslug}
\end{figure*}

}

\def\metric{Latency Walltime Inlet (ns)}\def\metricslug{latency-walltime-inlet-ns}\def\xvarslug{0-multithreading-1-multiprocessing}\def\yvarprefix{}\begin{figure*}[h]
  \centering

  \begin{subfigure}[b]{0.5\textwidth}
    \centering
    \includegraphics[width=\textwidth]{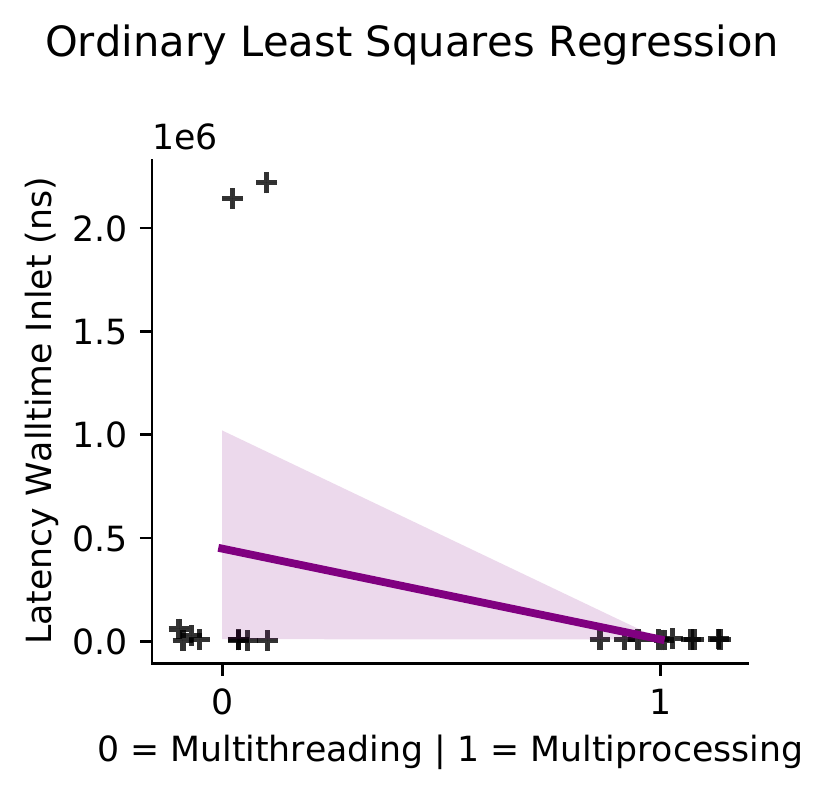}
    \caption{
      Ordinary least squares regression plot.
      Observations are means per replicate.
    }
    \label{fig:multithreading-vs-multiprocessing-regression-ols-\metricslug-complete-regression}
  \end{subfigure}%
  \begin{subfigure}[b]{0.5\textwidth}
    \centering
    \includegraphics[width=\textwidth]{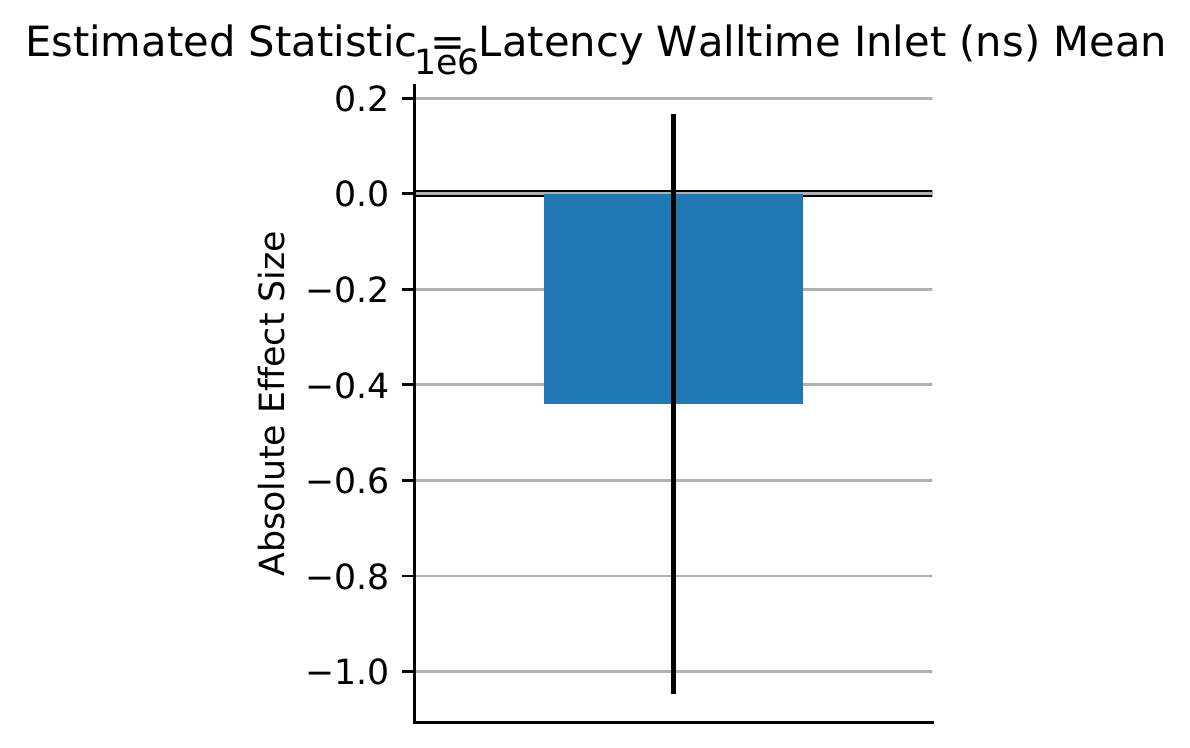}
    \caption{Estimated regression coefficient for ordinary least squares regression. Zero corresponds to no effect.}
    \label{fig:multithreading-vs-multiprocessing-regression-ols-\metricslug-complete-effect-size}
  \end{subfigure}

  \begin{subfigure}[b]{0.5\textwidth}
    \centering
    \includegraphics[width=\textwidth]{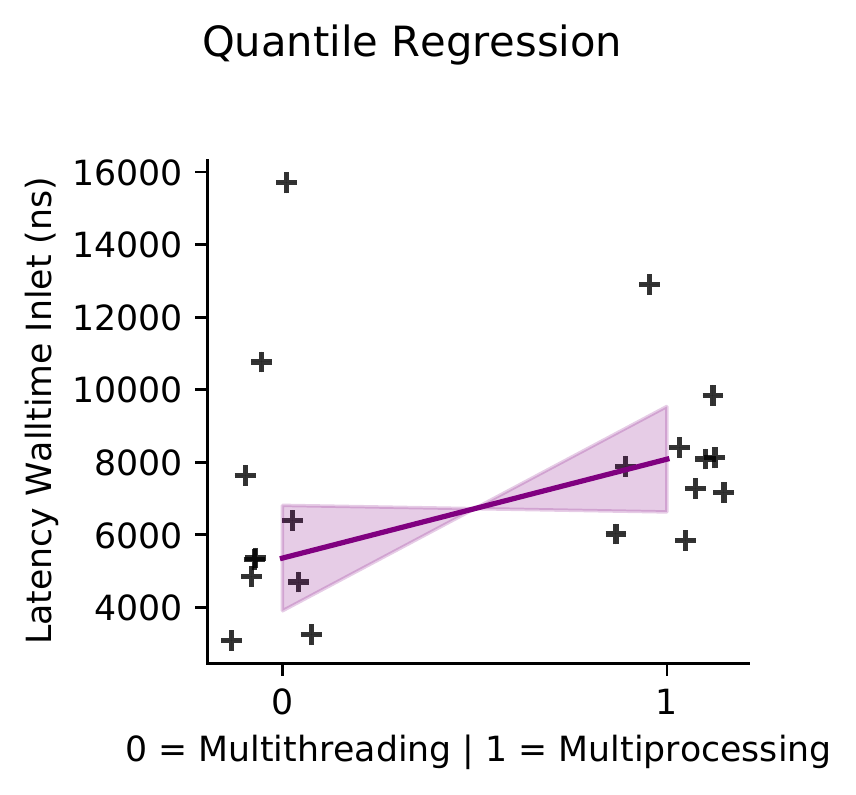}
    \caption{
      Quantile regression plot.
      Observations are medians per replicate.
    }
    \label{fig:multithreading-vs-multiprocessing-regression-quantile-\metricslug-complete-regression}
  \end{subfigure}%
  \begin{subfigure}[b]{0.5\textwidth}
    \centering
    \includegraphics[width=\textwidth]{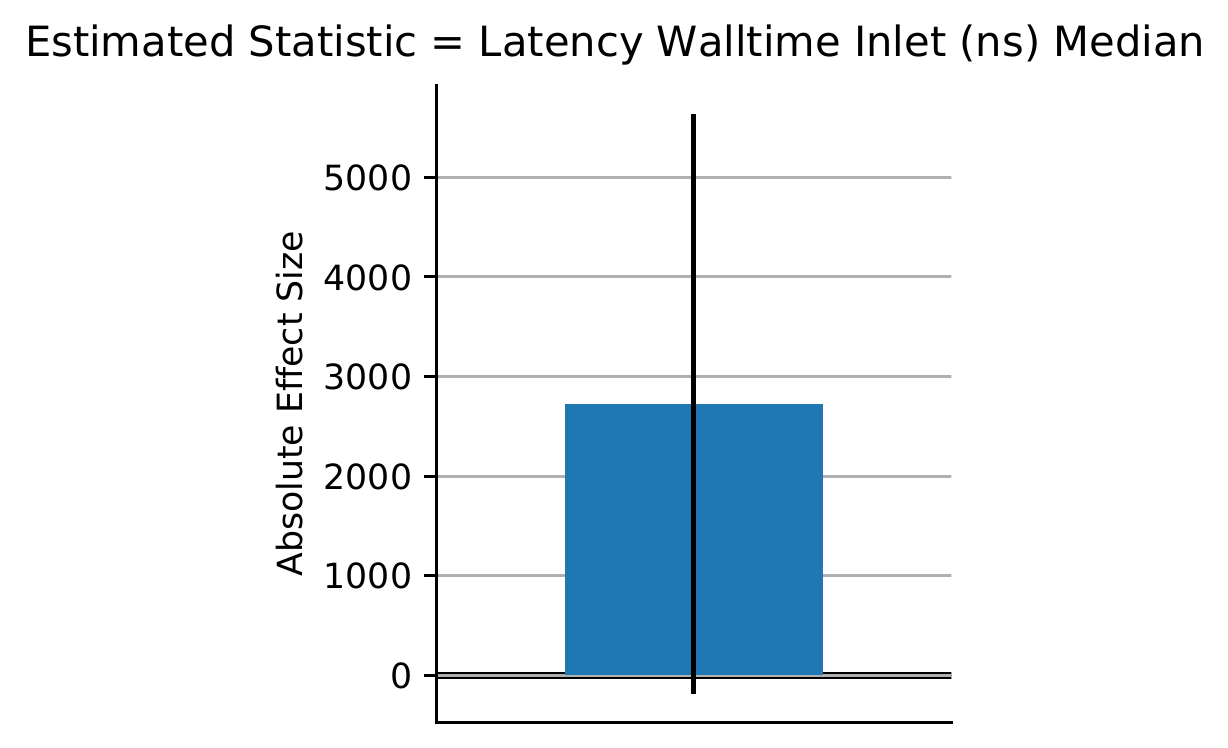}
    \caption{Estimated regression coefficient for quantile regression. Zero corresponds to no effect.}
    \label{fig:multithreading-vs-multiprocessing-regression-quantile-\metricslug-complete-effect-size}
  \end{subfigure}

  \caption{
  Regressions of \metric{} against categorically coded treatment for multithreading vs. multiprocessing experiment (Section \ref{sec:multithreading-vs-multiprocessing}).
  Lower is better.
  Ordinary least squares regression (top row) estimates relationship between categorical dependent variable and mean of response variable.
  Quantile regression (bottom row) estimates relationship between categorical independent variable and median of response variable.
  Error bands and bars are 95\% confidence intervals.
  }
  \label{fig:multithreading-vs-multiprocessing-regression-\metricslug}
\end{figure*}

\def\metric{Latency Simsteps Outlet}\def\metricslug{latency-simsteps-outlet}\def\xvarslug{0-multithreading-1-multiprocessing}\def\yvarprefix{}

\def\metric{Latency Walltime Outlet (ns)}\def\metricslug{latency-walltime-outlet-ns}\def\xvarslug{0-multithreading-1-multiprocessing}\def\yvarprefix{}

\def\metric{Delivery Clumpiness}\def\metricslug{delivery-clumpiness}\def\xvarslug{0-multithreading-1-multiprocessing}\def\yvarprefix{}

\def\metric{Simstep Period Inlet (ns)}\def\metricslug{simstep-period-inlet-ns}\def\xvarslug{0-multithreading-1-multiprocessing}\def\yvarprefix{}

\def\metric{Latency Simsteps Inlet}\def\metricslug{latency-simsteps-inlet}\def\xvarslug{0-multithreading-1-multiprocessing}\def\yvarprefix{}

\def\metric{Simstep Period Outlet (ns)}\def\metricslug{simstep-period-outlet-ns}\def\xvarslug{0-multithreading-1-multiprocessing}\def\yvarprefix{}

\def\metric{Delivery Failure Rate}\def\metricslug{delivery-failure-rate}\def\xvarslug{0-multithreading-1-multiprocessing}\def\yvarprefix{}

\providecommand{\tabmultithreadingvsmultiprocessingregressionmacro}[2]{
\def\regression{#1}\def\regressionslug{#2}

\providecommand{\dissertationelse}[2]{%
\ifdefined\DISSERTATION
#1
\else
#2
\fi
}

\makeatletter
  \providecommand\importpath{\import@path}
\makeatother

\pragmaonce

\def\replacesinglequote#1{%
  \IfSubStr{#1}{'}{%
    \StrSubstitute{#1}{'}{\,}}{#1}}

\pragmaonce

\providecommand\rotcol[1]{ \multicolumn{1}{@{}@{}}{} \adjustbox{angle=60,lap=\width}{\rule{0.4\textwidth}{0.8pt}} \hfill \makebox[3ex]{\adjustbox{angle=60,lap=\width}{\bfseries \hspace{1ex} #1}} \hfill }


\clearpage
{
\onecolumn

\begin{landscape}
\dissertationelse{\fontsize{8}{9}\selectfont}{\footnotesize}
\csvstyle{myTableStyle}{
  longtable=@{\hspace{-.5\arrayrulewidth}}c@{\hspace{-.5\arrayrulewidth}}|l|l|l|l|l|l|l|l|l|l|l|l|l|l|,
  table head=\caption{
    Full \regression{} results of quality of service metrics against categorically coded treatment for multithreading vs. multiprocessing experiment (Section \ref{sec:multithreading-vs-multiprocessing}).
    Significance level $p < 0.05$ used.
    Inf or NaN values may occur due to multicollinearity or due to inf or NaN observations.
  }\label{tab:multithreading-vs-multiprocessing-\regressionslug} \\
    \toprule%
    \multicolumn{1}{@{\hspace{-.5\arrayrulewidth}}l@{\hspace{-.5\arrayrulewidth}}}{} &%
    \rotcol{ Metric } &%
    \rotcol{ Statistic } &%
    \rotcol{ Significant Effect Sign } &%
    \rotcol{ Cpus Per Node } &%
    \rotcol{ Num Simels Per Cpu} &%
    \rotcol{ Num Processes} &%
    \rotcol{ Absolute Effect Size} &%
    \rotcol{ Absolute Effect Size 95\% CI Lower Bound} &%
    \rotcol{ Absolute Effect Size 95\% CI Upper Bound} &%
    \rotcol{ Relative Effect Size} &%
    \rotcol{ Relative Effect Size 95\% CI Lower Bound} &%
    \rotcol{ Relative Effect Size 95\% CI Upper Bound} &%
    \rotcol{$\mathbf{n}$} & %
    \rotcol{$\mathbf{p}$} %
    \vspace{-1ex} \\ \midrule\endfirsthead
    \caption*{\tablename{} \thetable{} (cont'd)} \\
    \toprule%
    \multicolumn{1}{@{\hspace{-.5\arrayrulewidth}}l@{\hspace{-.5\arrayrulewidth}}}{} &%
    \rotcol{ Metric } &%
    \rotcol{ Statistic } &%
    \rotcol{ Significant Effect Sign } &%
    \rotcol{ Cpus Per Node } &%
    \rotcol{ Num Simels Per Cpu} &%
    \rotcol{ Num Processes} &%
    \rotcol{ Absolute Effect Size} &%
    \rotcol{ Absolute Effect Size 95\% CI Lower Bound} &%
    \rotcol{ Absolute Effect Size 95\% CI Upper Bound} &%
    \rotcol{ Relative Effect Size} &%
    \rotcol{ Relative Effect Size 95\% CI Lower Bound} &%
    \rotcol{ Relative Effect Size 95\% CI Upper Bound} &%
    \rotcol{$\mathbf{n}$} & %
    \rotcol{$\mathbf{p}$} %
    \vspace{-1ex} \\ \midrule\endhead
    \bottomrule\endfoot,
  late after line=\\\hline,
  late after last line=\\\bottomrule,
  head to column names,
  filter=\equal{\csuse{RegressionModelSlug}}{\regressionslug},
}

\catcode`\%=12
\providecommand\filename{}
\renewcommand\filename{\detokenize{submodule/conduit-binder/binder/date=2021+project=72k5n/a=multithreading-vs-multiprocessing/outplots/a=multithreading_vs_multiprocessing_regression_results+async_mode=3+impl=proc_thread+nproc=1-2+nthread=1-2+proc=0-1+readability=latexcsvreader+replicate=0-9+slurm_nnodes=1+slurm_ntasks=2+subject=inlet_outlet+view=summary+ext=.csv}}
\catcode`\%=14

\csvreader[myTableStyle]{\importpath\filename}{}%
{ %
  &%
  \csuse{DependentVariable} &%
  \csuse{Statistic} &%
  \ifthenelse{\equal{\csuse{SignificantEffectSign}}{-}}{\cellcolor{red!25}}{}%
  \ifthenelse{\equal{\csuse{SignificantEffectSign}}{+}}{\cellcolor{green!25}}{}%
  \ifthenelse{\equal{\csuse{SignificantEffectSign}}{0}}{\cellcolor{gray!25}}{}%
  \csuse{SignificantEffectSign} &%
  \replacesinglequote{\csuse{CpusPerNode}} &%
  \replacesinglequote{\csuse{NumSimelsPerCpu}} &%
  \replacesinglequote{\csuse{NumProcessesPrettyprint}} &%
  \replacesinglequote{\csuse{AbsoluteEffectSize}} &%
  \replacesinglequote{\csuse{AbsoluteEffectSize95CILowerBound}} &%
  \replacesinglequote{\csuse{AbsoluteEffectSize95CIUpperBound}} &%
  \replacesinglequote{\csuse{RelativeEffectSize}} &%
  \replacesinglequote{\csuse{RelativeEffectSize95CILowerBound}} &%
  \replacesinglequote{\csuse{RelativeEffectSize95CIUpperBound}} &%
  \replacesinglequote{\csuse{n}} &%
  \replacesinglequote{\csuse{p}} %
}

\end{landscape}

}

\clearpage

}

\def\regression{Ordinary Least Squares Regression}\def\regressionslug{ordinary-least-squares-regression}

\providecommand{\dissertationelse}[2]{%
\ifdefined\DISSERTATION
#1
\else
#2
\fi
}

\makeatletter
  \providecommand\importpath{\import@path}
\makeatother

\pragmaonce

\def\replacesinglequote#1{%
  \IfSubStr{#1}{'}{%
    \StrSubstitute{#1}{'}{\,}}{#1}}

\pragmaonce

\providecommand\rotcol[1]{ \multicolumn{1}{@{}@{}}{} \adjustbox{angle=60,lap=\width}{\rule{0.4\textwidth}{0.8pt}} \hfill \makebox[3ex]{\adjustbox{angle=60,lap=\width}{\bfseries \hspace{1ex} #1}} \hfill }


\clearpage
{
\onecolumn

\begin{landscape}
\dissertationelse{\fontsize{8}{9}\selectfont}{\footnotesize}
\csvstyle{myTableStyle}{
  longtable=@{\hspace{-.5\arrayrulewidth}}c@{\hspace{-.5\arrayrulewidth}}|l|l|l|l|l|l|l|l|l|l|l|l|l|l|,
  table head=\caption{
    Full \regression{} results of quality of service metrics against categorically coded treatment for multithreading vs. multiprocessing experiment (Section \ref{sec:multithreading-vs-multiprocessing}).
    Significance level $p < 0.05$ used.
    Inf or NaN values may occur due to multicollinearity or due to inf or NaN observations.
  }\label{tab:multithreading-vs-multiprocessing-\regressionslug} \\
    \toprule%
    \multicolumn{1}{@{\hspace{-.5\arrayrulewidth}}l@{\hspace{-.5\arrayrulewidth}}}{} &%
    \rotcol{ Metric } &%
    \rotcol{ Statistic } &%
    \rotcol{ Significant Effect Sign } &%
    \rotcol{ Cpus Per Node } &%
    \rotcol{ Num Simels Per Cpu} &%
    \rotcol{ Num Processes} &%
    \rotcol{ Absolute Effect Size} &%
    \rotcol{ Absolute Effect Size 95\% CI Lower Bound} &%
    \rotcol{ Absolute Effect Size 95\% CI Upper Bound} &%
    \rotcol{ Relative Effect Size} &%
    \rotcol{ Relative Effect Size 95\% CI Lower Bound} &%
    \rotcol{ Relative Effect Size 95\% CI Upper Bound} &%
    \rotcol{$\mathbf{n}$} & %
    \rotcol{$\mathbf{p}$} %
    \vspace{-1ex} \\ \midrule\endfirsthead
    \caption*{\tablename{} \thetable{} (cont'd)} \\
    \toprule%
    \multicolumn{1}{@{\hspace{-.5\arrayrulewidth}}l@{\hspace{-.5\arrayrulewidth}}}{} &%
    \rotcol{ Metric } &%
    \rotcol{ Statistic } &%
    \rotcol{ Significant Effect Sign } &%
    \rotcol{ Cpus Per Node } &%
    \rotcol{ Num Simels Per Cpu} &%
    \rotcol{ Num Processes} &%
    \rotcol{ Absolute Effect Size} &%
    \rotcol{ Absolute Effect Size 95\% CI Lower Bound} &%
    \rotcol{ Absolute Effect Size 95\% CI Upper Bound} &%
    \rotcol{ Relative Effect Size} &%
    \rotcol{ Relative Effect Size 95\% CI Lower Bound} &%
    \rotcol{ Relative Effect Size 95\% CI Upper Bound} &%
    \rotcol{$\mathbf{n}$} & %
    \rotcol{$\mathbf{p}$} %
    \vspace{-1ex} \\ \midrule\endhead
    \bottomrule\endfoot,
  late after line=\\\hline,
  late after last line=\\\bottomrule,
  head to column names,
  filter=\equal{\csuse{RegressionModelSlug}}{\regressionslug},
}

\catcode`\%=12
\providecommand\filename{}
\renewcommand\filename{\detokenize{submodule/conduit-binder/binder/date=2021+project=72k5n/a=multithreading-vs-multiprocessing/outplots/a=multithreading_vs_multiprocessing_regression_results+async_mode=3+impl=proc_thread+nproc=1-2+nthread=1-2+proc=0-1+readability=latexcsvreader+replicate=0-9+slurm_nnodes=1+slurm_ntasks=2+subject=inlet_outlet+view=summary+ext=.csv}}
\catcode`\%=14

\csvreader[myTableStyle]{\importpath\filename}{}%
{ %
  &%
  \csuse{DependentVariable} &%
  \csuse{Statistic} &%
  \ifthenelse{\equal{\csuse{SignificantEffectSign}}{-}}{\cellcolor{red!25}}{}%
  \ifthenelse{\equal{\csuse{SignificantEffectSign}}{+}}{\cellcolor{green!25}}{}%
  \ifthenelse{\equal{\csuse{SignificantEffectSign}}{0}}{\cellcolor{gray!25}}{}%
  \csuse{SignificantEffectSign} &%
  \replacesinglequote{\csuse{CpusPerNode}} &%
  \replacesinglequote{\csuse{NumSimelsPerCpu}} &%
  \replacesinglequote{\csuse{NumProcessesPrettyprint}} &%
  \replacesinglequote{\csuse{AbsoluteEffectSize}} &%
  \replacesinglequote{\csuse{AbsoluteEffectSize95CILowerBound}} &%
  \replacesinglequote{\csuse{AbsoluteEffectSize95CIUpperBound}} &%
  \replacesinglequote{\csuse{RelativeEffectSize}} &%
  \replacesinglequote{\csuse{RelativeEffectSize95CILowerBound}} &%
  \replacesinglequote{\csuse{RelativeEffectSize95CIUpperBound}} &%
  \replacesinglequote{\csuse{n}} &%
  \replacesinglequote{\csuse{p}} %
}

\end{landscape}

}

\clearpage

\def\regression{Quantile Regression}\def\regressionslug{quantile-regression}

\clearpage

\subsection{With lac-417 vs. Sans lac-417}

This section provides full results from faulty hardware experiments discussed in Section \ref{sec:with-lac-417-vs-sans-lac-417}.

\providecommand{\figwithlacvssanslacdistributionmacro}[2]{
\def\metric{#1}\def\metricslug{#2}\begin{figure*}[h]
  \centering
  \begin{subfigure}[b]{0.5\textwidth}
    \centering
    \includegraphics[width=\textwidth]{submodule/conduit-binder/binder/date=2021+project=72k5n/a=with-lac-417-vs-sans-lac-417/teeplots/\metricslug/row=num-simels-per-cpu+transform=snapshot_diffs+viz=facet-boxplot-nofliers+x=0-with-lac-417-1-sans-lac-417+y=\metricslug+ext=}
    \caption{Distribution of \metric{} for each snapshot, without outliers.}
    \label{fig:with-lac-417-vs-sans-lac-417-distribution-\metricslug-nofliers}
  \end{subfigure}%
  \begin{subfigure}[b]{0.5\textwidth}
    \centering
    \includegraphics[width=\textwidth]{submodule/conduit-binder/binder/date=2021+project=72k5n/a=with-lac-417-vs-sans-lac-417/teeplots/\metricslug/row=num-simels-per-cpu+transform=snapshot_diffs+viz=facet-boxplot-withfliers+x=0-with-lac-417-1-sans-lac-417+y=\metricslug+ext=}
    \caption{Distribution of \metric{} for each snapshot, with outliers.}
    \label{fig:with-lac-417-vs-sans-lac-417-distribution-\metricslug-withfliers}
  \end{subfigure}
  \caption{Distribution of \metric{} for individual snapshot measurements for faulty hardware experiment (Section \ref{sec:with-lac-417-vs-sans-lac-417}). Lower is better.}
  \label{fig:with-lac-417-vs-sans-lac-417-distribution-\metricslug}
\end{figure*}

}

\def\metric{Latency Walltime Inlet (ns)}\def\metricslug{latency-walltime-inlet-ns}\begin{figure*}[h]
  \centering
  \begin{subfigure}[b]{0.5\textwidth}
    \centering
    \includegraphics[width=\textwidth]{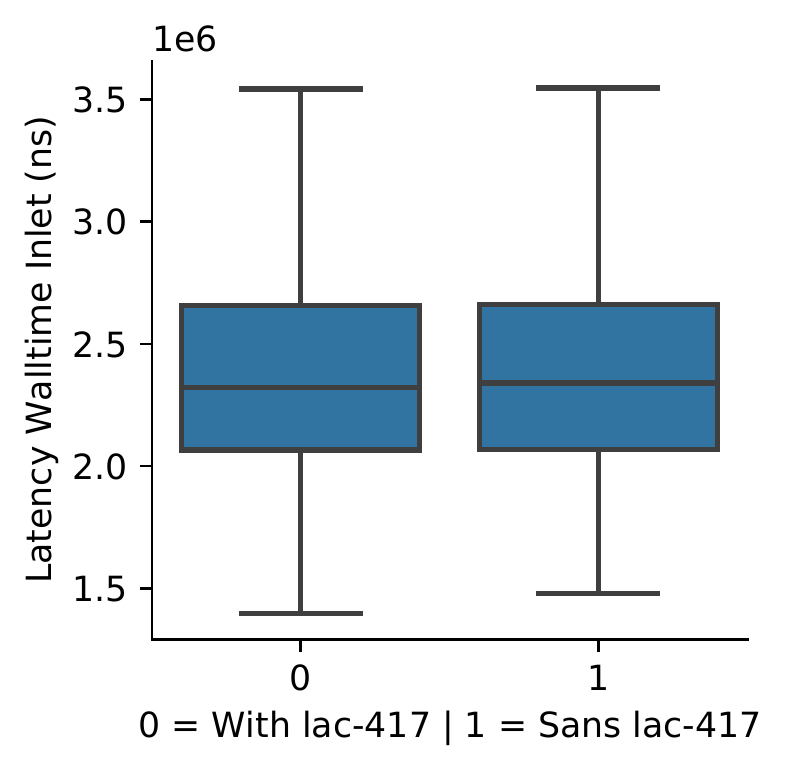}
    \caption{Distribution of \metric{} for each snapshot, without outliers.}
    \label{fig:with-lac-417-vs-sans-lac-417-distribution-\metricslug-nofliers}
  \end{subfigure}%
  \begin{subfigure}[b]{0.5\textwidth}
    \centering
    \includegraphics[width=\textwidth]{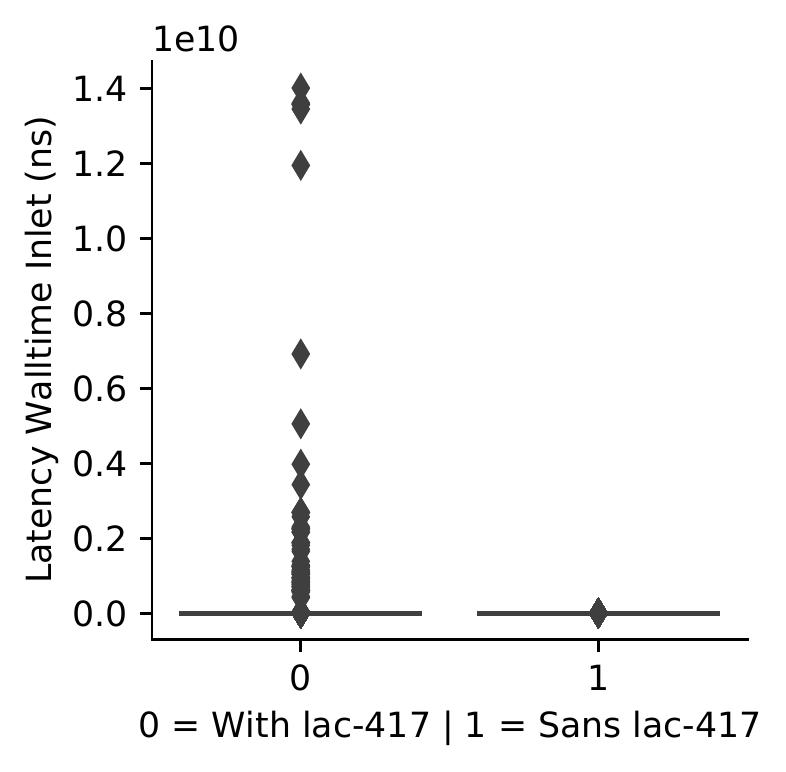}
    \caption{Distribution of \metric{} for each snapshot, with outliers.}
    \label{fig:with-lac-417-vs-sans-lac-417-distribution-\metricslug-withfliers}
  \end{subfigure}
  \caption{Distribution of \metric{} for individual snapshot measurements for faulty hardware experiment (Section \ref{sec:with-lac-417-vs-sans-lac-417}). Lower is better.}
  \label{fig:with-lac-417-vs-sans-lac-417-distribution-\metricslug}
\end{figure*}

\def\metric{Latency Simsteps Outlet}\def\metricslug{latency-simsteps-outlet}\begin{figure*}[h]
  \centering
  \begin{subfigure}[b]{0.5\textwidth}
    \centering
    \includegraphics[width=\textwidth]{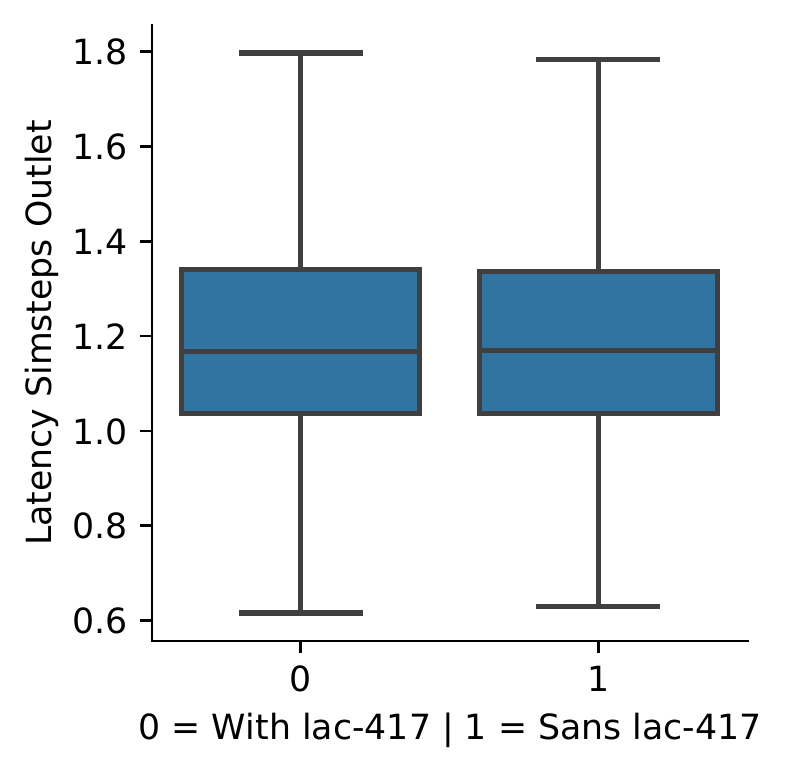}
    \caption{Distribution of \metric{} for each snapshot, without outliers.}
    \label{fig:with-lac-417-vs-sans-lac-417-distribution-\metricslug-nofliers}
  \end{subfigure}%
  \begin{subfigure}[b]{0.5\textwidth}
    \centering
    \includegraphics[width=\textwidth]{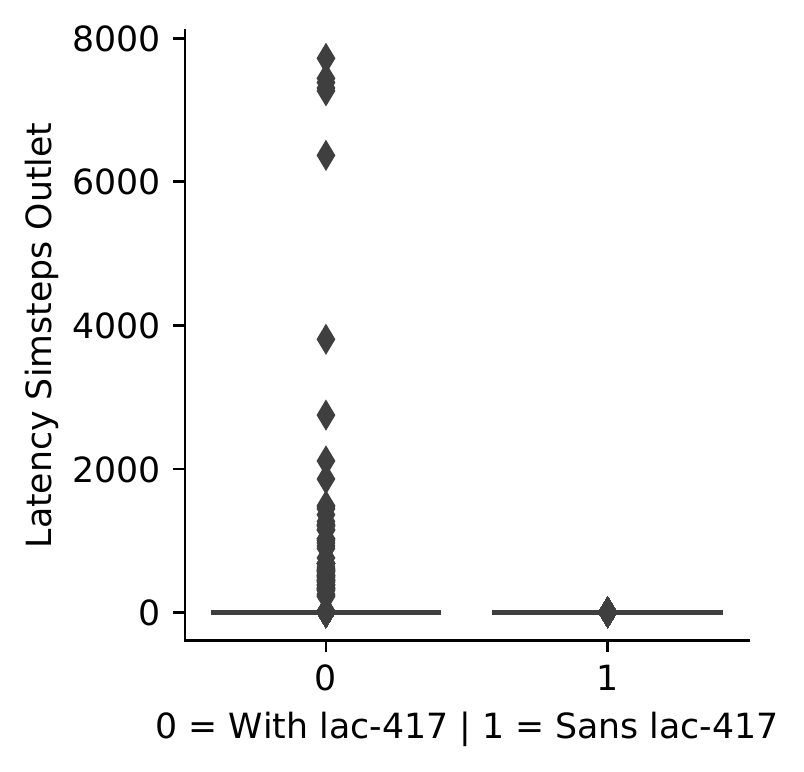}
    \caption{Distribution of \metric{} for each snapshot, with outliers.}
    \label{fig:with-lac-417-vs-sans-lac-417-distribution-\metricslug-withfliers}
  \end{subfigure}
  \caption{Distribution of \metric{} for individual snapshot measurements for faulty hardware experiment (Section \ref{sec:with-lac-417-vs-sans-lac-417}). Lower is better.}
  \label{fig:with-lac-417-vs-sans-lac-417-distribution-\metricslug}
\end{figure*}

\def\metric{Latency Walltime Outlet (ns)}\def\metricslug{latency-walltime-outlet-ns}\begin{figure*}[h]
  \centering
  \begin{subfigure}[b]{0.5\textwidth}
    \centering
    \includegraphics[width=\textwidth]{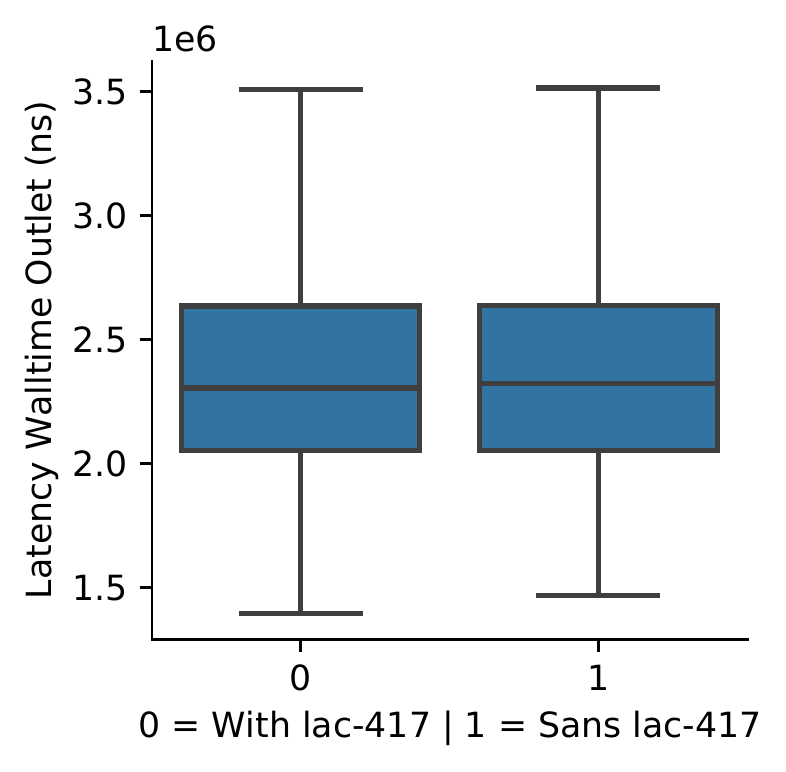}
    \caption{Distribution of \metric{} for each snapshot, without outliers.}
    \label{fig:with-lac-417-vs-sans-lac-417-distribution-\metricslug-nofliers}
  \end{subfigure}%
  \begin{subfigure}[b]{0.5\textwidth}
    \centering
    \includegraphics[width=\textwidth]{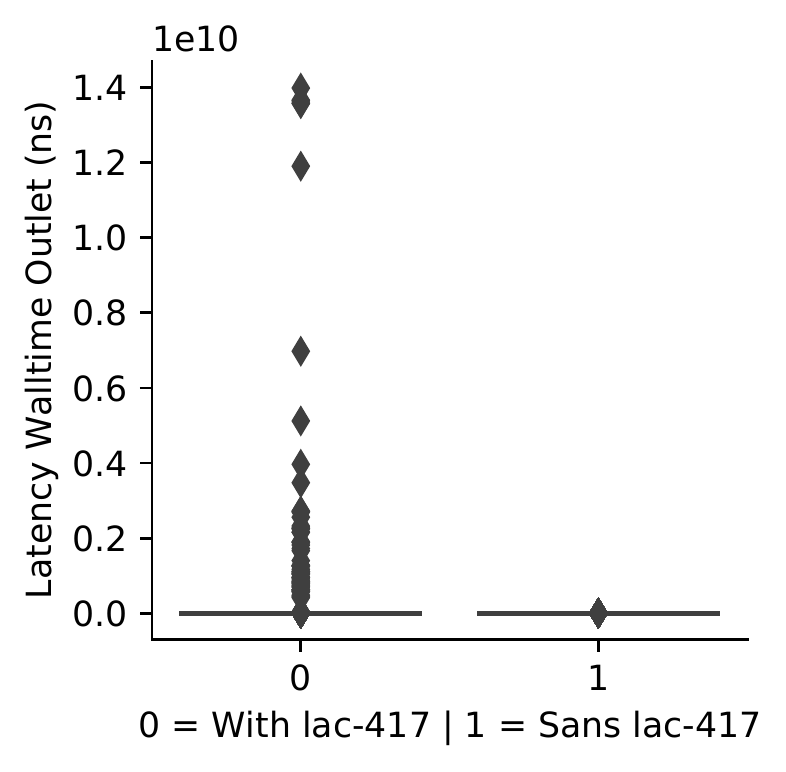}
    \caption{Distribution of \metric{} for each snapshot, with outliers.}
    \label{fig:with-lac-417-vs-sans-lac-417-distribution-\metricslug-withfliers}
  \end{subfigure}
  \caption{Distribution of \metric{} for individual snapshot measurements for faulty hardware experiment (Section \ref{sec:with-lac-417-vs-sans-lac-417}). Lower is better.}
  \label{fig:with-lac-417-vs-sans-lac-417-distribution-\metricslug}
\end{figure*}

\def\metric{Delivery Clumpiness}\def\metricslug{delivery-clumpiness}\begin{figure*}[h]
  \centering
  \begin{subfigure}[b]{0.5\textwidth}
    \centering
    \includegraphics[width=\textwidth]{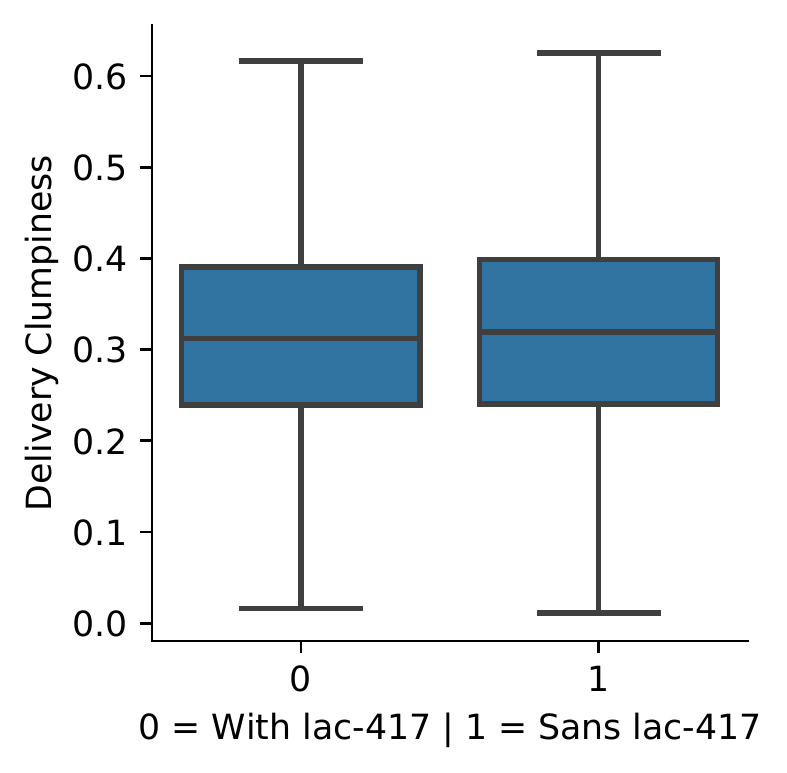}
    \caption{Distribution of \metric{} for each snapshot, without outliers.}
    \label{fig:with-lac-417-vs-sans-lac-417-distribution-\metricslug-nofliers}
  \end{subfigure}%
  \begin{subfigure}[b]{0.5\textwidth}
    \centering
    \includegraphics[width=\textwidth]{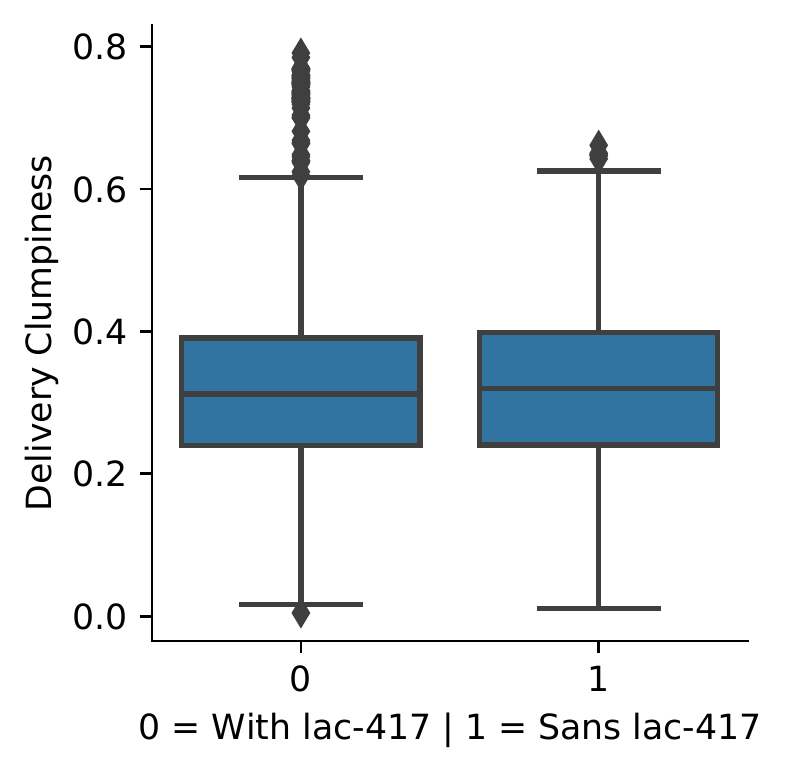}
    \caption{Distribution of \metric{} for each snapshot, with outliers.}
    \label{fig:with-lac-417-vs-sans-lac-417-distribution-\metricslug-withfliers}
  \end{subfigure}
  \caption{Distribution of \metric{} for individual snapshot measurements for faulty hardware experiment (Section \ref{sec:with-lac-417-vs-sans-lac-417}). Lower is better.}
  \label{fig:with-lac-417-vs-sans-lac-417-distribution-\metricslug}
\end{figure*}

\def\metric{Simstep Period Inlet (ns)}\def\metricslug{simstep-period-inlet-ns}\begin{figure*}[h]
  \centering
  \begin{subfigure}[b]{0.5\textwidth}
    \centering
    \includegraphics[width=\textwidth]{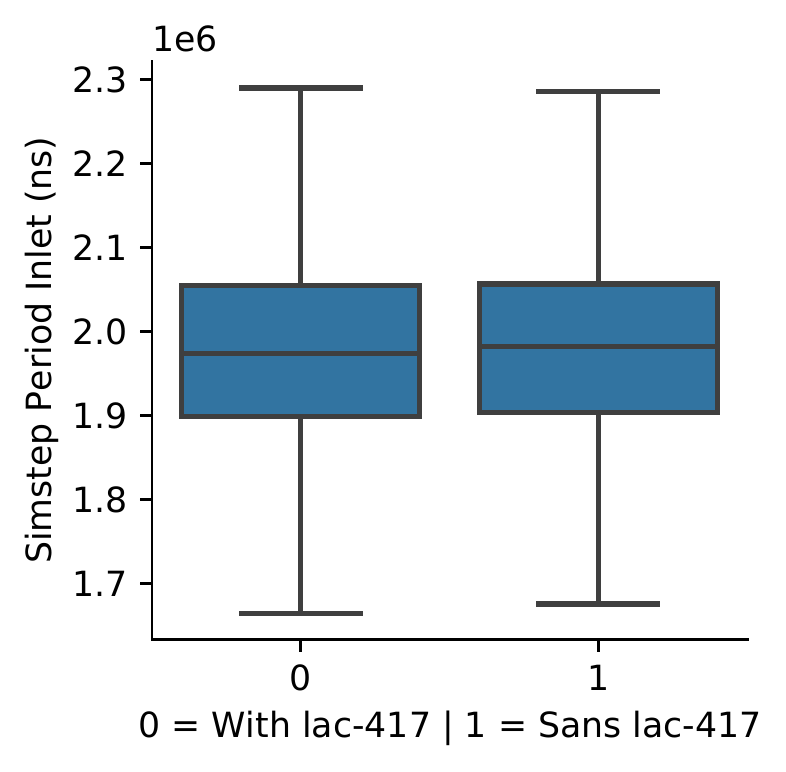}
    \caption{Distribution of \metric{} for each snapshot, without outliers.}
    \label{fig:with-lac-417-vs-sans-lac-417-distribution-\metricslug-nofliers}
  \end{subfigure}%
  \begin{subfigure}[b]{0.5\textwidth}
    \centering
    \includegraphics[width=\textwidth]{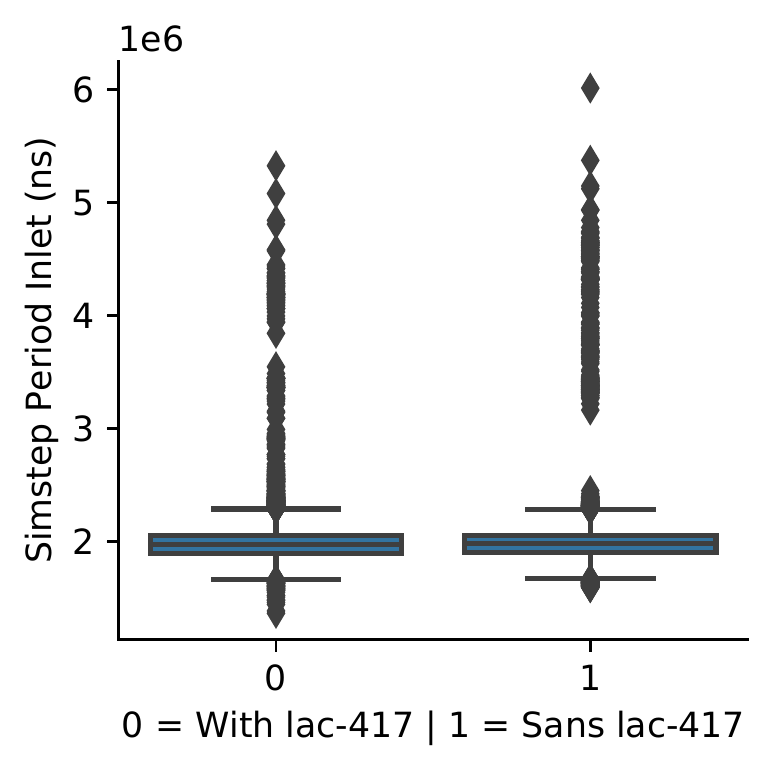}
    \caption{Distribution of \metric{} for each snapshot, with outliers.}
    \label{fig:with-lac-417-vs-sans-lac-417-distribution-\metricslug-withfliers}
  \end{subfigure}
  \caption{Distribution of \metric{} for individual snapshot measurements for faulty hardware experiment (Section \ref{sec:with-lac-417-vs-sans-lac-417}). Lower is better.}
  \label{fig:with-lac-417-vs-sans-lac-417-distribution-\metricslug}
\end{figure*}

\def\metric{Latency Simsteps Inlet}\def\metricslug{latency-simsteps-inlet}\begin{figure*}[h]
  \centering
  \begin{subfigure}[b]{0.5\textwidth}
    \centering
    \includegraphics[width=\textwidth]{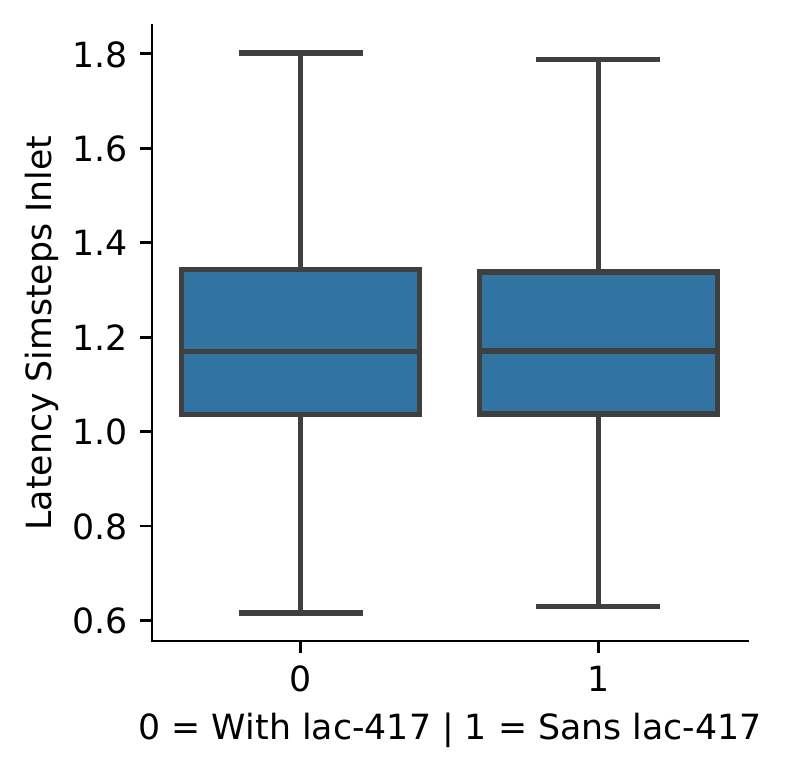}
    \caption{Distribution of \metric{} for each snapshot, without outliers.}
    \label{fig:with-lac-417-vs-sans-lac-417-distribution-\metricslug-nofliers}
  \end{subfigure}%
  \begin{subfigure}[b]{0.5\textwidth}
    \centering
    \includegraphics[width=\textwidth]{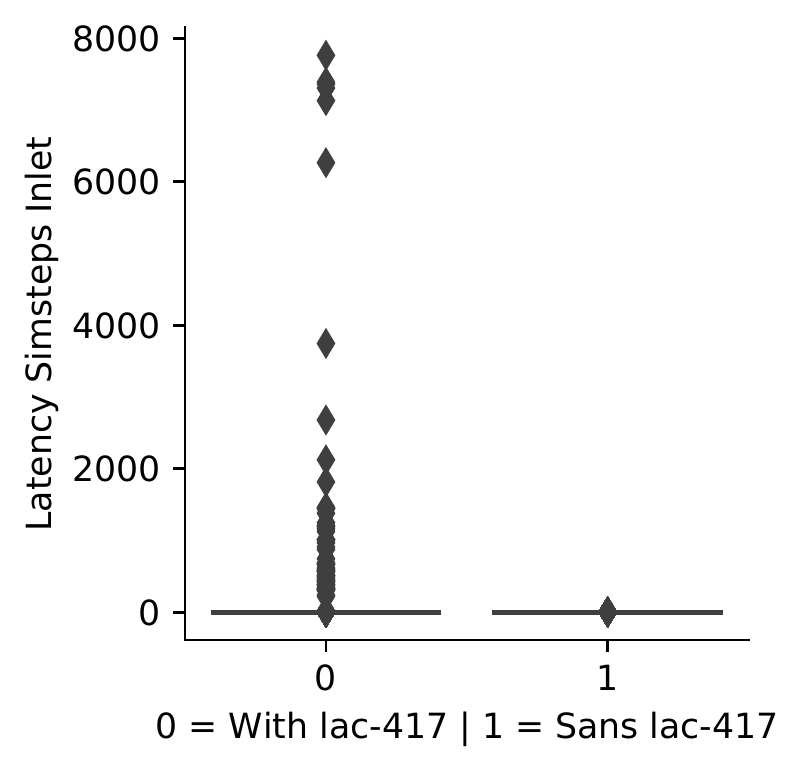}
    \caption{Distribution of \metric{} for each snapshot, with outliers.}
    \label{fig:with-lac-417-vs-sans-lac-417-distribution-\metricslug-withfliers}
  \end{subfigure}
  \caption{Distribution of \metric{} for individual snapshot measurements for faulty hardware experiment (Section \ref{sec:with-lac-417-vs-sans-lac-417}). Lower is better.}
  \label{fig:with-lac-417-vs-sans-lac-417-distribution-\metricslug}
\end{figure*}

\def\metric{Simstep Period Outlet (ns)}\def\metricslug{simstep-period-outlet-ns}\begin{figure*}[h]
  \centering
  \begin{subfigure}[b]{0.5\textwidth}
    \centering
    \includegraphics[width=\textwidth]{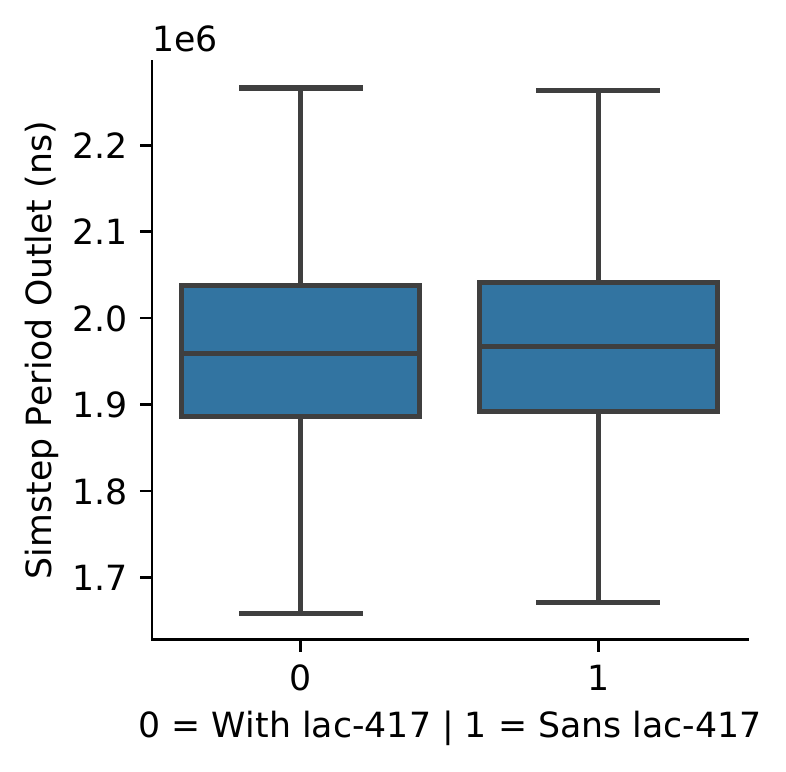}
    \caption{Distribution of \metric{} for each snapshot, without outliers.}
    \label{fig:with-lac-417-vs-sans-lac-417-distribution-\metricslug-nofliers}
  \end{subfigure}%
  \begin{subfigure}[b]{0.5\textwidth}
    \centering
    \includegraphics[width=\textwidth]{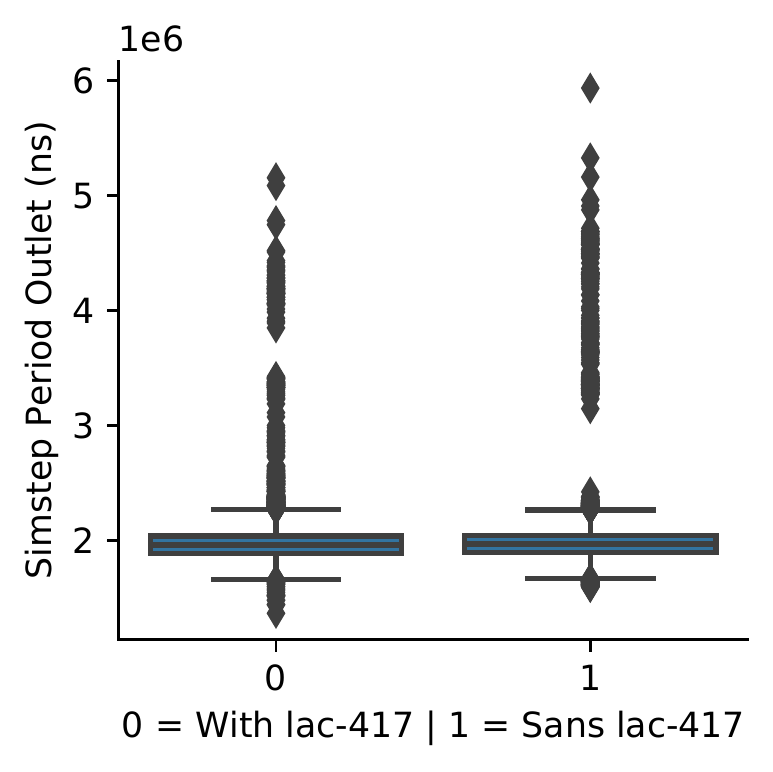}
    \caption{Distribution of \metric{} for each snapshot, with outliers.}
    \label{fig:with-lac-417-vs-sans-lac-417-distribution-\metricslug-withfliers}
  \end{subfigure}
  \caption{Distribution of \metric{} for individual snapshot measurements for faulty hardware experiment (Section \ref{sec:with-lac-417-vs-sans-lac-417}). Lower is better.}
  \label{fig:with-lac-417-vs-sans-lac-417-distribution-\metricslug}
\end{figure*}

\def\metric{Delivery Failure Rate}\def\metricslug{delivery-failure-rate}\begin{figure*}[h]
  \centering
  \begin{subfigure}[b]{0.5\textwidth}
    \centering
    \includegraphics[width=\textwidth]{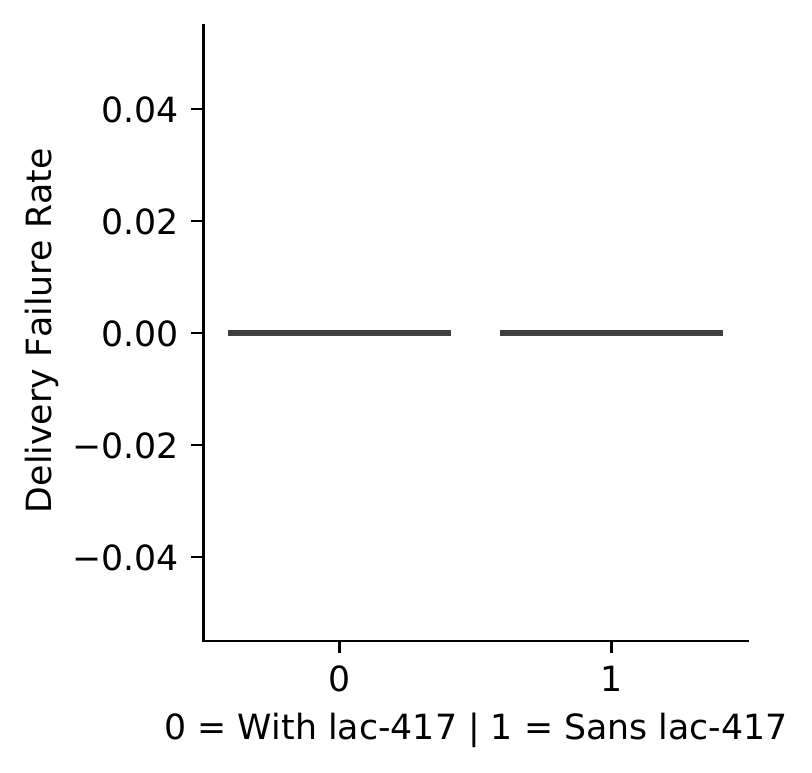}
    \caption{Distribution of \metric{} for each snapshot, without outliers.}
    \label{fig:with-lac-417-vs-sans-lac-417-distribution-\metricslug-nofliers}
  \end{subfigure}%
  \begin{subfigure}[b]{0.5\textwidth}
    \centering
    \includegraphics[width=\textwidth]{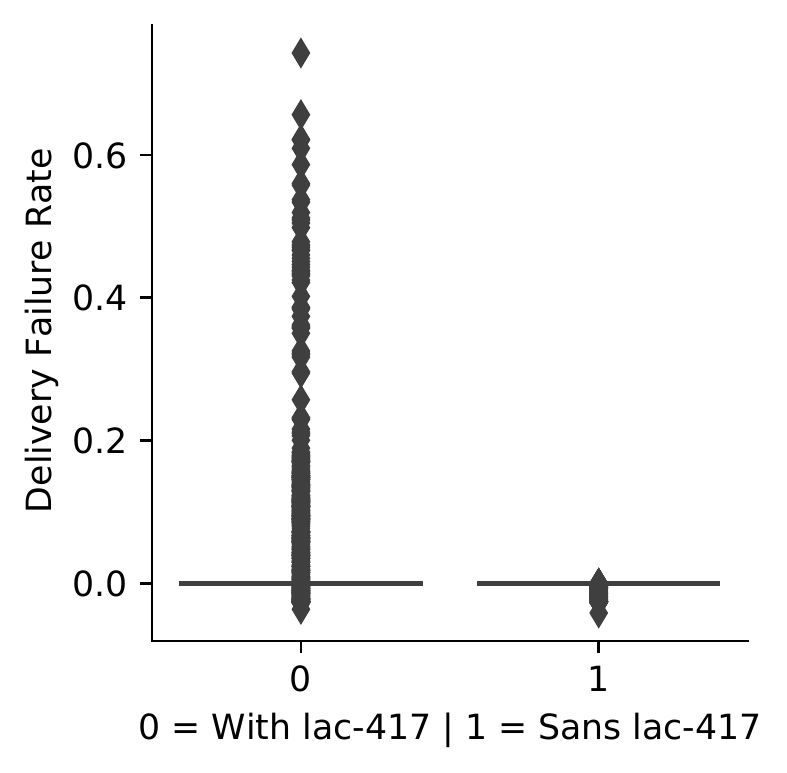}
    \caption{Distribution of \metric{} for each snapshot, with outliers.}
    \label{fig:with-lac-417-vs-sans-lac-417-distribution-\metricslug-withfliers}
  \end{subfigure}
  \caption{Distribution of \metric{} for individual snapshot measurements for faulty hardware experiment (Section \ref{sec:with-lac-417-vs-sans-lac-417}). Lower is better.}
  \label{fig:with-lac-417-vs-sans-lac-417-distribution-\metricslug}
\end{figure*}

\providecommand{\figwithlacvssanslacregressionmacro}[4]{
\def\metric{#1}\def\metricslug{#2}\def\xvarslug{#3}\def\yvarprefix{#4}\begin{figure*}[h]
  \centering

  \begin{subfigure}[b]{0.5\textwidth}
    \centering
    \includegraphics[width=\textwidth]{submodule/conduit-binder/binder/date=2021+project=72k5n/a=with-lac-417-vs-sans-lac-417/teeplots/\metricslug/marker=+regression=ordinary-least-squares-regression+row=num-simels-per-cpu+transform=snapshot_diffs-groupby_exec_instance-mean+viz=facet-unsplit-regression+x=\xvarslug+y=\yvarprefix\metricslug+ext=}
    \caption{
      Ordinary least squares regression plot.
      Observations are means per replicate.
    }
    \label{fig:with-lac-417-vs-sans-lac-417-regression-ols-\metricslug-complete-regression}
  \end{subfigure}%
  \begin{subfigure}[b]{0.5\textwidth}
    \centering
    \includegraphics[width=\textwidth]{submodule/conduit-binder/binder/date=2021+project=72k5n/a=with-lac-417-vs-sans-lac-417/teeplots/\metricslug/estimated-statistic=\metricslug-mean+row=num-simels-per-cpu+transform=fit_regression+viz=lambda+y=absolute-effect-size+ext=}
    \caption{Estimated regression coefficient for ordinary least squares regression. Zero corresponds to no effect.}
    \label{fig:with-lac-417-vs-sans-lac-417-regression-ols-\metricslug-complete-effect-size}
  \end{subfigure}

  \begin{subfigure}[b]{0.5\textwidth}
    \centering
    \includegraphics[width=\textwidth]{submodule/conduit-binder/binder/date=2021+project=72k5n/a=with-lac-417-vs-sans-lac-417/teeplots/\metricslug/marker=+regression=quantile-regression+row=num-simels-per-cpu+transform=snapshot_diffs-groupby_exec_instance-median+viz=facet-unsplit-regression+x=\xvarslug+y=\yvarprefix\metricslug+ext=}
    \caption{
      Quantile regression plot.
      Observations are medians per replicate.
    }
    \label{fig:with-lac-417-vs-sans-lac-417-regression-quantile-\metricslug-complete-regression}
  \end{subfigure}%
  \begin{subfigure}[b]{0.5\textwidth}
    \centering
    \includegraphics[width=\textwidth]{submodule/conduit-binder/binder/date=2021+project=72k5n/a=with-lac-417-vs-sans-lac-417/teeplots/\metricslug/estimated-statistic=\metricslug-median+row=num-simels-per-cpu+transform=fit_regression+viz=lambda+y=absolute-effect-size+ext=}
    \caption{Estimated regression coefficient for quantile regression. Zero corresponds to no effect.}
    \label{fig:with-lac-417-vs-sans-lac-417-regression-quantile-\metricslug-complete-effect-size}
  \end{subfigure}

  \caption{
  Regressions of \metric{} against categorically coded treatment for faulty hardware experiment (Section \ref{sec:with-lac-417-vs-sans-lac-417}).
  Lower is better.
  Ordinary least squares regression (top row) estimates relationship between categorical dependent variable and mean of response variable.
  Quantile regression (bottom row) estimates relationship between categorical independent variable and median of response variable.
  Error bands and bars are 95\% confidence intervals.
  }
  \label{fig:with-lac-417-vs-sans-lac-417-regression-\metricslug}
\end{figure*}

}

\def\metric{Latency Walltime Inlet (ns)}\def\metricslug{latency-walltime-inlet-ns}\def\xvarslug{0-with-lac-417-1-sans-lac-417}\def\yvarprefix{}\begin{figure*}[h]
  \centering

  \begin{subfigure}[b]{0.5\textwidth}
    \centering
    \includegraphics[width=\textwidth]{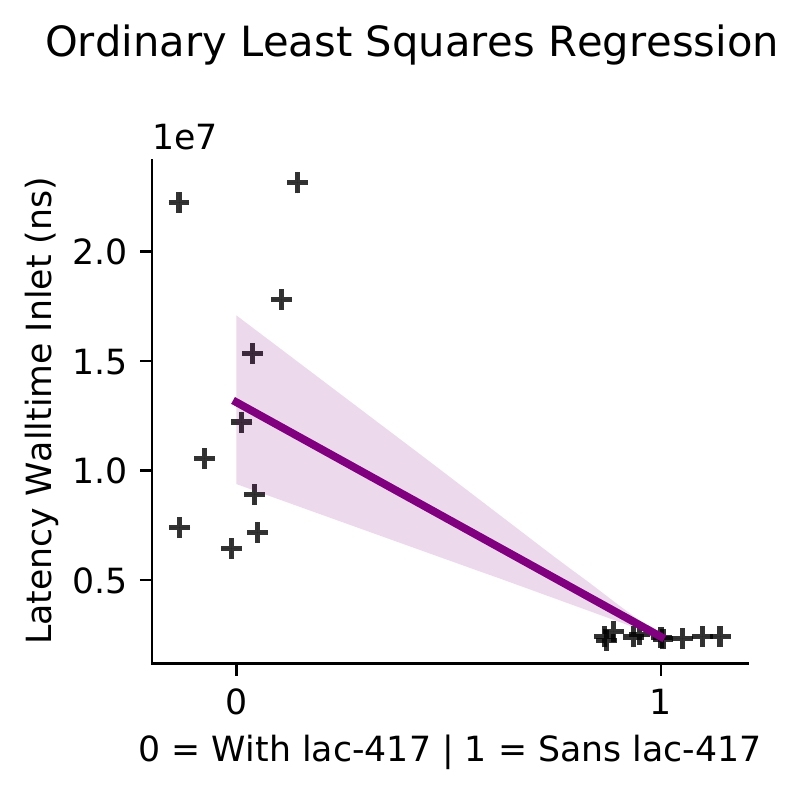}
    \caption{
      Ordinary least squares regression plot.
      Observations are means per replicate.
    }
    \label{fig:with-lac-417-vs-sans-lac-417-regression-ols-\metricslug-complete-regression}
  \end{subfigure}%
  \begin{subfigure}[b]{0.5\textwidth}
    \centering
    \includegraphics[width=\textwidth]{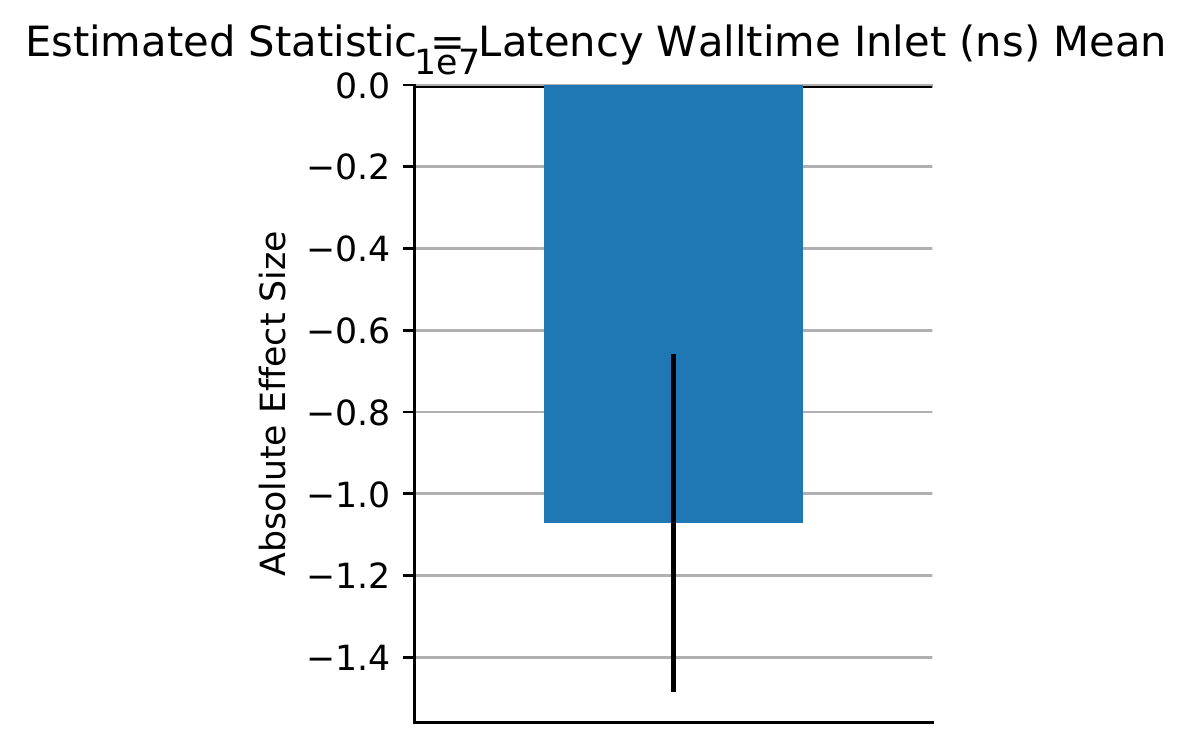}
    \caption{Estimated regression coefficient for ordinary least squares regression. Zero corresponds to no effect.}
    \label{fig:with-lac-417-vs-sans-lac-417-regression-ols-\metricslug-complete-effect-size}
  \end{subfigure}

  \begin{subfigure}[b]{0.5\textwidth}
    \centering
    \includegraphics[width=\textwidth]{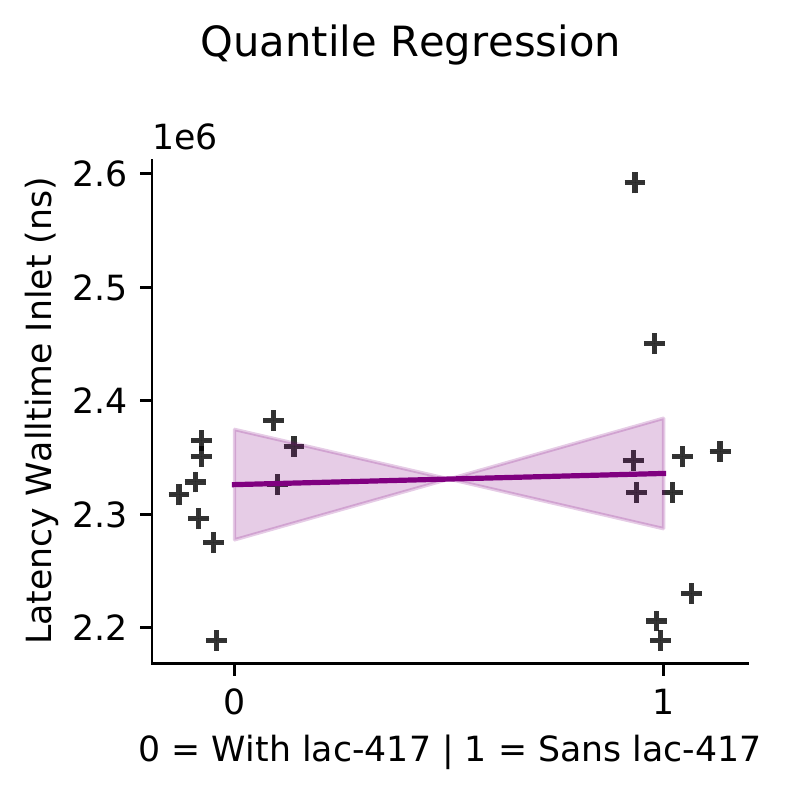}
    \caption{
      Quantile regression plot.
      Observations are medians per replicate.
    }
    \label{fig:with-lac-417-vs-sans-lac-417-regression-quantile-\metricslug-complete-regression}
  \end{subfigure}%
  \begin{subfigure}[b]{0.5\textwidth}
    \centering
    \includegraphics[width=\textwidth]{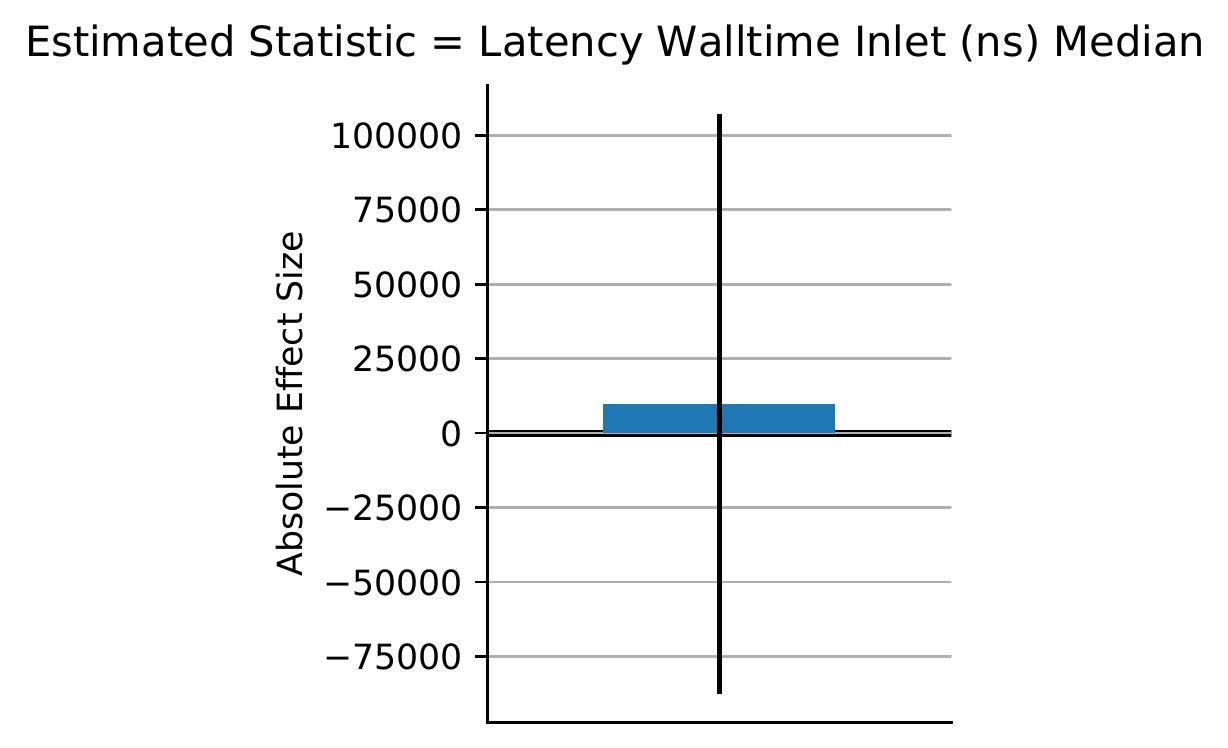}
    \caption{Estimated regression coefficient for quantile regression. Zero corresponds to no effect.}
    \label{fig:with-lac-417-vs-sans-lac-417-regression-quantile-\metricslug-complete-effect-size}
  \end{subfigure}

  \caption{
  Regressions of \metric{} against categorically coded treatment for faulty hardware experiment (Section \ref{sec:with-lac-417-vs-sans-lac-417}).
  Lower is better.
  Ordinary least squares regression (top row) estimates relationship between categorical dependent variable and mean of response variable.
  Quantile regression (bottom row) estimates relationship between categorical independent variable and median of response variable.
  Error bands and bars are 95\% confidence intervals.
  }
  \label{fig:with-lac-417-vs-sans-lac-417-regression-\metricslug}
\end{figure*}

\def\metric{Latency Simsteps Outlet}\def\metricslug{latency-simsteps-outlet}\def\xvarslug{0-with-lac-417-1-sans-lac-417}\def\yvarprefix{}

\def\metric{Latency Walltime Outlet (ns)}\def\metricslug{latency-walltime-outlet-ns}\def\xvarslug{0-with-lac-417-1-sans-lac-417}\def\yvarprefix{}

\def\metric{Delivery Clumpiness}\def\metricslug{delivery-clumpiness}\def\xvarslug{0-with-lac-417-1-sans-lac-417}\def\yvarprefix{}

\def\metric{Simstep Period Inlet (ns)}\def\metricslug{simstep-period-inlet-ns}\def\xvarslug{0-with-lac-417-1-sans-lac-417}\def\yvarprefix{}

\def\metric{Latency Simsteps Inlet}\def\metricslug{latency-simsteps-inlet}\def\xvarslug{0-with-lac-417-1-sans-lac-417}\def\yvarprefix{}

\def\metric{Simstep Period Outlet (ns)}\def\metricslug{simstep-period-outlet-ns}\def\xvarslug{0-with-lac-417-1-sans-lac-417}\def\yvarprefix{}

\def\metric{Delivery Failure Rate}\def\metricslug{delivery-failure-rate}\def\xvarslug{0-with-lac-417-1-sans-lac-417}\def\yvarprefix{}

\providecommand{\tabwithlacvssanslacregressionmacro}[2]{
\def\regression{#1}\def\regressionslug{#2}

\providecommand{\dissertationelse}[2]{%
\ifdefined\DISSERTATION
#1
\else
#2
\fi
}

\makeatletter
  \providecommand\importpath{\import@path}
\makeatother

\pragmaonce

\def\replacesinglequote#1{%
  \IfSubStr{#1}{'}{%
    \StrSubstitute{#1}{'}{\,}}{#1}}

\pragmaonce

\providecommand\rotcol[1]{ \multicolumn{1}{@{}@{}}{} \adjustbox{angle=60,lap=\width}{\rule{0.4\textwidth}{0.8pt}} \hfill \makebox[3ex]{\adjustbox{angle=60,lap=\width}{\bfseries \hspace{1ex} #1}} \hfill }


\clearpage
{
\onecolumn

\begin{landscape}
\dissertationelse{\fontsize{8}{9}\selectfont}{\footnotesize}
\csvstyle{myTableStyle}{
  longtable=@{\hspace{-.5\arrayrulewidth}}c@{\hspace{-.5\arrayrulewidth}}|l|l|l|l|l|l|l|l|l|l|l|l|l|l|,
  table head=\caption{
    Full \regression{} results of quality of service metrics against categorically coded treatment for faulty hardware experiment (Section \ref{sec:with-lac-417-vs-sans-lac-417}).
    Significance level $p < 0.05$ used.
    Inf or NaN values may occur due to multicollinearity or due to inf or NaN observations.
  }\label{tab:with-lac-417-vs-sans-lac-417-\regressionslug} \\
    \toprule%
    \multicolumn{1}{@{\hspace{-.5\arrayrulewidth}}l@{\hspace{-.5\arrayrulewidth}}}{} &%
    \rotcol{ Metric } &%
    \rotcol{ Statistic } &%
    \rotcol{ Significant Effect Sign } &%
    \rotcol{ Cpus Per Node } &%
    \rotcol{ Num Simels Per Cpu} &%
    \rotcol{ Num Processes} &%
    \rotcol{ Absolute Effect Size} &%
    \rotcol{ Absolute Effect Size 95\% CI Lower Bound} &%
    \rotcol{ Absolute Effect Size 95\% CI Upper Bound} &%
    \rotcol{ Relative Effect Size} &%
    \rotcol{ Relative Effect Size 95\% CI Lower Bound} &%
    \rotcol{ Relative Effect Size 95\% CI Upper Bound} &%
    \rotcol{$\mathbf{n}$} & %
    \rotcol{$\mathbf{p}$} %
    \vspace{-1ex} \\ \midrule\endfirsthead
    \caption*{\tablename{} \thetable{} (cont'd)} \\
    \toprule%
    \multicolumn{1}{@{\hspace{-.5\arrayrulewidth}}l@{\hspace{-.5\arrayrulewidth}}}{} &%
    \rotcol{ Metric } &%
    \rotcol{ Statistic } &%
    \rotcol{ Significant Effect Sign } &%
    \rotcol{ Cpus Per Node } &%
    \rotcol{ Num Simels Per Cpu} &%
    \rotcol{ Num Processes} &%
    \rotcol{ Absolute Effect Size} &%
    \rotcol{ Absolute Effect Size 95\% CI Lower Bound} &%
    \rotcol{ Absolute Effect Size 95\% CI Upper Bound} &%
    \rotcol{ Relative Effect Size} &%
    \rotcol{ Relative Effect Size 95\% CI Lower Bound} &%
    \rotcol{ Relative Effect Size 95\% CI Upper Bound} &%
    \rotcol{$\mathbf{n}$} & %
    \rotcol{$\mathbf{p}$} %
    \vspace{-1ex} \\ \midrule\endhead
    \bottomrule\endfoot,
  late after line=\\\hline,
  late after last line=\\\bottomrule,
  head to column names,
  filter=\equal{\csuse{RegressionModelSlug}}{\regressionslug},
}

\catcode`\%=12
\providecommand\filename{}
\renewcommand\filename{\detokenize{submodule/conduit-binder/binder/date=2021+project=72k5n/a=with-lac-417-vs-sans-lac-417/outplots/a=with_lac_417_vs_sans_lac_417_regression_results+async_mode=3+comp_work=0+impl=proc+nproc=256+nthread=1+proc=0-255+readability=latexcsvreader+replicate=0-9+slurm_nnodes=256+slurm_ntasks=256+subject=inlet_outlet+view=summary+ext=.csv}}
\catcode`\%=14

\csvreader[myTableStyle]{\importpath\filename}{}%
{ %
  &%
  \csuse{DependentVariable} &%
  \csuse{Statistic} &%
  \ifthenelse{\equal{\csuse{SignificantEffectSign}}{-}}{\cellcolor{red!25}}{}%
  \ifthenelse{\equal{\csuse{SignificantEffectSign}}{+}}{\cellcolor{green!25}}{}%
  \ifthenelse{\equal{\csuse{SignificantEffectSign}}{0}}{\cellcolor{gray!25}}{}%
  \csuse{SignificantEffectSign} &%
  \replacesinglequote{\csuse{CpusPerNode}} &%
  \replacesinglequote{\csuse{NumSimelsPerCpu}} &%
  \replacesinglequote{\csuse{NumProcessesPrettyprint}} &%
  \replacesinglequote{\csuse{AbsoluteEffectSize}} &%
  \replacesinglequote{\csuse{AbsoluteEffectSize95CILowerBound}} &%
  \replacesinglequote{\csuse{AbsoluteEffectSize95CIUpperBound}} &%
  \replacesinglequote{\csuse{RelativeEffectSize}} &%
  \replacesinglequote{\csuse{RelativeEffectSize95CILowerBound}} &%
  \replacesinglequote{\csuse{RelativeEffectSize95CIUpperBound}} &%
  \replacesinglequote{\csuse{n}} &%
  \replacesinglequote{\csuse{p}} %
}

\end{landscape}

}

\clearpage

}

\def\regression{Ordinary Least Squares Regression}\def\regressionslug{ordinary-least-squares-regression}

\providecommand{\dissertationelse}[2]{%
\ifdefined\DISSERTATION
#1
\else
#2
\fi
}

\makeatletter
  \providecommand\importpath{\import@path}
\makeatother

\pragmaonce

\def\replacesinglequote#1{%
  \IfSubStr{#1}{'}{%
    \StrSubstitute{#1}{'}{\,}}{#1}}

\pragmaonce

\providecommand\rotcol[1]{ \multicolumn{1}{@{}@{}}{} \adjustbox{angle=60,lap=\width}{\rule{0.4\textwidth}{0.8pt}} \hfill \makebox[3ex]{\adjustbox{angle=60,lap=\width}{\bfseries \hspace{1ex} #1}} \hfill }


\clearpage
{
\onecolumn

\begin{landscape}
\dissertationelse{\fontsize{8}{9}\selectfont}{\footnotesize}
\csvstyle{myTableStyle}{
  longtable=@{\hspace{-.5\arrayrulewidth}}c@{\hspace{-.5\arrayrulewidth}}|l|l|l|l|l|l|l|l|l|l|l|l|l|l|,
  table head=\caption{
    Full \regression{} results of quality of service metrics against categorically coded treatment for faulty hardware experiment (Section \ref{sec:with-lac-417-vs-sans-lac-417}).
    Significance level $p < 0.05$ used.
    Inf or NaN values may occur due to multicollinearity or due to inf or NaN observations.
  }\label{tab:with-lac-417-vs-sans-lac-417-\regressionslug} \\
    \toprule%
    \multicolumn{1}{@{\hspace{-.5\arrayrulewidth}}l@{\hspace{-.5\arrayrulewidth}}}{} &%
    \rotcol{ Metric } &%
    \rotcol{ Statistic } &%
    \rotcol{ Significant Effect Sign } &%
    \rotcol{ Cpus Per Node } &%
    \rotcol{ Num Simels Per Cpu} &%
    \rotcol{ Num Processes} &%
    \rotcol{ Absolute Effect Size} &%
    \rotcol{ Absolute Effect Size 95\% CI Lower Bound} &%
    \rotcol{ Absolute Effect Size 95\% CI Upper Bound} &%
    \rotcol{ Relative Effect Size} &%
    \rotcol{ Relative Effect Size 95\% CI Lower Bound} &%
    \rotcol{ Relative Effect Size 95\% CI Upper Bound} &%
    \rotcol{$\mathbf{n}$} & %
    \rotcol{$\mathbf{p}$} %
    \vspace{-1ex} \\ \midrule\endfirsthead
    \caption*{\tablename{} \thetable{} (cont'd)} \\
    \toprule%
    \multicolumn{1}{@{\hspace{-.5\arrayrulewidth}}l@{\hspace{-.5\arrayrulewidth}}}{} &%
    \rotcol{ Metric } &%
    \rotcol{ Statistic } &%
    \rotcol{ Significant Effect Sign } &%
    \rotcol{ Cpus Per Node } &%
    \rotcol{ Num Simels Per Cpu} &%
    \rotcol{ Num Processes} &%
    \rotcol{ Absolute Effect Size} &%
    \rotcol{ Absolute Effect Size 95\% CI Lower Bound} &%
    \rotcol{ Absolute Effect Size 95\% CI Upper Bound} &%
    \rotcol{ Relative Effect Size} &%
    \rotcol{ Relative Effect Size 95\% CI Lower Bound} &%
    \rotcol{ Relative Effect Size 95\% CI Upper Bound} &%
    \rotcol{$\mathbf{n}$} & %
    \rotcol{$\mathbf{p}$} %
    \vspace{-1ex} \\ \midrule\endhead
    \bottomrule\endfoot,
  late after line=\\\hline,
  late after last line=\\\bottomrule,
  head to column names,
  filter=\equal{\csuse{RegressionModelSlug}}{\regressionslug},
}

\catcode`\%=12
\providecommand\filename{}
\renewcommand\filename{\detokenize{submodule/conduit-binder/binder/date=2021+project=72k5n/a=with-lac-417-vs-sans-lac-417/outplots/a=with_lac_417_vs_sans_lac_417_regression_results+async_mode=3+comp_work=0+impl=proc+nproc=256+nthread=1+proc=0-255+readability=latexcsvreader+replicate=0-9+slurm_nnodes=256+slurm_ntasks=256+subject=inlet_outlet+view=summary+ext=.csv}}
\catcode`\%=14

\csvreader[myTableStyle]{\importpath\filename}{}%
{ %
  &%
  \csuse{DependentVariable} &%
  \csuse{Statistic} &%
  \ifthenelse{\equal{\csuse{SignificantEffectSign}}{-}}{\cellcolor{red!25}}{}%
  \ifthenelse{\equal{\csuse{SignificantEffectSign}}{+}}{\cellcolor{green!25}}{}%
  \ifthenelse{\equal{\csuse{SignificantEffectSign}}{0}}{\cellcolor{gray!25}}{}%
  \csuse{SignificantEffectSign} &%
  \replacesinglequote{\csuse{CpusPerNode}} &%
  \replacesinglequote{\csuse{NumSimelsPerCpu}} &%
  \replacesinglequote{\csuse{NumProcessesPrettyprint}} &%
  \replacesinglequote{\csuse{AbsoluteEffectSize}} &%
  \replacesinglequote{\csuse{AbsoluteEffectSize95CILowerBound}} &%
  \replacesinglequote{\csuse{AbsoluteEffectSize95CIUpperBound}} &%
  \replacesinglequote{\csuse{RelativeEffectSize}} &%
  \replacesinglequote{\csuse{RelativeEffectSize95CILowerBound}} &%
  \replacesinglequote{\csuse{RelativeEffectSize95CIUpperBound}} &%
  \replacesinglequote{\csuse{n}} &%
  \replacesinglequote{\csuse{p}} %
}

\end{landscape}

}

\clearpage

\def\regression{Quantile Regression}\def\regressionslug{quantile-regression}

\dissertationonly{\end{levelup}}

\end{document}